# JPEG XL Image Coding System

## CD WARNING

This document is not an ISO International Standard. It is distributed for review and comment. It is subject to change without notice and may not be referred to as an International Standard. Recipients of this draft are invited to submit, with their comments, notification of any relevant patent rights of which they are aware and to provide supporting documentation.

**Editor's note**: this is the Committee Draft of JPEG XL, technically identical to document wg1n84043. It is likely the final standard will differ in some ways from this draft.

**Authors**:

Alexander Rhatushnyak[a1†], Jan Wassenberg[b1], Jon Sneyers[c2], Jyrki Alakuijala[d1], Lode Vandevenne[e1], Luca Versari[f1], Robert Obryk[g1], Zoltan Szabadka[h1†], Evgenii Kliuchnikov[1], Iulia-Maria Comşa[1], Krzysztof Potempa[1], Martin Bruse[1], Moritz Firsching[1], Renata Khasanova[1†], Ruud van Asseldonk[1†], Sami Boukortt[1], Sebastian Gomez[1†], Thomas Fischbacher[1]

a lead on lossless mode
b editor, lead on decoding
c lead on modular mode
d manager, lead on image quality
e lead on integration
f lead on encoding
g lead on numerical algorithms
h lead on JPEG recompression

1 Google Research Zürich, Switzerland
2 Cloudinary, Israel
† Work done while the author was at Google

The ISO/IEC will provide a cover page and front matter.

# CONTENTS

**INTERNATIONAL STANDARD**

# JPEG XL Image Coding System

## 1 Scope

This draft International Standard defines a set of compression methods for coding one or more images of continuous-tone greyscale, or continuous-tone colour, or multichannel digital samples.

This International Standard:

- specifies decoding processes for converting compressed image data to reconstructed image data;
- specifies a codestream syntax containing information for interpreting the compressed image data;
- provides guidance on encoding processes for converting source image data to compressed image data.

## 2 Normative references

The following documents are referred to in the text in such a way that some or all of their content constitutes requirements of this document. For dated references, only the edition cited applies. For undated references, the latest edition of the referenced document (including any amendments) applies.

ISO/IEC 10918-1:1993(E), *Information technology — Digital compression and coding of continuous-tone still images: Requirements and guidelines*
ISO 10526:1999/CIE S005/E-1998, *CIE Standard Illuminants for Colourimetry*
IEC 61966-2-1:1999, *Multimedia systems and equipment — Colour measurement and management — Part 2-1: Colour management — Default RGB colour space — sRGB*
ISO 15076-1:2010, *Image technology colour management — Architecture, profile format and data structure — Part 1: Based on ICC.1:2010*
ISO/IEC 14882:2011, *Information technology — Programming languages — C++*
ISO/IEC/IEEE 60559:2011, *Information technology — Microprocessor Systems — Floating-Point arithmetic*
ISO/IEC FDIS 23091-2:2019(E), *Information technology — Coding-independent code points — Part 2: Video*

Rec. ITU-R BT.709-6, *Parameter values for the HDTV standards for production and international programme exchange*
Rec. ITU-R BT.2100-1, *Image parameter values for high dynamic range television for use in production and international programme exchange*
SMPTE ST 428-1, *D-Cinema Distribution Master Image Characteristics*

INTERNET ENGINEERING TASK FORCE (IETF). RFC 7932: *Brotli Compressed Data Format* [online].
Edited by J. Alakuijala and Z. Szabadka. July 2016 [viewed 2019-08-05].
Available at https://tools.ietf.org/html/rfc7932

## 3 Terms and definitions

For the purposes of this document, the following terms and definitions apply.
ISO and IEC maintain terminological databases for use in standardization at the following addresses:
— ISO Online browsing platform: available at https://www.iso.org/obp

— IEC Electropedia: available at http://www.electropedia.org/

## 3.1 Data storage

3.1.1
**bit**
binary digit (0 or 1)

3.1.2
**byte**
8 consecutive bits whose least-significant (0-based) bit index is divisible by 8

3.1.3
**big endian**
the bytes of a value representation occur in order of most to least-significant

3.1.4
**bitstream**
sequence of bytes from which bits are read

3.1.5
**JPEG marker**
two-byte code in which the first byte is 255 and the second byte is a value v with 1 <= v <= 254

3.1.6
**codestream**
bitstream plus associated JPEG markers

## 3.2 Inputs

3.2.1
**greyscale**
image representation in which each pixel is defined by a single sample representing luminance

3.2.2
**continuous-tone image**
image whose components have more than one bit per sample

3.2.3
**opsin**
photosensitive pigments in the human retina, whose dynamics are approximated by the XYB colour space

3.2.4
**burst**
sequences of images typically captured with identical settings

3.2.5
**animation**
series of pictures and timing delays to display as a video medium

3.2.6

**frame**
a single image, which may be part of a burst or animation

3.2.7
**preview**
lower-fidelity rendition of one of the frames, or a frame that represents the entire content of all frames

3.2.8
**ICC profile**
information for colour space transforms

## 3.3 Processes

3.3.1
**decoding process**
process which takes as its input a codestream and outputs a continuous-tone image

3.3.2
**decoder**
embodiment of a decoding process

3.3.3
**encoding process**
process which takes as its input a continuous-tone image and outputs compressed image data in the form of a codestream

3.3.4
**encoder**
embodiment of an encoding process

3.3.5
**lossless**
descriptive term for encoding and decoding processes in which the output of a decoding procedure is identical to the input to the encoding procedure

3.3.6
**lossy**
descriptive term for encoding and decoding processes which are not lossless

3.3.7
**upsampling**
procedure by which the spatial resolution of a channel is increased

3.3.8
**downsampling**
procedure by which the spatial resolution of a channel is reduced

3.3.9
**entropy encoding**
lossless procedure which converts a sequence of input symbols into a sequence of bits such that the average number of bits per symbol approaches the entropy of the input symbols

3.3.10
**entropy encoder**
embodiment of an entropy encoding procedure

3.3.11
**entropy decoding**
lossless procedure which recovers the sequence of symbols from the sequence of bits produced by the entropy encoder

3.3.12
**entropy decoder**
embodiment of an entropy decoding procedure

## 3.4 Image organization

3.4.1
**sample**
integer or real value representing one component of a pixel

3.4.2
**sample grid**
common coordinate system for all samples of an image, with top-left coordinates (0, 0), the first coordinate increasing towards the right, and the second increasing towards the bottom

3.4.3
**channel**
rectangular array of samples having the same designation, regularly aligned along a sample grid

3.4.4
**xsize**
width of a sample grid or a region

3.4.5
**ysize**
height of a sample grid or a region

3.4.6
**raster order**
access pattern from left to right in the top row, then in the row below and so on

3.4.7
**region**
rectangular range of samples

3.4.8
**naturally aligned**
positioning such that the top and left coordinates of a region are divisible by its (power of two) xsize and ysize, respectively

3.4.9
**tile**
naturally aligned square region covering up to 64 × 64 input pixels

### 3.4.10
**block**
naturally aligned square region covering up to 8 × 8 input pixels

### 3.4.11
**group**
naturally aligned square region covering up to 256 × 256 input pixels

### 3.4.12
**worker**
entity (e.g. thread of execution) capable of independently decoding a group

## 3.5 DCT

### 3.5.1
**coefficient**
input value to the inverse DCT

### 3.5.2
**quantization**
method of reducing the precision of individual coefficients

### 3.5.3
**varblock**
naturally aligned variable-size rectangular region of input pixels

### 3.5.4
**var-DCT**
lossy encoding of a frame that applies DCT to varblocks

### 3.5.5
**DC coefficient**
lowest frequency DCT coefficient, containing the average value of a block or the lowest-frequency component within the 8 × 8 region of a larger varblock

### 3.5.6
**AC coefficients**
all DCT coefficients except the DC coefficient

### 3.5.7
**pass**
subset of AC coefficients enabling decoding of successively higher resolutions

### 3.5.8
**DC group**
256 × 256 DC values from a naturally aligned square region covering up to 2048 × 2048 input pixels

# 4 Symbols and abbreviated terms

ICC: International Colour Consortium

IEC: International Electrotechnical Commission

ISO: International Organization for Standardization

ITU: International Telecommunication Union

JPEG: Joint Photographic Experts Group

JPEG1: reference to ISO/IEC 10918

RGB: additive colour model with red, green, blue components

LMS: colour space representing the response of cones in the human eye

XYB: absolute colour space based on gamma-corrected LMS, in which X is derived from the difference between L and M, Y is an average of L and M (behaves similarly to luminance), and B is derived from the S ("blue") channel

DCT: discrete cosine transform (DCT-II as specified in I.2)

IDCT: inverse discrete cosine transform (DCT-II as specified in I.2)

LLF: **N** / 8 × **N** / 8 square of lowest frequency coefficients of **N** × **N** DCT coefficients

LF: **N** / 4 × **N** / 4 square of lowest frequency coefficients of **N** × **N** DCT coefficients

HF: high frequency components of **N** × **N** DCT coefficients (all coefficients except LLF)

# 5 Conventions

## 5.1 Mathematical symbols

| | |
|---|---|
| \|x\| | absolute value of x, \|x\| equals –x for x < 0, otherwise x |
| [a, b], (c, d), [e, f) | closed or open or half-open intervals containing all integers or real numbers x (depending on context) such that a <= x <= b, c < x < d, e <= x < f. |
| {a, b, c} | ordered sequence of elements |
| π | the smallest positive zero of the sine function |

## 5.2 Functions

| | |
|---|---|
| sqrt(x) | square root, such that sqrt(x) × sqrt(x) == x and sqrt(x) >= 0. If x < 0, the codestream is ill-formed. |
| cos(r) | cosine of the angle r (in radians) |
| sin(r) | sine of the angle r (in radians) |
| erf(x) | error function of sigmoid shape |

log(x) natural logarithm of x. If x <= 0, the codestream is ill-formed.

log2(x) base-two logarithm of x. If x <= 0, the codestream is ill-formed.

ilog2(x) one-based index of the most significant 1-bit of x
EXAMPLE ilog2(1) == 1, ilog2(4) == 3

floor(x) the largest integer that is less than or equal to x

ceil(x) the smallest integer that is greater than or equal to x

abs(x) equivalent to |x|

sign(x) sign of x, 0 if |x| is 0, +1 if x is positive, -1 if x is negative

UnpackSigned(u) equivalent to u / 2 if u is even, and -(u + 1) / 2 if u is odd

clamp(x, lo, hi) equivalent to min({max({lo, x}), hi})

MakeBinary32(u) the number whose binary32 floating-point representation (as defined in ISO/IEC/IEEE 60559:2011) is u

signbit(x) equivalent to 1 if x is negative (including the floating-point value -0.0), otherwise 0

copysign(m, s) equivalent to $|m| \times (-1)^{signbit(s)}$

truncate(f) equivalent to copysign(floor(abs(f)), f)

len(a) length (number of elements) of array a

sum(a) sum of all elements of the array/tuple/sequence a

max(a) maximal element of the array/tuple/sequence a

min(a) smallest element of the array/tuple/sequence a

min_argmax(a) minimal number i such that a[i] == max(a)

## 5.3 Operators

### 5.3.1 Arithmetic operators

+ addition

− subtraction (as a binary operator) or negation (as a unary prefix operator)

++ increasing by one (before/after obtaining the value, if before/after a variable)

-- decreasing by one (before/after obtaining the value, if before/after a variable)

× multiplication

/ division of real numbers without truncation or rounding

$x^y$ exponentiation, x to the power of y

<< left shift: x << s is defined as $x \times 2^s$

>> right shift: x >> s is defined as $floor(x / 2^s)$

Umod a Umod d is the unique integer r in [0, d) for which a == r + q×d for a suitable integer q

Idiv a Idiv b is equivalent to truncate(a / b) - see 5.3.6

### 5.3.2 Bitwise arithmetic operators

~ bitwise NOT

& bitwise AND

^ bitwise XOR

| bitwise OR

### 5.3.3 Logical operators

|| logical OR with short-circuit evaluation

&& logical AND with short-circuit evaluation

! logical NOT

?: c ? t : f is defined as t if the condition c evaluates to true (or nonzero), otherwise f

### 5.3.4 Relational operators

> greater than

>= greater than or equal to

< less than

<= less than or equal to

== equal to

!= not equal to

### 5.3.5 Assignment operators

= variable = value assigns value to variable.

+= a += b is equivalent to a = a + b

-= a -= b is equivalent to a = a - b

×= a ×= b is equivalent to a = a × b

/= a /= b is equivalent to a = a / b

&= a &= b is equivalent to a = a & b

^= a ^= b is equivalent to a = a ^ b

|= a |= b is equivalent to a = a | b

### 5.3.6 Operator precedence

NOTE Operators are listed below in descending order of precedence. If several operators appear in the same line, they have equal precedence. When several operators of equal precedence appear at the same level in an expression, evaluation proceeds according to the associativity of the operator (either from right to left or from left to right).

| **Operators** | **Type of operation** | **Associativity** |
|---|---|---|
| () | expression | left to right |
| x++, x-- | postfix increment/decrement | left to right |

<table>
<tr><td>[]</td><td>indexing of arrays, indices start at 0 unless specified otherwise</td><td>left to right</td></tr>
<tr><td>.</td><td>member access</td><td>left to right</td></tr>
<tr><td>++x, --x</td><td>prefix increment/decrement</td><td>right to left</td></tr>
<tr><td>$x^y$</td><td>exponentiation</td><td>right to left</td></tr>
<tr><td>–</td><td>unary negation</td><td>right to left</td></tr>
<tr><td>!, ~</td><td>logical/bitwise NOT</td><td>right to left</td></tr>
<tr><td>×, /, Idiv, Umod</td><td>multiplication, division, integer division, remainder</td><td>left to right</td></tr>
<tr><td>+, −</td><td>addition and subtraction</td><td>left to right</td></tr>
<tr><td><<, >></td><td>left shift and right shift</td><td>left to right</td></tr>
<tr><td>< , >, <=, >=</td><td>relational</td><td>left to right</td></tr>
<tr><td>==, !=</td><td>equality comparison</td><td>left to right</td></tr>
<tr><td>&</td><td>bitwise AND</td><td>left to right</td></tr>
<tr><td>^</td><td>bitwise XOR</td><td>left to right</td></tr>
<tr><td>|</td><td>bitwise OR</td><td>left to right</td></tr>
<tr><td>&&</td><td>logical AND</td><td>left to right</td></tr>
<tr><td>||</td><td>logical OR</td><td>left to right</td></tr>
<tr><td>?:</td><td>ternary conditional</td><td>right to left</td></tr>
<tr><td>=</td><td>assignment</td><td>right to left</td></tr>
<tr><td>+=, -=, ×=, /=</td><td>compound assignment</td><td>right to left</td></tr>
<tr><td><<=, >>=, &=, ^=, |=</td><td>compound assignment</td><td>right to left</td></tr>
</table>

## 5.4 Pseudocode

This document describes functionality using pseudocode formatted as follows:

```
// Informative comment

var = u(8);  // Defined in A.3.1
if (var == 1) return;  // Stop executing this code snippet

[[Normative specification: if var == 0, the codestream is ill-formed]]

(out1, out2) = Function(var, kConstant);
```

Variables such as `var` are typically referenced by text outside the source code.

The semantics of this pseudocode shall be those of the C++ programming language as defined by ISO/IEC 14882:2011, with the following exceptions:

- Symbols from 5.1 and functions from 5.2 are allowed;
- Multiplication, division, remainder and exponentiation shall be expressed as specified in 5.3;
- Functions may return tuples which unpack to variables as in the above example;
- `[[ ]]` enclose normative directives specified using prose;

- `int` shall denote a 32 or 64-bit integer;
- All integers shall be stored using two's complement;
- Expressions, and variables whose types are omitted, shall be understood as real numbers.

Where unsigned integer wraparound and truncated division are required, Umod and Idiv (see 5.3.1) shall be used for those purposes. Where the size of a variable is important, this document may specify concrete `uint#_t`, where `#` is 8, 16, 32, or 64.

Numbers with a 0x prefix are in base 16 (hexadecimal), and apostrophe (') characters inside them shall be ignored.
EXAMPLE 0x0001'0000 == 65536.

# 6 Functional concepts

## 6.1 Image organization

A channel is defined as a rectangular array of (integer or real) samples regularly aligned along a sample grid of xsize sample positions horizontally and ysize sample positions vertically. The number of channels may be 1 to 4096 (see `num_extra_channels` in A.4.5).
A pixel is defined as a vector of dimension corresponding to the number of channels, and its component with index c (numbered from 0 to number of channels - 1) is the sample from channel c whose position matches that of the pixel. Thus, the number of components matches the number of channels.
An image is defined as the two-dimensional array of pixels, and its width is xsize and height is ysize.
Unless otherwise mentioned, channels shall be accessed in the following "raster order": left to right column within the topmost row, then left to right column within the row below the top, and so on until the rightmost column of the bottom row.

## 6.2 Group splitting

Channels shall be logically partitioned into naturally-aligned groups of 256 × 256 samples. The effective dimension of a group (i.e. how many pixels to read) may be smaller than 256 for groups on the right or bottom of the image. The decoder shall ensure the decoded image has the dimensions specified in SizeHeader by cropping at the right and bottom as necessary.

DC groups shall likewise consist of 256 × 256 DC samples, with the possibility of a smaller effective size on the right and bottom of the image.

If enabled, DC16 groups shall likewise consist of 256 × 256 DC16 samples, with the possibility of a smaller effective size on the right and bottom of the image.

Groups can be decoded independently. A 'table of contents' stores the size (in bytes) of each group to allow seeking to any group. An optional permutation allows groups to be arranged in arbitrary order.

## 6.3 Codestream and bitstream

A codestream is a bitstream plus associated JPEG markers. Other parts of, or amendments to, this International Standard may define a file format containing the codestream defined in this document.

NOTE A file format can add extensions, metadata and enable interoperability with existing systems.

A bitstream is a finite sequence of bytes that represent compressed image data and metadata. N bytes may also be viewed as 8 × N bits. The first 8 bits shall be the bits constituting the first byte, in least to most significant order, the next eight bits (again in least to most significant order) shall constitute the second byte, and so on. Unless otherwise specified, bits shall be read from the bitstream as specified in A.3.1.1 (u(n)).

Subsequent Annexes or subclauses may specify an element of the bitstream as byte-aligned. For such elements, the decoder shall take actions before and after reading the element as follows. Immediately before encountering the element, the decoder shall read a pu0() as specified in A.3.2.4. After finishing reading the element, the decoder shall read a pu0() as specified in A.3.2.4.

### 6.4 Multiple frames

A codestream may contain multiple frames. These may constitute an animation, a burst (arbitrary images with identical dimensions), or a still image with one or more overlays rendered on top of the first frame.

NOTE For any distinct frames f1 and f2, the decoder may decode f1 and f2 concurrently using two independent workers (which have their own state, including the decoded frame image). For every worker decoding a frame, that frame is referred to in this document as the current frame.

### 6.5 Mirroring

Some operations access samples where one or both of its coordinates, denoted `coord`, exceeds the valid range [0, `size`). The value of such samples shall be that of the input sample at the coordinate `coord` resulting from the following mirroring transform:

```
while (coord < 0 || coord >= size) {
  if (coord < 0) {
    coord = -coord - 1;
  } else {
    coord = 2 × size - 1 - coord;
  }
}
```

## 7 Encoder requirements

An encoder is an embodiment of the encoding process, which shall convert samples of a rectangular image into a codestream. This International Standard does not specify an encoding process, and any encoding process is acceptable as long as the codestream conforms to the codestream syntax specified in Annexes A to O.

## 8 Decoder requirements

A decoder is an embodiment of the decoding process. The decoder reconstructs sample values (arranged on a rectangular sampling grid) from a codestream as specified by Annexes A to O. These Annexes are normative in the sense that the decoder shall carry out all steps as described, or instead use alternate processes that achieve equivalent results. Equivalence and conformance requirements will be defined in other parts of, or amendments to, this International Standard.

## 9 Decoder overview

### 9.1 Summary

NOTE For a block diagram, design rationale and introduction to the coding tools, please refer to "JPEG XL next-generation image compression architecture and coding tools".

JPEG XL specifies four modes: modular (for responsive images), var-DCT (for lossy), lossless, and losslessly transcoded JPEG. Modular consists of entropy decoding followed by zero or more inverse transforms. Var-DCT operates in XYB colour space and performs DCT on varblocks, followed by prediction and adaptive quantization. Lossless uses a self-correcting predictor and context model. Lossless transcoding has a specialized context model for JPEG1 codestreams.

If the codestream starts with bytes {`0xFF`, `0xD8`, `0xFF`, `0xE0`}, the decoder shall decode a JPEG1 as specified in ISO/IEC 10918-1:1993(E). Otherwise, if the codestream starts with bytes {`0x0A`, `0x04`, `0x42`, `0xD2`, `0xD5`, `0x4E`}, the decoder shall reconstruct the original JPEG1 codestream as specified in Annex M. Otherwise, the codestream shall be structured as shown in Table 1; the syntax is described in Annex A.

**Table 1: Codestream overview**

| `condition` | `type` | `name` | `subclause` |
|---|---|---|---|
| | Signature | `signature` | A.4.1 |
| | SizeHeader | `size` | A.4.2.2 |
| | ImageMetadata | `metadata` | A.4.5 |
| `metadata.m2.have_preview` | PreviewHeader | `preview` | A.4.2.3 |
| `metadata.m2.have_animation` | AnimationHeader | `animation` | A.4.6 |
| `metadata.m2.have_preview` | Frame | `preview_frame` | C.1 |
| `for (i = 0; i == 0 \|\| !frames[i - 1].frame_header.animation_frame.is_last; ++i)` | Frame | `frames[i]` | C.1 |

The decoder can be viewed as a pipeline with multiple stages, in the order listed in the following subclauses, which refer to Annexes specifying their behaviour. Some of these stages may not participate in decoding some codestreams.

EXAMPLE The lossless mode does not involve AC predictions nor adaptive quantization.

The Annexes and their subclauses may begin with a condition on their use and decoders shall only apply such an Annex or subclause if its condition holds true.

## 9.2 Headers

The decoder shall read headers from the codestream as specified in Annex A.

## 9.3 ICC codec

The decoder shall synthesize or decode an ICC profile as specified in Annex B.

NOTE ICC profiles are standard display ICC profiles as defined by the International Color Consortium.

## 9.4 Frames

After headers and ICC codec, if present, the decoder shall read one or more frames as specified in Annex C. `frame_header.flags` determines which of the following stages (if any) shall be skipped.

## 9.5 Entropy decoding

For reading compressed data, the decoder shall use an entropy decoder as specified in Annex D.

## 9.6 Predictors

For channels which are residuals, the decoder shall add predictions based on previously reconstructed samples as specified in Annex E.

## 9.7 Adaptive dequantization

The decoder shall convert quantized 16-bit integers to coefficients as specified in Annex F.

## 9.8 Chroma from luma

For channels stored as differences versus a linear function of another channel, the decoder shall undo this decorrelation as specified in Annex G.

## 9.9 AC prediction

If AC coefficients are present, the decoder shall add to them predictions based on DC values as specified in Annex H.

## 9.10 Integral transform

The decoder shall perform or skip inverse integral transforms as specified in Annex I.

## 9.11 Loop filters

The decoder shall apply zero, one or two or loop filters (adaptive reconstruction, gaborish) as specified in Annex J.

## 9.12 Image features

The presence/absence of additional features (patches, dots, splines and noise) shall be indicated by `frame_header.flags`. The decoder shall draw these as specified in Annex K. Patches (if present) shall be rendered between adaptive reconstruction (if enabled) and gaborish (if enabled), whereas the others (if present) shall be rendered after gaborish (if enabled), in the listed order.

NOTE Dots and splines are special representations of point-like and curvilinear structures (which are common in images) that are more compact than a DCT-based representation.

## 9.13 Colour transform

The decoder shall perform a colour-space transform as specified in Annex L.

## 9.14 Lossless

The decoder shall decode lossless frames as specified in Annex O.

# Annex A
(normative)
**Headers**

**A**

## A.1 Introduction

The decoder shall read headers (all metadata listed in Table 1 prior to the first Frame) as follows. Each row of the table represents a bundle, which shall be initialized or read as specified in A.2. For each row, in top to bottom order, the decoder shall read the bundle if and only if its `condition` column is blank or evaluates to true.

NOTE Use cases for compact, compressed headers are given in "JPEG XL Short Headers" (see Annex Q).

## A.2 Bundles

A bundle shall consist of one or more conceptually related fields. A field may be a bundle, or an array/value of a type from A.3.1 or A.3.2. The graph of contained-within relationships between types of bundles shall be acyclic. This document specifies the structure of each bundle using tables structured as in Table A.1, with one row per field. If the `default` column is omitted, all of its entries shall be understood to be blank.

**Table A.1 — Structure of a table describing a bundle**

| `condition` | `type` | `default` | `name` |
|---|---|---|---|

If `condition` is blank or evaluates to true, the field shall be read from the codestream as specified below. Otherwise, the field shall instead be initialized to `default` as specified below. Fields whose `condition` is blank also have a blank `default` because it will not be used.

If `condition` is of the form `for(i = 0; condition; ++i)`, this shall be equivalent to replacing this row with a sequence of rows obtained from the current one by removing its condition and replacing the value of `i` in each column with the consecutive values assumed by the loop-variable `i` until `condition` evaluates to false. If `condition` is initially false, the current row has no effect.

Otherwise, if `name` ends with `[n]` then the field is a fixed-length array with `n` entries. Each of its elements shall be either read from the codestream or initialized to the same `default`, according to `condition` as specified above.

`name` may be referenced in the `condition` or `type` of a subsequent row, or the `condition` of the same row.

If a field is to be read from the codestream, the `type` entry determines how to do so. If it is a basic or derived type, it shall be read as described in A.3.1 or A.3.2. Otherwise `type` is a (nested) bundle denoted Nested, residing within a parent bundle Parent. Nested shall be read as if the rows of the table defining Nested were inserted (in top to bottom order) into the table defining Parent in place of the row that defined Nested. This principle shall apply recursively, corresponding to a depth-first traversal of all fields.

If a field is to be initialized to `default`, the `type` entry determines how to do so. If it is a bundle, then `default` shall be blank and each field of the bundle shall (recursively) be initialized to the `default` specified within the bundle's table. Otherwise, if `default` is not blank, the field shall be set to `default`, which shall be a valid value of the same `type`.

## A.3 Field types

### A.3.1 Basic types

#### A.3.1.1 u(n)

u(n) shall read n bits from the codestream, advance the position accordingly, and return the value in the range $[0, 2^n)$ represented by the bits. u(0) shall be interpreted as zero without reading any bits. The decoder shall first read the least-significant bit of the value, from the least-significant not yet consumed bit in the first not yet fully consumed byte of the codestream. The next bit of the value (in increasing order of significance) shall be read from the next (in increasing order of significance) bit of the same byte unless all its bits have already been consumed, in which case the decoder shall read from the least-significant bit of the next byte, and so on. When the decoder would consume a byte which is beyond the end of the codestream, the codestream is ill-formed.

#### A.3.1.2 pu()

pu() shall return u(n), where `n` shall be zero if the 0-based index of the next unread bit in the codestream `P` is a multiple of 8, otherwise `n` = 8 - (`P` Umod 8).

NOTE After pu(), the number of bits read from the codestream is divisible by 8.

#### A.3.1.3 U32(d0, d1, d2, d3)

U32(d0, d1, d2, d3) shall read an unsigned 32-bit `value` in $[0, 2^{32})$ as follows. The d0, d1, d2, d3 parameters each represent a distribution, i.e. an encoding of a range of values. Each distribution shall be one of Val(u) or Bits(n) or BitsOffset(n, offset), defined below.
The decoder shall first read a u(2) from the codestream indicating which distribution to use (0 selects d0, 2 selects d2 etc.).

Let d denote the chosen distribution. If d is Val(u), `value` shall be the unsigned 31-bit integer u.

NOTE The value of u is implicitly defined by Val(u) and not stored in the codestream. Similarly, d0..3 are determined by the type and also not stored in the codestream.

If d is Bits(n), `value` shall be read as u(n), where n <= 32. If d is BitsOffset(n, `offset`), `value` shall be (`offset` + v) Umod $2^{32}$, where v is the value read as u(n), with n <= 32.

EXAMPLE 1 For a field of type U32(Val(8), Val(16), Val(32), Bits(6)), the bits 10 (where 0 is the least-significant bit) result in `value = 32`.
EXAMPLE 2 For a field of type U32(Bits(2), Bits(4), Bits(6), Bits(8)), the bits 010111 result in `value = 7`.

#### A.3.1.4 U64()

U64() shall read an unsigned 64-bit `value` in $[0, 2^{64})$ using a single variable-length encoding. The decoder shall first read a u(2) selector s. If s == 0, `value` shall be 0. If s == 1, `value` shall be BitsOffset(4, 1) as defined in A.3.1.3. If s == 2, `value` shall be BitsOffset(8, 17). Otherwise s == 3 and `value` shall be read from a 12-bit component, zero or more 8-bit components, and zero or one 4-bit component as follows:

```
value = u(12);
shift = 12;
// Read 1-bit flag, stop if zero, read and add 4 or 8 bits
while (u(1) == 1) {
```

```
  // Last part only has 4 bits because we already read 60
  if (shift == 60) {
    value += u(4) << shift;
    break;
  }
  value += u(8) << shift;
  shift += 8;
}
```

EXAMPLE the largest possible value ($2^{64}$ - 1) is encoded as 73 consecutive 1-bits.

#### A.3.1.5 Varint()

Varint() shall read an unsigned integer `value` of up to 63 bits as follows:

```
value = 0;
shift = 0;
// Read a byte, add its 7 lower bits, stop if upper bit zero
while (1) {
  b = u(8);
  value += (b & 127) << shift;
  if (b <= 127) break;
  shift += 7;
  if (shift >= 63) [[the codestream is ill-formed]];
}
```

#### A.3.1.6 SL(n)

SL(n) shall read an unsigned value in $[0, 2^n)$ using $[1, 2 \times n)$ bits as follows.
A sequence of bits shall be read as a series of u(1), until it is long enough to conclude the resulting value.
Values less than $2^{n-2}$ are encoded as $\{1, b_0, \ldots 1, b_m, 0\}$ bit sequence, where $b_i$ is the i-th lowest binary digit of the value, and m is less than n - 2; greater values are encoded as $\{1, b_0, \ldots 1, b_{n-2}, b_{n-1}\}$ bit sequence.
EXAMPLES (with `n == 4`):

- {0} bit sequence encodes 0x0
- {1, 1, 0} bit sequence encodes the value 0x1
- {1, 0, 1, 1, 0} bit sequence encodes the value 0x2
- {1, 1, 1, 1, 0} bit sequence encodes the value 0x3
- {1, 0, 1, 0, 1, 1, 0} bit sequence encodes the value 0x4
- {1, 0, 1, 0, 1, 0, 1} bit sequence encodes the value 0x8
- {1, 1, 1, 1, 1, 1, 1} bit sequence encodes the value 0xF

#### A.3.1.7 GL(g, n)

GL(g, n) shall read an unsigned value in $[0, 2^{g \times n})$ using $[1, (g + 1) \times n]$ bits as follows.
A sequence of bits shall be read as an alternating series of u(1) and u(n), until it is long enough to conclude the resulting value.
Values less than $2^{g \times (n - 1)}$ are encoded as $\{1, b_0, \ldots b_{g-1}, \ldots 1, b_{m-g}, \ldots b_{m-1}, 0\}$ bit sequence, where $b_i$ is the i-th lowest binary digit of the value, m is divisible by g, and m is less than or equal to $(g - 1) \times n$; greater values are encoded as $\{1, b_0, \ldots b_{g-1}, \ldots 1, b_{g \times (n-1)}, \ldots b_{g \times n - 1}\}$ bit sequence.
EXAMPLES (with g == 4, n == 3):

- {0} bit sequence encodes 0x0
- {1, 1, 1, 0, 1, 0} bit sequence encodes the value 0xD
- {1, 1, 1, 0, 1, 1, 1, 0, 1, 0, 0} bit sequence encodes the value 0xAD
- {1, 1, 1, 0, 1, 1, 1, 0, 1, 0, 1, 1, 0, 1, 1} bit sequence encodes the value 0xBAD

### A.3.1.8 U8()

U8() shall read an unsigned integer `value` in the range [0, 256) using [1, 12) bits as follows:

```
if (u(1) == 0) {
  value = 0;
} else {
  n = u(3);
  value = u(n) + (1 << n);
}
```

EXAMPLES

- {0} bit sequence encodes the value 0x00
- {1, 0, 0, 0} bit sequence encodes the value 0x01
- {1, 0, 0, 1, 0} bit sequence encodes the value 0x02
- {1, 0, 0, 1, 1} bit sequence encodes the value 0x03
- {1, 1, 1, 0, 0, 0, 0, 0, 1, 0} bit sequence encodes the value 0x42
- {1, 1, 1, 1, 1, 1, 1, 1, 1, 1, 1} bit sequence encodes the value 0xFF

### A.3.1.9 bu(n)

bu(n) shall read n bits from the bit buffer defined below. bu(0) shall be interpreted as zero without reading any bits. Otherwise, the n bits shall represent a value in [0, $2^n$).

The bu(n) decoder has a separate buffer that shall be initialized as specified in subclauses that reference bu(). Initialization is performed as follows:

```
buffer = u(16);
buffer_len = 16;
```

Reading `value` shall be performed as specified by the following:

```
if (buffer_len < n) {
  buffer |= (u(16) << buffer_len);
  buffer_len += 16;
}
value = buffer & ((1 << n) - 1);
buffer >>= n;
buffer_len -= n;
```

### A.3.1.10 F16()

F16() shall read a real `value` in [-65504, 65504] as follows:

```
bits16 = u(16);
sign = bits16 >> 15;
biased_exp = (bits16 >> 10) & 0x1F;
mantissa = bits16 & 0x3FF;
```

```
if (biased_exp == 0) {
  value = mantissa / (1 << 24);
} else {
  biased_exp32 = biased_exp + (127 - 15);
  mantissa32 = mantissa << (23 - 10);
  bits32 = (sign << 31) | (biased_exp32 << 23) | mantissa32;
  value = MakeBinary32(bits32);
}
```

The codestream is ill-formed if `biased_exp` is 31.

NOTE This type is intended to match binary16 from ISO/IEC/IEEE 60559:2011, with the exception that NaN and infinities are disallowed.

### A.3.2 Derived types

#### A.3.2.1 Bool()

Bool() shall read a boolean value as u(1) ? true : false.

#### A.3.2.2 Enum(`EnumTable`)

Enum(`EnumTable`) shall read an enumerated type as follows. The decoder shall obtain v = U32(Val(0), Val(1), BitsOffset(4, 2), BitsOffset(6, 18)). If v exceeds 63 or is not a `value` defined by the table in the subclause titled `EnumTable`, the codestream is ill-formed. Such tables are structured according to Table A.2, with one row per unique `value`.

**Table A.2 — Structure of a table describing an enumerated type**

| `name` | `value` | `meaning` |
|---|---|---|

An enumerated type shall be interpreted as having the `meaning` of the row whose `value` is v. `name` (or where ambiguous, `EnumTable.name`) is an arbitrary identifier for purposes of referring to the same `meaning` elsewhere in this document. `name` begins with a k prefix, e.g. `kRGB`.

#### A.3.2.3 S32(d0, d1, d2, d3)

S32(d0, d1, d2, d3) shall read a signed `value` in $[-2^{31}, 2^{31})$ as follows. The decoder shall obtain v = U32(d0, d1, d2, d3). If the least-significant bit of v is zero, `value` shall be v >> 1. Otherwise, if v == 0xFFFF'FFFF, `value` shall be 0x8000'0000. Otherwise, `value` shall be the negation of the signed 32-bit integer whose two's complement representation is ((v >> 1) + 1).

#### A.3.2.4 pu0()

The decoder shall read a pu(). If its value is not equal to 0, the codestream is ill-formed. Thus, the effect is skip to the next byte boundary if not already at a byte boundary, and require all skipped bits to have value 0.

## A.4 Headers

### A.4.1 Signature

| condition | type | default | name |
|---|---|---|---|
| | u(8) | | ff |
| | u(8) | | type |

ff shall be 255.

type shall be 10.

### A.4.2 Image dimensions

#### A.4.2.1 AspectRatio

| ratio | Meaning |
|---|---|
| 0 | xsize coded separately |
| 1 | xsize = ysize |
| 2 | xsize = floor(ysize × 12 / 10) |
| 3 | xsize = floor(ysize × 4 / 3) |
| 4 | xsize = floor(ysize × 3 / 2) |
| 5 | xsize = floor(ysize × 16 / 9) |
| 6 | xsize = floor(ysize × 5 / 4) |
| 7 | xsize = floor(ysize × 2 / 1) |

If the ratio field of SizeHeader or PreviewHeader != 0, their xsize field shall be computed from ysize and ratio as defined in this table.

#### A.4.2.2 SizeHeader

| condition | type | default | name |
|---|---|---|---|
| | Bool() | | small |
| small | u(5) | 0 | ysize_div8_minus_1 |
| !small | U32(Bits(9), Bits(13), Bits(18), Bits(30)) | 0 | ysize_minus_1 |
| | u(3) | | ratio |
| small && ratio == 0 | u(5) | 0 | xsize_div8_minus_1 |
| !small && ratio == 0 | U32(Bits(9), Bits(13), Bits(18), Bits(30)) | 0 | xsize_minus_1 |

ysize shall be computed as small ? (ysize_div8_minus_1 + 1) × 8 : ysize_minus_1 + 1.

xsize shall be computed as follows. If ratio == 0, then xsize = small ? (xsize_div8_minus_1 + 1) × 8 : xsize_minus_1 + 1. Otherwise, ratio is in the range [1, 8) and xsize shall be computed as defined in the AspectRatio subclause.

NOTE These encodings cannot represent a zero-width or zero-height image.

Unlike other bundles, SizeHeader provides a guarantee on when the decoder can access its fields. In a streaming use case, it can be helpful for the decoder to know the image dimensions before having received the entire codestream. Once at least 94 bits of the codestream are accessible, or the entire codestream is known to have been transmitted, the decoder can decode SizeHeader and access its fields.

#### A.4.2.3 PreviewHeader

NOTE The presence of this bundle is signaled by metadata.m2.have_preview.

| condition | type | default | name |
|---|---|---|---|
| | Bool() | | div8 |
| div8 | U32(Val(15), Val(31), Bits(5), BitsOffset(9, 32)) | 0 | ysize_div8_minus_1 |
| !div8 | U32(Bits(6), BitsOffset(8, 64), BitsOffset(10, 320), BitsOffset(12, 1344)) | 0 | ysize_minus_1 |
| | u(3) | | ratio |
| div8 && ratio == 0 | U32(Val(15), Val(31), Bits(5), BitsOffset(9, 32)) | 0 | xsize_div8_minus_1 |
| !div8 && ratio == 0 | U32(Bits(6), BitsOffset(8, 64), BitsOffset(10, 320), BitsOffset(12, 1344)) | 0 | xsize_minus_1 |

ysize shall be computed as div8 ? (ysize_div8_minus_1 + 1) × 8 : ysize_minus_1 + 1.

xsize shall be computed as follows. If ratio == 0, then xsize = div8 ? (xsize_div8_minus_1 + 1) × 8 : xsize_minus_1 + 1. Otherwise, ratio is in the range [1, 8) and xsize shall be computed as defined in the AspectRatio subclause. If xsize or ysize exceeds 4096, the codestream is ill-formed.

NOTE This encoding cannot represent a zero-width or zero-height preview.

### A.4.3 ColourEncoding

#### A.4.3.1 Customxy

| condition | type | default | name |
|---|---|---|---|
| | S32(Bits(19), BitsOffset(19, 524288), BitsOffset(20, 1048576), BitsOffset(21, 2097152)) | | x |
| | S32(Bits(19), BitsOffset(19, 524288), BitsOffset(20, 1048576), BitsOffset(21, 2097152)) | | y |

x and y shall be the coordinates of a point on the CIE xy chromaticity diagram, scaled by $10^6$. The unscaled coordinates may be outside [0, 1] for imaginary primaries.

#### A.4.3.2 ColourSpace

| name | value | meaning |
|---|---|---|
| kRGB | 0 | Tristimulus RGB, with various white points and primaries |
| kGrey | 1 | Luminance, with various white points; colour_encoding.primaries shall be ignored |
| kXYB | 2 | XYB (opsin); colour_encoding.white_point shall be kD65, colour_encoding.primaries shall be ignored, colour_encoding.have_gamma shall be true, and colour_encoding.gamma shall be 3333333 |
| kUnknown | 3 | The colour space shall not be one of the other meanings in this table |
| kXYZ | 4 | CIE XYZ; colour_encoding.white_point shall be kE and colour_encoding.primaries shall be ignored |

#### A.4.3.3 WhitePoint

| name | value | meaning |
|---|---|---|
| kD65 | 1 | CIE Standard Illuminant D65: 0.3127, 0.3290 |
| kCustom | 2 | Custom white point stored in colour_encoding.white |
| kE | 10 | CIE Standard Illuminant E (equal-energy): 1/3, 1/3 |
| kDCI | 11 | DCI-P3 from SMPTE RP 431-2: 0.314, 0.351 |

The meaning column shall be interpreted as CIE xy chromaticity coordinates.

NOTE The values are defined by ISO/IEC FDIS 23091-2:2019(E).

#### A.4.3.4 Primaries

| name | value | meaning |
|---|---|---|
| kSRGB | 1 | 0.639998686, 0.330010138; 0.300003784, 0.600003357; 0.150002046, 0.059997204 |
| kCustom | 2 | Custom red/green/blue primaries stored in colour_encoding.red/green/blue |
| k2100 | 9 | As specified in Rec. ITU-R BT.2100-1 |
| kP3 | 11 | As specified in SMPTE RP 431-2 |

NOTE The values are defined by ISO/IEC FDIS 23091-2:2019(E).

The meaning column shall be interpreted as three CIE xy chromaticity coordinates: red; green; blue, respectively.

NOTE The xy coordinates of kSRGB are quantized and match the values that would be stored in an ICC profile.

#### A.4.3.5 TransferFunction

| name | value | meaning |
|---|---|---|
| k709 | 1 | As specified in Rec. ITU-R BT.709-6 |
| kUnknown | 2 | The transfer function shall not be one of the other meanings in this table |
| kLinear | 8 | The gamma exponent is 1 |
| kSRGB | 13 | As specified in IEC 61966-2-1 sRGB |
| kPQ | 16 | As specified in Rec. ITU-R BT.2100-1 (PQ) |
| kDCI | 17 | As specified in SMPTE ST 428-1 |
| kHLG | 18 | As specified in Rec. ITU-R BT.2100-1 (HLG) |

NOTE The values are defined by ISO/IEC FDIS 23091-2:2019(E).

#### A.4.3.6 RenderingIntent

| name | value | meaning |
|---|---|---|
| kPerceptual | 0 | As specified in ISO 15076-1:2010 (vendor-specific) |
| kRelative | 1 | As specified in ISO 15076-1:2010 (media-relative) |
| kSaturation | 2 | As specified in ISO 15076-1:2010 (vendor-specific) |

| kAbsolute | 3 | As specified in ISO 15076-1:2010 (ICC-absolute) |
|---|---|---|

NOTE The values are defined by ISO 15076-1:2010.

#### A.4.3.7 ColourEncoding

In this subclause, CS shall denote ColourSpace, WP shall denote WhitePoint, PR shall denote Primaries, TF shall denote TransferFunction, use_desc shall expand to !all_default && !opaque_icc, and not_xy shall expand to colour_space != kXYZ && colour_space != kXYB.

| condition | type | default | name |
|---|---|---|---|
| | Bool() | | all_default |
| !all_default | Bool() | false | received_icc |
| received_icc | Bool() | false | opaque_icc |
| use_desc | Enum(ColourSpace) | kRGB | colour_space |
| use_desc && not_xy | Enum(WhitePoint) | kD65 | white_point |
| use_desc && white_point == WP.kCustom | Customxy | | white |
| use_desc && not_xy && colour_space != kGray | Enum(Primaries) | PR.kSRGB | primaries |
| use_desc && primaries == PR.kCustom | Customxy | | red |
| use_desc && primaries == PR.kCustom | Customxy | | green |
| use_desc && primaries == PR.kCustom | Customxy | | blue |
| use_desc && not_xy | Bool() | false | have_gamma |
| use_desc && have_gamma | u(24) | 0 | gamma |
| use_desc && !have_gamma && not_xy | Enum(TransferFunction) | TF.kSRGB | transfer_function |
| use_desc && colour_space != kGray && not_xy | Enum(RenderingIntent) | kRelative | rendering_intent |

received_icc shall indicate whether the source image was accompanied by an ICC profile.

opaque_icc shall be true if and only if received_icc && (transfer_function == TF.kUnknown || colour_space == CS.kUnknown). If it is true, the colour space shall be characterized by the ICC profile from Annex B. Otherwise, the other fields fully describe the colour space.

colour_space shall be kGrey if the input image was greyscale.

have_gamma == false shall indicate the transfer function is defined by transfer_function. Otherwise, the transfer function shall be characterized by the OETF exponent colour_encoding.gamma / $10^7$, and the codestream is ill-formed if this exponent is not in (0, 1).

The applicability of ColourEncoding is specified in FrameEncoding.

### A.4.4 Extensions

This bundle shall be the last field in bundles (ImageMetadata2, FrameHeader, LoopFilter) that may be extended in future.

| condition | type | default | name |
|---|---|---|---|
| | U64() | | extensions |
| extensions != 0 | U64() | 0 | extension_bits |
| extensions != 0 | u(extension_bits) | 0 | |

An extension consists of zero or more bits whose interpretation shall be established by amendments to this document. Each extension shall be identified by an ext_num in the range [0, 63), assigned in increasing sequential order. extensions shall be a bit array whose i-th bit (least-significant == 0) indicates whether the extension with ext_num = i is present. Extensions that are present shall be stored in ascending order of their ext_num. extension_bits shall indicate the total number of bits (after extension_bits has been read) used for storing extensions. The decoder shall choose a value of ExtensionEnd in [0, 64] and support all extensions whose ext_num < ExtensionEnd, and no others.

NOTE This definition allows the decoder to support no extensions at all.

The decoder shall record pos_begin (the total number of bits read subsequent to reading extension_bits), read any extensions whose ext_num < ExtensionEnd that are present, record pos_end (the total number of bits read subsequent to reading supported and present extensions), and skip the next extension_bits - (pos_end - pos_begin) bits. Skipping n bits shall be equivalent to reading a u(n) and ignoring its value.

### A.4.5 ImageMetadata

Holds information that applies to all Frames (including the preview); decoders shall use it to interpret pixel values and restore the original image.

| condition | type | default | name |
|---|---|---|---|
| | Bool() | | all_default |
| !all_default | Bool() | false | have_icc |
| !all_default | U32(Val(8), Val(16), Val(32), Bits(5)) | 8 | bits_per_sample |
| !all_default | ColourEncoding | | colour_encoding |
| !all_default | U32(Val(0), Val(8), Val(16), Bits(4)) | 0 | alpha_bits |
| !all_default | U32(Val(5), Val(20), Val(80), BitsOffset(10, 1)) | 5 | target_nits_div50 |
| !all_default | ImageMetadata2 | | m2 |

have_icc shall be set if and only if the codestream stores an ICC profile. If so, it shall be decoded as specified in Annex B. If colour_encoding.opaque_icc == true and have_icc == false, the codestream is ill-formed.

bits_per_sample shall be the number of bits per channel of the original image and shall not exceed 32.

colour_encoding shall indicate the colour image encoding of the original image. This may differ from the encoding of the stored image in lossy modes.

alpha_bits shall be 0 if the original image and therefore the stored image lack an alpha channel. Otherwise, it shall be the number of valid bits in the alpha channel, and at most 16.

target_nits_div50 indicates the intended display luminance in candelas per square meter, divided by 50.

NOTE the maximum representable value of target_nits_div50 is 1024, corresponding to 51200 nits.

m2 contains additional metadata with a separate all_default flag.

### A.4.6 AnimationHeader

Meta-information about an animation or image sequence that shall be present in FrameHeader if metadata.m2.have_animation.

| condition | type | default | name |
|---|---|---|---|
| | Bool() | | composite_still |
| !composite_still | U32(Val(99), Val(999), Bits(6), Bits(18)) | 0 | tps_numerator_minus_1 |
| !composite_still | U32(Val(0), Val(1000), Bits(8), Bits(10)) | 0 | tps_denominator_minus_1 |
| !composite_still | U32(Val(0), Bits(3), Bits(16), Bits(32)) | 0 | num_loops |
| !composite_still | Bool() | false | have_timecodes |

composite_still shall indicate whether there is a single frame followed by zero or more frames with frame_header.animation_frame.duration == 0.

NOTE this document defines the interval between presenting the current and next frame in units of ticks.

(tps_numerator_minus_1 + 1) / (tps_denominator_minus_1 + 1) shall indicate the number of ticks per second.

num_loops shall be the number of times to repeat the animation (0 shall be interpreted as infinity).

have_timecodes shall indicate whether AnimationFrame.timecode is present in the codestream.

### A.4.7 ExtraChannelInfo

| condition | type | default | name |
|---|---|---|---|
| | U32(Val(0), Val(1), Val(2), Bits(6)) | | meaning |
| meaning == 1 | F16() | 0 | red |
| meaning == 1 | F16() | 0 | green |
| meaning == 1 | F16() | 0 | blue |
| meaning == 1 | F16() | 0 | solidity |

meaning shall be 0 if the semantics of an extra channel are unknown but without impact on the decoded image. meaning shall be 1 if the extra channel represents a spot colour, in which case red, green, blue shall indicate its colour encoded as sRGB, where in-gamut samples are in the range [0, 255] and out-of-gamut samples are allowed, and solidity in [0, 1] indicates the overall blending factor, with 0 corresponding to fully translucent (invisible) and 1 corresponding to fully opaque. meaning shall be 2 to denote selection masks, which indicate a region of interest plus the degree of membership, where zero means not selected.

Extra channels shall be interpreted as specified in L.7.

### A.4.8 ImageMetadata2

| condition | type | default | name |
|---|---|---|---|
| | Bool() | | all_default |
| !all_default | Bool() | false | have_preview |
| !all_default | Bool() | false | have_animation |
| !all_default | u(3) | 0 | orientation_minus_1 |
| !all_default | U32(Val(0), Val(8), Val(16), Bits(4)) | 0 | depth_bits |
| !all_default | U32(Val(0), Val(3), Val(4), BitsOffset(3, 1)) | 0 | depth_shift |
| !all_default | U32(Val(0), Bits(4), BitsOffset(8, 16), BitsOffset(12, 1)) | 0 | num_extra_channels |
| num_extra_channels > 0 | U32(Val(0), Val(8), Val(16), Bits(4)) | 0 | extra_channel_bits |
| | ExtraChannelInfo | | extra_channel_info[ num_extra_channels] |
| !all_default | Extensions | | extensions |

have_preview shall indicate whether the codestream includes a preview and preview_frame.

have_animation shall indicate whether the codestream includes animation and FrameHeader includes an animation_frame.

orientation_minus_1 shall indicate the interpretation of the stored image versus the original image as follows:

| **orientation_minus_1** | **Side of original stored in first row** | **Side of original stored in first column** | **Transform to apply to the stored image** |
|---|---|---|---|
| 0 | top | left | none |
| 1 | top | right | flip horizontally |
| 2 | bottom | right | rotate 180° |
| 3 | bottom | left | flip vertically |
| 4 | left | top | transpose (rotate 90° clockwise then flip horizontally) |
| 5 | right | top | rotate 90° clockwise |
| 6 | right | bottom | flip horizontally then rotate 90° clockwise |
| 7 | left | bottom | rotate 90° counterclockwise |

NOTE **orientation_minus_1** + 1 matches the values used by JEITA CP-3451C (Exif version 2.3).

depth_bits shall be 0 if the original image and therefore the stored image lack a depth channel. Otherwise, it shall the number of valid bits in the depth channel, and at most 16.

depth_shift shall be the base-2 logarithm of the downsampling factor of the dimensions of the depth image (if any) with respect to the main image dimensions defined in size. If 1 << depth_shift exceeds the dimensions of a group, the codestream is ill-formed.

EXAMPLE depth_shift == 3 implies 8 × 8 downsampling, e.g. a 2 × 2 depth image for a 16 × 16 image.

num_extra_channels shall be the number of additional image channels. If it is nonzero, all extra channel samples shall be in [0, 1 << extra_channel_bits), and extra_channel_info shall signal the semantics (or lack thereof) of each extra channel.

# Annex B
(normative)
# ICC Profile

**B**

## B.1 General
If !`metadata.have_icc`, the decoder shall synthesize an ICC profile with the `colour_space`, `white_point`, `primaries`, `transfer_function` and `rendering_intent` specified in `metadata.colour_encoding`. Otherwise, the decoder shall decode an ICC profile as specified in the remainder of this Annex.

## B.2 Data stream

The decoder shall ensure byte alignment as specified in 6.3, and decompress the ICC profile stream with Brotli (D.2). The decoder shall reconstruct the ICC profile from the decompressed stream as described below, and resume decoding starting at the end of the Brotli stream, as indicated by the Brotli decoder.

## B.3 The encoded ICC stream

The decoder shall decompress the ICC stream from the following high level format:

- `output_size` (the size of the decoded ICC profile) = Varint()
- `commands_size` = Varint()
- `commands_size` bytes: commands stream
- all remaining bytes: data stream

The decoder shall maintain the current position within the command and data stream, initially at their beginning, and increment each when a byte from the corresponding stream is read. If the decoder reads from a stream whose position is beyond the last byte of that stream, the codestream is ill-formed.

The decoder shall decode an ICC header (B.4), ICC tag list (B.5) and main content (B.6). The decoded ICC profile shall be the concatenation of these three parts.

## B.4 ICC Header

The header is the first of three concatenated parts that the decoder shall output.

The header size in bytes shall be:

```
header_size = min(128, output_size)
```

To decode the header, the decoder shall read `header_size` bytes from the data stream and for each byte at position `i` shall compute the result as follows.

A predicted value `p` shall be computed as:

- If `i` equals 0, 1, 2 or 3: `p` shall be set to byte `i` of `output_size` encoded as an unsigned 32-bit integer in big endian order
- If `i` equals 41, 42 or 43 and the earlier output header byte corresponding to `i == 40` matches ASCII value 'A', then `p` shall be set to respectively ASCII value 'P', 'P', 'L'

- If `i` equals 41, 42 or 43 and the earlier output header byte corresponding to `i == 40` matches ASCII value 'M', then `p` shall be set to respectively ASCII value 'S', 'F', 'T'
- If `i` equals 42 or 43 and the earlier output header bytes corresponding to `i == 40` and `i == 41` matches respectively ASCII values 'S' and 'G', then `p` shall be set to respectively ASCII value 'I', ' '
- If `i` equals 42 or 43 and the earlier output header bytes corresponding to `i == 40` and `i == 41` matches respectively ASCII values 'S' and 'U', then `p` shall be set to respectively ASCII value 'N', 'W'
- If `i` equals 80 + j for j in [0, 4), then `p` shall be equal to the earlier output header bytes corresponding to `i` == 4 + j
- Otherwise, `p` shall be set to the value corresponding to the following options:
    - if `i` equals 8: 4
    - if `i` equals 12 to 23: the corresponding ASCII value at index `i - 12` from the 12-character string "mntrRGB XYZ ", where there is one space after "RGB" and one space after "XYZ"
    - if `i` equals 36 to 39: the corresponding ASCII value at index `i - 36` from the 4-character string "acsp"
    - if `i` equals 70: 246
    - if `i` equals 71: 214
    - if `i` equals 73: 1
    - if `i` equals 78: 211
    - if `i` equals 79: 45
    - otherwise: 0

A byte `e` shall then be read from the data stream and header byte `i` shall be output as `(p + e) & 255`.

If `output_size` is smaller than or equal to 128, then the above procedure has produced the full output, the decoder shall stop decoding and return the result. In that case, if not all bytes of the commands stream and data stream have been read at this point, the codestream is ill-formed.

## B.5 ICC tag list

The ICC tag list is the second of three concatenated output parts that the decoder shall decode.

The decoder shall resume at the stream positions reached after decoding the previous part (the header). Since the encoded header size is constant and no commands were used, this means at this point it is still at the start of the commands stream, and at position 128 in the data stream.

The tag list shall be decoded as follows:

1. If the end of the command stream is reached, the decoder shall stop decoding and return the result.
    - If not all bytes of the data stream have been read at this point, or the output size does not match `output_size`, the codestream is ill-formed
2. Read `v` = Varint() from the command stream and set `num_tags` to `v` - 1
3. If `num_tags == -1`, output nothing, stop reading the tag list, and proceed to B.6
4. Append `num_tags` to the output as a big endian unsigned 32-bit integer
5. If the end of the command stream is reached, the decoder shall stop decoding and return the result
    - If not all bytes of the data stream have been read at this point, or the output size does not match `output_size`, the codestream is ill-formed
6. Read 1 byte `command` from the command stream
7. Set `tagcode` to `command & 63`
8. If `tagcode` == 0, decoding the tag list is done, and the decoding shall continue as specified in B.6
9. Read or compute `tagstart`:
    - if (`command & 64`) == 0, set tagstart to the previous value of tagstart + previous value of tagsize (if there are no previous values yet, set it to `num_tags` × 12 + 128)
    - else read tagstart = Varint() from the command stream
10. Read or compute `tagsize`:

- If (`command & 128`) == 0, the decoder shall set tagsize to the previous tagsize (if there is no previous tagsize yet, this shall be 0)
   - else read tagsize = Varint() from the command stream
11. if `tagcode` == 1, the decoder shall append the following 3 groups of 4 bytes to the output:
   - 4 bytes, which shall read from the data stream
   - tagstart as big endian 32-bit integer
   - tagsize as big endian 32-bit integer
12. if `tagcode` == 2, the decoder shall append the following 9 groups of 4 bytes to the output:
   - the string "rTRC"
   - tagstart as big endian 32-bit integer
   - tagsize as big endian 32-bit integer
   - the string "gTRC"
   - tagstart as big endian 32-bit integer
   - tagsize as big endian 32-bit integer
   - the string "bTRC"
   - tagstart as big endian 32-bit integer
   - tagsize as big endian 32-bit integer
13. if `tagcode` == 3, the decoder shall append the following 9 groups of 4 bytes to the output:
   - the string "rXYZ"
   - tagstart as big endian 32-bit integer
   - tagsize as big endian 32-bit integer
   - the string "gXYZ"
   - (tagstart + tagsize) as big endian 32-bit integer
   - tagsize as big endian 32-bit integer
   - the string "bXYZ"
   - (tagstart + 2 × tagsize) as big endian 32-bit integer
   - tagsize as big endian 32-bit integer
14. if `tagcode` is in the range [4, 21), the decoder shall append the following 3 groups of 4 bytes to the output, in top to bottom order:
   - 4 bytes from a string chosen from the following list using a 0-based index given by `tagcode - 4`:
     {"cprt", "wtpt", "bkpt", "rXYZ", "gXYZ", "bXYZ", "kXYZ", "rTRC", "gTRC", "bTRC", "kTRC", "chad", "desc", "chrm", "dmnd", "dmdd", "lumi"}
   - tagstart as big endian 32-bit integer
   - tagsize as big endian 32-bit integer
15. If `tagcode` has any other value not handled above, the codestream is ill-formed.
16. The decoder shall go back to step 5 to read the next command

## B.6 Main content

The main content is the third of three concatenated output parts that the decoder shall decode. The decoder shall resume at the stream positions reached after decoding the previous part (the ICC tag list).

In this subclause, `Shuffle(bytes, width)` denotes the following operation. Inscribe `bytes` in raster order into a matrix with `width` columns, and as many rows as needed. The last row may have missing elements if `len(bytes)` is not a multiple of `width`. Then transpose the matrix, and overwrite `bytes` with elements of the matrix read in raster order, skipping any missing elements.

The decoder shall decode as specified below:

1. If the end of the command stream is reached, the decoder shall stop decoding and return the result.
   - If not all bytes of the data stream have been read at this point, or the output size does not match `output_size`, the codestream is ill-formed
2. Read 1 byte `command` from the command stream

3. If the `command` is 1, append bytes to the output as follows:
   - read `num` = [Varint]() from the command stream
   - read num bytes from the data stream and append them to the output stream
4. If the `command` is 2 or 3, append shuffled bytes to the output as follows:
   - read `num` = [Varint]() from the command stream
   - read `num` bytes from the data stream
   - apply `Shuffle` with width 2 (if command is 2) or width 4 (if command is 4) to all the `num` bytes.
   - append the resulting `num` bytes to the output stream
5. If `command` is 4, append a series of Nth order predicted bytes to the output stream as follows:
   - read byte `flags` from the command stream
   - set `width` to `(flags & 3) + 1`
   - if the width is 3, the codestream is ill-formed
   - set `order` to `(flags & 12) >> 2`
   - if the order is 3, the codestream is ill-formed
   - compute or read `stride`:
     - if (flags & 16) is 0, set `stride` to `width`
     - else set `stride` = [Varint]() from the command stream
     - if `stride` × 4 >= amount of bytes already output, the codestream is ill-formed
   - read `num` = [Varint]() from the command stream
   - read `num` bytes from the data stream into `v`.
   - if `width` is 2 or 4, `Shuffle(v, width)`.
   - run an Nth-order predictor on `num` bytes as follows (with the bytes representing numbers of the given width, but `num` is not required to be a multiple of width):
     - Read an N-element array `prev` of previous numbers with N = `order` + 1, where each number is encoded as `width` big-endian bytes. `prev`[i] shall be read from (`stride` × (i + 1)) bytes before the current position, where i is in the range [1, N]. The bytes read are those appended to the output from predicting the previous bytes, or output by previous commands before that
     - compute the predicted number `p` with the following formula: for order 0: `prev`[0]. For order 1: 2 × `prev`[0] - `prev`[1]. For order 2: 3 × `prev`[0] - 3 × `prev`[1] + `prev`[2]
     - read `width` bytes from the data stream into `e`
     - For j in the range [0, `width`), ending short if `num` is reached, output (`e`[`width` - 1 - j] + `p`[`width` - 1 - j]) & 255. In this notation, `p`[0] is the least-significant byte.
6. If `command` is 10, append an XYZ section as follows:
   - append the 4 ASCII characters "XYZ " to the output
   - append 4 bytes with value 0 to the output
   - read 12 bytes from the data stream and append them to the output
7. If `command` is in the range [16, 24), output a common type signature as follows:
   - output 4 bytes from a string chosen from the following list of strings, the string chosen from the index given by `command - 16`: {"XYZ ", "desc", "text", "mluc", "para", "curv", "sf32", "gbd "}
   - append 4 bytes with value 0 to the output
8. If `command` has any other value not handled above, the codestream is ill-formed.
9. The decoder shall go back to step 1 to read the next command

**Annex C**
(normative)
**Frames**

**C**

## C.1 General

Subsequent to headers (Annex A) and ICC profile, if present (Annex B), the codestream shall consist of one or more frames. Each Frame shall be byte-aligned as specified in 6.3 and have the structure shown in Table C.1.

**Table C.1: Frame structure**

| condition | type | name |
|---|---|---|
| | FrameHeader | frame_header |
| frame_header.encoding == kPasses | LoopFilter | loop_filter |
| | TOC | toc |
| frame_header.encoding == kPasses | DcGlobal | dc_global |
| frame_header.flags & kProgressive16 | Dc16Group | dc_group[num_dc16_groups] |
| frame_header.encoding == kPasses | DcGroup | dc_group[num_dc_groups] |
| frame_header.encoding == kPasses \|\| frame_header.encoding == kLossless | AcGlobal | ac_global |
| frame_header.encoding == kPasses | AcPass | ac_pass[frame_header.passes.num_passes] |
| frame_header.encoding != kModular | PassGroup | group_pass[num_groups × frame_header.passes.num_passes] |
| frame_header.encoding == kModular | Modular | modular |

The dimensions of a frame in pixels (xsize and ysize) shall be size.xsize and size.ysize if !frame_header.animation_frame.have_crop, otherwise frame_header.animation_frame.xsize and frame_header.animation_frame.ysize. If xsize >= size.xsize or ysize >= size.ysize, the codestream is ill-formed.

xsize and ysize shall be interpreted as the dimensions of the sample grid, not its rotated/mirrored interpretation (which is indicated by metadata.m2.orientation).

In this document, num_groups shall be ceil(xsize / kGroupDim) × ceil(ysize / kGroupDim), where kGroupDim is the width and height of a group (6.2). num_dc_groups shall be ceil(xsize / (kGroupDim×8)) × ceil(ysize / (kGroupDim×8)). num_dc16_groups shall be ceil(xsize / (kGroupDim×16)) × ceil(ysize / (kGroupDim×16)).

## C.2 FrameHeader

### C.2.1 Overview

Each FrameHeader shall have the structure shown in Table C.2.

**Table C.2: FrameHeader structure**

| condition | type | default | name |
|---|---|---|---|
| | Bool() | | all_default |
| !all_default && metadata.m2.have_animation | AnimationFrame | | animation_frame |
| !all_default | Enum(FrameEncoding) | kPasses | encoding |
| encoding == kLossless | LosslessMode | | lossless |
| !all_default && encoding == kPasses | Passes | | passes |
| !all_default && encoding == kPasses | U64() | 0 | flags |
| !all_default | Extensions | | extensions |

### C.2.2 FrameEncoding

| name | value | meaning |
|---|---|---|
| kLossless | 0 | Indicates the frame is stored in mathematically-lossless mode using either a single channel or three channels, with either 8-bit or 16-bit unsigned integer precision per channel as indicated by LosslessMode. |
| kPasses | 1 | Indicates the frame is stored in Var-DCT mode |
| kModular | 2 | Indicates the frame is stored in full-frame modular mode |
| kModularGroup | 3 | Indicates the frame is stored in modular mode |

This document specifies the meaning of **name** where it is referenced. Furthermore, all encodings except kModular split each frame into groups.

All encodings except kPasses store pixels in the colour space described by the ICC profile from Annex B. For kPasses, the pixels shall be in XYB colour space.

### C.2.3 AnimationFrame

| condition | type | default | name |
|---|---|---|---|
| | U32(Val(0), Val(1), Bits(8), Bits(32)) | | duration |
| duration > 0 | u(2) | 1 | dispose_mode |
| | u(2) | | blend_mode |
| animation.have_timecode | u(32) | 0 | timecode |
| | Bool() | | have_crop |
| have_crop | U32(Bits(8), BitsOffset(11, 256), BitsOffset(14, 2304), BitsOffset(30, 18688)) | 0 | x0 |
| have_crop | U32(Bits(8), BitsOffset(11, 256), BitsOffset(14, 2304), BitsOffset(30, 18688)) | 0 | y0 |
| have_crop | U32(Bits(8), BitsOffset(11, 256), BitsOffset(14, 2304), BitsOffset(30, 18688)) | 0 | xsize |
| have_crop | U32(Bits(8), BitsOffset(11, 256), BitsOffset(14, 2304), BitsOffset(30, 18688)) | 0 | ysize |
| | Bool() | | is_last |

duration (in units of ticks, see AnimationHeader) is the intended period of time between presenting the current frame and the next one. If the total number of pixels in frames whose duration sum to dt seconds (at least 1 tick) exceeds 25 × size.xsize × size.ysize / dt, the decoder may increase duration. If duration is zero, the decoder shall not present the current frame until all zero-duration frames following it have been decoded.

timecode shall indicate the SMPTE timecode of the current frame, or 0. The decoder shall interpret groups of 8 bits from most-significant to least-significant as hour, minute, second, and frame. If timecode is nonzero and less than or equal to that of a previous frame whose duration is nonzero, the bitstream is ill-formed.

If have_crop, the decoder shall consider the current frame to have dimensions xsize × ysize, and update the rectangle of the previous frame whose top-left corner in the sample grid is x0, y0 with the current frame using the given blend_mode. If x0 + xsize >= size.xsize or y0 + ysize >= size.ysize, the codestream is ill-formed. If !have_crop, the frame shall have the same dimensions as the image.

The following values for blend_mode are defined; if the value is 3, the codestream is ill-formed.

| name | value | description |
|---|---|---|
| kReplace | 0 | Each sample in channel c (including alpha, if present) is overwritten with the corresponding new sample in channel c |
| kAdd | 1 | Each new sample in channel c (including alpha, if present) is added to the corresponding previous sample in channel c |
| kBlend | 2 | Each new sample is alpha-blended over the corresponding previous sample, that is:<br>alpha = old_alpha + new_alpha × ($2^{alpha_bits}$ - 1 - old_alpha)/($2^{alpha_bits}$ - 1)<br>rgb = (old_alpha × old_rgb + new_alpha × new_rgb × ($2^{alpha_bits}$ - 1 - old_alpha)/($2^{alpha_bits}$ - 1)) / alpha |

The dispose_mode indicates what the new base image will be, on top of which the next frame will be rendered. The following values are defined; if the value is 3, the codestream is ill-formed.

| name | value | description |
|---|---|---|
| kPrevious | 0 | The new base is equal to the *previous* frame, as it was before it was updated with the current frame. |
| kCurrent | 1 | The new base is equal to the *current* frame, i.e. the update is preserved. |
| kKeyframe | 2 | The new base is an all-zeroes, fully transparent frame. That is, the next frame will be a "keyframe" since it can be rendered without first rendering the preceding frames. |

The decoder shall cease decoding after the current frame if is_last.

### C.2.4 LosslessMode

This bundle shall indicate properties of the lossless coding when frame_header.encoding == kLossless.

| condition | type | default | name |
|---|---|---|---|
| | Bool() | | greyscale |
| | Bool() | | bits16 |

The lossless decoder for the current frame shall use a greyscale mode if greyscale, otherwise a coloured mode. The lossless decoder for the current frame shall use a 16-bit mode if bits16, otherwise an 8-bit mode.

### C.2.5 Passes

| condition | type | default | name |
|---|---|---|---|
| | U32(Val(1), Val(2), Val(3), BitsOffset(3, 4)) | | num_passes |
| num_passes != 1 | U32(Val(0), Val(1), Val(2), BitsOffset(1, 3)) | 0 | num_downsample |
| for (i = 0; i < num_passes - 1; ++i) | u(2) | 0 | shift[i] |
| num_passes != 1 | U32(Val(1), Val(2), Val(4), Val(8)) | 1 | downsample [num_downsample] |
| num_passes != 1 | U32(Val(0), Val(1), Val(2), Bits(3)) | 0 | last_pass [num_downsample] |

num_passes shall indicate the number of passes into which the frame is partitioned.

num_downsample shall indicate the number of (downsample, last_pass) pairs between 0 and 4, inclusive. If it matches or exceeds num_passes, the codestream is ill-formed.

shift[i] shall indicate the amount by which the AC coefficients of the pass with index i in the range [0, num_passes) shall be left-shifted immediately after entropy decoding. The last pass shall behave as if it had a shift value of 0.

downsample[i] (i in the range [0, num_downsample)) shall indicate a downsampling factor in x and y direction.

last_pass[i] (i in the range [0, num_downsample)) shall indicate the zero-based index of the final pass that shall be decoded in order to achieve a downsampling factor of least downsample[i]. The codestream is ill-formed if this is at least num_passes.

In addition to the (downsample, last_pass) pairs that are explicitly encoded in the bitstream, the decoder shall behave as if a final pair equal to (1, num_passes - 1) were present.

### C.2.6 Flags

| name | value | meaning |
|---|---|---|
| kNoise | 1 | Enable noise feature stage |
| kSkipDots | 2 | Disable dot feature stage |
| kPatches | 4 | Enable patch feature stage |
| kSplines | 16 | Enable spline feature stage |
| kProgressive16 | 32 | Enable 16× downsampled progressive mode |
| kSkipPredictLF | 128 | Disable low-frequency prediction |

Other values than listed above shall only be used if they are defined in amendments to this document or subsequent parts.

In this subclause and references to frame_header.flags, a flag with a value from the table above shall be considered set when (frame_header.flags & value) != 0. The decoder shall skip the respective processing steps in the current frame for each flag that is set and whose name starts with kSkip. The decoder shall enable the respective processing steps in the current frame for each flag that is set and whose name does not start with kSkip.

EXAMPLE if frame_header.flags == 18, the decoder shall enable spline rendering and skip dots.

## C.3 TOC

### C.3.1 Overview

The TOC (Table of Contents) is an array of entries.

If frame_header.encoding == kModular, then the TOC has exactly six entries. These correspond to the (relative) byte offsets for partially decoding the frame. These offsets provide hints that allow the decoder to determine in advance how many bytes of the bitstream it needs to have access to in order to render a lower resolution/quality image preview. The first entry is for a LQIP (low-quality image placeholder), the second entry for a scale 1:16 image, the third entry for a scale 1:8 image, the fourth entry for a scale 1:4 image, the fifth entry for a scale 1:2 image, and finally the sixth entry for the full-resolution image.

If `frame_header.encoding != kModular`, then each TOC entry indicates the size in bytes of a section (defined below). Each section shall be aligned to a byte boundary, as defined in 6.3. If `num_groups` == 1, there shall be a single TOC entry and section containing all frame data structures. Otherwise, there shall be one entry for each of the following sections, in the order they are listed: DcGlobal, one per Dc16Group in raster order (if `kProgressive16` is set in `frame_header.flags`), one per DcGroup in raster order, one for AcGlobal, followed by AcPass data for all the passes, and `num_groups` × `frame_header.passes.num_passes` PassGroup. The first PassGroup shall be the groups (in raster order) of the first pass, followed by all groups (in raster order) of the second pass (if present), and so on.

The decoder shall first read `permuted_toc = u(1)`. If and only if its value is 1, the decoder shall read `permutation` from a single ANS stream with `8` clustered distributions, as specified in D.3.5, with `size` equal to the number of entries and `skip = 0`.

### C.3.2 Decoding permutations

The decoder shall compute a sequence `lehmer` containing `size` elements as follows. Let `GetContext(x)` be defined as `min(8, ceil(log2(x + 1)))`. The decoder shall first decode an ANS symbol (as specified in D.3.3) `log_end` using distribution `D[GetContext(size)]`. Then, it shall read a `u(log_end)` value `b` and compute `end` as `(1 << log_end) + b - 1 + skip`. Then, `(end - skip)` elements of the `lehmer` sequence shall be produced as follows. For each element, a symbol shall be read from the ANS steam using the distribution `D[prev_elem]`, where `prev_elem` is `s[i - 1]` if `i > 0`, or `0` otherwise. After reading this symbol, a `u(s[i])` value `bp` shall be read and the following element shall be set in the `lehmer` sequence:

```
lehmer[skip + i] = (1 << s[i]) + bp - 1
```

All other elements of the sequence `lehmer` shall be `0`.

The decoder shall maintain a sequence of elements `temp`, initially containing the numbers `[0, size)` in increasing order, and a sequence of elements `permutation`, initially empty. Then, for each integer `i` in the range `[0, size)`, the decoder shall append to `permutation` element `temp[lehmer[i]]`, then remove it from `temp`, leaving the relative order of other elements unchanged. `permutation` shall be the decoded permutation.

### C.3.3 Decoding TOC

The sequence of TOC entries shall be byte-aligned (6.3). The decoder shall read each TOC entry in order of increasing index via U32(Bits(10), BitsOffset(14, 1024), BitsOffset(22, 17408), BitsOffset(30, 4211712)). If any TOC entry is zero, the codestream is ill-formed.
The decoder shall then compute an array `group_offsets`, whose first element is 0 and subsequent `group_offsets[i]` are the sum of all TOC entries [0, `i`).

If and only if `permuted_toc`, the decoder shall permute `group_offsets` according to `permutation`, such that `group_offsets[i]` is the one whose index was previously `group_offsets[permutation[i]]`.

Let `P` be the index of the byte containing the next unread bit after decoding the TOC. When decoding a group with index `i`, the decoder shall read from the codestream starting at the byte with index `P + group_offsets[i]`.

## C.4 DcGlobal

### C.4.1 DC channel dequantization weights

The decoder shall first read a u(1). If its value is `0`, the DC dequantization weights shall be their default values of `{4096, 512, 256}`. Otherwise, the decoder shall read three F16(), representing the X, Y and B dequantization weights respectively. $mX_{DC}$, $mY_{DC}$ and $mB_{DC}$ shall then be obtained by multiplying each channel weight by the value of `1 / ((global_scale_minus_1 + 1) × (quant_dc_minus_1 + 1))`, as decoded in C.4.2.

### C.4.2 Quantizer

| condition | type | default | name |
|---|---|---|---|
| | U32(Bits(11), BitsOffset(11, 2048), BitsOffset(12, 4096), BitsOffset(15, 8192)) | | global_scale_minus_1 |
| | U32(Val(15), Bits(5), Bits(8), Bits(16)) | | quant_dc_minus_1 |

These fields are referenced in C.4.1 and in C.5.5.

### C.4.3 DC channel correlation factors

The decoder shall read a u(1). If its value is 0, the DC colour correlation factors x_factor_dc and b_factor_dc shall be 128. Otherwise, the decoder shall first read x_factor_dc as u(8) and then b_factor_dc as u(8).

### C.4.4 Subpixel position decoding

#### C.4.4.1 Overview

Dots and splines refer to pixel locations (x, y) with 0 <= x < xsize and 0 <= y < ysize. The decoder shall read n_points positions (x, y) from the bit stream using a QuadTree. A QuadTree is defined as a tree where each node has either four or zero children. The nodes with zero children are called leaf nodes. The QuadTree represents a recursive sub-division of the image into four parts, that stores how many point positions are located within each of the four divisions.

#### C.4.4.2 Decoding the QuadTree

Each Node in the QuadTree is composed of the following attributes:

- first_x, last_x, first_y, last_y: Rectangle of the image that the node represents.
- n_points: Number of points located in the rectangle represented by the node.
- is_leaf: Whether the node is a leaf.
- children: An array of size four with references to the children nodes.

A node shall be considered a leaf node if its rectangle contains at most kLeafDots = 2 points or both sides of the rectangle are at most kLeafSize = 1. If a node is not a leaf, the decoder shall recursively partition its rectangle into four parts, read the number of points from the first three children from the (already initialized) arithmetic decoder, and deduce the fourth number of points, as specified by the following:

```
Node DecodeQTreeBAE(int n_points, int first_x, int last_x,
        int first_y, int last_y) {
  Node node;
  int xsize = last_x - first_x + 1;
  int ysize = last_y - first_y + 1;
  node.n_points = n_points;
  node.first_x = first_x;
  node.last_x = last_x;
  node.first_y = first_y;
  node.last_y = last_y;
  if (n_points <= kLeafDots || (xsize <= kLeafSize && ysize <= kLeafSize)) {
    node.is_leaf = true;
```

```
    return node;
  }

  int x_split = (first_x + last_x) Idiv 2;
  int y_split = (first_y + last_y) Idiv 2;
  int child_n[4];
  for (i = 0; i < 3; i++) {
    child_n[i] = GetUniformInt(n_points + 1);
    n_points -= child_n[i];
  }
  child_n[3] = n_points;  // The remaining points

  node.children[0] =
      DecodeQTreeBAE(child_n[0], first_x, x_split, first_y, y_split);
  node.children[1] =
      DecodeQTreeBAE(child_n[1], first_x, x_split, y_split + 1, last_y);
  node.children[2] =
      DecodeQTreeBAE(child_n[2], x_split + 1, last_x, first_y, y_split);
  node.children[3] =
      DecodeQTreeBAE(child_n[3], x_split + 1, last_x, y_split + 1, last_y);

  node.is_leaf = false;
  return node;
}
```

If the variable `n_points` becomes negative, the codestream is ill-formed. `GetUniformInt` is defined in D.5.6. `qtree`, the root of the QuadTree, represents the entire image, and is obtained using

```
n_points = GetUniformInt(xsize × ysize / 4);
qtree = DecodeQTreeBAE(n_points, 0, xsize - 1, 0, ysize - 1);
```

After the QuadTree is decoded, the positions shall be obtained by traversing the leaves of the QuadTree with the order determined by a Depth First Search (DFS) visit. For each non-leaf node, the children are visited in ascending order. For each leaf, the position of the point in the node rectangle shall be obtained by decoding an integer uniformly distributed in `[0, xsize × ysize)` from the arithmetic coder. Subsequently, the absolute position of the point in the image shall be obtained by adding the coordinates of the top left corner `(first_x, first_y)` of the rectangle from the QuadTree. The behaviour of the position decoder is described below:

```
void DecodeDotsBAE(const QTNode node) {
  if (!node.is_leaf) {
    for (i = 0; i < node.children.count; ++i) {
      DecodeDotsBAE(node.children[i]);
    }
  } else {
    for (i = 0; i < node.n_points; i++) {
      int xsize = node.last_x + 1 - node.first_x;
      int ysize = node.last_y + 1 - node.first_y;
      int pos = GetUniformInt(xsize × ysize);
      int y = (pos Idiv xsize) + node.first_y;
      int x = (pos Umod xsize) + node.first_x;
      [[Append {x, y} to list of dot positions.]]
```

```
    }
  }
}
```

### C.4.5 Dots

If the `kSkipDots` flag in `frame_header.flags` is set, this subclause shall be skipped for that frame. Otherwise, the dot dictionary contains a set of dots. Each dot has the following unsigned integer attributes:

- `x, y`: coordinates of the center of the ellipse times `5`
- `sigma_x, sigma_y`: quantized standard deviation
- `angle`: quantized angle
- `sign_intensity`: 3 bit integer representing the sign of the intensity on each channel; a zero means positive and one means negative
- `intensity`: array of 3 integers representing the quantized unsigned intensity
- `is_circle`: boolean flag indicating whereas the ellipse is a circle; the circle is a special case of an ellipse with `sigma_x = sigma_y` and `angle = 0`.

To decode a set of quantized dots from the bit stream, the decoder shall:

1) Initialise an arithmetic decoder from the code stream as specified in D.5
2) Decode the positions `{x, y}` as specified by C.4.4.2
3) Decode the rest of the parameters of the ellipse

Steps 1 and 2 are specified in previous subclauses. For the third step, the decoder shall decode first the histograms and subsequently the dot information. Histograms shall be decoded as follows, where `DecodeHist(nbins, sum)` is specified in D.5.7:

```
signHist = DecodeHist(8, n_points);
sigmaHist = DecodeHist(kSigmaQ, n_points);
sigmaDelta = DecodeHist(kSigmaQ, n_points);
for (j = 0; j < 3; j++) {
  intHist[j] = DecodeHist(kIntensityQ[j], n_points);
}
```

There is one histogram for `sigma_x` and `sigma_y`. The constants `kSigmaQ = 16`, `kAngleQ = 8` and `kIntensityQ = {10, 36, 10}` specify how many quantization levels are used for the standard deviation, angle and intensity per channel respectively. The angle shall be assumed to be uniformly distributed.

Each time an integer is decoded using a histogram, the histogram shall be updated by removing the decoded element. The following code describes the behaviour of the dot dictionary decoder. `GetHistInt(hist, U)` and `GetUniformInt(U)` obtain an integer in `[0, U)` distributed according to the histogram `hist` or uniformly, respectively, as specified in D.5.6 and D.5.7.

```
elem.sigma_x = GetHistInt(sigmaHist, kSigmaQ);
sigmaHist[elem.sigma_x] -= 1;

sDelta = GetHistInt(sigmaDelta, kSigmaQ);
elem.sigma_y = elem.sigma_x - sDelta;
sigmaDelta[sDelta] -= 1;

if (elem.sigma_x == elem.sigma_y) {
  elem.is_circle = 1;
```

```
  elem.angle = 0;
} else {
  elem.is_circle = 0;
  elem.angle = GetUniformInt(kAngleQ);
}

elem.sign_intensity = GetHistInt(signHist);
signHist[elem.sign_intensity] -= 1;

// Decode the intensity
for (j = 0; j < 3; j++) {
  elem.intensity[j] = GetHistInt(intHist[j]);
  intHist[j][elem.intensity[j]] -= 1;
}
```

The quantized dots shall subsequently be dequantized as specified in K.2.1.

## C.4.6 **Patches**

### C.4.6.1 Overview

If the `kPatches` flag in `frame_header` is not set, this subclause shall be skipped for that frame.
Otherwise, the patch dictionary contains a set of small image patches, to be added to the decoded image in specified locations.

The decoder shall read a set of `7` clustered distributions `D`, according to D.3.5, and shall then read the following values from a single ANS stream as specified in D.3.3.

Let ReadHybridVarLenUint(x) denote decoding a symbol `sym` from the ANS stream using the distribution D[x], and then calling DecodeHybridVarLenUint(`sym`) as defined in D.3.6.

### C.4.6.2 Decoding patch types

Each patch type has the following attributes:

- `xsize`, `ysize`: dimensions of the image patch; if either exceeds 16, the codestream is ill-formed.
- `sample(x, y, c)`: the sample from channel `c` at the position `(x, y)` within the patch
- `count`: the number of distinct positions
- `x`, `y`: arrays of `count` positions where this patch shall be added to the image

The decoder shall first set `num_patches` = ReadHybridVarLenUint(0), specified in D.3.6. For `i` in the range [0, `num_patches`) in ascending order, the decoder shall decode `patch[i]` as specified by the following:

```
patch[i].xsize = ReadHybridVarLenUint(1);
patch[i].ysize = ReadHybridVarLenUint(1);
if (i != 0) {
  patch[i].xsize = UnpackSigned(patch[i].xsize) + patch[i - 1].xsize;
  patch[i].ysize = UnpackSigned(patch[i].ysize) + patch[i - 1].ysize;
}

for (c = 0; c < 3; c++) {
  int ctx = 2 + c;
  for (iy = 0; iy < patch[i].ysize; iy++) {
```

```
    for (ix = 0; ix < patch[i].xsize; ix++) {
      pred = 0;
      if (ix != 0) {
        pred += patch[i].sample(ix - 1, iy, c);
      }
      if (iy != 0) {
        pred += patch[i].sample(ix, (iy - 1), c);
      }
      if (ix != 0 && iy != 0) pred = pred Idiv 2;
    }
    patch[i].sample(ix, iy, c) = UnpackSigned(ReadHybridVarLenUint(ctx)) + pred;
  }
}
```

#### C.4.6.3 Decoding patch positions

For `i` in the range [0, `num_patches`) in ascending order, the decoder shall read patch positions as specified below.

```
patch[i].count = ReadHybridVarLenUint(6) + 1;
for (j = 0; j < patch[i].count; j++) {
  patch[i].x[j] = ReadHybridVarLenUint(5);
  patch[i].y[j] = ReadHybridVarLenUint(5);
  if (j != 0) {
    patch[i].x[j] = UnpackSigned(patch[i].x[j]) + patch[i].x[j - 1];
    patch[i].y[j] = UnpackSigned(patch[i].y[j]) + patch[i].y[j - 1];
  }
  [[If the xsize × ysize region with top-left coordinates x, y, is not
  fully contained within the frame, the codestream is ill-formed]]
}
```

### C.4.7 **Splines**

#### C.4.7.1 Overview

Spline data is present in the codestream if and only if `kSplines` is set in `frame_header.flags`. If so, it shall consist of a quadtree stream followed by an ANS stream.

#### C.4.7.2 Quadtree stream

The first substream is a quadtree encoding (as described in C.4.4.2) of the first control point of each spline.

#### C.4.7.3 Splines stream

Directly following the quadtree stream is an ANS stream. The decoder shall first read five clustered distributions from it as described in D.3.5, and set `quantization_adjustment` = UnpackSigned(ReadHybridVarLenUint(0)) (C.4.6). The decoder shall then read the number of splines `num_splines` = ReadHybridVarLenUint(1) + 1. The following functions will be used below.

```
int DecodeVarLenUint(symbol) {
  return (1 << symbol) + u(symbol) - 1;
}

int DecodeVarLenInt(symbol) {
  return UnpackSigned(DecodeVarLenUint(symbol));
```

```
}

DecodeDoubleDelta(starting_value, delta[n]) {
  [[Append starting_value to the list of decoded values]]
  current_value = starting_value;
  current_delta = 0;
  for (i = 0; i < n; ++i) {
    current_delta += delta[i];
    current_value += current_delta;
    [[Append current_value to the list of decoded values]]
  }
}
```

Let Read32Ints() denote the sequence of integers obtained by repeating the following 32 times: reading one symbol from D[4] and transforming it into an integer using `DecodeVarLenInt`.

Then, for each of `num_splines` splines:

- the number of control points (including the starting point included in the quadtree) shall be obtained as ReadHybridVarLenUint(2) + 1.
- The result of applying double delta encoding to the control point coordinates (separately for *x* and *y*) and then dropping the very first control point shall be read as interleaved values (*x1*, *y1*, *x2*, *y2*…) via UnpackSigned(ReadHybridVarLenUint(3)). That is, for a starting point (sp_x, sp_y) obtained from the quadtree, DecodeDoubleDelta(sp_x, {x1, x2…}) and DecodeDoubleDelta(sp_y, {y1, y2}}) give the x and y coordinates of the spline's control points, respectively.
- A sequence of thirty-two integers representing the quantized DCT32 coefficients of the X channel of the spline along its arc length shall be obtained by Read32Ints().
- A sequence of thirty-two integers representing the quantized DCT32 coefficients of the Y channel of the spline along its arc length shall be obtained by Read32Ints().
- A sequence of thirty-two integers representing the quantized and partially decorrelated (see below) coefficients of the B channel of the spline along its arc length symbols shall be obtained by Read32Ints().
- A sequence of thirty-two integers representing the quantized DCT32 coefficients of the σ parameter of the spline (defining its apparent thickness) along its arc length shall be obtained by Read32Ints().

The order of the decoded starting points from the quadtree shall be the same as the order of the respective splines to which they belong in the ANS stream.

### C.4.7.4 Dequantization

After decoding, the DCT32 coefficients of the X, Y, B and σ values shall be divided by their respective predefined quantization weights and by `quantization_adjustment >= 0 ? 1 + quantization_adjustment / 8 : 1 / (1 + quantization_adjustment / 8)`. The quantization weights are distinct for each frequency band of each channel and their values are obtained using the following formula:

```
QuantizationWeight(channel, k) = 1 / (DCTScale(k) × kChannelWeight[channel]))
```

Where:

- `DCTScale` is defined in K.4;
- `channel` is in the range [0, 4), corresponding respectively to X, Y, B and σ;
- `kChannelWeight[4]` shall be `{0.025, 0.45, 0.42, 2}`.

### C.4.7.5 Decorrelation

Before quantization of B, but after that of Y, B is partially decorrelated from Y by subtracting dequantized Y × `kYToBRatio` (see K.2.1) from it. The decoder shall recorrelate B by adding that quantity back to it.

### C.4.8 Noise synthesis parameters

The decoder shall read noise parameters if and only if `kNoise` is set in `frame_header.flags`.
The 16 LUT values representing the noise synthesis parameters at different intensity levels shall be decoded sequentially from index 0 to 15. Each LUT value shall be decoded by reading a u(10) quantized value and dividing it by `(1 << 10)`.

### C.4.9 Custom opsin absorbance matrix

The decoder shall read a u(1) which, if `1`, indicates that the default inverse opsin absorbance matrix and opsin biases from L.2 shall be used. Otherwise, the decoder shall replace the default inverse opsin absorbance matrix with the 9 values read using F16() in raster order, and the default opsin biases with the 3 values read using F16() for R, G, and B, respectively.

## C.5 DcGroup

### C.5.1 Overview

In this subclause, `xsize` and `ysize` shall refer to the size of the current DC group, and all coordinates shall be relative to the top-left corner of the DC group.

The decoder shall read a u(2) that represents the `dc_preset` to use (controlling the `extra_levels_xyb` that will be used during decoding of DC coefficients, as specified in C.5.2). It shall then read `all_default_cmap` (referenced in C.5.3) as `u(1)`.

Each DC group shall be decoded as a single ANS stream, as specified in D.3. Thus, the decoder shall read `126` clustered distributions for the stream, as specified in D.3.5.

### C.5.2 DC coefficients

The decoder shall obtain three unsigned 8-bit channels with `ceil(ysize / 8)` rows and `ceil(xsize / 8)` columns, as well as a three auxiliary channels `DcQuantField`, as follows.

The decoder shall use the following values of `extra_levels_xyb` for each `dc_preset` value:

| `dc_preset` | `extra_levels_xyb` |
|---|---|
| `0` | `{0, 0, 0}` |
| `1` | `{0, 1, 1}` |
| `2` | `{1, 2, 1}` |
| `3` | `{2, 3, 2}` |

In this subclause, `GetContextForDC`, a slightly modified version of `GetContext` from E.2 shall be used:

```
uint8_t GetContextForDC(c) {
  if (nc == 0) {
    residual = (me + 1) >> 1;
    if (extra_levels_xyb[c] == 1) {
      if (residual > 2) residual = ((residual + 1) >> 2) + 2;
    }
```

```
    if (extra_levels_xyb[c] == 2) {
      if (residual > 3) {
        if (residual < 6) residual = 3;
        else residual = ((residual + 2) >> 3) + 3;
      }
    }
    if (extra_levels_xyb[c] == 3) {
      if (residual > 3) {
        if (residual < 6) residual = 3;
        else if (residual < 8) residual = 4;
        else residual = ((residual + 4) >> 4) + 4;
      }
    }
    return c × 12 + min(residual, 8) - 1;
  }
  return c × 12 + 8 + ceil(log2(9 - nc));
}
```

The decoder shall use the adaptive predictor specified in E.2, using `XBPredictor` as a sub-predictor for each of the 3 channels and `PackSigned` as an error metric.

During the execution of the predictor, for the samples (in order Y, then X, then B) of each pixel in raster order, the decoder shall read a single ANS symbol `sym`, using D[`GetContextForDC(c)`], and then use the procedure specified in D.3.6 to decode the `residual` value. The `delta` value shall then be obtained from the `residual` value, as specified by `PackSigned` in E.2.2.1. For each channel, the following steps shall then be performed:

- If `|delta|` is `0`, the value of `DcQuantField` corresponding to the current sample shall be set to the per-channel DC dequantization multiplier (as specified in F.2) divided by `1 << extra_levels_xyb[c]`, and `d` shall be 0.
- If `|delta|` is not `0` but less than or equal to `extra_levels_c`, then `d = copysign(1 << (|delta| - 1), delta)`, and the value of `DcQuantField` corresponding to the current sample shall be set to the per-channel DC dequantization multiplier (as specified in F.2) divided by `1 << (extra_levels_xyb[c] - |d| + 1)`.
- Otherwise, `d = copysign((delta - extra_levels_xyb[c]) << extra_levels_xyb[c], delta)`, and the value of `DcQuantField` corresponding to the current sample shall be set to the per-channel DC dequantization multiplier (as specified in F.2).

For each channel, the decoded value shall be `ev + d`, where `ev` is the *expected value* for channel `c` (see E.3).

### C.5.3 AC colour correlation

The decoder shall obtain two real channels (with `ceil(ysize/64)` rows and `ceil(xsize/64)` columns), denoted `XFromY` and `BFromY`, as follows.

If `all_default_cmap` is `1`, the decoder shall not read any symbol from the ANS stream, and shall fill both `XFromY` and `BFromY` with `128`.

Otherwise, the decoder shall use the adaptive predictor specified in E.2, using `YPredictor` as a sub-predictor for each of the two channels and `PackSignedRange` with `min = 0` and `max = 255` as an error metric. Channel `0` shall correspond to the `XFromY` values, and channel `1` to the `BFromY` values.

During the execution of the predictor, for each sample and each channel, the decoder shall read a single ANS symbol `sym`, using D[`36 + GetContext(c)`], and then use the procedure specified in D.3.6 to decode the `residual` value. It shall then compute the *decoded value* for that sample as specified by `PackSigned`, except that `ev` shall be `128` if `x == 0 && y == 0`, and shall set the corresponding value in the colour correlation channels.

### C.5.4 DctSelect

The decoder shall obtain an unsigned 8-bit channel with `ceil(ysize / 8)` rows and `ceil(xsize / 8)` columns, denoted `DctSelect`, as follows.

The decoder shall use the adaptive predictor specified in E.2, using `YPredictor` as a sub-predictor for the only channel and `PackSigned` as an error metric.

During the execution of the predictor, for each sample, the decoder shall read a single ANS symbol `sym`, using D[`60 + ev × 2 + (nc > 5 ? 1 : 0)`], where `nc` is defined in E.2. If `sym` does not correspond to a valid varblock type, or if the corresponding varblock is not naturally aligned, or covers $8 \times 8$ regions that are out-of-bounds (the `DctSelect` sample is stored at the coordinates of the top-left $8 \times 8$ region of the varblock), the codestream is ill-formed. Otherwise, the decoder shall produce `sym` as a decoded value and store it in the `DctSelect` field. The decoder shall also skip this procedure for the non-top-left blocks that are covered by the decoded varblock.

The mapping between transform types (defined in I.2) and numerical values shall be the following:

| Transform type | Numerical value |
|---|---|
| DCT8×8 | 0 |
| IDENTITY | 1 |
| DCT2×2 | 2 |
| DCT4×4 | 3 |
| DCT16×16 | 4 |
| DCT32×32 | 5 |
| DCT16×8 | 6 |
| DCT8×16 | 7 |
| DCT32×8 | 8 |
| DCT8×32 | 9 |
| DCT32×16 | 10 |
| DCT16×32 | 11 |
| DCT4×8 | 12 |
| DCT8×4 | 13 |
| AFV | 14 |

### C.5.5 Quantization field

The decoder shall obtain a real channel with `ceil(ysize / 8)` rows and `ceil(xsize / 8)` columns, denoted `AcMul`, as follows.

The decoder shall use the adaptive predictor specified in E.2, using `XBPredictor` as a sub-predictor for the only channel and `PackSignedRange` with `min = 0` and `max = 255` as an error metric.

During the execution of the predictor, for each sample that corresponds to a top-left block in a varblock, the decoder shall read a single ANS symbol `sym`, using D[`90 + GetContext(0)`], and then use the procedure specified in D.3.6 to decode the `residual` value. The decoder shall then compute the *decoded value* for that block, as well as all the other blocks covered by the varblock, as specified by the error metric. The decoded value shall be clamped to the `[0, 256)` range. Finally, the values in the `AcMul` field shall be set to `(1 << 16) / ((global_scale_minus_1 + 1) × (1 + dv))`, where `dv` is the *decoded value* for the corresponding block and `global_scale_minus_1` is the value read in C.4.2.

### C.5.6 Adaptive reconstruction parameters

The decoder shall unconditionally obtain an unsigned 8-bit channel with `ceil(ysize / 8)` rows and `ceil(xsize / 8)` columns, denoted `Sharpness` as follows.

The decoder shall use the adaptive predictor specified in E.2, using `YPredictor` as a sub-predictor for the only channel and `PackSigned` as an error metric.

During the execution of the predictor, for each sample that corresponds to a top-left block in a varblock, the decoder shall read an ANS symbol `sym`, using D[`102 + ev × 2 + (nc > 5 ? 1 : 0)`]. The *decoded value* for this block, as well as the corresponding `Sharpness` value, shall then be `sym`. The decoder shall then read a symbol `s` for each other block in the varblock, using D[`118 + sym`], in raster order. `s` shall be both the *decoded value* and the `Sharpness` value for those blocks.

## C.6 AcGlobal

### C.6.1 Dequantization matrices

The dequantization matrices shall be used as multipliers for the AC coefficients, as specified in F.3. They are defined by the channel, the transform type and the index of the coefficient inside the varblock. The parameters that define the dequantization matrices shall be read from the stream as follows.

First, a u(1) shall be read. If this is 0, all matrices have their default encoding as specified in C.6.1.1. Otherwise, each matrix for the 15 types of quantization shall be decoded individually in order of ascending index. Each parameter in this subclause shall be read using F16() from A.3.1.

For each matrix, the *encoding mode* shall be read as a u(3). There are 8 possible encoding modes: `Library (0)`, `Identity (1)`, `DCT2 (2)`, `DCT4 (3)`, `DCT4x8 (4)`, `AFV (5)`, `DCT (6)`, and `Copy (7)`. The `Identity`, `DCT2`, `DCT4`, `DCT4x8` and `AFV` modes are only valid for defining matrices that contain 64 coefficients.

If the mode is `Library`, the matrix encoding shall be the default encoding from C.6.1.1. If the mode is `Identity`, `3 × 3` parameters shall be read in raster order. If the mode is `DCT2`, the `3 × 6` parameters shall be read in raster order. If the mode is `DCT4`, the `3 × 2` parameters shall be read in raster order and further DCT parameters shall be read as described below. If the mode is `DCT`, only the DCT parameters shall be read, as described below. If the mode is `Copy`, the next `ceil(log2(current_index))` bits shall specify the index of the previously-decoded matrix encoding that shall be used for this matrix, where `current_index` is the 0-based index of the matrix currently being decoded. If the mode is `DCT4x8`, the decoder shall read three parameters (for X, Y, B channels, respectively) of `dct4x8multipliers`, and further parameters as specified below.
If the mode is `AFV`, the decoder shall read, for each of the three channels `c`, in order:

- Two parameters `dct4multipliers[c][0]` and `dct4multipliers[c][1]`.
- Eight parameters `afv_weights[c][0]` to `afv_weights[c][7]`

The DCT parameters shall be specified by `num_bands` distance and eccentricity bands. Reading these parameters shall be performed as specified below. First, a u(4) shall be read from the stream, representing `(num_bands - 2)`. Then, for each channel, `num_bands` parameters shall be read; these represent the *distance bands* `db` for the transform, and control how the matrix changes with the distance from coefficient `(0, 0)`. Similarly, next, a `u(3)` shall be read from the stream, representing `(num_ecc_bands - 1)`, and then, for each channel, `num_ecc_bands` parameters; these represent the *eccentricity bands* `eb`, that control how the matrix changes with the distance from the diagonal.

Once the above parameters are read, the dequantization matrices shall be defined by the element-by-element ratio of a `numerator` matrix, that only depends on the transform type, and a `weights` matrix, that only depends on the encoding parameters. The `numerators` matrix is identical for all channels, whereas the `weights` matrix is computed per channel.

The number of coefficients `num_coef` present in the final dequantization matrix shall be `64` for `DCT2`, `DCT4`, `DCT4x8`, `AFV` and the `Identity` transforms (defined in I.2), and `X × Y` for a transform $\mathrm{DCT}_{Y \times X}$ with `X` and `Y` in `{8, 16, 32}`.

The `numerators` matrix, containing `num_coef` elements, shall be computed given the DCT scaling factors defined by function `I` in I.2.6.1. For AFV, $\mathrm{DCT}_{2 \times 2}$ and the `Identity` transform, `numerators` shall be filled with `1`. Otherwise, it shall be computed as follows, with `dct4factor` set to `0.5` if the corresponding DCT dimension is `4`, and `1` otherwise (if the corresponding DCT dimension is one of `{8, 16, 32}`):

```
numerators[i] = I(X, i × dct4factor Umod Y] ×
                I(Y, i × dct4factor Idiv Y] / (X × Y)
```

Next, a `weights` matrix of dimensions `X × Y` shall be computed for each channel using the following procedure.

First, a `distance_bands` array, of length `num_bands`, shall be obtained from the encoding's distance bands `db`:

```
distance_bands[0] = db[0];
for (i = 1; i < num_bands; i++) {
  distance_factor = db[i] > 0 ? (db[i] + 1) : (1 / (1 - db[i]));
  distance_bands[i + 1] = distance_factor × distance_bands[i];
}
```

Similarly, an `eccentricity_bands` array, of length `num_ecc_bands + 1`, shall be obtained from the encoding's eccentricity bands `eb`:

```
eccentricity_bands[0] = 1;
for (i = 0; i < num_ecc_bands; i++) {
  eccentricity_factor = eb[i] > 0 ? (eb[i] + 1) : (1 / (1 - eb[i]));
  eccentricity_bands[i + 1] = eccentricity_factor × eccentricity_bands[i];
}
```

For a coefficient in position `(x, y)`, the distance and eccentricity values shall be computed as:

```
distance(x, y) = sqrt((x / X)² + (y / Y)²);
```

```
eccentricity(x, y) = |x / X - y / Y| / distance(x, y);
```

The previously-computed bands shall be used as interpolation points to compute the distance and eccentricity multipliers. Given a raw value `pos` of either distance or eccentricity, a scaled position shall be computed and a geometric interpolation shall be performed (with `max = sqrt(2)` for distance values, and `max = 1` for eccentricity values, and where `len` is the length of the corresponding `distance_bands` or `eccentricity_bands` array):

```
interpolate(pos, max, bands, len) {
  scaled_pos = pos × (len - 1) / max;
  scaled_index = floor(scaled_pos);
  frac_index = scaled_pos - scaled_index;
  A = bands[scaled_index];
  B = bands[scaled_index + 1];
  interpolated_value = A × (B / A)^frac_index;
  return interpolated_value;
}
```

The resulting entry in the matrix shall be the product of the distance and eccentricity multipliers.

For encoding mode `DCT`, the `weights` matrix for $\mathrm{DCT}_{X \times Y}$ is the matrix of size `X × Y` computed by the procedure above.

For encoding mode `DCT4`, the `weights` matrix shall be obtained by copying into position `(x, y)` the value in position `(x Idiv 2, y Idiv 2)` in the `4 x 4` matrix computed by the above procedure; coefficients `(0, 1)` and `(1, 0)` shall be divided by the per-channel parameter of index `0`, and the `(1, 1)` coefficient shall be divided by the per-channel parameter of index `1`.

For encoding mode `DCT2`, per-channel parameters shall be copied in position `(x, y)` as follows:

- `0`: positions `(0, 1)`, `(1, 0)`
- `1`: position `(1, 1)`
- `2`: all positions in rectangle from `(2, 0)` to `(4, 2)`, and symmetric
- `3`: all positions in rectangle from `(2, 2)` to `(4, 4)`
- `4`: all positions in rectangle from `(4, 0)` to `(8, 4)`, and symmetric
- `5`: all positions in rectangle from `(4, 4)` to `(8, 8)`

For encoding mode `Identity`, coefficient `(1, 1)` shall be equal to the per-channel parameter of index `2`. Coefficients `(0, 1)` and `(1, 0)` shall be equal to the per-channel parameter of index `1`, and all other coefficients to the per-channel parameter of index `0`. Coefficient `(0, 0)` is invalid and shall not be used by the decoder.

For encoding mode `DCT4x8`, the `weights` matrix shall be obtained by copying into position `(x, y)` the value in position `(x, y Idiv 2)` in the `8 × 4` matrix computed by the above procedure; coefficient `(0, 1)` shall be divided by the per-channel parameter.

If encoding mode `DCT4x8` is used for the `DCT8x4` transform, the decoder shall obtain the `weights` matrix as in the `DCT4x8` mode, and then transpose it, per channel.

For encoding mode `AFV`, the decoder shall obtain the `weights` according to the following algorithm, per channel.

```
set_weight(weights, x, y, val) {
    weights(2 × x, 2 × y) = val;
```

```
    weights(2 × x, 2 × y + 1) = val;
    weights(2 × x + 1, 2 × y) = val;
    weights(2 × x + 1, 2 × y + 1) = val;
}
kFreqs[] = {
    0,
    0,
    0.8517778890324296,
    5.37778436506804,
    0,
    0,
    4.734747904497923,
    5.449245381693219,
    1.6598270267479331,
    4,
    7.275749096817861,
    10.423227632456525,
    2.662932286148962,
    7.630657783650829,
    8.962388608184032,
    12.97166202570235,
};
lo = 0.8517778890324296;
hi = 12.1198851366699204;
bands[4] = { afv_weights[4] };

for (i = 1; i < 4; i++)
  bands[i] = bands[i - 1] × (
      afv_weights[i + 4] > 0 ?
        (afv_weights[i + 1] + 1) :
        (1 / (1 - afv_weights[i + 1]));

set_weight(0, 0, afv_weights[0]);
set_weight(0, 1, afv_weights[1]);
set_weight(1, 0, afv_weights[2]);
set_weight(1, 1, afv_weights[3]);

for (y = 0; y < 4; y++) {
  for (x = 0; x < 4; x++) {
    if (x < 2 && y < 2) continue;
    val = interpolate(kFreqs[y × 4 + x] - lo, hi, bands, 4);
    set_weight(x, y, val);
  }
}

weights(1, 0) ×= dct4multipliers[0];
weights(0, 1) ×= dct4multipliers[0];
```

```
weights(1, 1) *= dct4multipliers[1];
```

Finally, each element at position `(x, y)` in the final dequantization matrices for channel `c` shall be computed as:

```
val(x, y) = numerators(x, y) / weights(x, y)
```

If any of the resulting values is non-positive or infinity, the codestream is ill-formed.

#### C.6.1.1 Default values for each dequantization matrix

The dequantization matrix encodings defined here shall be used as the default encodings for each kind of dequantization matrix.
For $DCT_{8x8}$, there shall be the following `6` distance bands `{{5.0, 0.0, -0.4, -0.4, -0.4, -2.0}, { 1.1, 0.0, -0.3, -0.3, -0.3, -0.3}, { 1.0, -3.0, 0.0, 0.0, -1.0, -2.0}}`. There shall be 3 eccentricity bands (all `0`).
For `Identity`, the default per-channel weights shall be:

```
{{221.50801256717435, 7087.151589594002, 4466.9924939259863},
 {29.535505378287006, 1454.6663024982124, 1368.041230099175},
 {9.2887405354132238, 23.982311358829531, 11.201853748207084}}
```

For `DCT2`, the default per-channel weights shall be:

```
{3840.0, 2560.0, 1280.0, 640.0, 480.0, 300.0},
{960.0, 640.0, 320.0, 180.0, 140.0, 120.0},
{640.0, 320.0, 128.0, 64.0, 32.0, 16.0}
```

For `DCT4`, there shall be the following `4` distance bands `{{32.0, -0.4, 0.0, 0.0}, {6.0, 0.0, 0.0, -0.0}, {4.0, -5.0, -1.0, -1.0}}`, `2` eccentricity bands (all `0`), and the following per-channel parameters: `{{1.0, 1.0}, {0.7, 0.7}, {1.0, 1.0}}`.

For $DCT_{16x16}$, there shall be the following `7` distance bands `{{3.1965020925735868, -1.8000777393353804, -0.39424529824571225, -0.339093774457103443, -0.6350101832695744, -0.90177264050827612, -1.5162099239887414}, {0.77917081615440487, -0.57424582104194355, -0.90745813428471001, -0.64925837484843441, -0.35865440981033403, -0.21322389111877305, -0.27615025315725483}, {0.22259376988644336, -3.0531423165804414, -0.2, -0.30687130033378396, -0.42708730624733904, -1.4856834539296244, -5.9209142884401604}}`, and 3 eccentricity bands (all `0`).

For $DCT_{32x32}$, there shall be the following `8` distance bands `{{0.47968775359573942, -1.025, -0.98, -0.9012, -0.4, -0.48819395464, -0.421064, -0.27}, {0.22295421389982905, -0.8041958212306401, -0.7633036457487539, -0.55660379990111464, -0.49785304658857626, -0.43699592683512467, -0.40180866526242109, -0.27321683125358037}, {0.11607457694128877, -3.060733579805728, -2.0413270132490346, -2.0235650159727417,

-0.5495389509954993, -0.4,
-0.4, -0.3}}, and 4 eccentricity bands (all 0).

For $DCT_{16x8}$ and $DCT_{8x16}$, there shall be the following 7 distance bands {{5.0, -0.7, -0.7, -0.2, -0.2, -0.2, -0.5}, {1.0, -0.5, -0.5, -0.5, -0.2, -0.2, -0.2}, {0.35, -1.4, -0.2, -0.5, -0.5, -1.5, -3.6}} and 3 eccentricity bands (all 0).

For $DCT_{32x8}$ and $DCT_{8x32}$, there shall be the following 8 distance bands
{{0.90897201669289496, -2.523575079652864,
1.5606102630460328, -0.6828050717528521,
-14.770016895841904, -31.365240527393656,
0.5659292541666433, 26.789646874558219},
{0.3603042187746216, -0.11724474917497789,
-0.85426302204085935, -0.5404181207241352,
-0.45637619443533173, -0.82707923060149569,
-1.09760551067614, -54.16057009483842},
{0.25176400522276937, -10.060949304689093,
-4.5316627381323995, -5.4569229849460807,
-10.549754675878864, 1.7512906265092971,
-1.6854706115725824, 3.3092762331901451}}
and the following 4 eccentricity bands {{-0.40751133465833289, 2.2055690721475925,
-1.4434547377534228, 2.0590289434445102},
{0.12452072219659419, 0.065578935001406033,
-0.14598888885596958, 0.12114139909222557},
{0.0586589025020419, -2.2654958780302183,
-4.6996221931893647, 1.8254498073137184}}.

For $DCT_{32x16}$ and $DCT_{16x32}$, there shall be the following 8 distance bands {{1.1280371559733287, -0.98868, -0.658, -0.42026, -0.22712, -0.2206, -0.226, -0.2},
{0.38961075971152237, -0.62525308982767058,
-0.83770786552491361, -0.79014862079498627,
-0.2692727459704829, -0.38272769465388551,
-0.22924222653091453, -0.20719098826199578},
{0.098, -2.41, -0.23, -0.2, -0.4, -0.8, -0.4, -0.5}} and 4 eccentricity bands, all 0.

For $DCT_{4x8}$ and $DCT_{8x4}$, there shall be the following 4 distance bands
{{17.142628543359718, -0.16269623020744692,
-0.31194253026666783, -0.2151140670773547},
{4.6225628588812384, -0.82630200888366945,
-1.0675229603596517, -0.17845290869168118},
0.81188546639482984, -0.6594385811273854,
-0.050082094097871593, -2.6843722511996204}} and 2 eccentricity bands, all 0. Per-channel multipliers shall be {{1.0}, {1.0}, {1.0}}.

For AFV mode the default afv_weights shall be
{{6368.834817724610816928, 232.1096058872860158351,
415.6635474797509090044, 220.0720486370546864556,
329.8189627387656012564, -0.9919974139384578,
0, -0.60},
{1608.323575150649711025, 67.2724358236086,
89.06657498438612, 64.07248898359674,
125.6541247697168638611, -2.2833978673775697,
-0.14, -0.2485},
{195.9278285747304246438, 21.841306388764952,

18.5279466567289584105, 12.573868715940659,
15.89916083965558, -0.41,
-0.1608, 0}}, with the following per-channel multipliers: {{1.0, 1.0}, {1.0, 1.0}, {1.0, 1.0}}.

### C.6.2 Number of AC decoding presets

The decoder shall read num_ac_presets_minus_1 as u(ceil(log2(num_groups - 1))). The number of coefficient orders and histograms in each pass num_ac_presets shall be num_ac_presets_minus_1 + 1.

NOTE The number of bits used to represent this number is not a constant because the set of histograms and orders to be used is decided on a per-group basis.

### C.6.3 Lossless mode

If the frame_header.encoding is kLossless, the decoder shall proceed as specified in O.2.

## C.7 AcPass

### C.7.1 AC coefficient order

The data described here shall be read num_ac_presets times, once for each preset in ascending order.

The decoder shall first read a 10-bit bitmask used_orders using U32(Val(0x337), Val(0x33F), Val(0), Bits(10)). If used_orders is not 0, the decoder shall read a set of 8 clustered distributions D, according to subclause D.3.5. The decoder shall then repeat the following procedure for each value b in [0, 10) for which (used_orders & (1 << b)) != 0, in ascending order of b.

Let dcts be any DctSelect value whose order corresponds to the order that is currently being decoded, as defined in I.2.5. The decoder shall decode a permutation nat_ord_perm from a *single* ANS stream (containing permutations for all the orders) as specified in C.3.2, with size the number of coefficients covered by dcts and skip = size / 64.

Finally, the decoder shall compute order[i] as natural_coeff_order[nat_ord_perm[i]], where natural_coeff_order is the natural coefficient order defined by dcts, as specified in I.2.5.

If (used_orders & (1 << b)) == 0, the decoder shall instead define order[i] as natural_coeff_order[i], for each i.

### C.7.2 AC coefficient histograms

The decoder shall read an array of num_ac_presets histograms, concatenated in increasing index order, once for each preset. For each preset, the decoder shall read a set of 3780 distributions D from the bitstream as specified in D.3.5.

## C.8 PassGroup

### C.8.1 Overview

In all subsections of C.8, xsize and ysize shall refer to the size of the current group (at most kGroupDim × kGroupDim), and all coordinates shall be relative to the top-left corner of the group.

### C.8.2 Alpha

This part shall only be present in the first *pass*, and if `metadata.alpha_bits` != 0 and `frame_header.encoding != kModularGroup`. The alpha channel data corresponding to the group is encoded using the modular image sub-bitstream (C.9), where the number of channels is 1 and dimensions correspond to `xsize` × `ysize`.

### C.8.3 Depth

This part shall only be present in the first *pass*, and if `metadata.m2.depth_bits` != 0 and `frame_header.encoding != kModularGroup`. The depth channel data corresponding to the group is encoded using the modular image sub-bitstream (C.9), where the number of channels is 1 and dimensions correspond to $\mathtt{ceil(xsize\ /\ 2^{metadata.m2.depth_shift})} \times \mathtt{ceil(ysize\ /\ 2^{metadata.m2.depth_shift})}$.

### C.8.4 Extra channels

This part shall only be present in the first *pass*, and if `metadata.m2.num_extra_channels` is nonzero and `frame_header.encoding != kModularGroup`.
The decoder shall follow the procedure described in the rest of this subclause to decode the area of each extra channel corresponding to the current group, independently. Extra channels appear in the bitstream in increasing order of index. They are encoded using the modular image sub-bitstream (C.9), where the number of channels is 1 and dimensions correspond to `xsize` × `ysize`.

### C.8.5 AC coefficients

A `u(ceil(log2(num_ac_presets)))` shall indicate the histogram and coefficient order to be used for this group. These are chosen from the `num_ac_presets` possibilities of this *pass*.

To compute context values for the number of nonzeros in a varblock and for each coefficient, the decoder shall use the following procedures (where `c` is the current channel, and `s` is the value of the `DctSelect` field for the current varblock):

```
BlockContext(c, s) {
  if (s == 0) return 10 × c;
  if (s < 4) return 10 × c + 1;
  return 10 × c + s - 2;
}
```

```
GetNonZerosContext(c, s, predicted, num_blocks) {
  return BlockContext(c, s) + 30 × ceil(log2(num_blocks × predicted + 1));
}
```

```
CoeffFreqContext[64] = {
  0,  1,  2,  3,  4,  4,  5,  5,  6,  6,  7,  7,  8,  8,  8,  8,
  9,  9,  9,  9,  10, 10, 10, 10, 10, 10, 10, 10, 11, 11, 11, 11,
  11, 11, 11, 11, 12, 12, 12, 12, 12, 12, 12, 12, 13, 13, 13, 13,
  13, 13, 13, 13, 14, 14, 14, 14, 14, 14, 14, 14, 14, 15, 15, 15,
};
```

```
CoeffNumNonzeroContext[63] = {
            0,  16,  32,  32,  47,  47,  47,  47,  62,  62,  62,  62,
    62,     62,  77,  77,  77,  77,  77,  77,  77,  77,  91,  91,  91,
    91,     91,  91,  91,  91,  91,  91,  91,  91,  91,  91,  91, 104,
    104,   104, 104, 104, 104, 104, 104, 104, 104, 104, 104, 104, 104,
    104,   104, 104, 104, 104, 104, 104, 104, 104, 104, 104, 104,
};

GetCoefficientContext(c, s, k, non_zeros, num_blocks, size) {
 non_zeros = (non_zeros + num_blocks - 1) Idiv num_blocks;
 k = k Idiv num_blocks;
 return CoeffNumNonzeroContext[non_zeros - 1] + CoeffFreqContext[k] +
   BlockContext(c, s) × 115 + 330;
}
```

For each block of the group, `PredictedNonZeros(x, y)` shall be the following function, where `NonZeros` is defined below:

- if `x == y == 0`, `32`.
- if `x == 0 && y != 0`, `NonZeros(x, y - 1)`.
- if `x != 0 && y == 0`, `NonZeros(x - 1, y)`.
- if `x != 0 && y != 0`, `(NonZeros(x, y - 1) + NonZeros(x - 1, y) + 1) >> 1`.

After selecting the histogram and coefficient order, the decoder shall read symbols from an ANS stream, as specified in D.3.3. The decoder shall proceed by decoding varblocks of channels X, Y, then B in raster order. For the purposes of this ordering, each varblock corresponds to its top-left block. For each varblock of size `X × Y` in the image, with a `DctSelect` value of `s`, covering `num_blocks = X × Y / 64` blocks, and of total size `size = X × Y`, the decoder shall first read a symbol `sym` using D[`GetNonZerosContext(c, s, PredictedNonZeros(x, y), num_blocks)`]. The decoder shall then compute `non_zeros = (1 << sym) + u(sym)`.

The decoder shall then compute the `NonZeros(x, y)` field for each block in the current varblock as follows. Let `i` be the difference between `x` and the x coordinate of the top-left block of the current varblock, and `j` the difference between `y` and the y coordinate of the top-left block of the current varblock. Then let `cur = j × X + i`. `NonZeros(x, y)` shall then be

```
((cur + 1) × non_zeros) Idiv num_blocks - (cur × non_zeros) Idiv num_blocks
```

Finally, for `k` in the range [`num_blocks`, `size`), the decoder shall read a symbol `sym` from the bitstream, using D[`GetCoefficientContext(c, s, k, non_zeros, num_blocks, size)`], and then use the procedure specified in D.3.6 to decode the `ucoeff` value. The decoder shall then set the quantized AC coefficient in the position corresponding to index `k` in the coefficient order for the current `DctSelect` value to UnpackSigned(`ucoeff`). If `ucoeff != 0`, the decoder shall decrease `non_zeros` by `1`. If `non_zeros` reaches `0`, the decoder shall stop decoding further coefficients for the current block.

If this is not the first *pass*, the decoder shall add the decoded AC coefficients to the previously-decoded ones.

### C.8.6 Lossless group data

If the `frame_header.encoding` is `kLossless`, the decoder shall proceed as specified in Annex O.

### C.8.7 Modular group data

This part shall only be present if `frame_header.encoding == kModularGroup`. The pixel data corresponding to the group is encoded using the modular image sub-bitstream (C.9), where the dimensions correspond to `xsize` × `ysize`.

## C.9 Modular image sub-bitstream

### C.9.1 Overview

This Annex describes the modular image sub-bitstream, which encodes the pixels of a frame or group if `frame_header.encoding == kModular` or `kModularGroup`, respectively. It is also used to encode additional channels (alpha, depth and extra channels) in the other frame encodings.

The modular image sub-bitstream encodes an arbitrary number N of channels. The dimensions are implicit (i.e. are already known at decode time). We denote the channels as a vector: `channel[0], channel[1], ... , channel[N - 1]`. Each channel has a width and height (`channel[i].width, channel[i].height`); initially (before transformations are taken into account), the dimensions of all channels (with the possible exception of the Depth channel, if present) are identical. Channels also have a horizontal and vertical subsampling factor, denoted by `channel[i].hshift` and `channel[i].vshift`, which are initially set to zero (except for the Depth channel, if present, where both are initialized to `metadata.m2.depth_shift`). Finally, channels have a quantization factor, denoted by `channel[i].q`, which is only relevant if the Quantization transformation is used. `channel[i](x, y)` denotes the value of the sample in column `x` and row `y` of channel number `i`.

Unless otherwise specified, the number of channels is equal to `(metadata.colour_encoding.colour_space == kGrey ? 1 : 3) + (alpha_bits > 0 ? 1 : 0) + (depth_bits > 0 ? 1 : 0) + num_extra_channels` and the dimensions correspond to `xsize` × `ysize`, except for the depth channel which has dimensions $\text{ceil}(\text{xsize} / 2^{\text{metadata.m2.depth_shift}}) \times \text{ceil}(\text{ysize} / 2^{\text{metadata.m2.depth_shift}})$. The channel order is as follows: Grey or (Red, Green, Blue), followed by Alpha (if `alpha_bits > 0`), followed by Depth (if `depth_bits > 0`) and finally the extra channels (if any) in ascending order of index.

Transformations can be applied to the array of channels, which can result in changes to the number of channels and their dimensions. If the dimensions correspond to a power-of-two subsampling of the original image dimensions, this is denoted by `hshift` and `vshift`, where the horizontal subsampling factor is $2^{\text{hshift}}$ and the vertical subsampling factor is $2^{\text{vshift}}$. If the dimensions are not related to the original dimensions, the `hshift` and `vshift` values are set to -1. The sub-bitstream starts with an encoding of the series of transformations that was applied, so the decoder can anticipate the corresponding changes in the number of channels and their dimensions (so channel numbers and dimensions do not need to be explicitly signalled) and can afterwards apply the appropriate inverse transformations.
The first `nb_meta_channels` channels are used to store information related to transformations that require extra information (for example a colour palette). Initially, `nb_meta_channels` is set to zero, but transformations can increment this value.

### C.9.2 Image decoding

The following header information shall be decoded:

| condition | type | value |
|---|---|---|
| | Varint() | max_extra_properties |
| | Varint() | nb_transforms |
| | TransformInfo | transform[nb_transforms] |

The array of channels shall be initialized according to the previously specified number of channels. Their dimensions and subsampling factors shall be derived from and the series of transforms and their parameters (see C.9.4).

The decoder shall then decode channel data:

```
for (i = skipchannels; i < channel.size(); i++) {
  if (channel[i].width == 0 || channel[i].height == 0) continue;
  [[channel decoding as described in C.9.3]]
}
```

Finally the series of inverse transformations shall be applied as described in C.9.4.

#### C.9.2.1 TransformInfo

| condition | type | value |
|---|---|---|
| | Varint() | (nb_parameters << 4) + tr_id |
| | Varint() | parameter[nb_parameters] |

### C.9.3 Channel decoding

The encoding of a single channel starts with the following channel header:

| condition | type | default | value |
|---|---|---|---|
| | Varint() | | (predictor << 2) + entropy_coder |
| | Varint() | | 1 - channel[i].min |
| channel[i].min == 1 | Varint() | | channel[i].min - 1 |
| | Varint() | | channel[i].max - channel[i].min |
| transform contains Quantize or AdaptiveQuantize and !(channel[i].min == 0 && channel[i].max == 0) | Varint() | 1 | channel[i].q |

NOTE: the channel minimum value channel[i].min shall be signalled in a single Varint() if it is negative or zero (channel[i].min <= 0), while if it is strictly positive then first a zero is signalled (which implies channel[i].min == 1) and then the actual channel minimum value minus one is signalled.

If `channel[i].min == channel[i].max`, then the decoder shall set all pixels in the channel to this value, and immediately continues to decode the next channel.

Otherwise, the decoder shall initialize a range encoder and decode a MA tree as described in D.7.3. If `entropy_coder == 0`, then signal_initialization = true; otherwise signal_initialization = false;

Then the actual channel data shall be decoded as follows. If `entropy_coder == 0`, MABEGABRAC is used. If `entropy_coder == 1`, MABrotli is used. If `entropy_coder == 2`, MAANS is used.

### C.9.3.1 MABEGABRAC channel decoding

```
ch_zero = max(channel[i].min, min(channel[i].max, 0));
for (y = 0; y < channel[i].height; y++) {
  for (x = 0; x < channel[i].width; x++) {
    [[properties = (see D.7.2)]]
    k = MA(properties);
    top = (y > 0 ? channel[i](x, y - 1) : ch_zero);
    left = (x > 0 ? channel[i](x - 1, y) : ch_zero);
    topleft = (x > 0 && y > 0 ? channel[i](x - 1, y - 1) : left);
    if (predictor == 0) predicted = ch_zero;
    if (predictor == 1) predicted = (left + top) Idiv 2;
    if (predictor == 2) predicted = median(left + top - topleft, left, top);
    if (predictor == 3) predicted = left;
    if (predictor == 4) predicted = top;
    diff = BEGABRAC_k(channel[i].min - predicted, channel[i].max - predicted);
    channel[i](x, y) = predicted + diff;
  }
}
```

### C.9.3.2 MABrotli and MAANS channel decoding

First, the number of bytes per pixel and the pixel encoding is determined as follows:

```
maxval = channel[i].max - channel[i].min;
if (predictor == 0 && (channel[i].max <= 0 || channel[i].min >= 0)) {
  signed = false;
} else {
  signed = true;
  maxval *= 2;
}
bytes_per_pixel = 1;
if (maxval > 0xff) bytes_per_pixel++;
if (maxval > 0xffff) bytes_per_pixel++;
if (maxval > 0xffffff) bytes_per_pixel++;
```

In the case of MABrotli, a single Brotli stream is then decoded. The decoded stream is split as follows: for each leaf node in the MA tree (in depth-first order), bytes_per_pixel substreams are obtained by repeatedly reading a Varint() to obtain the substream length and taking that many bytes from the decoded stream.

In the case of MAANS, first all the substream lengths are decoded as follows: for each leaf node in the MA tree (in depth-first order), bytes_per_pixel times a Varint() is read.
The decoded stream corresponding to MA leaf node *k* and byte number *b* is denoted with $\texttt{stream}_{k,b}$.
The channel data is then reconstructed as follows:

```
ch_zero = max(channel[i].min, min(channel[i].max, 0));
for (y = 0; y < channel[i].height; y++) {
  for (x = 0; x < channel[i].width; x++) {
    [[properties = (see D.7.2)]]
    k = MA(properties);
    top = (y > 0 ? channel[i](x, y - 1) : ch_zero);
    left = (x > 0 ? channel[i](x - 1, y) : ch_zero);
    topleft = (x > 0 && y > 0 ? channel[i](x - 1, y - 1) : left);
    if (predictor == 0) predicted = ch_zero;
    if (predictor == 1) predicted = (left + top) Idiv 2;
    if (predictor == 2) predicted = median(left + top - topleft, left, top);
    if (predictor == 3) predicted = left;
    if (predictor == 4) predicted = top;
    diff = stream_k,0.next_byte();
    if (bytes_per_pixel > 1) diff += stream_k,1.next_byte() << 8;
    if (bytes_per_pixel > 2) diff += stream_k,2.next_byte() << 16;
    if (bytes_per_pixel > 3) diff += stream_k,3.next_byte() << 24;
    if (signed) {
      if (diff & 1) diff = (diff + 1) >> 1;
      else diff = - (diff >> 1);
      channel[i](x, y) = diff + predicted;
    } else {
      channel[i](x, y) = diff + channel[i].min;
    }
  }
}
```

### C.9.4 Transformations

The following table lists the transformations that can be applied:

| `tr_id` | Name | `nb_meta_channels` | Summary of channels/dimensions impact | Defined in section |
|---|---|---|---|---|
| 0 | YCbCr | no change | no change | L.3 |
| 1 | YCoCg | no change | no change | L.4 |
| 2 | SubtractGreen | no change | no change | L.5 |
| 3 | Palette | +1 | channels `parameter[0] + 1` until `parameter[1]` are removed; one meta-channel is inserted in the beginning | L.6 |
| 4 | Subsample | no change | number of channels does not change; depending on parameters, the width and/or height of one or more channels is divided by a power of two and the hshift/vshift is incremented accordingly | C.9.4.1 |
| 5 | Squeeze | no change | number of channels and their dimensions change, depending on parameters | I.3 |
| 6 | Quantize | no change | no change | C.9.4.2 |
| 7 | AdaptiveQuantize | +1 | no change if `nb_parameters < 2`, otherwise channels are duplicated; one meta-channel is inserted in the beginning | C.9.4.3 |

### C.9.4.1 Subsample

This transform has a number of parameters that is either 0, 1, or a multiple of 4. The parameters are interpreted as follows if there are at least 4: `[ [begin] [end] [h_shift] [v_shift] ]+`.
Channels with index in the range `[begin, end]` shall be upsampled horizontally by the factor $2^{h_shift}$ and vertically by the factor $2^{v_shift}$.
If there is only one parameter, then it is interpreted as an abbreviation, as follows:

| Single parameter | Shorthand for | Description |
|---|---|---|
| 0 | 1 2 1 1 | 4:2:0 chroma subsampling (assuming channels 1 and 2 are chroma) |
| 1 | 1 2 1 0 | 4:2:2 chroma subsampling |
| 2 | 1 2 0 1 | 4:4:0 chroma subsampling |
| 3 | 1 2 2 0 | 4:1:1 chroma subsampling |

If there are no parameters, this is an abbreviation for 1 2 1 1 (just like the single parameter 0).
The effect on the channel vector (when decoding the transformation header) shall be as follows:

```
for (c = begin; c <= end; c++) {
  channel[c].width += (1 << h_shift) - 1;
  channel[c].width <<= srh;
  channel[c].height += (1 << v_shift) - 1;
  channel[c].height <<= srv;
  channel[c].hshift += h_shift;
  channel[c].vshift += v_shift;
}
```

The inverse Subsample transform shall use the following upsampling method. For factor two horizontal (vertical) upsampling, the channel width (height) is multiplied by two, and every pixel `B` with left (top) neighbor `A` and right (bottom) neighbor `C` is replaced by two new pixels with values `(A + 3 × B + 1) >> 2` and `(3 × B + C + 2) >> 2`.

For higher factors the factor two upsampling shall be repeated as often as necessary. Channel dimensions shall not exceed the image dimensions, and the border pixels are replicated for the purpose of the computation.

### C.9.4.2 Quantize

The inverse Quantize transform shall multiply all the pixel values in all channels by the quantization factor of that channel, `channel[i].q`.

### C.9.4.3 AdaptiveQuantize

This transformation has one obligatory parameter, `aq_shift`, and one optional parameter, `store_remainders`, which has a default value of zero.

When this transform is used, one meta-channel is inserted in the beginning of the channel vector, with the following properties:

```
width = ((image_width - 1) >> aq_shift) + 1;
height = ((image_height - 1) >> aq_shift) + 1;
hshift = vshift = aq_shift;
```

If `store_remainders == 1`, then all the non-meta channels shall be duplicated:

```
endc = channel.size();
for (i = nb_meta_channels; i < endc; i++) {
  channel.push_back(channel[i].copy());
}
```

The inverse transformation uses the values in the meta-channel as weights for the quantization factor for the corresponding region. The inverse AdaptiveQuantize transform shall behave as follows:

```
denominator = channel[0].q;
first = nb_meta_channels;
last = channel.size() - 1;
if (store_remainders == 1) {
  offset = first + (last - first + 1) Idiv 2;
last = offset - 1;
} else {
  offset = first;
}
for (c = first; c <= last; c++) {
  if (aq_shift < channel[c].hshift) continue;
  if (aq_shift < channel[c].vshift) continue;
  q = channel[offset + c - first].q;
  for (y = 0; y < channel[c].height; y++) {
    for (x = 0; x < channel[c].width; x++) {
            w = channel[0]((x << channel[c].hshift) >> aq_shift,
                           (y << channel[c].vshift) >> aq_shift);
            aq = q × w / denominator;
            channel[c](x, y) ×= aq;
            if (store_remainders == 1)
              channel[c](x, y) += channel[offset + c - first];
    }
```

```
  }
}
if (store_remainders == 1) } [[Remove all channels with index >= offset]];
[[Remove the first channel]]
```

### C.10 LoopFilter

Immediately after decoding FrameHeader, the decoder shall read the following bundle if and only if `frame_header.encoding == kPasses`, otherwise its fields shall be initialized to their defaults.

| condition | type | default | name |
|---|---|---|---|
| | Bool() | true | all_default |
| !all_default | Bool() | true | gab |
| !all_default && gab | Bool() | false | gab_custom |
| gab_custom | F16() | 0.115169525 | gab_x_weight1 |
| gab_custom | F16() | 0.061248592 | gab_x_weight2 |
| gab_custom | F16() | 0.115169525 | gab_y_weight1 |
| gab_custom | F16() | 0.061248592 | gab_y_weight2 |
| gab_custom | F16() | 0.115169525 | gab_b_weight1 |
| gab_custom | F16() | 0.061248592 | gab_b_weight2 |
| !all_default | Bool() | true | epf |
| !all_default && epf | Bool() | false | epf_weight_custom |
| epf_weight_custom | F16() | 0.3 | epf_weight_min |
| !all_default && epf | Bool() | false | epf_scale_custom |
| epf_scale_custom | F16() | 899.676941 | epf_scale_x |
| epf_scale_custom | F16() | 111.965157 | epf_scale_y |
| epf_scale_custom | F16() | 45.275791 | epf_scale_b |
| !all_default && epf | Bool() | false | epf_sharp_custom |
| for (i = 0; i < 8 && epf_sharp_custom; ++i) | F16() | i / 7 | epf_sharp_lut[i] |
| !all_default && epf | Bool() | false | epf_sigma_custom |
| epf_sigma_custom | F16() | 70 | epf_range_max |
| epf_sigma_custom | F16() | 12.6 | epf_range_mul |
| epf_sigma_custom | F16() | 20 | epf_quant_mul |
| epf_sigma_custom | U32(Val(0), Val(170), Val(672), Bits(10)) | 170 | epf_sigma_max |
| !all_default | Bool() | false | qco |
| qco | Bool() | false | qco_custom |
| qco_custom | F16() | 0.5 | qco_interval_mul |
| !all_default | Extensions | | extensions |

epf, qco, and gab shall determine whether edge-preserving filter, quantization constraint, and Gaborish shall be applied as specified in J.2, J.3, and J.4. The other fields shall parameterize any enabled filter(s).

## C.11 Dc16Group

### C.11.1 Overview

If the kProgressive16 flag in frame_header.flags is not set, this subclause shall be skipped.
In this subclause, xsize and ysize shall refer to the size of the current DC16 group, and all coordinates shall be relative to the top-left corner of the DC16 group.
Each DC16 group shall be decoded as a single ANS stream, as specified in D.3. Thus, the decoder shall read 36 clustered distributions D for the stream.

### C.11.2 DC16 coefficients

The decoder shall obtain three unsigned 8-bit channels with ceil(ysize / 16) rows and ceil(xsize / 16) columns.

The decoder shall use the adaptive predictor specified in E.2, using XBPredictor as a sub-predictor for each of the 3 channels and PackSigned as an error metric.

During the execution of the predictor, for the samples (in order Y, then X, then B) of each pixel in raster order, the decoder shall read a single ANS symbol sym, using D[GetContext(c)], and then use the procedure specified in D.3.6 to decode the residual value. The decoded value shall then be obtained from the residual value, as specified by PackSigned in E.2.2.1.

When decoding DC coefficients (as specified in C.5.2), the decoded DC coefficient at position (x, y) shall be increased by the decoded DC16 coefficient in position (x >> 1, y >> 1).

# Annex D
(normative)
# Entropy decoding

## D.1 General

Each subclause in this Annex specifies an entropy decoder. Other Annexes and their subclauses indicate which of these entropy decoders shall be used for various elements of the codestream.

NOTE These entropy decoders enable various size and speed tradeoffs.

## D.2 Brotli

Brotli shall be decoded as specified in RFC 7932.

## D.3 ANS

### D.3.1 Overview

This subclause describes a symbol-based stream that shall be decoded to a sequence of 8-bit symbols, each of which is taken to belong to a specific probability distribution `D`. `D` shall be represented as an array of values in `[0, 1 << 12]`, one per symbol, that sum to `1 << 12`. Indexing `D` with an out-of-bounds value shall result in a `0` value.

### D.3.2 Alias mapping

For a given probability distribution `D` of symbols in the range `[0, 256)` and a symbol `x` from D, the *alias mapping* of `x` under `D` (`AliasMapping(D, x)`) shall be the output of the following procedure, that produces a *symbol* in `[0, 256)` and an *offset* in `[0, 1 << 12)`.

Let `max_symbol` be the highest-index symbol in `D` with nonzero probability. `table_size` shall be `256`, and `bucket_size` shall be `(1 << 12) / table_size`, which is also a power of two. In order to set up the AliasMapping table, the decoder shall create an auxiliary table `T`, an array of `table_size` sequences called *buckets*; bucket `i` is assigned the (possibly empty) sequence of *(symbol, offset)* pairs `{i, k}` for `k` in `[0, D[i])`, in this order. We say that a bucket is *overfull* if strictly more than `bucket_size` pairs belong to it, and that it is *underfull* if strictly less than `bucket_size` pairs belong to it.

The decoder shall maintain a stack of overfull buckets and also a stack of underfull buckets, initialized by pushing overfull and underfull buckets into their respective stack in order of increasing `i`. Note that either both stacks are empty, or none is.

While the stacks are not empty, the decoder shall remove bucket `u` from the underfull stack, and bucket `o` from the overfull one, then move as many pairs from `o` as needed to make `u` reach exactly `bucket_size` pairs (note that the symbol of those pairs will always be the same). The pairs to be removed from `o` shall be the highest-offset ones, and they shall be added to `u` at the end of existing pairs, preserving their relative order. After this step, `u` will have exactly `bucket_size` pairs, but `o` may be overfull, underfull, or neither; in the first two cases, it shall be added to the corresponding stack.

At the end of this procedure, all buckets shall be concatenated together, in increasing order of index and preserving relative order inside a given bucket. The *alias mapping* of value `x` in `[0, 1 << 12)` shall then be the *(symbol, offset)* pair with index `x` in this sequence of pairs.

### D.3.3 ANS symbol decoding

The ANS decoder shall keep an internal 32-bit state. Upon creation of the decoder (immediately before reading from a new ANS stream), this state shall be initialized with a u(32) from the bitstream. After the decoder reads the last symbol in a given stream, the state shall be `0x130000`; if this condition does not hold, the codestream is ill-formed.

The decoder shall decode a symbol belonging to a given distribution `D` as specified by the following:

```
index = state & 0xFFF;
(symbol, offset) = AliasMapping(D, index);
state = D[symbol] × (state >> 12) + offset;
```

If this update procedure causes the `state` to become smaller than `1 << 16`, the decoder shall read the update bits `upd` as a u(16) from the bitstream, and update the state to `(state << 16) | upd`.

`symbol` shall then be returned as the next decoded symbol.

### D.3.4 ANS distribution encoding

To decode a distribution `D`, which is a 256-element array whose elements represent symbol probabilities and sum to 1 << 12, the decoder shall first read `simple` as u(1).

If `simple == 1`, the decoder shall read number of symbols as `ns = u(1) + 1`. If `ns` is `1`, the decoder shall decode `x = U8()`: the resulting distribution shall be `D[x] = 1 << 12`, and `D[y] = 0` for `x != y`. If `ns` is `2`, the decoder shall decode `v1 = U8()` and `v2 = U8()`. If `v1 == v2`, the codestream is ill-formed. Otherwise, the decoder shall then read the distribution value of the `v1` symbol as a u(12); the `v2` symbol shall have a distribution value `1 << 12` minus the distribution value of the `v1` symbol, and all other distribution values shall be `0`.

If `simple == 0`, the decoder shall read `flat` as u(1). If `flat == 1`, the decoder shall read `alphabet_size = U8() + 1`. The first `(1 << 12) Umod alphabet_size` values of `D` shall be `floor((1 << 12) / alphabet_size) + 1`, the next `alphabet_size - ((1 << 12) Umod alphabet_size)` shall be `floor((1 << 12) / alphabet_size)`, and the rest shall be `0`.

If `simple == 0` and `flat == 0`, the decoder shall read up to `3` bits, stopping at the first occurence of `0`; let `len` be the number of `1` bits that were read. The decoder shall then read another `u(len)` representing the lower bits `b` of the `shift` value, computed as `b + (1 << len) - 1`. Then, it shall set `alphabet_size = U8() + 3`. If `alphabet_size > 256`, the codestream is ill-formed.

Let `kLogCountLut` be the following Huffman decoding table:

```
kLogCountLut[128][2] = {
       {3, 10}, {7, 12}, {3, 7}, {4, 3}, {3, 6}, {3, 8}, {3, 9}, {4, 5},
       {3, 10}, {4, 4},  {3, 7}, {4, 1}, {3, 6}, {3, 8}, {3, 9}, {4, 2},
       {3, 10}, {5, 0},  {3, 7}, {4, 3}, {3, 6}, {3, 8}, {3, 9}, {4, 5},
       {3, 10}, {4, 4},  {3, 7}, {4, 1}, {3, 6}, {3, 8}, {3, 9}, {4, 2},
       {3, 10}, {6, 11}, {3, 7}, {4, 3}, {3, 6}, {3, 8}, {3, 9}, {4, 5},
       {3, 10}, {4, 4},  {3, 7}, {4, 1}, {3, 6}, {3, 8}, {3, 9}, {4, 2},
       {3, 10}, {5, 0},  {3, 7}, {4, 3}, {3, 6}, {3, 8}, {3, 9}, {4, 5},
       {3, 10}, {4, 4},  {3, 7}, {4, 1}, {3, 6}, {3, 8}, {3, 9}, {4, 2},
       {3, 10}, {7, 13}, {3, 7}, {4, 3}, {3, 6}, {3, 8}, {3, 9}, {4, 5},
       {3, 10}, {4, 4},  {3, 7}, {4, 1}, {3, 6}, {3, 8}, {3, 9}, {4, 2},
       {3, 10}, {5, 0},  {3, 7}, {4, 3}, {3, 6}, {3, 8}, {3, 9}, {4, 5},
```

```
    {3, 10}, {4, 4},  {3, 7}, {4, 1}, {3, 6}, {3, 8}, {3, 9}, {4, 2},
    {3, 10}, {6, 11}, {3, 7}, {4, 3}, {3, 6}, {3, 8}, {3, 9}, {4, 5},
    {3, 10}, {4, 4},  {3, 7}, {4, 1}, {3, 6}, {3, 8}, {3, 9}, {4, 2},
    {3, 10}, {5, 0},  {3, 7}, {4, 3}, {3, 6}, {3, 8}, {3, 9}, {4, 5},
    {3, 10}, {4, 4},  {3, 7}, {4, 1}, {3, 6}, {3, 8}, {3, 9}, {4, 2},
};
```

For each symbol in the alphabet, in increasing order, the decoder shall read a u(7) index `h` at first without advancing its position in the bitstream, and then advance the bitstream position by `kLogCountLut[h][0]`. It shall then compute `logcount` as `kLogCountLut[h][1]`. If `logcount` is `13`, then the decoder shall read `r = U8()`, indicating that the current symbol and the next `r + 3` shall have the same distribution value as the previous one (or `0`, if the current symbol is the first). In that case, if the index of the current symbol incremented by `r + 3` exceeds alphabet_size, the codestream is ill-formed. Thus, decoding the index `h` and `logcount` of the next `r + 3` symbols shall be skipped. During this procedure, the decoder also keeps track of `omit`, the symbol with the highest `logcount` (excluding `logcount`s of `13`), breaking ties by choosing the symbol of lowest index. If the symbol that follows `omit` has a `logcount` value of `13`, the codestream is ill-formed.

For each symbol that is not run-length coded and not the `omit` symbol, the decoder shall then proceed as follows:

- If `logcount` is 0, the distribution value shall be `0`.
- If `logcount` is 1, the distribution value shall be `1`.
- Otherwise, let `nbits = min(logcount - 1, shift - ((13 - logcount) >> 1))`. The distribution value shall then be `(u(nbits) << (logcount - 1 - nbits)) + (1 << (logcount - 1))`.

Finally, the `omit` symbol shall have a distribution value such that the sum of all distribution values is `1 << 12`. If the required distribution value would be non-positive, the codestream is ill-formed.

### D.3.5 Distribution clustering

In most cases, the probability distributions that symbols belong to can be clustered together. This subclause specifies how this clustering shall be read by the decoder. When referring to a distribution to be used for ANS decoding, unless otherwise specified, the distribution to be used shall be the distribution of the corresponding cluster.

Let `num_distributions` be the number of non-clustered distributions. The output of this procedure shall be an array with `num_distributions` entries, with values in the range [`0`, `num_clusters`). The codestream is ill-formed if any integer in `[0, num_clusters)` is not present in this array. Position `i` in this array indicates that distribution `i` is merged into the corresponding cluster.

After decoding of the clustering map, the decoder shall decode the `num_clusters` histograms, one for each distribution, appearing in the bitstream in increasing index order, as specified in D.3.4.

The decoder shall first read a u(1) `is_simple` indicating a simple clustering. If `is_simple` is `1`, it shall then decode a u(2) representing the number of bits per entry `nbits`. For each non-clustered distribution, the decoder shall read a `u(nbits)` value that indicates the cluster that the given distribution belongs to.

Otherwise, if `is_simple` is `0`, the decoder shall read a single distribution `D` from the bitstream, as specified in D.3.4. For each non-clustered distribution `i`, the decoder shall read an ANS symbol `sym` using distribution `D`. If this symbol is less than `8`, the corresponding value for position `i` shall be `(1 << sym) + u(sym) - 1`. Otherwise, the decoder shall compute `r = (1 << (sym - 8)) + u(sym - 8) + 1`, fill the current and the next `r` positions with `0`, and skip the next `r` distributions. Finally, the decoder shall apply an inverse move-to-front transform to the cluster mapping, as follows:

```
MTF(uint8_t v[256], index) {
  value = v[index];
  for (i = index; i; --i) v[i] = v[i - 1];
  v[0] = value;
}

InverseMoveToFrontTransform(uint8_t clusters[num_distributions]) {
  uint8_t mtf[256];
  for (i = 0; i < 256; ++i) {
    mtf[i] = i;
  }
  for (i = 0; i < num_distributions; ++i) {
    uint8_t index = clusters[i];
    clusters[i] = mtf[index];
    if (index) MTF(mtf, index);
  }
}
```

### D.3.6 Hybrid integer coding

The following procedure reads a 32-bit unsigned integer from ANS-encoded streams given a 8-bit token.

```
DecodeHybridVarLenUint(uint8_t token) {
  if (token < 16) {
    return token;
  }
  n = 4 + ((token - 16) >> 1);
  return (1 << n) + ((token & 1) << (n - 1)) + u(n - 1);
}
```

## D.4 sANS

### D.4.1 Overview

This subclause describes a symbol-based stream that shall be decoded to a sequence of symbols in the range [0, 18), each of which is taken to belong to a specific probability distribution `D`. `D` shall be represented as an array of integer values in the range $[0, 2^{10} + 1)$, one per symbol, that sum to $2^{10}$.
Auxiliary function `S(D, i)` returns the sum of the first `i + 1` elements of `D`.
`alphabet_size`, the length of `D`, is always 18.

### D.4.2 Symbol resolving

Resolving a symbol means finding a symbol which, according to the given distribution `D`, is represented by the given value `x`. More precisely, `ResolveSymbol(D, x)` returns the smallest non-negative value `s` such that `S(D, s) > i`.

### D.4.3 sANS symbol decoder

The sANS decoder shall keep an internal 32-bit unsigned integer `state`; upon creation of the decoder, this state shall be initialized by reading `state` as u(32). Whenever reading from sANS token-stream is (explicitly) completed, its integrity shall be checked: if `state != 0x130000`, the codestream is ill-formed.

To decode a symbol t belonging to a given distribution D, the following procedure shall be used:

```
x = state & 0x3FF;
t = ResolveSymbol(D, x);
o = x - S(D, t) - D[t];
state = o + D[t] × (state >> 10);
if (state < 0x10000) state = (state << 16) | u(16);
```

### D.4.4 sANS distribution stream

Initially all alphabet_size elements of D are assigned 0. To decode a distribution D, at first the decoder shall read is_simple as u(1). If is_simple is 1, then only 1 or 2 symbols have non-zero weight, and distribution D shall be decoded as follows:

```
max_bits = ilog2(alphabet_size - 1);
num_tokens = u(1) + 1;
v1 = u(max_bits);
if (num_tokens == 1) {
  D[v1] = 1 << 10;
  return;
}
v2 = u(max_bits);
if (v1 == v2) [[ the codestream is ill-formed]];
weight = u(10);
D[v1] = weight;
D[v2] = 1024 - weight;
```

Otherwise (is_simple == 0), the number of non-zero weight symbols num_symbols is in the range [3, alphabet_size + 1), and shall be read using the fixed prefix code specified in Table D.1. Reading the prefix code means fetching bits using u(1) until the left-to-right concatenation of fetched bits matches exactly some code in the table; the result shall be the corresponding value.

**Table D.1 – prefix code for number of non-zero weight symbols**

| value | code | value | code |
|---|---|---|---|
| 10 | 000 | 8 | 11100 |
| 11 | 001 | 17 | 11101 |
| 12 | 010 | 5 | 111100 |
| 13 | 011 | 6 | 111101 |
| 14 | 100 | 7 | 111110 |
| 15 | 101 | 18 | 1111110 |
| 9 | 1100 | 3 | 11111110 |
| 16 | 1101 | 4 | 11111111 |

Distribution D is decoded as follows:

- `weight_code` is initialized as an empty array of unsigned integers
- `num_symbols` times:
  - `code` is decoded using predefined prefix code specified in Table D.2
  - `code` is appended to `weight_code`
- `remainder_index = min_argmax(weight_code)`
- `total_count = 0`
- for each `code` element of `weight_code`:
  - let `index` be the index of the considered array element
  - if `index == remainder_index`, loop iteration is finished
  - if `code == 0`, loop iteration is finished
  - if `code == 1`:
    - `D[index] = 1`
    - `total_count += 1`
    - loop iteration is finished
  - `bits = floor(code / 2)`
  - `shift = code - 1 - bits`
  - `extra = u(bits)`
  - `weight = (1 << (code - 1)) + (extra << shift)`
  - `D[index] = weight`
  - `total_count += weight`
- if `total_count >= 1024`, the codestream is ill-formed
- `D[reminder_index] = 1024 - total_count`
- decoding of distribution is finished

**Table D.2 – prefix code for weight bit-length**

| value | code |
|---|---|
| 6 | 00 |
| 4 | 010 |
| 5 | 011 |
| 7 | 100 |
| 8 | 101 |
| 1 | 1100 |

| value | code |
|---|---|
| 2 | 1101 |
| 3 | 1110 |
| 0 | 11110 |
| 9 | 111110 |
| 10 | 111111 |

## D.5 Binary arithmetic coding

### D.5.1 Base binary arithmetic decoder

The binary arithmetic decoder is a source of 1-bit values. When the decoder is queried to read the value the expectancy of value 0 is supplied. The decoder state consists of unsigned integer values: `low`, `value`, and `high`. Furthermore, the decoder instance is specialized with two parameters: an unsigned integer `precision_bits` and the function `GetWord()` that returns a 16-bit unsigned integer.
During initialization and reading the value, the decoder shall use the subroutine `Normalize()` defined as follows:

```
mask = 0xFFFFFFFF;
value = ((value << 16) | GetWord()) & mask;
low = (low << 16) & mask;
high = ((high << 16) | 0xFFFF) & mask;
```

The decoder is initialized as follows:

```
low = 0;
high = 0;
value = 0;
Normalize();
Normalize();
```

The function ReadBit returns a bit given the probability that it is zero, as specified by the following:

```
ReadBit(prob_zero) {
  diff = high - low;
  split = low + ((diff × prob_zero) >> precision_bits);
  if (value > split) {
    low = split + 1;
    bit = 1;
  } else {
    high = split;
    bit = 0;
  }
  if (((low ^ high) >> 16) == 0) {
    Normalize();
  }
  return bit;
}
```

### D.5.2 8-bit binary arithmetic decoder

8-bit Binary Arithmetic Decoder (precision_bits == 8) is used for decoding losslessly compressed JPEG streams. GetWord() of this specialization is equal to u(16).

### D.5.3 16-bit binary arithmetic decoder

16-bit Binary Arithmetic Decoder (precision_bits == 16) is used to decode the dot and spline dictionaries. This specialization has state and extra initialization state, performed before the base decode initialization:

```
nbits = u(4);
nwords = u(nbits);
uint16_t stream[nwords];
for (i = 0; i < nwords; i++) {
  stream[i] = u(16);
}
cur_pos = 0;
```

GetWord() of this specialization returns the next unused element of stream, or 0, as specified by the following:

```
GetWord() {
  if (curr_pos < nwords) {
    word = stream[curr_pos];
    ++curr_pos;
    return word;
  } else {
```

```
    return 0;
  }
}
```

### D.5.4 Binary arithmetic decoding of integers

An unsigned integer shall be decoded one bit at a time, using a cumulative distribution function (cdf) to determine the probability of the next bit being zero. `cdf(low, m, high)` shall denote

```
(1 << 16) × P(low <= x < m) / P(low <= x < high)
```

The function `ReadInt(bits)` shall read an integer given the number of bits and `cdf`, as specified by the following:

```
ReadInt(bits) {
  uint64_t low = 0;
  uint64_t high = 1 << bits;
  for (i = 0; i < bits; i++) {
    uint64_t m = (low + high) Idiv 2;
    prob_zero = cdf(low, m, high);
    bit = ReadBit(prob_zero);
    if (bit == 0) {
      high = m;
    } else {
      low = m;
    }
  }
  return low;
}
```

The next two subclauses specify how to compute `cdf`.

### D.5.5 Integers distributed with a given histogram

Given a histogram represented as an array `hist` of size `U`, the `cdf(low, m, high)` for integers in the interval `[0, U)` shall be computed as specified below. The probability of obtaining a number `x` is given by `p(x) = hist[x] / sum(hist)`. The `cdf` is a sum of the elements of the histogram, for the numerator between `[low, m)` and for the denominator between `[low, high)`:

```
uint32_t num = 0;
uint32_t den = 0;
if (high > U) high = U;
for (i = low; i < high; i++) {
  uint32_t x = (i < U) ? hist[i] : 0;
  if (i < m) num += x;
  den += x;
}
if (den == 0) return 0;
return (num × 0xFFFF) Idiv den;
```

### D.5.6 Uniformly distributed integers

The `cdf(low, m, high)` for a uniform distribution in the interval `[0, U)` is equivalent to that of a histogram with a value of one in the interval `[0, U)` and zero otherwise. It shall be computed as specified by the following:

```
uint32_t num = min(m, U) - min(low, U);
```

```
uint32_t den = min(high, U) - min(low, U);
if (den == 0) return 0;
return (num × 0xFFFF) Idiv den;
```

GetUniformInt(U) shall denote ReadInt(ceil(log2(U))) with the cdf computed as specified above.

#### D.5.7 Decoding histograms

The decoder shall decode a histogram, represented as an array hist, as specified by the following:

```
s = sum(hist);
for (i = 0; i < len(hist) - 1; i++) {
  hist[i] = GetUniformInt(s + 1);
  s -= hist[i];
  if (s < 0) [[the codestream is ill-formed]];
}
hist[len(hist) - 1] = s;
```

GetHistInt(hist, U) shall denote ReadInt(ceil(log2(len(hist)))) with the above cdf.

### D.6 Histogram code

#### D.6.1 Huffman histogram stream

Huffman histogram stream stores metadata required for decoding Huffman symbol stream.
See RFC 7932 section 3.5 ("Complex Prefix Codes") for format encoding description and definitions. Alphabet size mentioned in RFC is explicitly specified as parameter alphabet_size when histogram is being decoded.

#### D.6.2 Huffman symbol stream

Huffman symbol stream stores a sequence of unsigned integers (symbols). To read a symbol, the decoder shall read a series of bits via u(1), until the bits concatenated in left-to-right order exactly match one of the prefix codes associated with the symbol.
See RFC 7932 section 3.2 ("Use of Prefix Coding in the Brotli Format") for a detailed description of how to build the prefix codes from Huffman histogram.

### D.7 Adaptive binary range coding (ABRAC)

This entropy decoder is the basic building block from which the next two entropy decoders (BEGABRAC and MABEGABRAC) are constructed. It encodes individual bits using a range coder and chance adaptation.

The range coder has a state consisting of two unsigned 32-bit integers, range and low, which are initialized as follows:

```
range = 1 << 24;
low = u(8) << 16;
low |= u(8) << 8;
low |= u(8);
```

Chances correspond to the likelihood of the bit being 1, and are represented using 12-bit unsigned integers, i.e. they are between 1 and 4095 and have an implicit denominator of 4096.
To decode a single bit with a given chance **c**, the decoder performs the following computation:

```
get_bit(c) {
```

```
  nc = (range × c) >> 12;
  if (low >= range - nc) {
    bit = 1;
    low -= range - nc;
    range = nc ;
  } else {
    bit = 0;
    range -= nc;
  }
  while (range <= 1 << 16) {
    low <<= 8;
    low |= u(8);
    range <<= 8;
  }
  return bit;
}
```

To decode bits with adaptive chances, the chance **ac** is initialized to 2048 (50%) or some other value (as specified below), and it shall be updated each time a bit value has been decoded:

```
get_adaptive_bit(ac) {
  if (get_bit(ac)) {
    ac += (4096 - ac) >> 5;
    return 1;
  } else {
    ac -= ac >> 5;
    return 0;
  }
}
```

### D.7.1 Bounded-Exp-Golomb ABRAC (BEGABRAC)

Based on ABRAC, integers are decoded using multiple adaptive bit coders. The number of bit chances needed, depends on the maximum range of the integers to be encoded. The following adaptive chances are defined (where N denotes the maximum bit depth of the numbers being encoded):

ac_zero
ac_sign
$ac_exponent_0$, $ac_exponent_1$, …, $ac_exponent_{N-2}$
$ac_mantissa_0$, $ac_mantissa_1$, …, $ac_mantissa_{N-1}$

The decoder is assumed to know bounds lower, upper on the number x to be decoded, such that x is in the range [lower, upper]. It is assumed that zero is always a possible value, hence lower <= 0 <= upper.

The procedure for decoding a BEGABRAC-encoded integer is as follows:

```
get_begabrac(lower, upper) {
  if (get_adaptive_bit(ac_zero)) return 0;
  if (lower < 0) {
    if (upper > 0) {
```

```
      sign = get_adaptive_bit(ac_sign);
    } else {
      sign = 0;
    }
  } else {
    sign = 1;
  }
  max = sign == 1 ? upper : -lower;
  exp = 0;
  while (exp < ilog2(max) - 1) {
    if (get_adaptive_bit(ac_exponent[exp]) == 1) break;
    exp++;
  }
  v = 1 << exp;
  for (i = exp - 1; i >= 0; i--) {
    one = v | (1 << i);
    if (one > max) continue;
    if (get_adaptive_bit(ac_mantissa[i]) == 1) v = one;
  }
  if (sign == 1) return v;
  else return -v;
}
```

The initialization of the adaptive chances is parametrized by providing the zero bit chance **init_ac_zero**:

```
init_begabrac(init_ac_zero) {
  ac_zero = init_ac_zero;
  ac_sign = 2048;
  c = 4096 - ac_zero;
  for (i = 0; i < N - 1; i++) {
    if (c < 256) c = 256;
    if (c > 3840) c = 3840;
    ac_exponent[i] = 4096 - c;
    c = (c × c + 2048) >> 12;
  }
  for (i = 0; i < N; i++) ac_mantissa[i] = 1024;
}
```

### D.7.2 Meta-Adaptive BEGABRAC (MABEGABRAC)

The Meta-Adaptive context model uses a vector of integer numbers, which are called properties. To determine which context to use, a decision tree is used. The full structure of the tree is known in advance by the decoder; it is explicitly signalled in the codestream (C.9.4 describes where it is signalled and D.7.3 describes how it is signalled) so encoders can tune this tree to obtain a context model that is relevant for the actual image content. The inner nodes (decision nodes) of a MA tree contain a test of the form property[k] > value. If this test evaluates to true, then the left branch is taken, otherwise the right branch is taken. Eventually a leaf node is reached, which corresponds to a context to be used to encode a symbol; i.e. each leaf node contains a set of adapting BEGABRAC probabilities.

The properties which are always used in the context model are given in the table below. In addition to these properties, a variable number of ‘previous channel’ properties are also used. The maximum number of such extra properties (max_extra_properties) is obtained from the modular bitstream header. The extra properties get the smallest property index numbers; the number *n* in the table below represents the actual number of such extra properties.

Assuming the value to be decoded corresponds to the sample at column x of row y of channel number c, the following values are used to define the properties:

| | | |
|---|---|---|
| n: | initially equal to max_extra_properties | |
| ch_min: | the lowest value occurring in channel[c] (signalled) | |
| ch_max: | the highest value occurring in channel[c] (signalled) | |
| channel[c].zero: | ch_min | if (ch_min > 0) |
| | ch_max | if (ch_max < 0) |
| | 0 | otherwise |
| left: | pixel value at row y, column x - 1 | if x > 0 |
| | channel[c].zero | otherwise |
| top: | pixel value at row y - 1, column x | if y > 0 |
| | left | otherwise |
| topleft: | pixel value at row y - 1, column x - 1 | if x > 0 and y > 0 |
| | left | otherwise |
| topright: | pixel value at row y - 1, column x + 1 | if y > 0 and x + 1 < channel[c].width |
| | top | otherwise |
| leftleft: | pixel value at row y, column x - 2 | if x > 1 |
| | left | otherwise |

| Property | Value | Range |
|---|---|---|
| *n*+0 | abs(top) | 0 .. max(abs(ch_min), abs(ch_max)) |
| *n*+1 | abs(left) | 0 .. max(abs(ch_min), abs(ch_max)) |
| *n*+2 | top | ch_min .. ch_max |
| *n*+3 | left | ch_min .. ch_max |
| *n*+4 | y | 0 .. channel[c].height - 1 |
| *n*+5 | x | 0 .. channel[c].width - 1 |
| *n*+6 | left+top-topleft | 2 × ch_min - ch_max .. 2 × ch_max - ch_min |
| *n*+7 | topleft+topright-top | 2 × ch_min - ch_max .. 2 × ch_max - ch_min |
| *n*+8 | left-topleft | ch_min - ch_max .. ch_max - ch_min |
| *n*+9 | topleft-top | ch_min - ch_max .. ch_max - ch_min |
| *n*+10 | top-topright | ch_min - ch_max .. ch_max - ch_min |
| *n*+10 | top-toptop | ch_min - ch_max .. ch_max - ch_min |
| *n*+11 | left-leftleft | ch_min - ch_max .. ch_max - ch_min |

The extra properties are defined as follows:

```
k = 0;
for (i = c - 1; i >= 0 && k < max_extra_properties; i--) {
  if (channel[i].min == channel[i].max) continue;
  if (channel[i].width == 0 || channel[i].height == 0) continue;
  if (channel[i].hshift < 0) continue;
  ry = (y << channel[c].vshift) >> channel[i].vshift;
  rx = (x << channel[c].hshift) >> channel[i].hshift;
  if (ry >= channel[i].height) ry = channel[i].height - 1;
  if (rx >= channel[i].width) rx = channel[i].width - 1;
  rv = channel[i](rx, ry);
  rleft = (rx > 0 ? channel[i](rx - 1, ry) : channel[i].zero);
  rtop = (ry > 0 ? channel[i](rx, ry - 1) : rleft);
  rtopleft = (rx > 0 && ry > 0 ? channel[i](rx - 1, ry - 1) : rleft);
  rp = median(rleft + rtop - rtopleft, rleft, rtop);
  property[k + 0] = abs(rv);
  property[k + 1] = rv;
  property[k + 2] = abs(rv - rp);
  property[k + 3] = rv - rp;
  k += 4;
}
n = k;
```

The ranges of the properties each time they are assigned above are summarized in the following table:

| Property | Value | Range |
|---|---|---|
| *k*+0 | abs(rv) | 0 .. max(abs(channel[i].min), abs(channel[i].max)) |
| *k*+1 | rv | channel[i].min .. channel[i].max |
| *k*+2 | abs(rv - rp) | 0 .. abs(channel[i].min - channel[i].max) |
| *k*+3 | rv - rp | channel[i].min-channel[i].max .. channel[i].max-channel[i].min |

To decode a MABEGABRAC-encoded number, the MA tree is traversed using the above properties, and then the corresponding BEGABRAC context is used.

### D.7.3 MA tree encoding

The MA tree itself is encoded using 4 BEGABRAC contexts, all initialized with ac_init_zero = 1024. It is encoded recursively in a depth-first way. To decode it, it is important to keep track of the valid range of each property. These ranges are initialized as in the table above, and narrowed down based on the ancestor decision nodes. The decoder starts at the root node of the tree.

```
ZERO_INIT[11] = {4, 128, 512, 1024, 1536, 2048, 2560, 3072, 3584, 3968, 4088};
SIGN_INIT[7] = {512, 1024, 1536, 2048, 2560, 3072, 3584};
decode_subtree(node, ranges) {
```

```
  property = BEGABRAC1(0, n + 12) - 1;
  if (property < 0) {
    node.type = leaf_node;
    if (signal_initialization) {
      zc = BEGABRAC2(-5, 5) + 5;
      node.init_begabrac(ZERO_INIT[zc]);
      if (zc < 3) node.ac_sign = SIGN_INIT[BEGABRAC3(-3, 3) + 3];
    }
  } else {
    node.type = decision_node;
    node.property = property;
    v = BEGABRAC4(ranges[property].min, ranges[property].max - 1);
    node.value = v;
    ranges_left = ranges;
    ranges_left[property].min = v + 1;
    decode_subtree(node.left_child, ranges_left);
    ranges_right = ranges;
    ranges_right[property].max = v;
    decode_subtree(node.right_child, ranges_right);
  }
}
```

The leaf nodes of the MA tree are numbered in depth-first order, that is, in the order in which they are decoded. We say MA(properties) == k if the result of traversing the MA tree with given property values properties is the leaf node with index k.

# Annex E
(normative)
# Predictors

## E.1 Overview

The decoder shall support two kinds of predictors. Adaptive predictors (E.2) may be parameterised for use cases such as `DctSelect`. The weighted, self-correcting predictor (E.3) shall be used when `frame_header.encoding == kLossless`.

## E.2 Adaptive

Lossless prediction shall proceed through each sample of the channel to be decoded in raster order. All channels of the image shall be processed independently, in the order defined by each reference to this subclause.

The lossless predictor is parameterised by:

- an ordered set of 8 sub-predictors, for each channel, used for computing the *expected value* of each sample;
- an error metric, used to choose between sub-predictors;
- a decoder that produces the *decoded value* from the *expected value* and information computed from previously decoded samples.

Both the *expected value* and the *decoded value* shall be 16-bit signed integers. The output of the error metric shall be a 16-bit unsigned integer.

For each channel, each sub-predictor shall compute an *expected value* for sample **c** by taking as input the decoded values at samples **w** (the left sample on the same row), **l** (the left sample on the row above), **n** (the sample directly above) and **r** (the right sample on the row above), as shown in the figure below. If one of the sub-predictors requires the use of a sample that is outside the image bounds, that sub-predictor shall be marked as *unavailable* for the given sample.

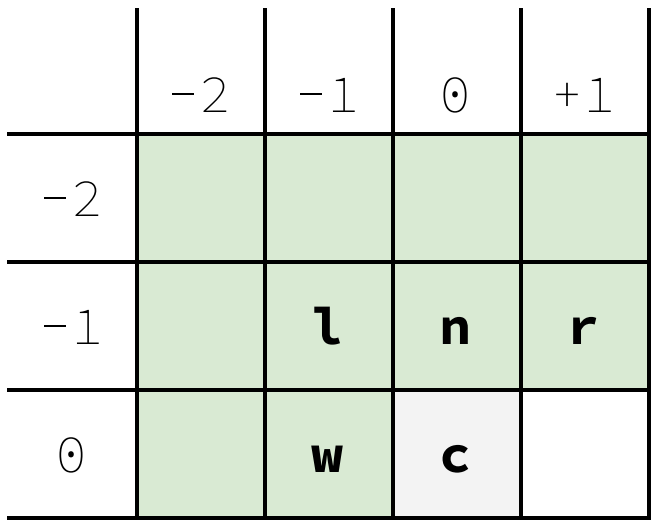

The *error* of an available sub-predictor is defined as the result of applying the error metric on the pair of the computed (*expected value*, *decoded value*) for sample **c**.

An *error estimate* for each (sample, sub-predictor) pair shall be computed as the maximum of the *error* of that sub-predictor for the **n**, **l** and **w** samples, excluding those samples that are out-of-bounds or for which that sub-predictor is unavailable. If no such error can be computed, the value of the *error estimate* for that sample shall be `(1 << 16) - 2`. If the sub-predictor is unavailable for that sample, then its *error estimate* shall be `(1 << 16) - 1`.

For each sample, the *expected value* `ev` shall be computed as the output of the sub-predictor for which the *error estimate* is the lowest; ties shall be broken by choosing the first predictor that realizes this minimum. If no predictor is available for that sample, the *expected value* shall be 0.

For each sample, the decoder is given the `x` and `y` coordinates, and shall compute the minimum `me` across all *error estimates* and the number `nc` of sub-predictors that have an *error estimate* of 0. The decoder shall then produce corresponding *decoded values* given the `ev`, `me` and `nc` values of each channel.

### E.2.1 Sub-predictors

All sub-predictors described here shall be computed as if all intermediate quantities were signed 32-bit integers. The following subclauses describe ordered (top to bottom) sets of sub-predictors that are used. In this subclause, the average of two values `a` and `b` shall be `(a + b) Idiv 2`.

#### E.2.1.1 Luminance sub-predictors (YPredictor)

- average of (average of **n** and **w**), and **r**
- average of **w** and **n**
- average of **n** and **r**
- average of **w** and **l**
- average of **n** and **l**
- **w**
- clamp((**n** + **w** - **l**), min(**n**, **w**, **l**), max(**n**, **w**, **l**))
- **n**

#### E.2.1.2 Chrominance sub-predictors (XBPredictor)

- clamp((**n** + **w** - **l**), min(**n**, **w**, **l**), max(**n**, **w**, **l**))
- average of **w** and **n**
- **n**
- average of **r** and **n**
- **w**
- average of **w** and **l**
- **r**
- average of (average of **n** and **w**) and **r**

### E.2.2 Error metrics

#### E.2.2.1 PackSigned

Let `d` be the difference between the *expected* and the *decoded* value, clamped to the range of a 16-bit integer. Then the error shall be `2 × d` if `d` is non-negative, and `-2 × d - 1` otherwise.

If this error metric is used, the *actual value* of a sample shall be obtained from `ev` and a *residual* `r` by the following procedure. `delta` shall be UnpackSigned(`r`). The *actual value* shall then be `ev + delta`.

#### E.2.2.2 PackSignedRange

This error metric is parameterised by two 16-bit integers `min` and `max` (where `min < max`). Let `ev` be the *expected value* and `dv` be the *decoded value*. If `ev == min`, the error shall be `dv - min`. If `ev == max`, the error shall be `max - dv`. Otherwise, this error metric shall be equivalent to PackSigned.

If this error metric is used, the *actual value* of a sample shall be obtained from `ev` and a *residual* `r` by the following procedure. `delta` shall be computed as

- If `ev == min`, `delta` shall be `r`.
- If `ev == max`, `delta` shall be `-r`.
- Otherwise, `delta` shall be computed as for PackSigned.

The *actual value* shall then be `ev + delta`.

### E.2.3 Context computation

Unless otherwise specified, the predictor shall compute a context value (i.e. index for the D array) from `me` and `nc` as follows, where `c` represents the channel that the sample belongs to:

```
GetContext(c) {
  if (nc == 0) {
    return c × 12 + min((me + 1) >> 1, 8) - 1;
  }
  return c × 12 + 8 + ceil(log2(9-nc));
}
```

## E.3 Weighted

### E.3.1 Overview

The weighted prediction shall be performed using the following parameters, obtained as described in subclauses O.1 and O.3:

- `xsize` and `ysize`, the dimensions of the current image group to decode
- the LosslessMode: 8-bit greyscale, 8-bit colour, 16-bit greyscale or 16-bit colour
- clustered distributions to read symbols from as specified in D.3.3.
- `max_err_shift`, `max_err_round`, `num_contexts`, `channel_index` and `max_value`
- for 8-bit modes only: `prediction_mode: 0`, `1` and `2`: `PM_Regular`, `PM_West` and `PM_North`. This value is not used for 16-bit modes, they have only one prediction mode.

A value `PBits` shall be set to `3` in 8-bit modes and to `0` in 16-bit modes, `PRound` to `((1 << PBits) >> 1)` and `PRoundm1` to `max(0, PRound - 1)`.

Unless noted otherwise, computations and variable types in this subclause use signed 32-bit integers.

The decoding of individual samples shall then proceed sequentially in raster order, with `x` the column index of the current sample (starting from 0) and `y` the row index of the current sample (starting from 0). For greyscale modes, this shall be done once for the grey samples. For colour modes, this shall be done three times: once independently for each of the three channels (in ascending order of index), with the corresponding input data streams, `prediction_mode` and `max_value` for each channel.

For each sample, the predictor specified by the image mode and prediction mode shall be called, implemented as indicated in the subclauses E.3.3, and E.3.4 below. It shall return a `prediction` and a `max_error`, as well as 4 sub-predictor values $\texttt{prediction}_i$, with i in range `[0, 4)`.

The `max_error` shall be incremented by `max_err_round`, then right-shifted by `max_err_shift` positions, then incremented by `num_channels × channel_index` to obtain the value of the `context`.

The decoder shall read a single ANS symbol, using D[context], and then use the procedure specified in D.3.6 to decode a value d. It shall be a value in range [0, 255] for 8-bit modes and in range [0, 65535] for 16-bit modes.

We shall then compute true_error, $sub_error_i$ and quantized_error for the current sample and store them for use in predictions of next samples:

The true_value of that sample shall then be computed with the signed least-significant bit shift as indicated in E.3.2:

```
true_value = restoreTrueValue((prediction + PRoundm1) >> Pbits, d);
```

The true_error for that sample shall be calculated as the difference between the prediction and the true_value of the sample:

```
true_error_CURRENT_SAMPLE = prediction - (true_value << PBits);
```

Each mode uses multiple sub-predictors $prediction_i$ to produce the final prediction (see E.3.3 and E.3.4). The error of a sub-predictor that produced a $prediction_i$ shall be defined as:

```
sub_error_i,CURRENT_SAMPLE = diff2error(prediction_i, true_value);
```

Here, diff2error shall be defined as the following function in an 8-bit mode:

```
int diff2error(int prediction, int truePixelValue) {
  prediction = (prediction + PRound) >> Pbits;
  return min(101, abs(512 - truePixelValue + prediction));
}
```

diff2error shall be defined as the following function in a 16-bit mode:

```
int diff2error(int prediction, int truePixelValue) {
  return min(0x3fbf, abs(prediction - truePixelValue));
}
```

For 8-bit modes, the quantized_error for the current sample shall be computed as follows, where quantizeError shall take a value in range [0, 256) and yield an even value in range [0, 14]:

```
quantized_error_CURRENT_SAMPLE = quantizeError(d);

int quantizeError(error) {
  error = (error + 1) >> 1;
  res = (error >= 4 ? 4 : error);
  if (error >= 6) res = 5;
  if (error >= 9) res = 6;
  if (error >= 15) res = 7;
  return res × 2;
}
```

For 16-bit modes, the quantized_error for the current sample shall be computed as follows, where quantizeError shall take a value in range [0, 65536) and yield an even or odd value in range [0, 13]:

```
quantized_error_CURRENT_SAMPLE = quantizeError(abs(prediction - true_value));

int quantizeError(error) {
  int res = 0;
  if (error >= 256) return min(13, 8 + quantizeError(error >> 8));
  if (error >= 16) res = 4, error >>= 4;
  if (error >= 4) res += 2, error >>= 2;
  return (res + min(error, 2));
}
```

The true_value corresponds to the sample for the given channel in the resulting image, and shall be further processed by the lossless decoder to apply colour transforms and expand per-channel palette as described in O.2.

### E.3.2 Signed least-significant bit shift

Given an encoded residual value d decoded from the stream and a value prediction predicted as described in the next subsections, the true value of a sample shall be computed as follows.

Here, m is max_value of the current channel for 16-bit or 255 for 8-bit.

```
int restoreTrueValue(int prediction, int d) {
  int v;
  if (prediction > (m >> 1)) {
    if (d > (m - prediction) × 2)
      v = m - d;
    else if (d & 1)
      v = prediction - 1 - (d >> 1);
    else
      v = prediction + (d >> 1);
  } else {
    if (d > prediction × 2)
      v = d;
    else if (d & 1)
      v = prediction - 1 - (d >> 1);
    else
      v = prediction + (d >> 1);
  }
  return v;
}
```

### E.3.3 8 bit prediction

The computation of predictors is based on six samples already scanned that are closest to the current sample - the NW (NorthWest), N (North), NE (NorthEast), W (West) samples, NN (North of North if in third or further row, or same as N when in second row) and WW (West of West if in third or further column, or same as W otherwise), as shown in the image below:

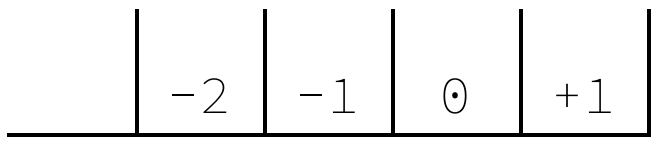

| | | | | |
|---|---|---|---|---|
| -2 | | | NN | |
| -1 | | NW | N | NE |
| 0 | WW | W | C | |

The variables N, NW, NE, W, NN, and WW shall be defined as the true values of their corresponding samples left shifted by PBits, and these symbols in subscripts refer to the corresponding variable for the sample at that location (e.g. $true_error_{W}$).
The variables x and y shall be the location of the current sample and xsize and ysize the dimensions of the current image group to decode, as described in E.3.1. When x or y are near the borders at 0 or xsize and ysize respectively, some neighbour samples may not exist. Unless otherwise noted, the following replacements shall be done for non-existing samples:

- If no NE sample exists, NE shall be set to N and subscripts NE, e.g. $true_error_{NE}$, shall correspond to N.
- If no NN sample exists, NN shall be set to N and subscripts NN, e.g. $true_error_{NN}$, shall correspond to N.
- If no WW sample exists, WW shall be set to W and subscripts WW, e.g. $true_error_{WW}$, shall correspond to W.

In each prediction mode, there shall be a set of sub-predictors whose predicted values and errors shall be used. The 4 sub-predictors shall be calculated differently depending on prediction_mode, as indicated below, and in another way in the top row or left column, as indicated further below.

The values $prediction_{i}$ with subscript i in the range [0, 4) shall refer to these 4 sub-predictors while prediction without subscript shall refer to the final computed prediction for the sample value.

When computing a sample that is not in the first row or first column of the image, the sub-predictors, prediction and max_error shall be computed as follows:

If prediction_mode is PM_Regular, the 4 sub-predictions and weight multipliers shall be defined as follows:

```
prediction_0 = W - (true_error_W + true_error_N + true_error_NW);
prediction_1 = N - (true_error_W + true_error_N + true_error_NE) Idiv 4;
prediction_2 = W + NE - N;
prediction_3 = N + ((N - NN) × 23 Idiv 32) +((W - NW) Idiv 16) -
               (true_error_NE × 3 + true_error_NW × 4 + 7) >> 5;

weight_mult_0 = {34, 31, 33, 36, 39, 42, 43, 43};
weight_mult_1 = {36, 37, 37, 40, 44, 46, 47, 42};
weight_mult_2 = {32, 32, 32, 32, 32, 32, 32, 32};
weight_mult_3 = {28, 24, 24, 24, 23, 23, 25, 32};
weight_mult_NE = {0, 15, 19, 16, 12, 12, 11, 11};

sum_weights_shift = 3;

NW_mult = 20;
```

If prediction_mode is PM_West, the 4 sub-predictions and weight multipliers shall be defined as follows:

```
prediction_0 = W - ((true_error_W + true_error_N + true_error_NW) × 9) Idiv 32;
prediction_1 = N - ((true_error_W + true_error_N + true_error_NE) × 171) Idiv 512;
```

```
prediction_2 = W + NE - N;
prediction_3 = N + ((N - NN) >> 1) +
    (((W - NW) × 19 - true_error_NW × 13)) Idiv 64;

weight_mult_0 = {27, 33, 40, 43, 52, 59, 63, 65};
weight_mult_1 = {31, 31, 34, 36, 43, 45, 43, 28};
weight_mult_2 = {32, 32, 32, 32, 32, 32, 32, 32};
weight_mult_3 = {31, 31, 29, 28, 26, 28, 32, 43};
weight_mult_NE = {0, 21, 19, 13, 14, 24, 26, 35};

sum_weights_shift = 1;

NW_mult = 1;
```

If prediction_mode is PM_North, the 4 sub-predictions and weight multipliers shall be defined as follows:

```
prediction_0 = N -((true_error_W + true_error_N + true_error_NW + true_error_NE) Idiv 4);
prediction_1 = W - ((true_error_W × 2 + true_error_NW) >> 2);
prediction_2 = W + NE - N;
prediction_3 = N + (((N - NN) × 47) Idiv 64) - (true_error_N >> 2);

weight_mult_0 = {43, 38, 35, 34, 35, 33, 28, 23};
weight_mult_1 = {23, 21, 24, 27, 29, 31, 31, 31};
weight_mult_2 = {32, 32, 32, 32, 32, 32, 32, 32};
weight_mult_3 = {27, 23, 26, 29, 30, 35, 40, 51};
weight_mult_NE = {0, 29, 34, 29, 13, 13, 11, 9};

sum_weights_shift = 1;

NW_mult = 23;
```

The weights $weight_i$ for each of the 4 sub-predictions shall be computed as follows, based on the sub_error values already computed for sub-predictors of earlier samples. Here $sub_error_{i,LOCATION}$ shall be taken to have value 0 if the corresponding sample does not exist at that location.

```
mxe = max(quantized_error_W, quantized_error_N, quantized_error_NW);
if (x + 1 < xsize) mxe = max(mxe, quantized_error_NE);
if (x > 1) mxe = max(mxe, quantized_error_WW);

if (prediction_mode == PM_North) {
  err_sum_i = (((sub_error_i,N + sub_error_i,W) × 3) >> 1) + sub_error_i,NW + sub_error_i,WW +
sub_error_i,NE;
} else {
  err_sum_i = sub_error_i,N + sub_error_i,W + sub_error_i,NW + sub_error_i,WW + sub_error_i,NE;
}
weight_i = error2weight(err_sum_i) × weight_mult_i[mxe >> 1];
```

The function error2weight shall be defined as follows:

```
uint16_t error2weight(int err_sum) {
  return floor(150 × 512 / (58 + err_sum × sqrt(err_sum_i + 50)));
}
```

The prediction shall then be computed based on the sub-predictions i and their corresponding weight multipliers and shifts as follows:

```
sum_weights = sum_i=0..3(weight_i);
weighted_sum_predictions = sum_i=0..3(prediction_i × weight_i);
prediction = weighted_sum_predictions + (sum_weights >> sum_weights_shift)) Idiv
             sum_weights;

// if true_error_N, true_error_W and true_error_NW have the same sign
if (((true_error_N ^ true_error_W) | (true_error_N ^ true_error_NW)) > 0) {
  prediction = clamp(prediction, 0, max_value << Pbits);
} else {
  prediction = clamp(prediction, min(W, N, NE), max(W, N, NE));
}
```

The final max_error shall be the value of mxe, post-processed as follows:

```
max_error = mxe;
if (max_error != 0) {
  if ((true_error_W + true_error_N) × 40 + true_error_NW × NW_mult +
     true_error_NE × weight_mult_NE[mxe >> 1] <= 0)
       ++max_error;
} else if (N == W && N == NE) {
  max_error =
      (((true_error_W + true_error_N) | true_error_NE | true_error_NW) == 0 ? 16 : 1);
}
```

In the special case where the current pixel is in the first row or the first column, where some or all of the sub-predictors are not available, the sub-predictions, prediction and max_error shall instead be computed as follows::

When in both the first row and first column (the top left sample):

```
prediction_0 = prediction_1 = prediction_2 = prediction_3 = 27 << PBits;
prediction = clamp(prediction_0, 0, max_value << PBits);
max_error = 14;
```

Otherwise, when in the first row:

```
prediction_0 = W + (((W - WW) × 5) Idiv 16);
prediction_1 = prediction_2 = prediction_3 = W;
prediction = clamp(prediction_0, 0, max_value << PBits);
max_error = max(quantized_error_W, quantized_error_WW);
```

Otherwise, when in the first column:

```
prediction_0 = (N × 7 + NE + 4) >> 3;
```

```
prediction_1 = prediction_2 = prediction_3 = N;
prediction = prediction_0;
max_error = max(quantized_error_N, quantized_error_NE);
```

### E.3.4 16 bit prediction

The variables and subscripts `N`, `NW`, `NE`, `W`, `NN`, `WW` shall be defined the same way as for 8-bit modes as described in E.3.3, including their replacements if a sample does not exist.

Each individual prediction shall be computed as follows.

```
prediction_0 = N - ((true_error_W + true_error_N) × 3) Idiv 4;
prediction_1 = W - ((true_error_W + true_error_N + true_error_NW) × 5) Idiv 16;
prediction_2 = W + (((NE - N) × 13 + 8) >> 4);
prediction_3 = N - ((true_error_N + true_error_NW + true_error_NE) × 7) Idiv 32;
```

A set of weight multipliers shall be defined as follows:

```
mult_weight_0 = {30, 0, 28, 0, 4, 0, 4, 0, 8, 11, 11, 14, 15, 21, 21, 21, 22, 23};
mult_weight_1 = {33, 33, 33, 33, 33, 33, 33, 33, 33, 33, 33, 33, 33, 33, 33, 33, 33,
33};
mult_weight_2 = {31, 0, 31, 0, 31, 0, 34, 0, 35, 32, 33, 33, 33, 31, 28, 25, 22, 9};
mult_weight_3 = {30, 0, 32, 0, 56, 0, 57, 0, 51, 51, 50, 44, 44, 35, 34, 31, 29,
25};
```

The 4 predictor weights `weight_i` shall each be computed as follows. Here, the `sub_error_i,LOCATION` shall be taken to have value `0` if the corresponding sample does not exist at that location.

```
mxe = max(quantized_error_W, quantized_error_N, quantized_error_NW);
if (x + 1 < xsize) mxe = max(mxe, quantized_error_NE);
if (x > 1) mxe = max(mxe, quantized_error_WW);

err_sum_i = (((sub_error_i,N + sub_error_i,W) × 9) >> 3) + sub_error_i,NW + sub_error_i,WW +
sub_error_i,NE;

weight_i = error2weight_i(err_sum_i) × weight_mult_i[mxe];
weight_1 += 1;
```

Here, error2weight shall be defined as the following function:

```
uint16_t error2weight(int err_sum) {
  return (err_sum == 0) ? 0xffff : ((180 × 256) Idiv err_sum);
}
```

Note that in the case of the first row, first column and last column only a subset of these will be available.

The `prediction` shall then be computed based on the sub-predictions `i` and their corresponding `weights` as follows, using signed 64-bit integer precision to handle potential overflow:

```
sum_weights = sum_i=0..3(weight_i);
weighted_sum_predictions = sum_i=0..3(prediction_i × weight_i);
```

```
prediction = (weighted_sum_predictions + (sum_weights >> 2)) Idiv sum_weights;
```

The final `prediction` shall be clamped between `min(N + 1, W, NE)` and `max(N - 1, W, NE)`.

The final `quantized_error` shall be the value of `mxe`, post-processed as follows:

```
if (mxe && mxe <= 8) {
  if (true_error_W + true_error_N × 2 + true_error_NW + true_error_NE < 0) --mxe;
}
quantized_error = mxe;
```

In the special case where the current pixel is in the first row or the first column, where some or all of the sub-predictors are not available, the sub-predictions, prediction and max_error shall instead be computed as follows:
When in both the first row and first column (the top left sample):

```
prediction_0 = prediction_1 = prediction_2 = prediction_3 = 3584;
prediction = clamp(prediction_0, 0, max_value);
max_error = 17;
```

Otherwise, when in the first row:

```
prediction_0 = prediction_1 = prediction_2 = prediction_3 = W + (W - WW Idiv 4);
prediction = clamp(prediction_0, 0, max_value)
max_error = max(quantized_error_W, quantized_error_WW);
```

Otherwise, when in the first column:

```
prediction_1 = (N × 3 + NE + 2) >> 2;
prediction_0 = prediction_2 = prediction_3 = N;
prediction = prediction_1;
max_error = max(quantized_error_N, quantized_error_NE);
```

# Annex F
(normative)
# Adaptive quantization

## F.1 Overview

In var-DCT mode, quantized DC and AC coefficients qX, qY and qB shall be converted into values dX, dY and dB as specified in F.2 and F.3, respectively.

## F.2 DC dequantization

Let $mX_{DC}$, $mY_{DC}$ and $mB_{DC}$ be the three per-channel DC dequantization multipliers, as decoded in C.4.2.
For each DC group, the dequantization process shall be influenced by the number of extra_levels_xyb that have been used in that part of the image and for that channel, as described in C.5.2.
To compute dequantized values, the following formula shall be used:

```
dX = mX_DC × qX / (1 << extra_levels_X);
dY = mY_DC × qY / (1 << extra_levels_Y);
dB = mB_DC × qB / (1 << extra_levels_B);
```

After dequantizing DC coefficients, the decoder shall apply the following adaptive smoothing algorithm, unless kSkipAdaptiveDCSmoothing is set in frame_header.flags.

For each DC sample of the image that is not in the first row, last row, first column, or last column, the decoder shall compute a weighted average wa of the 9 samples that are in neighbouring rows and columns, using a weight of 0.05226273532324128 for the current sample s, 0.20345139757231578 for horizontally/vertically adjacent samples (taxicab distance 1), and 0.0334829185968739 for diagonally adjacent (taxicab distance 2) samples. Let dcq be the dequantization range for the current sample, as defined in C.5.2. Then, the smoothed value shall be wa if (s - 0.5 × dcq <= wa && wa <= s + 0.5 × dcq), and shall be s otherwise.

After applying this smoothing, the decoder shall use the process described in I.2.7 to compute the top-left X / 8 × Y / 8 coefficients of each varblock of size X × Y, using the corresponding X / 8 × Y / 8 samples from the dequantized DC image.

## F.3 AC dequantization

Every quantized AC coefficient quant shall first be bias-adjusted using the following procedure, depending on its channel (0 for X, 1 for Y or 2 for Z):

```
if (|quant| <= 1) quant ×= (1 - kOneBias[channel]);
else quant -= 0.145 / quant;
```

kOneBias shall be {0.05465007330715401, 0.07005449891748593, 0.04993510337343655}.
The resulting quant shall then be multiplied by a per-block multiplier, the value of AcMul at the coordinates of the 8 × 8 region containing the current sample.
The final dequantized value shall then be obtained by multiplying the result by a multiplier defined by the channel, the transform type and the index of the coefficient inside the varblock, as specified in C.6.1.

# Annex G
(normative)
# Chroma from Luma

This Annex shall only apply to var-DCT mode.

Each X, Y and B sample shall be reconstructed from the dequantized samples dX, dY and dB using a linear chroma from luma model. The reconstruction shall use the per-tile colour correlation coefficient multipliers kX and kB to restore the correlation between the X/B and the Y channel. The reconstruction shall then be obtained as:

```
Y = dY;
X = dX + kX × Y;
B = dB + kB × Y;
```

The correlation coefficients kX and kB shall be computed as follows:

```
kX = (x_factor - 128) / 256;
kB = 0.935669 + (b_factor - 128) / 128;
```

For DC coefficients, x_factor and b_factor shall be x_factor_dc and b_factor_dc. For AC coefficients, x_factor and b_factor shall be values from XFromY and BFromY, respectively, at the coordinates of the 64 × 64 region containing the current sample.

# Annex H
(normative)
# AC prediction

This Annex shall only apply to var-DCT mode, and shall be skipped if `kSkipPredictLf` is set in `frame_header.flags`.

Before applying inverse integral transforms, but after performing adaptive DC smoothing (as described in F.2) and LLF computation, the decoder shall modify the top-left `X / 4 × Y / 4` coefficients of each `X × Y` varblock as follows.

For each group of size `xsize` × `ysize`, the decoder shall compute an image `DC2x2` of dimensions `(xsize × 2 + 4) × (ysize × 2 + 4)` as follows. For each varblock of size `X × Y` of the group, the decoder shall use the procedure in I.2.6 (with `N = 4`, using only the top `X / 8 × Y / 8` coefficients as an input) to compute a corresponding `X / 4 × Y / 4` area in `DC2x2`, starting from coordinate `(2, 2)` in the `DC2x2` image. Then, for each dequantized DC sample of value `v` in a `1x1` border around the group (DC samples that would be out of image bounds shall be replicated by mirroring, as defined in 6.5), the decoder shall set the corresponding 4 border pixels in the `DC2x2` image to `v`.

The decoder shall then apply the following adaptive smoothing procedure to the resulting `DC2x2` image. For each sample of the `DC2x2` image that is not in the first two rows, last two rows, first two columns, or last two columns, the decoder shall compute a weighted average `wa` of the 9 samples that are in neighbouring rows and columns, using a weight of 0.12 for the current sample s, 0.2 for horizontally/vertically adjacent samples (taxicab distance 1), and 0.02 for diagonally adjacent (taxicab distance 2) samples. Let `dcq` be the DC dequantization multiplier for the current channel, as defined in C.4.2. Then, the smoothed value shall be `wa` if `(s - 1.7 × dcq <= wa && wa <= s + 1.7 × dcq)`, and `s` otherwise.

Finally, for each varblock of size `X × Y` within the group, the decoder shall use the procedure in I.2.7 to compute the top-left corner of `X / 4 × Y / 4` coefficients from the corresponding area in `DC2x2`, and shall increase all coefficients by the computed value, excluding the top-left `X / 8 × Y / 8` corner.

# Annex I
(normative)
# Integral transforms

## I.1 Overview

In var-DCT mode, the decoder shall apply DCT as specified in I.2.

In modular mode, the decoder shall apply Squeeze as specified in I.3.

## I.2 Variable DCT

### I.2.1 One-dimensional DCT and IDCT

One-dimensional discrete cosine transforms and their inverses (DCT) shall be computed as follows. The input shall be a vector of size `s`, and `s` shall be a power of two. If `s == 8`, the forward DCT transform shall be:

```
c1 = 1 / sqrt(2);
c2 = cos(3 × π / 8);
c3 = 1 / (2 × cos(3 × π / 8));
c4 = sqrt(2) × cos(3 × π / 8);

t00 = i0 + i7;
t01 = i0 - i7;
t02 = i3 + i4;
t03 = i3 - i4;
t04 = i2 + i5;
t05 = i2 - i5;
t06 = i1 + i6;
t07 = i1 - i6;
t08 = t00 + t02;
t09 = t00 - t02;
t10 = t06 + t04;
t11 = t06 - t04;
t12 = t07 + t05;
t13 = t01 + t07;
t14 = t05 + t03;
t15 = t11 + t09;
t16 = t14 - t13;
t17 = c1 × t15;
t18 = c1 × t12;
t19 = c2 × t16;
t20 = t01 + t18;
t21 = t01 - t18;
t22 = c3 × t13 + t19;
t23 = c4 × t14 + t19;
```

```
o0 = t08 + t10;
o1 = t20 + t22;
o2 = t09 + t17;
o3 = t21 - t23;
o4 = t08 - t10;
o5 = t21 + t23;
o6 = t09 - t17;
o7 = t20 - t22;
```

Where `i0` to `i7` represent the inputs and `o0` to `o7` the outputs.
The inverse (IDCT) transform for `s == 8` shall be

```
c1 = sqrt(2);
c2 = 1 / cos(3 × π / 8);
c3 = 2 × cos(3 × π / 8);
c4 = 2 × sqrt(2) × cos(3 × π / 8);

t00 = i0 + i4;
t01 = i0 - i4;
t02 = i6 + i2;
t03 = i6 - i2;
t04 = i7 + i1;
t05 = i7 - i1;
t06 = i5 + i3;
t07 = i5 - i3;
t08 = t04 + t06;
t09 = t04 - t06;
t10 = t00 + t02;
t11 = t00 - t02;
t12 = t07 - t05;
t13 = c3 × t12;
t14 = c1 × t03 + t02;
t15 = t01 - t14;
t16 = t01 + t14;
t17 = c2 × t05 + t13;
t18 = c4 × t07 +  t13;
t19 = t08 + t17;
t20 = c1 × t09 + t19;
t21 = t18 - t20;
o0 = t10 + t08;
o1 = t15 - t19;
o2 = t16 + t20;
o3 = t11 + t21;
o4 = t11 - t21;
o5 = t16 - t20;
o6 = t15 + t19;
```

```
o7 = t10 - t08;
```

Transformations with `s == 2`, `s == 4`, `s == 16` and `s == 32` shall use the DCT defined in "A low multiplicative complexity fast recursive DCT-2 algorithm", as generated by Electronic Insert I.1 "DCT-II / DCT-III code generator" (see Annex R).

### I.2.2 Two-dimensional DCT and IDCT

In this subclause, `Transpose` is a matrix transpose and `ColumnIDCT/ColumnDCT` performs a one dimensional IDCT/DCT on each column vector of the given matrix as defined in I.2.1. The DCT transform of a square varblock `DCT_2D(varblock)` shall be:

```
dct1 = ColumnDCT(varblock);
dct1_t = Transpose(dct1);
result = ColumnDCT(dct1_t);
```

If the varblock is not a square, `DCT_2D(varblock)` shall be:

```
dct1 = ColumnDCT(varblock);
dct1_t = Transpose(dct1);
dct2 = ColumnDCT(dct1_t);
result = Transpose(dct2);
```

NOTE The final `Transpose` ensures the DCT image has the same dimensions as the original varblock.

Similarly, the inverse DCT (IDCT) of a square varblock `IDCT_2D(coefficients)` shall be:

```
dct1 = ColumnIDCT(coefficients);
dct1_t = Transpose(dct1);
result = ColumnIDCT(dct1_t)
```

If the varblock is not square, `IDCT_2D(coefficients)` shall be:

```
dct1 = ColumnIDCT(varblock);
dct1_t = Transpose(dct1);
dct2 = ColumnIDCT(dct1_t);
result = Transpose(dct2);
```

The AFV transform uses the following orthonormal basis (here and in the rest of the subclause, we will assume that a `4 × 4` matrix is stored in an array such that entry `(x, y)` corresponds to index `4 × y + x`:

```
AFVBasis[16][16] = {{0.25, 0.25, 0.25, 0.25, 0.25, 0.25, 0.25, 0.25, 0.25, 0.25,
0.25, 0.25, 0.25, 0.25, 0.25, 0.25,},
{0.876902929799142, 0.2206518106944235, -0.10140050393753763, -0.1014005039375375,
0.2206518106944236, -0.10140050393753777, -0.10140050393753772,
-0.10140050393753763, -0.10140050393753758, -0.10140050393753769,
-0.1014005039375375, -0.10140050393753768, -0.10140050393753768,
-0.10140050393753759, -0.10140050393753763, -0.10140050393753741,},
{0.0, 0.0, 0.40670075830260755, 0.44444816619734445, 0.0, 0.0, 0.19574399372042936,
0.2929100136981264, -0.40670075830260716, -0.19574399372042872, 0.0,
0.11379074460448091, -0.44444816619734384, -0.29291001369812636,
```

```
-0.1137907446044814, 0.0,},
{0.0, 0.0, -0.21255748058288748, 0.3085497062849767, 0.0, 0.4706702258572536,
-0.1621205195722993, 0.0, -0.21255748058287047, -0.16212051957228327,
-0.47067022585725277, -0.1464291867126764, 0.3085497062849487, 0.0,
-0.14642918671266536, 0.4251149611657548,},
{0.0, -0.7071067811865474, 0.0, 0.0, 0.7071067811865476, 0.0, 0.0, 0.0, 0.0, 0.0,
0.0, 0.0, 0.0, 0.0, 0.0, 0.0,},
{-0.4105377591765233, 0.6235485373547691, -0.06435071657946274,
-0.06435071657946266, 0.6235485373547694, -0.06435071657946284,
-0.0643507165794628, -0.06435071657946274, -0.06435071657946272,
-0.06435071657946279, -0.06435071657946266, -0.06435071657946277,
-0.06435071657946277, -0.06435071657946273, -0.06435071657946274,
-0.0643507165794626,},
{0.0, 0.0, -0.4517556589999482, 0.15854503551840063, 0.0, -0.04038515160822202,
0.007418226379242387 5, 0.39351034269210167, -0.45175565899994635,
0.00741822637924435 1, 0.1107416575309343, 0.08298163094882051, 0.15854503551839705,
0.3935103426921022, 0.0829816309488214, -0.45175565899994796,},
{0.0, 0.0, -0.304684750724869, 0.5112616136591823, 0.0, 0.0, -0.290480129728998,
-0.06578701549142804, 0.304684750724884, 0.2904801297290076, 0.0,
-0.23889773523344604, -0.5112616136592012, 0.06578701549142545,
0.23889773523345467, 0.0,},
{0.0, 0.0, 0.3017929516615495, 0.25792362796341184, 0.0, 0.16272340142866204,
0.09520022653475037, 0.0, 0.3017929516615503, 0.09520022653475055,
-0.16272340142866173, -0.35312385449816297, 0.25792362796341295, 0.0,
-0.3531238544981624, -0.6035859033230976,},
{0.0, 0.0, 0.40824829046386274, 0.0, 0.0, 0.0, 0.0, -0.4082482904638628,
-0.4082482904638635, 0.0, 0.0, -0.40824829046386296, 0.0, 0.4082482904638634,
0.408248290463863, 0.0,},
{0.0, 0.0, 0.1747866975480809, 0.0812611176717539, 0.0, 0.0, -0.3675398009862027,
-0.307882213957909, -0.17478669754808135, 0.3675398009862011, 0.0,
0.4826689115059883, -0.08126111767175039, 0.30788221395790305,
-0.48266891150598584, 0.0,},
{0.0, 0.0, -0.21105601049335784, 0.18567180916109802, 0.0, 0.0,
0.49215859013738733, -0.38525013709251915, 0.21105601049335806,
-0.49215859013738905, 0.0, 0.17419412659916217, -0.18567180916109904,
0.3852501370925211, -0.1741941265991621, 0.0,},
{0.0, 0.0, -0.14266084808807264, -0.3416446842253372, 0.0, 0.7367497537172237,
0.24627107722075148, -0.08574019035519306, -0.14266084808807344,
0.24627107722075137, 0.14883399227113567, -0.04768680350229251,
-0.3416446842253373, -0.08574019035519267, -0.047686803502292804,
-0.14266084808807242,},
{0.0, 0.0, -0.13813540350758585, 0.3302282550303788, 0.0, 0.08755115000587084,
-0.07946706605909573, -0.4613374887461511, -0.13813540350758294,
-0.07946706605910261, 0.49724647109535086, 0.12538059448563663, 0.3302282550303805,
-0.4613374887461554, 0.12538059448564315, -0.13813540350758452,},
{0.0, 0.0, -0.17437602599651067, 0.0702790691196284, 0.0, -0.2921026642334881,
0.3623817333531167, 0.0, -0.1743760259965108, 0.36238173335311646,
```

```
0.29210266423348785, -0.4326608024727445, 0.07027906911962818, 0.0,
-0.4326608024727457, 0.34875205199302267,},
{0.0, 0.0, 0.11354987314994337, -0.07417504595810355, 0.0, 0.19402893032594343,
-0.435190496523228, 0.21918684838857466, 0.11354987314994257, -0.4351904965232251,
0.5550443808910661, -0.25468277124066463, -0.07417504595810233, 0.2191868483885728,
-0.25468277124066413, 0.1135498731499429,},
};
```

The AFV DCT transform `AFV_DCT` of a square `4 × 4` varblock from `samples` to `coefficients` shall be computed as:

```
for (i = 0; i < 16; ++i) {
  scalar = 0;
  for (j = 0; j < 16; ++j) {
    scalar += samples[j] × AFVBasis[i][j];
  }
  coefficients[i] = scalar;
}
```

The AFV IDCT transform `AFV_IDCT` of a square `4 × 4` varblock from `coefficients` to `samples` shall be computed as:

```
for (i = 0; i < 16; ++i) {
  sample = 0;
  for(j = 0; j < 16; ++j) {
    sample += coefficients[j] × AFVBasis[j][i];
  }
  samples[i] = sample;
}
```

### I.2.3 Samples to coefficients

#### I.2.3.1 General

The partitioning of the image into varblocks shall be defined by `DctSelect`. The size of each varblock shall be identified by the `DctSelect` value at the coordinates of the top-left 8 × 8 region of the varblock. The `DctSelect` values shall be “Identity” or “DCT `N` × `M`”, with `N` and `M` taking values 2, 4, 8, 16 or 32.

The DCT coefficients shall be obtained using `coefficients = DCT_2D(varblock)` for varblock with sizes 8 × 8, 8 × 16, 8 × 32, 16 × 8, 16 × 16, 32 × 8, 32 × 16, 32 × 32, 8 × 4, 4 × 8.

#### I.2.3.2 DCT 2 × 2

The DCT 2 × 2 and DCT 4 × 4 shall be computed on 8 × 8 varblocks. Let `AuxDCT2x2(block, S)` denote:

```
num_2x2 = S / 2;
for (y = 0; y < num_2x2; y++) {
  for (x = 0; x < num_2x2; x++) {
    c00 = block(x × 2, y × 2);
    c01 = block(x × 2 + 1, y × 2);
```

```
    c10 = block(x × 2, y × 2 + 1);
    c11 = block(x × 2 + 1, y × 2 + 1);
    r00 = (c00 + c01 + c10 + c11) / 4;
    r01 = (c00 + c01 - c10 - c11) / 4;
    r10 = (c00 - c01 + c10 - c11) / 4;
    r11 = (c00 - c01 - c10 + c11) / 4;
    result(x, y) = r00;
    result(num_2x2 + x, y) = r01;
    result(x, y + num_2x2) = r10;
    result(num_2x2 + x, y + num_2x2) = r11;
  }
}
```

Where `block` is a matrix of size `S × S`, and `S` is a power of two. If `block` is larger than `S × S` the function is applied to the top `S × S` cells and the rest of the cells shall be copied to the answer. The `result` (an 8 × 8 matrix containing the resulting coefficients) of DCT 2 × 2 shall be:

```
temp8x8 = AuxDCT2x2(block, 8);
top4x4 = AuxDCT2x2(temp8x8, 4);
result = AuxDCT2x2(top4x4, 2);
```

### I.2.3.3 DCT 4 × 4

The DCT 4 × 4 shall be computed on 8 × 8 blocks by dividing the block into 4 sub-matrices each of size 4 × 4. For each sub-matrix, represented by the variable `sub_coeff`, an 8 × 8 array of coefficients shall be computed using:

```
sub_coeff = DCT_2D(sub_block);
for (iy = 0; iy < 4; iy++) {
  for (ix = 0; ix < 4; ix++) {
    coefficients(x + ix × 2, y + iy × 2) = sub_coeff(ix, iy);
  }
}
```

Where `x` and `y` are indices taking values zero or one indicating the position of each sub-block. The top 2 × 2 cells of the `coefficients` matrix represent the DC coefficients of each sub-block, and the rest of the cells represent the AC coefficients. The DC coefficients shall be transformed by a final 2 × 2 DCT using:

```
coefficients = AuxDCT2x2(coefficients, 2);
```

### I.2.3.4 Identity

When `DctSelect` is "identity", the 8 × 8 input matrix `pixels` shall be transformed to the 8 × 8 matrix `coefficients` as follows:

```
for (y = 0; y < 2; y++) {
  for (x = 0; x < 2; x++) {
    block_dc = 0;
    for (iy = 0; iy < 4; iy++) {
      for (ix = 0; ix < 4; ix++) {
        block_dc += sample(x × 4 + ix, y × 4 + iy);
      }
```

```
    }
    block_dc /= 16;
    for (iy = 0; iy < 4; iy++) {
      for (ix = 0; ix < 4; ix++) {
        if (ix == 1 && iy == 1) continue;
        coefficients(x + ix × 2, y + iy × 2) =
            sample(x × 4 + ix, y × 4 + iy) - sample(x × 4 + 1, y × 4 + 1);
      }
    }
    coefficients(x + 2, y + 2) = coefficients(x, y);
    coefficients(x, y) = block_dc;
  }
}
coefficients = AuxDCT2x2(coefficients, 2);
```

#### I.2.3.5 DCT 8 × 4

The DCT 8 × 4 `coefficients` shall be computed on 8 × 8 varblocks of `samples` by dividing them into two 8 × 4 vertical blocks `samples_8x4`, using the `DCT_2D` function defined in I.2.2:

```
for (x = 0; x < 2; x++) {
  coefficients_8x4 = DCT_2D(samples_8x4[x]);
  for (iy = 0; iy < 8; iy++) {
     for (ix = 0; ix < 4; ix++) {
        coefficients(x + ix × 2, iy) = coefficients_8x4(ix, iy);
     }
  }
}
coef0 = coefficients(0, 0);
coef1 = coefficients(1, 0);
coefficients(0, 0) = (coef0 + coef1) × 0.5;
coefficients(1, 0) = (coef0 - coef1) × 0.5;
```

#### I.2.3.6 DCT 4 x 8

The DCT 4 × 8 `coefficients` shall be computed on 8 × 8 varblocks of `samples` by dividing them into two 4 × 8 horizontal blocks `samples_4x8`, using the `DCT_2D` function defined in I.2.2:

```
for (y = 0; y < 2; y++) {
  coefficients_4x8 = DCT_2D(samples_4x8[y]);
  for (iy = 0; iy < 4; iy++) {
     for (ix = 0; ix < 8; ix++) {
        coefficients(ix, y + iy × 2) = coefficients_4x8(ix, iy);
     }
  }
}
coef0 = coefficients(0, 0);
coef1 = coefficients(0, 1);
```

```
coefficients(0, 0) = (coef0 + coef1) × 0.5;
coefficients(0, 1) = (coef0 - coef1) × 0.5;
```

#### I.2.3.7 AFV

The AFV `coefficients` shall be computed on 8 × 8 varblocks of `samples` by dividing them into four 4 × 4 square blocks and using the `AFV_DCT` function defined in I.2.2:

```
for (y = 0; y < 2; y++) {
  for (x = 0; x < 2; x++) {
      for (iy = 0; iy < 4; iy++) {
        for (ix = 0; ix < 4; ix++) {
          samples_afv(x ? 3 - ix : ix, y ? 3 - iy : iy) =
              sample(x × 4 + ix, y × 4 + iy);
        }
      }
      coeffs_afv = AFV_DCT(samples_afv);
      for (iy = 0; iy < 4; iy++) {
        for (ix = 0; ix < 4; ix++) {
           coefficients(x + ix × 2, y + iy × 2) = coeffs_afv(ix, iy);
        }
      }
  }
}
coef1 = coefficients(0, 0) × 0.25;
coef2 = coefficients(1, 0) × 0.25;
coef3 = coefficients(0, 1) × 0.25;
coef4 = coefficients(1, 1) × 0.25;
coefficients(0, 0) = (coef1 + coef2 + coef3 + coef4) × 0.25;
coefficients(1, 0) = (coef1 + coef2 - coef3 - coef4) × 0.25;
coefficients(0, 1) = (coef1 - coef2 + coef3 - coef4) × 0.25;
coefficients(1, 1) = (coef1 - coef2 - coef3 + coef4) × 0.25;
```

### I.2.4 Coefficients to samples

#### I.2.4.1 General

The pixels shall be obtained using `samples = IDCT_2D(coefficients)` for any varblock with sizes 8 × 8, 8 × 16, 8 × 32, 16 × 8, 16 × 16, 32 × 8, 32 × 16, 32 × 32, 8 × 4, 4 × 8.

#### I.2.4.2 IDCT 2 × 2

The IDCT 2 × 2 is computed from 8 × 8 blocks. Let `AuxIDCT2x2(block, S)` denote the following:

```
num_2x2 = S / 2;
for (y = 0; y < num_2x2; y++) {
  for (x = 0; x < num_2x2; x++) {
    c00 = block(x, y);
    c01 = block(num_2x2 + x, y);
```

```
    c10 = block(x, y + num_2x2);
    c11 = block(num_2x2 + x, y + num_2x2);
    r00 = c00 + c01 + c10 + c11;
    r01 = c00 + c01 - c10 - c11;
    r10 = c00 - c01 + c10 - c11;
    r11 = c00 - c01 - c10 + c11;
    result(x × 2, y × 2) = r00;
    result(x × 2 + 1, y × 2) = r01;
    result(x × 2, y × 2 + 1) = r10;
    result(x × 2 + 1, y × 2 + 1) = r11;
  }
}
```

Where `block` is a matrix of size `S × S`, and `S` is a power of two. In case `block` is larger than `S × S`, the `AuxIDCT2x2` function modifies only the top `S × S` cells. In that case, the decoder shall copy the rest of the cells from `block` to `result`. The inverse DCT 2 × 2 shall be

```
block = AuxIDCT2x2(block, 2);
block = AuxIDCT2x2(block, 4);
result = AuxIDCT2x2(block, 8);
```

### I.2.4.3 IDCT 4 × 4

The IDCT 4 × 4 shall be computed on the 8 × 8 coefficient matrix as follows:

```
dcs = AuxIDCT2x2(coefficients, 2);
for (y = 0; y < 2; y++) {
  for (x = 0; x < 2; x++) {
    for (iy = 0; iy < 4; iy++) {
      for (ix = 0; ix < 4; ix++) {
        if (ix == 0 && iy == 0) continue;
        block(ix, iy) = coefficients(x + ix × 2, y + iy × 2);
      }
    }
    block(0, 0) = dcs(x, y);

    sample(x, y) = IDCT_2D(block);
  }
}
```

Where `block` is a 4 × 4 matrix, and `sample` is the answer represented here as a 2 × 2 × 4 × 4 4-index array. The decoder shall re-shape the `sample` array to an 8 × 8 matrix using `result(4 × i + k, 4 × j + L) = sample(i, j, k, L)`.

### I.2.4.4 Identity

When DctSelect is "identity", the 8 × 8 input matrix `coefficients` shall be transformed to the 8 × 8 matrix `samples` as follows:

```
dcs = AuxIDCT2x2(coefficients, 2);
```

```
for (y = 0; y < 2; y++) {
  for (x = 0; x < 2; x++) {
    block_dc = dcs(x, y);
    residual_sum = 0;
    for (iy = 0; iy < 4; iy++) {
      for (ix = 0; ix < 4; ix++) {
        if (ix == 0 && iy == 0) continue;
        residual_sum += coefficients(x + ix × 2, y + iy × 2);
      }
    }
    sample(4 × x + 1, 4 × y + 1) = block_dc - residual_sum / 16.0;
    for (iy = 0; iy < 4; iy++) {
      for (ix = 0; ix < 4; ix++) {
        if (ix == 1 && iy == 1) continue;
        sample(x × 4 + ix, y × 4 + iy) =
            coefficients(x + ix × 2, y + iy × 2) + sample(4 × x + 1, 4 × y + 1);
      }
    }
    sample(4 × x, 4 × y) =
        coefficients(x + 2, y + 2) + sample(4 × x + 1, 4 × y + 1);
  }
}
```

### I.2.4.5 IDCT 8 × 4

In the case of DCT 8 × 4, the 8 × 8 `samples` shall be reconstructed from the 8 × 8 `coefficients` by dividing them into two 8 × 4 vertical blocks `samples_8x4`, as follows. A temporary 8 × 4 matrix `coeffs_8x4` shall be used. The `IDCT_2D` function is defined in I.2.2.

```
coef0 = coefficients(0, 0);
coef1 = coefficients(1, 0);
dcs = {coef0 + coef1,
       coef0 - coef1};
for (x = 0; x < 2; x++) {
  coeffs_8x4(0, 0) = dcs[x];
  for (iy = 0; iy < 8; iy++) {
    for (ix = 0; ix < 4; ix++) {
      if (ix == 0 && iy == 0) continue;
        coeffs_8x4(ix, iy) = coefficients(x + ix × 2, iy);
    }
  }
  samples_8x4[x] = IDCT_2D(coeffs_8x4);
}
```

### I.2.4.6 IDCT 4 × 8

In the case of DCT 4 × 8, the 8 × 8 `samples` shall be reconstructed from the 8 × 8 `coefficients` by dividing them into two 4 × 8 horizontal blocks `samples_4x8`, as follows. A temporary 8 × 4 matrix `coeffs_4x8` shall be used. The `IDCT_2D` function is defined in I.2.2.

```
coef0 = coefficients(0, 0);
coef1 = coefficients(0, 1);
dcs = {coef0 + coef1,
       coef0 - coef1};
for (y = 0; y < 2; y++) {
  coeffs_4x8(0, 0) = dcs[x]
  for (iy = 0; iy < 4; iy++;) {
    for (ix = 0; ix < 8; ix++) {
      if (ix == 0 && iy == 0) continue;
        coeffs_4x8(ix, iy) = coefficients(ix, y + iy × 2);
    }
  }
  samples_4x8[y] = IDCT_2D(coeffs_4x8);
}
```

#### I.2.4.7 AFV

In the case of AFV, the 8 × 8 `samples` shall be reconstructed from the 8 × 8 `coefficients` by dividing them into four 4 × 4 square blocks, as follows. A temporary 8 × 4 matrix `coeffs_afv` shall be used. `AFV_IDCT` is defined in I.2.2.

```
coef1 = coefficients(0, 0) × 4.0;
coef2 = coefficients(1, 0) × 4.0;
coef3 = coefficients(0, 1) × 4.0;
coef4 = coefficients(1, 1) × 4.0;
dcs = {coef1 + coef2 + coef3 + coef4,
       coef1 + coef2 - coef3 - coef4,
       coef1 - coef2 + coef3 - coef4,
       coef1 - coef2 - coef3 + coef4};
for (y = 0; y < 2; y++) {
  for (x = 0; x < 2; x++) {
    coeffs_afv(0, 0) = dcs[y × 2 + x];
    for (iy = 0; iy < 4; iy++) {
      for (ix = 0; ix < 4; ix++) {
        if (ix == 0 && iy == 0) continue;
        coeffs_afv(ix, iy) = coefficients(x + ix × 2, y + iy × 2);
      }
    }
    samples_afv = AFV_IDCT(coeffs_afv);
    for (iy = 0; iy < 4; iy++) {
      for (ix = 0; ix < 4; ix++) {
        sample(x × 4 + ix, y × 4 + iy) =
             samples_afv(x ? 3 - ix : ix, y ? 3 - iy : iy);
      }
    }
  }
}
```

### I.2.5 Natural ordering of the DCT coefficients

For every `DctSelect` value (Identity, DCT 2 × 2, etc), the natural order of the coefficients shall be computed as follows. The varblock size `(bxsize, bysize)` for a DctSelect value with name "DCT `N` × `M`" is `bxsize = max(8, N)` and `bysize = max(8, M)`, respectively. The varblock size for all other transforms is `bxsize = bysize = 8`. The natural ordering of the DCT coefficients is defined as a vector order of cell positions `(x, y)` between `(0, 0)` and `(bxsize, bysize)`, described below. The number of elements in the vector `order` is therefore `bxsize × bysize`, and shall be computed using:

```
order = concatenate(LLF, HF);
```

Where `concatenate(LLF, HF)` is a function that creates an array containing all the elements of `LLF` in their original order, followed by the elements of `HF` also in their original order. `LLF` is a vector of lower frequency components, containing cells `(x, y)` satisfying

```
(x < (bxsize / 8)) && (y < (bysize / 8))
```

The cells `(x, y)` that do not satisfy the condition described above, shall go in the higher frenciencies vector `HF`.

The rest of this subclause specifies how to order the elements within each of the arrays `LLF` and `HF`. The pairs `(x, y)` in the `LLF` vector shall be sorted in ascending order according to the value `y × bxsize + x`.

For the pairs `(x, y)` in the `HF` vector, the decoder shall compute first the value of the variables `key1` and `key2` defined by:

```
cx = bxsize / 8;
cy = bysize / 8;
max_dim = max(cx, cy);
scaled_y = y × max_dim / cy;
scaled_x = x × max_dim / cx;
key1 = scaled_x + scaled_y;
if (key1 Umod 2 == 0) {
  key2 = scaled_x - scaled_y;
} else {
  key2 = scaled_y - scaled_x;
}
```

Subsequently, the decoder shall sort the `(x, y)` pairs on the vector `HF` in ascending order according to the value `key1`. In case of a tie, the decoder shall also sort in ascending order according to the value `key2`.

The *order ID* of each transform shall be defined as follows:

- For DCT8×8, the order ID shall be 0.
- For IDENTITY, DCT2×2, DCT4×4, DCT4×8, DCT8×4 and AFV, the order ID shall be 1.
- For DCT16×16, the order ID shall be 2.
- For DCT32×32, the order ID shall be 3.
- For DCT16×8, the order ID shall be 4.
- For DCT8×16, the order ID shall be 5.
- For DCT32×8, the order ID shall be 6.
- For DCT8×32, the order ID shall be 7.
- For DCT32×16, the order ID shall be 8.
- For DCT16×32, the order ID shall be 9.

### I.2.6 Downsampled image from LF coefficients

#### I.2.6.1 General

This subclause specifies how to obtain an `N` times downsampled image from the LF coefficients for each `DctSelect` value, where `N` can be 4 or 8. The input will be a varblock represented as a matrix of size `bxsize × bysize` containing the DCT coefficients, from which the top `bxsize / N × bysize / N` cells are called the LF coefficients. If the caller requests a `4×` downsampled image, but only provides the top `bxsize / 8 × bysize / 8` cells, the missing cells shall be filled with `0`. The possible values for `bxsize` and `bysize` are 8, 16 and 32.

In this subclause, `ScaleF` shall be:

```
ScaleF(N, n, x) = sqrt(n × N) × D(N, x) × I(n, x) × C(N, n, x);
```

With subroutines:

```
I8(int N, int u) {
  eps = (u == 0) ? sqrt(0.5) : 1;
  return sqrt(2.0 / N) × eps × cos(u × π / (2.0 × N));
}

D8(int N, int u) { return 1 / (N × I8(N, u)); }

I(int N, int u) {
  if (N == 8)
    return I8(N, u);
  else
    return D8(N, u);
}

D(int N, int u) {
  if (N == 8)
    return D8(N, u);
  else
    return I8(N, u);
}

C(int N, int n, int x) {
  if (n > N) return 1 / C(n, N, x);
  if (n == N) return 1;
  else return cos(x × π / (2 × N)) × C(N / 2, n, x);
}
```

#### I.2.6.2 DctSelect DCT 8 × 8, 8 × 16, 8 × 32, 16 × 8, 16 × 16, 32 × 8, 32 × 16, 32 × 32

`output` shall be a matrix of samples downsampled by a factor of N as follows:

```
cx = bxsize / N;
cy = bysize / N;
```

```
for (y = 0; y < bysize; y++) {
  for (x = 0; x < bxsize; x++) {
    llf(x, y) = input(x, y) × ScaleF(bysize, cy, y) × ScaleF(bxsize, cx, x);
  }
}
output = IDCT_2D(llf);
```

#### I.2.6.3 DctSelect Identity, DCT 2 × 2, 4 × 4, 8 × 4, 4 × 8, AFV

The `input` is an 8 × 8 matrix, and the `output` shall be a 1 × 1 or 2 × 2 matrix for the 8× and 4× downsamples respectively. All the cells of the `output` matrix have the value of `input[0][0]`.

### I.2.7 Low frequency coefficients from downsampled image

#### I.2.7.1 General

This subclause specifies how to obtain the low frequency DCT coefficients from an `N` times downsampled image for each `DctSelect` value, where `N` can be 4 or 8. This process is the inverse of the downsampling process from I.2.6.

The input will be a matrix with `bxsize / N` columns and `bysize / N` rows containing the DC coefficients, or equivalently, the downsampled image. The possible values for `bxsize` and `bysize` are 8, 16, and 32.

The output will be a matrix with `bxsize / N` columns and `bysize / N` rows. If the caller requires a matrix of a larger size, the decoder shall zero-initialize the other elements.

#### I.2.7.2 DctSelect 8 × 8, 8 × 16, 8 × 32, 16 × 8, 16 × 16, 32 × 8, 32 × 16, 32 × 32

The `output` matrix shall be computed using

```
cx = bxsize / 8;
cy = bysize / 8;
C = DCT_2D(input)
for (y = 0; y < cy y++) {
  for (x = 0; x < cx; x++) {
    output(x, y) = C(x, y) × ScaleF(cy, bysize, y) × ScaleF(cx, bxsize, x);
  }
}
```

#### I.2.7.3 DctSelect Identity, DCT 2 × 2, 4 × 4, 8 × 4, 4 × 8, AFV

The `output` matrix shall be computed using:

```
if (N == 4) {
  output = AuxDCT2x2(input, 2);
} else {
  output = input;
}
```

where `N` is the downsampling factor, which may be 4 or 8, `AuxDCT2x2` computes a 2 × 2 DCT transform as described in Section I.2.3.2 for Identity, DCT 2 × 2, DCT 4 × 4 and as follows for the other `DctSelect` options. For DCT 8 x 4:

```
coef1 = input(0, 0);
coef2 = input(1, 0);
coef3 = input(0, 1);
coef4 = input(1, 1);
dc0 = (coef1 + coef3) / 2;
ac0 = (coef1 - coef3) / 2;
dc1 = (coef2 + coef4) / 2;
ac1 = (coef2 - coef4) / 2;
output(0, 0) = (dc0 + dc1) / 2;
output(1, 0) = (dc0 - dc1) / 2;
output(0, 1) = ac0 × ScaleF(2, 8, 1);
output(1, 1) = ac1 × ScaleF(2, 8, 1);
```

For DCT 8 x 4:

```
coef1 = input(0, 0);
coef2 = input(1, 0);
coef3 = input(0, 1);
block11 = input(1, 1);
dc0 = (coef1 + coef2) / 2;
ac0 = (coef1 - coef2) / 2;
dc1 = (coef3 + coef4) / 2;
ac1 = (coef3 - coef4) / 2;
output(0, 0) = (dc0 + dc1) / 2;
output(0, 1) = (dc0 - dc1) / 2;
output(1, 0) = ac0 × ScaleF(2, 8, 1);
output(1, 1) = ac1 × ScaleF(2, 8, 1);
```

For AFV: same as for Identity.

## I.3 Squeeze

### I.3.1 Parameters

The squeeze transform consists of a series of horizontal and vertical squeeze steps. The sequence of squeeze steps to be applied is defined by the transformation parameters; if no parameters are given, default parameters are used which are derived from the image dimensions and number of channels.

The number of parameters shall be a multiple of three. They are interpreted as follows: [ [step_ID] [begin] [end] ]*.

The step_ID is defined as follows:

| step_ID | Description |
|---|---|
| 0 | Vertical squeeze steps, insert residual channels immediately after squeezed channels |
| 1 | Horizontal squeeze steps, insert residual channels immediately after squeezed channels |
| 2 | Vertical squeeze steps, insert residual channels at the very end of the channel list |
| 3 | Horizontal squeeze steps, insert residual channels at the very end of the channel list |

The input channels are replaced by the squeezed channels, and the residual channels are inserted at the position determined by the step_ID. To determine the new channel vector (i.e. when interpreting the transformation description, before channel data decoding starts), the steps are applied in the order in which they are specified. After the channel data has been decoded, to apply the inverse transformation, the steps are applied in reverse order.

The channel vector is modified as follows:

| **step_ID** | **Channel vector modification** |
|---|---|
| 0 | `offset = end + 1;`<br>`for (c = begin, i = 0; c <= end; c++, i++) {`<br>`  h = channel[c].height;`<br>`  channel[c].height = (h + 1) Idiv 2;`<br>`  channel[c].vshift++;`<br>`  residu = channel[c].copy();`<br>`  residu.height = h Idiv 2;`<br>`  [[Insert residu into channel at index offset + i]]`<br>`}` |
| 1 | `offset = end + 1;`<br>`for (c = begin, i = 0; c <= end; c++, i++) {`<br>`  w = channel[c].width;`<br>`  channel[c].width = (w + 1) Idiv 2;`<br>`  channel[c].hshift++;`<br>`  residu = channel[c].copy();`<br>`  residu.width = w Idiv 2;`<br>`  [[Insert residu into channel at index offset + i]]`<br>`}` |
| 2 | Same as step_ID == 0, except offset = channel.size() |
| 3 | Same as step_ID == 1, except offset = channel.size() |

The inverse transform does the following:

| **step_ID** | **Inverse transform** |
|---|---|
| 0 | `offset = end + 1;`<br>`for (c = begin; c <= end; c++) {`<br>`  Channel output;`<br>`  output.height = channel[c].height;`<br>`  output.height += channel[offset].height;`<br>`  output.width = channel[c].width;`<br>`  vert_isqueeze(channel[c], channel[offset], output);`<br>`  channel[c] = output;`<br>`  [[Remove the channel with index offset]]`<br>`}` |
| 1 | `offset = end + 1;`<br>`for (c = begin; c <= end; c++) {`<br>`  Channel output;`<br>`  output.width = channel[c].width;` |

```
    output.width += channel[offset].width;
    output.height = channel[c].height;
    horiz_isqueeze(channel[c], channel[offset], output);
    channel[c] = output;
    [[Remove the channel with index offset]]
  }
```

| | |
|---|---|
| 2 | Same as step_ID == 0, except offset = channel.size() + begin - end - 1 |
| 3 | Same as step_ID == 1, except offset = channel.size() + begin - end - 1 |

The default parameters (the case when the number of parameters is zero) are defined as follows.

```
first = nb_meta_channels;
last = channel.size() - 1;
count = first - last + 1;
w = channel[first].width;
h = channel[first].height;
if (count > 2 && channel[first + 1].width == w && channel[first + 1].height == h) {
  parameters.push_back(3);
  parameters.push_back(first + 1);
  parameters.push_back(first + 2);
  parameters.push_back(2);
  parameters.push_back(first + 1);
  parameters.push_back(first + 2);
}
if (h >= w && h > 8) {
  parameters.push_back(0);
  parameters.push_back(first);
  parameters.push_back(last);
  h = (h + 1) Idiv 2;
}
while (w > 8 || h > 8) {
  if (w > 8) {
    parameters.push_back(1);
    parameters.push_back(first);
    parameters.push_back(last);
    w = (w + 1) Idiv 2;
  }
  if (h > 8) {
    parameters.push_back(0);
    parameters.push_back(first);
    parameters.push_back(last);
    h = (h + 1) Idiv 2;
  }
}
```

If the image has at least 3 channels and they have the same dimensions, then first the default parameters correspond to a horizontal followed by a vertical squeeze on channels 1 and 2, putting the residuals at the very end. These channels presumably contain chroma, so this effectively corresponds to 4:2:0 chroma subsampling if these residuals are truncated. If the number of channels is at least 3 and there are no meta-channels, the default parameter sequence starts with 3,1,2,  2,1,2.

Then an alternating sequence of horizontal and vertical squeeze steps is applied to all channels. The sequence starts with a horizontal step if the image width is (strictly) larger than the image height, and it starts with a vertical step otherwise. The sequence of squeeze steps ends when the dimensions of the first squeezed channel (presumably luma) are smaller than 8 × 8 (both width and height are at most 8).

### I.3.2 Horizontal inverse squeeze step

This step takes two input channels and replaces them with one output channel. The first input channel has dimensions $W_1$ x H; the second input channel has dimensions $W_2$ x H, where either $W_1 = W_2$ or $W_1 = W_2 + 1$.
The output channel has dimensions $(W_1 + W_2)$ x H.

The output channel is reconstructed as specified by the following:

```
horiz_isqueeze(input_channel_1, input_channel_2, output_channel) {
  for (y = 0; y < H; y++) {
    for (x = 0; x < W₂; x++) {
      avg = input_channel_1(x, y);
      residu = input_channel_2(x, y);
      next_avg = (x + 1 < W₁ ? input_channel_1(x + 1, y) : avg);
      left = (x > 0 ? output_channel((x << 1 ) - 1, y) : avg);
      diff = residu + tendency(left, avg, next_avg);
      first = (2 × avg + diff - sign(diff) × (diff & 1)) >> 1;
      second = first - diff;
      output_channel(2 × x, y) = first;
      output_channel(2 × x + 1, y) = second;
    }
    if (W₁ > W₂) {
      output_channel(2 × W₂) = input_channel_1(W₂);
    }
  }
}
```

The `tendency` function is specified by the following:

```
tendency(A, B, C) {
  X = (4 × A - 3 × C - B + 6) Idiv 12;
  if (A >= B && B >= C) {
    if (X - (X & 1) > 2 × (A - B)) X = 2 × (A - B) + 1;
    if (X + (X & 1) > 2 × (B - C)) X = 2 × (B - C);
    return X;
  } else if (A <= B && B <= C) {
    if (X + (X & 1) < 2 × (A - B)) X = 2 × (A - B) - 1;
    if (X - (X & 1) < 2 × (B - C)) X = 2 × (B - C);
    return X;
  } else {
```

```
    return 0;
  }
}
```

### I.3.3 Vertical inverse squeeze step

This step takes two input channels and replaces them with one output channel. The first input channel has dimensions W x $H_1$; the second input channel has dimensions W x $H_2$, where either $H_1 = H_2$ or $H_1 = H_2 + 1$.
The output channel has dimensions W x ($H_1 + H_2$).

The output channel is reconstructed as specified by the following:

```
vert_isqueeze(input_channel_1, input_channel_2, output_channel) {
  for (y = 0; y < H2; y++) {
    for (x = 0; x < W; x++) {
      avg = input_channel_1(x, y);
      residu = input_channel_2(x, y);
      next_avg = (y + 1 < H1 ? input_channel_1(x, y + 1) : avg);
      top = (y > 0 ? output_channel(x, (y << 1) - 1) : avg);
      diff = residu + tendency(top, avg, next_avg);
      first = (2 × avg + diff - sign(diff) × (diff & 1)) >> 1;
      second = first - diff;
      output_channel(x,2 × y) = first;
      output_channel(x,2 × y + 1) = second;
    }
  }
  if (H1 > H2) {
    for (x = 0; x < W; x++) {
      output_channel(x, 2 × H2) = input_channel_1(x, H2);
    }
  }
}
```

The `tendency` function is defined as above.

# Annex J
(normative)
# Loop filters

## J.1 Overview

The decoder shall support two kinds of loop filters: adaptive reconstruction and Gaborish.

If `loop_filter.epf`, an edge-preserving adaptive filter shall be applied to the entire frame immediately after decoding all groups, as defined in J.2. If `loop_filter.qco` is also set, a quantization constraint shall subsequently be applied as defined in J.3. The combination of edge-preserving filter and optional quantization constraint is referred to as adaptive reconstruction.

If enabled, patches shall subsequently be added as defined in K.3.

If `loop_filter.gab`, a convolution shall subsequently be applied to the entire frame as defined in J.4.

## J.2 Edge-preserving filter

### J.2.1 Overview

For each reference pixel, the filter outputs a pixel which is a weighted sum of all pixels in a 7×8 window. Each weight shall be computed as an exponentially decreasing function of an L1 distance metric. The distance for each weight shall be computed over two 3×4 patches whose origins are the reference pixel, and the pixel corresponding to the weight, located within the 7×8 window around the reference pixel.

In this subclause, `lf` denotes `loop_filter` and `intensity_multiplier3` denotes the cube root of `metadata.target_nits_div50` × 50 / 250.

### J.2.2 Input image

The distances shall be computed from an unsigned 8-bit guide image derived from the input image. For each pixel of the guide image, its samples `g[c]` (for `c` in the range [0, 3)) shall be computed from the input samples `in[c]` as follows:

```
min[c] = (c != 0 ? 0 : -0.03) × intensity_multiplier3;
scaled[c] = (in[c] - min[c]) × lf.epf_scale[c];
g[c] = clamp(scaled[c], 0, 255);
```

The filter may reference guide or input pixels whose positions lie outside the valid bounds. The decoder shall behave as if every such access were redirected to valid input pixels, as defined in 6.5.

### J.2.3 Distances

Distances shall be computed as sums of absolute differences over patches of 3×4 samples. The origin of a patch shall be one sample below and to the right of the top-left sample of the patch. Distances shall be computed for every sample, denoted the reference. In channel `c`, the distance `PatchDist(cx, cy, c)` between patches whose origins are the reference and one of its neighbors at offset (`cx`, `cy`) shall be computed as follows:

```
patch_dist = 0;
for (iy = -1; iy <= 1; ++iy) {
```

```
  for (ix = -1; ix <= 2; ++ix) {
    patch_dist += abs(guide(ix, iy, c) - guide(cx + ix, cy + iy, c);
  }
}
```

guide(dx, dy, c) shall be the sample in channel c of the guide image at an offset of (dx, dy) from the reference.

For each pixel, the patchwise L1 distances sad[0..55] between the reference and its 55 neighbors (plus itself) shall be computed as follows:

```
for (i = 0; i < 56; ++i) sad[i] = 0;

for (c = 0; c < 3; ++c) {
  for (cy = -3; cy <= 3; ++cy) {
    for (cx = -3; cx <= 4; ++cx) {
      i = (cy + 3) × 8 + cx + 3;
      sad[i] += PatchDist(cx, cy, c);
    }
  }
}
```

NOTE The distance between two patches whose origins are the reference pixel is sad[27] == 0.

### J.2.4 Weights

The convolution kernel for each reference pixel shall be defined by the 56 weights for its neighbors and the reference pixel, computed using an exponentially decreasing function Weight(sad, mul) defined as follows:

```
prescaled = min(sad, 507) << 6;
high16 = ((prescaled × mul) >> 16) + (127 << (23 - 16));
weight = MakeBinary32(high16 << 16);
```

The mul parameter determines the rate of decay and shall be set to mul_table[sigma], both defined below. Let dcts, quantization_width and sharpness denote the values in DctSelect, AcMul and Sharpness at the coordinates of the 8 × 8 region containing the reference pixel.

range shall be the largest sample within the varblock containing the reference pixel, of the channel all_r. all_r shall be a channel with samples equal to max({r[0], r[1], r[2]}). r[c] is the sample from channel c (0 for X, 1 for Y, 2 for B), equal to the largest value minus the smallest value encountered in the same (naturally aligned) 8 × 8 block of channel c of an image scaled. For each input sample p, the corresponding sample of scaled shall be as defined in J.2.2.

quant shall be 1 / quantization_width.

sigma shall be computed as follows:

```
range = min(range, lf.epf_range_max);
sigma = range × lf.epf_range_mul × lf.epf_sharp_lut[sharpness];
sigma = min(sigma, quant × epf_quant_mul);
sigma = truncate(min(sigma, lf.epf_sigma_max));
```

mul_table shall be initialized as follows:

```
gap = 4;
mul = -32768;
for (sigma = 16; sigma <= 672; sigma += gap) {
  weight = 0.0;
  for (; mul < 0; ++mul) {
    weight = Weight(sigma >> kSigmaShift, mul);
    if (weight > 0.5) break;
  }
  mul_table[sigma] = mul;
}
for (sigma = 16; sigma < 672; sigma += gap) {
  mul_step = (mul_table_[sigma + gap] - mul_table_[sigma]) / gap;
  for (i = 1; i < gap; ++i) {
    mul_table_[sigma + i] = mul_table_[sigma] + i × mul_step;
  }
}
```

### J.2.5 Weighted average

Given `sad[]` and `sigma` for the reference pixel, the convolution shall be computed as a weighted sum of neighboring pixels as follows.

```
for (i = 0; i < 56; ++i) {
  weights[i] = Weight(sad[i], mul_table[sigma]);
  if (weights[i] < lf.epf_weight_min) weights[i] = 0;
}
weighted_sum[3] = {0.0, 0.0, 0.0};
sum_weights[3] = {0.0, 0.0, 0.0};
i = 0;
for (cy = -3; cy <= 3; ++cy) {
  for (cx = -3; cx <= 4; ++cx) {
    weight = weights[i++];
    for (c = 0; c < 3; ++c) {
      weighted_sum[c] += Sample(cx, cy, c) × weight;
      sum_weights[c] += weight;
    }
  }
}
for (c = 0; c < 3; ++c) filtered_sample[c] = weighted_sum[c] / sum_weights[c];
```

`Sample(dx, dy, c)` shall be the sample from channel c of the padded input image at an offset (`dx`, `dy`) from the reference pixel.

## J.3 Quantization constraint

For every varblock, the decoder shall clamp the filtered pixels to the quantization intervals in DCT space. For every 8 × 8 block that `DctSelect` indicates is the top-left of a (possibly larger than 8 × 8) varblock `b`, the decoder shall construct a temporary varblock, whose samples are the values after convolution (J.2), minus the original samples (at sample locations covered by `b`). The decoder shall compute `dct`, the DCT of the temporary varblock as specified in I.2.3, and then obtain DC from LLF in-place as specified in I.2.6. The decoder shall clamp each coefficient v within `dct` as follows:

```
v = clamp(v, -lf.qco_interval_mul × interval, lf.qco_interval_mul × interval);
```

Where `interval` depends on whether the coefficient is LLF (i.e. its row is less than the height of b / 8 and column is less than the width of b / 8) or HF. For HF, `interval` is the quantization matrix entry for that coefficient divided by the value of `AcMul` (for the varblock b). Otherwise, `interval` is the `DcQuantField` for that coefficient, as defined in C.5.2.

The decoder shall then inverse-DCT transform the clamped coefficients as specified in I.2.4 after obtaining LLF coefficients from DC in-place as specified in I.2.7. The decoder shall obtain each output sample by adding the IDCT result (which was a filter residual) to the corresponding sample of the filter input.

## J.4 Gaborish

The decoder shall apply a convolution to each channel of the entire image with the following symmetric 3×3 kernels. The unnormalized weights for the center, its 4-neighbors (taxicab distance 1) and the four corner (taxicab distance 2) shall be, respectively, 1, `loop_filter.gab_C_weight1` $\times 10^{-6}$ and `loop_C_filter.gab_weight2` $\times 10^{-6}$, where `C` is `x`, `y`, or `b` according to the channel currently being processed. These weights shall be rescaled uniformly before convolution, such that the nine kernel weights sum to 1.

When the convolution references input pixels whose locations are outside the bounds of the original image, the decoder shall apply the mirroring logic in 6.5 to redirect such accesses to valid input pixels.

# Annex K
(normative)
# Image features

## K.1 Overview

Each frame may include alternate representations of image contents (dots, splines, patches and noise). If any of them are enabled (as indicated by `frame_header.flags`), the decoder shall draw them on top of the frame as follows.

## K.2 Dots

If the `kSkipDots` flag in `frame_header` is set, this subclause shall be skipped for that frame. Otherwise, all dots shall be added after the Gaborish filter (J.4) is applied or skipped, as follows.

Each dequantized dot is modeled as a 2D Gaussian function with the following attributes:

- `x`, `y`: Position of the dot (mean of the Gaussian)
- `var_x`: The variance along the long axis (principal eigenvalue)
- `var_y`: The variance along the short axis (secondary eigenvalue)
- `angle`: Angle in radians between the long axis (principal eigenvector) and the x axis of the image (pointing to the right) in counter-clockwise direction.
- `intensity`: Array of three values specifying the intensity in each of the channels (X, Y, B).

The quantized dot attributes are specified in C.4.5. Note that quantized dots differ from dequantized dots in that the former encode the sign and magnitude of the ellipse colour separately, and include a boolean flag `is_circle` used to represent circular dots.

### K.2.1 Dots dequantization

This subclause describes how to obtain a dequantized dot from a quantized dot `quantized[i]`, where `i` is in the range [0, `n_points`). The position shall be obtained by dividing the quantized position by `kPosMult = 5`:

```
ellipse[i].x = quantized[i].x / kPosMult;
ellipse[i].y = quantized[i].y / kPosMult;
```

`deq = LinDequant(x, xMin, xMax, levels)` shall be computed as

```
delta = (xMax - xMin) / levels;
deq = delta × (x + 0.5) + xMin;
```

The angle and variance shall be dequantized as specified by the following:

```
if (quantized[i].is_circle) {
  ellipse[i].angle = 0.0;
  radius = LinDequant(quantized[i].sigma_x, kMinSigma, kMaxSigma, qSigma);
  stdev_x = radius;
  stdev_y = radius;
} else {
  ellipse[i].angle = LinDequant(quantized[i].angle, 0, π, qAngle);
  stdev_x = LinDequant(quantized[i].sigma_x, kMinSigma, kMaxSigma, qSigma);
  stdev_y = LinDequant(quantized[i].sigma_y, kMinSigma, kMaxSigma, qSigma);
}
```

```
ellipse[i].var_x = stdev_x × stdev_x;
ellipse[i].var_y = stdev_y × stdev_y;
```

With `kMinSigma = 0.1`, `kMaxSigma = 3.1`, `qSigma = 16`, `qAngle = 8`. The intensity shall be dequantized:

```
for (c = 0; c < 3; c++) {
  ellipse[i].intensity[c] = kColourQuant[c] × (quantized[i].intensity[c] + 0.5);
  if (quantized[i].sign_intensity & (1 << c)) {
    ellipse[i].intensity[c] = ellipse[i].intensity[c];
  }
}
ellipse[i].intensity[2] += kYToBRatio × ellipse[i].intensity[1];
```

The constant `kYToBRatio = 0.935669` represents the correlation between the Y and B channel.

### K.2.2 Adding dots

Each `ellipse[i]` shall be added to the decoded image by modulating all samples in a 5 × 5 window centered on `floor(x), floor(y)`. Let `ix`, `iy` denote the x, y coordinates of every such sample, and `c` the colour channel. The per-sample modulation is specified by the following, where `e = ellipse[i]`:

```
rx =  cos(e.angle) × (ix - e.x) + sin(e.angle) × (iy - e.y);
ry = -sin(e.angle) × (ix - e.x) + cos(e.angle) × (iy - e.y);
md = (rx × rx / e.var_x) + (ry × ry / e.var_y);
image(ix, iy, c) += e.intensity[c] × exp(-0.5 × md);
```

## K.3 Patches

If the `kPatches` flag in `frame_header` is not set, this subclause shall be skipped for that frame. Otherwise, all patches shall be added after adaptive reconstruction (J.2 / J.3) and before gaborish (J.4), as follows.
The decoder shall dequantize and add each `patch[i]` (from C.4.6), where `i` is in the range [0, `num_patches`):

```
image(pos.x + ix, pos.y + iy, c) +=
  kChannelDequant[c] × patch[i].sample(ix, iy, c);
```

`image(x, y, c)` represents the sample from channel `c` at position `(x, y)`. The values `ix, iy` shall be iterated between zero and `(patch[i].xsize, patch[i].ysize)` respectively. `kChannelDequant[3] = {0.01615, 0.08875, 0.1922}` represents the quantization level.

## K.4 Splines

If the `kSplines` flag in `frame_header` is not set, this subclause shall be skipped for that frame. Otherwise, all (centripetal Catmull-Rom) splines shall be rendered after dots (K.2), via pixel-by-pixel addition, as follows.
In this subclause, the following definitions apply:

```
Mirror(center, point) = center + (center - point);

DCTScales32(k) = D(32, k);  // from I.2.6.1
DCTScale(k) = DCTScales32(k) × (k == 0 ? 1 / (4 × sqrt(2)) : 0.25);
ContinuousIDCT(dct, t) = [[sum for k = [0, 32):
  DCTScale(k) × dct[k] × cos(k × (π / 32) × (t + 0.5))]];
```

Given the decoded control point positions of the splines as well as the dequantized and recorrelated DCT32 coefficients of their X, Y, B and σ values along their arc length, compliant decoders shall, for each spline *S*:

1. determine an ordered sequence *C* of points constructed in the following manner:
    a. if *S* has only one control point, then *C* is the singleton with that control point;
    b. otherwise:
        i. let *E* be a sequence of control points constructed in the following manner:
            1. add `Mirror(`first control point of S, second control point of S`)` to *E*;
            2. add all the control points of *S* to *E*;
            3. add `Mirror(`last control point of S, second-to-last control point of S`)` to *E*;
        ii. for each sliding window (*p*[0], *p*[1], *p*[2], *p*[3]) of *E*:
            1. add *p*[1] to *C*;
            2. let *t*[0] = 0 and, for *k* from 1 to 3 (inclusive), `t[k] = t[k - 1] + ((p[k].x - p[k - 1].x)² + (p[k].y - p[k - 1].y)²)^0.25`;
            3. for *t* from *t*[1] (exclusive) to *t*[2] (exclusive) by steps of at most `(t[2] - t[1]) / 16` in size:
                a. let *A*[*k*] = *p*[*k*] + ((*t* − *t*[*k*]) / (*t*[*k* + 1] − *t*[*k*])) × (*p*[*k* + 1] − *p*[*k*]) for *k* from 0 to 2 (inclusive);
            4. let *B*[*k*] = *A*[*k*] + ((*t* − *t*[*k*]) / (*t*[*k* + 2] − *t*[*k*])) × (*A*[*k* + 1] − *A*[*k*]) for *k* from 0 to 1 (inclusive);
                a. add *B*[0] + ((*t* − *t*[1]) / (*t*[2] − *t*[1])) × (*B*[1] − *B*[0])) to *C*;
        iii. add the second-to-last element of *E* (the last original control point) to *C*;
2. determine an ordered sequence *P* of points obtained by joining the consecutive points of *C* with straight lines, yielding an approximation *S′* of *S*, and going through it with an arc length of 1 between each pair of consecutive points, adding the last element of *C* if it would not have been included otherwise;
3. determine the total arc length *L* of *S′*;
4. for each point *p* of *P*:
    a. determine the position *l* of *p* along the arc length of *S′*, which may be computed during step 2;
    b. determine the arc length *d* of *p* from the previous point in *P* along *S′*, or set *d* to 1 if *p* is the first point (note: this shall be 1 for every point of *P* except possibly the last one);
    c. compute the effective value of X, Y, B and σ at that point of the spline using the formula: `value = ContinuousIDCT(dct, 31 × l / L)`
    d. for each channel *c* in {X, Y, B}:
        i. for each pixel (x, y) within -2 × σ² × log(0.1) of *p*:
            1. Let `s2s` denote `sqrt(2) × σ`.
            2. add `((value×d×σ) / 4) × (erf((p.x - x + 0.5) / s2s) - erf((p.x - x - 0.5) / s2s)) × (erf((p.y - y + 0.5) / s2s) - erf((p.y - y - 0.5) / s2s))` to channel *c* of pixel (x, y).

## K.5 Noise

### K.5.1 Overview

If the `kNoise` flag in `frame_header` is not set, this subclause shall be skipped for that frame. Otherwise, noise shall be added to each 256 × 256 pixel group after rendering splines (K.4) as follows. The decoder shall generate pseudorandom channels of the same size as a group, modulate them, and add their samples to the corresponding sample of the group.

### K.5.2 Pseudorandom number generator

Three pseudorandom channels denoted RR, RG, RB shall be generated in left to right order as follows. Rows in a channel shall be generated in top to bottom order. Each row shall be generated as follows.

For every segment of up to 16 samples, from left to right, the decoder shall generate an array of eight pseudorandom 64-bit variables `batch[i]`, where i is [0, 8). Given 64-bit arrays `s0[i]` and `s1[i]`, the decoder shall compute `batch[i]` as follows:

```
s1[i] ^= s1[i] << 23;
batch[i] = s1[i] + s0[i];
s1[i] ^= s0[i] ^ (s1[i] >> 18) ^ (s0[i] >> 5);
```

`s0[i]` and `s1[i]` constitute the internal state of a XorShift128Plus generator, and their updated values shall be used to compute the next `batch[i]`. `s0` and `s1` shall be (re)initialized as follows prior to generating random numbers for a 256 × 256 pixel group:

```
s0[0] = SplitMix64(seed + 0x9E3779B97F4A7C15);
s1[0] = SplitMix64(s0[0]);
for (i = 1; i < 8; ++i) {
  s0[i] = SplitMix64(s1[i - 1]);
  s1[i] = SplitMix64(s0[i]);
}
```

`seed` shall be $((y_0 \ll 32) + x_0)$, where $(x_0, y_0)$ are the coordinates of the top-left pixel of the current group.

SplitMix64 shall be the following function of the 64-bit current state `z`:

```
z = (z ^ (z >> 30)) × 0xBF58476D1CE4E5B9;
z = (z ^ (z >> 27)) × 0x94D049BB133111EB;
z ^= z >> 31;
```

The resulting batch[i] shall be interpreted as a 16-element array of 32-bit values bits[], where bits[0] are the lower 32 bits of batch[0], bits[1] are the upper 32 bits of batch[0], bits[2] are the lower 32 bits of batch[1], and so forth. The next 16 (or the number of remaining samples in the row, whichever is less) samples s[j] shall be generated as follows:

```
rand12 = MakeBinary32((bits[j] >> 9) | 0x3F800000);
s[j] = rand12 - 1;
```

After all rows are thus initialized, each channel shall be convolved (with boundary conditions as defined in 6.5) by the following Laplacian-like 5×5 kernel, where blank weights shall be zero:

| | | | | |
|---|---|---|---|---|
| | | 1 | | |
| | | | | |
| 1 | | -4 | | 1 |
| | | | | |
| | | 1 | | |

### K.5.3 Additive noise

The following procedure shall be applied to each pixel.

$A_R$, $A_G$, $A_B$ shall denote the corresponding sample from each of the preceding pseudorandom channels RR/RG/RB, further scaled by multiplication with `0.22`.

The strength of the noise shall be obtained from as follows. First, the input quantities $In_R$ and $In_G$ shall be obtained from the X and Y samples of the input pixel:

```
In_R = Y + X;
In_G = Y - X;
```

These values shall be scaled using the formula:

```
In_scaled = In × 14;
```

The result shall be split into an integer part `in_int` = floor($In_{scaled}$) and a fractional part `in_frac` = $In_{scaled}$ - `in_int`. The integer part indicates the intensity and shall be clamped to the interval `[0, 14]`.

Using the LUT values decoded as described in C.4.8, the strength `S` shall be defined, for each of R and G, as:

```
S = LUT[in_int] × (1 - in_frac) + LUT[in_int + 1] × in_frac;
```

The strength shall be clamped between 0 and 1.
Using the additive noise `A` and the strength `S`, the final noise values shall be obtained as:

```
N_R = 0.01 × A_R × S_R + 0.99 × A_B × S_R;
N_G = 0.01 × A_G × S_G + 0.99 × A_B × S_G;
```

The noise shall be added to the X, Y and B samples of each input pixel as follows:

```
X += N_R - N_G;
Y += N_R + N_G;
B += 0.935669 × (N_R + N_G);
```

# Annex L
(normative)
# Colour transforms

## L.1 Overview

In var-DCT mode, the samples are stored in XYB colour space and shall be transformed to RGB as defined in L.2. Other modes shall interpret samples according to the colour encoding indicated by `metadata.colour_encoding`.

Modular modes may use decorrelating colour transforms which shall be reversed as specified in L.3 to L.6. Extra channels shall be rendered as specified in L.7.

## L.2 XYB

If `metadata.colour_encoding.colour_space == kXYB`, the decoder shall return the X, Y, B samples unchanged. Otherwise, each X, Y, B sample shall be converted to an RGB colour encoding with linear transfer function and sRGB white point and primaries as specified below.

First, the colour space basis shall be changed:

$$R_{gamma} = Y + X;$$
$$G_{gamma} = Y - X;$$
$$B_{gamma} = B;$$

Then, the gamma compression shall be undone as follows, where the $(R_{bias}, G_{bias}, B_{bias})$ values shall default to (-0.00100549, -0.00100549, -0.00094081):

$$R_{mix} = R_{gamma}{}^{3} + R_{bias};$$
$$G_{mix} = G_{gamma}{}^{3} + G_{bias};$$
$$B_{mix} = B_{gamma}{}^{3} + B_{bias};$$

R, G, B column vector shall be obtained by multiplying the 3×3 row-major inverse opsin absorbance matrix with the column vector $(R_{mix}, G_{mix}, B_{mix})$. The default inverse opsin absorbance matrix (in raster order) is:

```
(2542.56835937500000, -2326.53173828125000,  44.17734527587891,
 -698.04852294921875,  1012.71777343750000, -76.22974395751953,
 -827.36480712890625,   709.83502197265625, 451.38488769531250)
```

NOTE Both the $_{bias}$ and $_{mix}$ terms above may be overridden as specified in C.4.9.

Finally, the R, G and B values shall be divided by `metadata.target_nits_div50` × 50 / 250.

NOTE The expected range of the resulting R, G and B samples is [0, 255], but some samples may lie outside this range (e.g. if the input gamut is larger than sRGB).

## L.3 YCbCr

This transformation operates on the first three channels. These channels must have the same dimensions. For every pixel position in these channels, the values (Y, Cb, Cr) are replaced by (R, G, B) values, as specified by the following, where $S = 2^{bit_depth-1}$:

```
R = Y + 1.402 × (Cr - S);
```

```
G = Y - 0.344136 × (Cb - S) - 0.714136 × (Cr - S);
B = Y + 1.772 × (Cb - S);
```

### L.4 YCoCg

This transformation operates on the first three channels after the meta-channels. These channels must have the same dimensions. For every pixel position in these channels, the values (Y, Co, Cg) are replaced by (R, G, B) values, as specified by the following:

```
B = Y + ((1 - Cg) >> 1) - (Co >> 1);
G = Y - ((-Cg) >> 1);
R = Co + B;
```

### L.5 SubtractGreen

This transformation operates on the first three channels after the meta-channels. These channels must have the same dimensions. For every pixel position in these channels, the values (G2, R2, B2) are replaced by (R, G, B) values, as specified by the following, where $S = 2^{bit_depth-1}$:

```
G = clamp(G2, 0, 2^bit_depth - 1);
R = R2 - S + G Umod 2^bit_depth;
B = B2 - S + G Umod 2^bit_depth;
```

### L.6 Palette

This transformation has three obligatory parameters: `begin_channel, end_channel, nb_colours`.

It replaces the channels channel[begin_channel] ... channel[end_channel], which shall have the same dimensions, with two new channels:

- one meta-channel which is inserted at the very beginning of the channel vector and has dimensions `width = nb_colours` and `height = (begin_channel - end_channel + 1)` and `hshift = -1` (which indicates that the dimensions are not related to the image dimensions). This channel represents all the colours of the palette.
- one channel (at the same position in the vector as the original channels, same dimensions) which encodes the same pixels as the original channels but using only one channel and palette indexes.

The decoder restores the original pixels as follows:

```
first = begin_channel + 1;
last = end_channel + 1;
nb = last - first + 1;
for (i = first + 1; i <= last; i++)
  [[Insert a copy of channel[first] at index i]];
for (y = 0; y < channel[first].height; y++) {
  for (x = 0; x < channel[first].width; x++) {
    index = channel[first](x, y) - channel[first].minval;
      for (c = 0; c < nb; c++)
        channel[first+c](x, y) = channel[0](index, c);
  }
}
```

```
[[Remove the first channel]]
```

## L.7 Extra channel rendering

Some extra channels are suitable for rendering into the main image. Decoders that wish to render an output image (instead of returning samples and extra channels to the application), shall do so as follows.

At the final step in the decode process (when the inverse colour transform has been applied), the decoded image is either in linear RGB with sRGB primaries (when the inverse XYB transform was applied) or in the colour space defined by the image metadata. In both cases the in-gamut range of the pixel data is [0, 255].

The decoder shall then render the extra channel with index `n` (in ascending order of `n`) as follows:

- If `metadata.m2.extra_channel_info[n].meaning` is zero or any other even number, nothing is done.
- If `metadata.m2.extra_channel_info[n].meaning == 1`:
  For every (R, G, B) sample in the image and corresponding sample S in `extra_channel[n]`:
  - mix = S × `metadata.m2.extra_channel_info[n].solidity` / ($2^{\texttt{metadata.m2.extra_channel_bits}}$ - 1)
  - R = mix × `metadata.m2.extra_channel_info[n].red` + (1-mix) × R
  - G = mix × `metadata.m2.extra_channel_info[n].green` + (1-mix) × G
  - B = mix × `metadata.m2.extra_channel_info[n].blue` + (1-mix) × B

# Annex M
(normative)
# Lossless JPEG1 recompression

## M.1 Overview

The lossless JPEG1 decoder may be viewed as a pipeline with two stages:

- parsing input bitstream and decoding DCT coefficients, quantization tables and other auxiliary data
- reconstruction of the original JPEG1 bitstream

The second stage is optional and may be replaced with direct rasterization to an image, if desired.

In this Annex, the following definitions apply:

### M.1.1

**grid**
a 2-dimensional analogue of an array; `a[x, y]` means addressing an element of grid `a` at row `y` and column `x`; for the sake of simplicity, addressing elements with coordinates outside of bounding rectangle (`x < 0`, or `y < 0`, or `x >= width`, or `y >= height`) may be allowed

### M.1.2

**dct_block**
an array with 64 elements corresponding to DCT coefficients of a (8 × 8) block

### M.1.3

**histo**
an array of unsigned integers, used as sANS "distribution" for entropy decoding

### M.1.4

**set**
an unordered collection of elements

## M.2 Entities

### M.2.1 QuantTable

Quantization table-specification (see ISO/IEC 10918-1:1993(E) / B.2.4.1).

| **name** | **type** | **description** |
|---|---|---|
| `values` | `dct_block` of `int` | multipliers used for dequantization; all elements shall be positive after decoding of entity is finished |
| `precision` | `int` | • 0 means `values` elements shall be less than `256`<br>• 1 means that `values` elements shall be less than `65536` |
| `index` | `int` | the index of this quantization table in original JPEG1 stream (from DQT marker); valid values are in the range [0, 4) |
| is_last | boolean | true if this table is the last entry within its (DQT) marker |

### M.2.2 HuffmanCode

Huffman table-specification (see ISO/IEC 10918-1:1993(E) / B.2.4.2).

| name | type | description |
|---|---|---|
| counts | array of 17 int | bit length histogram |
| values | array of 257 int | symbol values sorted by increasing bit lengths |
| slot_id | int | concatenation of **Tc** and **Th** values from ISO/IEC 10918-1:1993(E) / B.2.4.2 |
| is_last | boolean | true if this table is the last entry within its (DHT) marker |

### M.2.3 Component

Core data structure that holds colour channel data (DCT coefficients).

| name | type | description |
|---|---|---|
| id | int | identifier of the colour channel; valid values are in the range [0, 256); most frequently used values are: 0x1, 0x2, 0x3, 'R', 'G', 'B' |
| h_samp_factor | int | horizontal sampling factor; valid values are in the range [1, 16) |
| v_samp_factor | int | vertical sampling factor; valid values are in the range [1, 16) |
| quant_idx | int | the index of the M.2.1 entity to be used for dequantization of coeffs |
| coeffs | grid of dct_block of int | quantized DCT coefficients |

### M.2.4 ComponentScanInfo

Association between Component and HuffmanCode.

| name | type | description |
|---|---|---|
| comp_idx | int | Component index |
| dc_tbl_idx | int | index of HuffmanCode for DC; valid values are in the range [0, 4) |
| ac_tbl_idx | int | index of HuffmanCode for AC; valid values are in the range [0, 4) |

### M.2.5 ExtraZeroRunInfo

| name | type | description |
|---|---|---|
| block_idx | int | scan block index |
| num_extra_zero_runs | int | number of extra ZRL (RRRRSSSS == 0xF0) symbols to be emitted before EOB (RRRRSSSS == 0x00) (see ISO/IEC 10918-1:1993(E) / F.1.2.2.1) |

### M.2.6 ScanInfo

Scan header (see ISO/IEC 10918-1:1993(E) / B.2.3) plus extra information.
Parameters used for progressive scans are named the same as in the JPEG1 spec.

| name | type | description |
|---|---|---|
| Ss | int | start of spectral band in zig-zag sequence |
| Se | int | end of spectral band in zig-zag sequence |
| Ah | int | successive approximation bit position, high |
| Al | int | successive approximation bit position, low |
| component_scans | array of ComponentScanInfo | HuffmanCode associations for components (colour channels) |
| reset_points | set of int | scan block indexes where the JPEG1 stream has an out-of-order EOB |
| extra_zero_runs | array of ExtraZeroRunInfo | data about extra ZRL (RRRRSSSS = 0xF0) in JPEG1 stream (see ISO/IEC 10918-1:1993(E) / F.1.2.2.1) |

### M.2.7 Stream

Structure used to store information required for byte-precise JPEG1 data reconstruction.

| name | type | description |
|---|---|---|
| width | int | image width, in pixels |
| height | int | image height, in pixels |
| version | int | • 0 means that regular image is represented by this structure<br>• 1 means that raw unmodified image data is stored in original_jpg field |
| restart_interval | int | value stored in DRI marker; 0 if original image does not contain that marker; after serialization of corresponding DRI marker the bit stream is restarted (padded) every restart_interval MCUs and RSTn marker is injected |
| app_data | array of array of byte | elements are APP*n* markers with first byte (0xFF) stripped; (in other words, the first byte of each item is 0xE*n*) |
| com_data | array of array of byte | elements are COM markers with the first two bytes (0xFF, 0xFE) stripped |
| quant | array of QuantTable | collection of quantization tables and corresponding metadata |
| huffman_code | array of HuffmanCode | collection of Huffman codes and corresponding metadata |
| components | array of Component | image data (DCT coefficients) |
| scan_infos | array of ScanInfo | metadata for interlaced JPEG1 |
| marker_order | array of byte | second byte of markers as they shall appear reconstructed JPEG1; 0xFF value is used to denote series of unrecognized markers |
| inter_marker_data | array of array of byte | elements store unrecognized markers contents (see marker_order) |
| tail_data | array of byte | raw original JPEG1 stream contents after the EOI marker |
| original_jpg | array of byte | raw unmodified JPEG1 stream data (see version) |
| padding_bits | array of int | concatenated bitstream padding bits (see also restart_interval); if empty, then all padding bits are considered to be ISO/IEC 10918-1:1993(E) / F.1.2.3 compliant |

### M.2.8 BinaryProbabilityTracker

This structure is used for adaptive entropy decoding. The structure is initialized with "a priori" probability value P. Default value (i.e. if it is not explicitly specified) of P is 134.

| name | type | description | initial value |
|---|---|---|---|
| prob | int | current probability value | P |
| total | int | probability numerator | 3 |
| count | int | probability denominator | 3 × P |

Before the next bit is queried from the entropy source, prob Idiv 256 is used as an estimate that bit value is 0. After actual bit value is calculated, BinaryProbabilityTracker instance is updated as follows:

- total += 1
- count += (c == 0) ? 256 : 1
- prob = (((1 << 17) Idiv total) × count) >> 17
- if total == 254, then:
  - total >>= 1
  - count >>= 1

### M.2.9 ComponentStateDC

ComponentStateDC is initialized with width and height values, corresponding to component dimensions – number of columns and rows of DCT blocks.

| name | type | description |
|---|---|---|
| is_zero | BinaryProbabilityTracker | instance is initialized with P = 135 |
| is_empty | array of 3 BinaryProbabilityTracker | elements are initialized with P = 74 |
| sign | array of 9 BinaryProbabilityTracker | elements are initialized with P = 128 |
| first_extra | array of 10 BinaryProbabilityTracker | elements are initialized with P = 150 |
| prev_is_nonempty | grid of int | width columns, height rows; initial value of all elements is 0; elements outside of bounding rectangle always have implicit value 1 |
| prev_abs | grid of int | width columns, height rows; initial value of all elements is 0; elements outside of bounding rectangle always have implicit value 0 |
| prev_sign | grid of int | width columns, height rows; initial value of all elements is 0; elements outside of bounding rectangle always have implicit value 0 |

### M.2.10 ComponentStateAC

ComponentStateAC is initialized with width and height values, corresponding to component dimensions – number of columns and rows of DCT blocks.

| name | type | description |
|---|---|---|
| context_offset | int | |
| is_zero | array of 11 dct_block of BinaryProbabilityTracker | |
| sign | array of 19 dct_block of BinaryProbabilityTracker | |
| first_extra | array of 10 dct_block of BinaryProbabilityTracker | elements are initialized with P = 158 |
| num_nonzeros | array of 32 array of 63 BinaryProbabilityTracker | |
| prev_is_nonempty | grid of int | width columns, height rows; initial value of all elements is 0; elements outside of bounding rectangle always have implicit value 1 |
| prev_num_nonzeros | grid of int | width columns, height rows; initial value of all elements is 0; elements outside of bounding rectangle always have implicit value 0 |
| prev_abs | grid of dct_block of int | width columns, height rows; initial value of all elements is 0 |
| prev_sign | grid of dct_block of int | width columns, height rows; initial value of all elements is 0 |
| order | dct_block of int | |
| mult_row | dct_block of int | |
| mult_col | dct_block of int | |

is_zero elements are initialized as follows:

- let bucket be the index dct_block element
- let k be the index of BinaryProbabilityTracker element in the dct_block
- base is assigned is_zero_base[k] (see Table M.1)
- considered BinaryProbabilityTracker is initialized with P = base + 9 × (bucket - 7)

First 8 × 64 elements of sign are initialized with P = 108; next 64 elements are initialized with P = 128; remaining elements are initialized with P = 148.

num_nonzeros elements are initialized with P equal to the corresponding values from num_nonzeros_base (see Table M.2).

**Table M.1 — is_zero_base table**

Defined in is_zero_base.md (see Annex R).

**Table M.2 — num_nonzeros_base table**

Defined in num_nonzeros_base.md (see Annex R).

### M.2.11 State

| name | type | description |
|---|---|---|
| schemes | array of int | context assignment scheme index, one per colour component |
| context_map | array of int | mapping from context to entropy_codes elements |
| entropy_codes | array of histo | elements are sANS distributions |
| block_state | array of boolean | "is empty" flags, one per DCT block |

#### M.2.11.1 DCT Block Order

Elements in State.block_state correspond to the following DCT blocks order:

<table><tr><td>first MCU row<table><tr><td>first colour component<table><tr><td>first row (inside MCU row)<table><tr><td>first block of the row (corresponds to block_state[0])</td></tr><tr><td>second block of the row (corresponds to block_state[1])</td></tr><tr><td>...</td></tr></table></td></tr><tr><td>...</td></tr></table></td></tr><tr><td>...</td></tr></table></td></tr><tr><td>...</td></tr></table>

Depending on the horizontal and vertical subsampling MCU rows of different colour components may contain different number of rows and columns.

### M.2.12 Constants

`num_avg_ctx = 9`
`zigzag` is an array that maps the "zig-zag" sequence (described in ISO/IEC 10918-1:1993(E) / 4.3) to index in `dct_block`; in other words, the first 6 elements of `zigzag` are {0, 1, 8, 16, 9, 2}.

### M.2.13 Aliases

After corresponding members of `Stream` and `Component` are parsed, other values shall be derived from them:

- (per `Stream`) `max_h_samp_factor` – maximum of `components` `h_samp_factor` values
- (per `Stream`) `max_v_samp_factor` – maximum of `components` `v_samp_factor` values
- (per `Stream`) `MCU_rows = ceil(height / (max_v_samp_factor × 8))`
- (per `Stream`) `MCU_cols = ceil(width / (max_h_samp_factor × 8))`
- (per `Stream`) `has_zero_padding_bit` – `true` if any element of `padding_bits` is 0, otherwise `false`
- (per `Component`) `width_in_blocks = h_samp_factor × MCU_cols`
- (per `Component`) `height_in_blocks = v_samp_factor × MCU_rows`
- (per `Component`) `num_blocks = width_in_blocks × height_in_blocks`

## M.3 Subroutines

### M.3.1 A(ba, t)

`A(ba, t)` denotes the result of querying of 1-bit value from the 8-bit Binary Arithmetic Decoder `ba`, with the 0-expectancy `prob` supplied by `BinaryProbabilityTracker` `t`; after value is decoded, `t` is updated with the result.

### M.3.2 E(ans, D)

`E(ans, D)` denotes the result of querying of a value from sANS Symbol Decoder `ans`, with distribution `D`.

## M.4 Stream structure

### M.4.1 Sections and subsections

Thi section is a top level entity in the losslessly compressed JPEG stream.
Every section starts with a `marker` byte marker followed by `Varint()` value.
Marker byte requirements:

- `(marker & 0x81) == 0`
- `marker > 2`

If the marker byte does not meet the requirements, the codestream is ill-formed.
Lowest 3 bits of marker byte represent `wire_type`, and highest 5 bits represent `tag`. According to the marker byte restrictions, `wire_type` shall be 0 or 2, and `tag` shall be in the range [1, 16).
At each level, `tag` of all the sibling (sub-)sections shall be different, otherwise the codestream is ill-formed.
`wire_type` defines the kind of section:

- 0 means "Varint" sections – the decoded `value` is the only content of the section; section finishes immediately after it
- 2 means "Length-delimited" sections – the decoded `value` is the length of the section payload, in bytes; further section parsing shall use exactly `value` bytes of input bitstream, otherwise the codestream is ill-formed

Subsections follow the same rules as sections but they are themselves reside in the payload of "Length-delimited" sections.
Decoder may ignore the `value` of "Varint" (sub-)sections, and skip parsing of payload of "Length-delimited" (sub-)sections, if their meaning is not described in this document.

NOTE Overall sections / subsections encoding is a subset of "Protocol Buffers" serialization. It should be possible to parse top level structure of valid stream using "Protocol Buffers" implementation. However, that way it is impossible to fully validate the stream, because information about order of fields, and value overwrites is lost. Corresponding Protocol Buffer descriptor is presented in Table M.3.

**Table M.3 – Protocol Buffer descriptor of top-level structure of losslessly compressed JPEG stream**

Defined in `brn_proto.md` (see Annex R).

### M.4.2 Stages

This stream consists of a number of sections. Parsing the stream shall be performed in several stages.
On the first stage, the decoder expects Signature section. If unexpected sections appear, the codestream is ill-formed.
Before the second stage `stream` is initialized as an instance of `Stream` entity.
On the second stage, the decoder expects Header section. If unexpected sections appear, the codestream is ill-formed.
If, after parsing Header section, `stream.version == 1`, then the next section shall be Fallback section, and it shall be the last section in the stream. If Fallback section appears under any other circumstances, the codestream is ill-formed.
Before the third stage `state` is initialized as an instance of `State` entity. `state` is used for storing and sharing intermediate information.
On the third stage remaining sections may appear in an arbitrary order, once the following precedence requirements are met:

- JPEG1 Internals section ≻ Quant Data section
- Quant Data section ≻ DC Data section
- Histogram Data section ≻ DC Data section
- DC Data section ≻ AC Data section

All the sections described in the following subclauses (except for the Fallback section) are required for a regular JPEG stream reconstruction.

### M.4.3 Signature section

Signature section allows quick stream type detection.
`marker == 0x0A` (`tag` is 1, `wire_type` is "Length-delimited").
Payload shall be 4 bytes long and shall contain the following sequence: {0x42, 0xD2, 0xD5, 0x4E}.

### M.4.4 Header section

Header section contains metadata.
`marker == 0x12` (`tag` is 2, `wire_type` is "Length-delimited").
Payload: subsections.
All subsections shall have different marker bytes, otherwise the codestream is ill-formed.
All possible subsections not mentioned in subclauses are considered "reserved" and their contents may be ignored by the decoder.
Only Version And Component Count subsection is mandatory. If it is not present, the codestream is ill-formed.
After every subsection is parsed, its data is transformed, validated and stored in the `stream` instance.
`stream.version = version_and_component_count_code >> 2`
If `stream.version > 1`, the codestream is ill-formed.
If `stream.version == 1` (fallback mode), then the contents of all other subsections shall be ignored.
If `stream.version == 0` (regular mode), then Width, Height and Subsampling subsections are rendered mandatory. If any of them has not appeared after section parsing is complete, the codestream is ill-formed.
`width` and `height` values are required to be in the range [1, 65536); if any value is out of range, the codestream is ill-formed. `stream.width` and `stream.height` are assigned corresponding values.
`number_of_components = (version_and_component_count_code & 3) + 1`
`stream.components` is resized to contain exactly `number_of_components` elements.

`subsampling_code` is interpreted as a concatenation of `number_of_components` pairs of unsigned 4-bit integers $\{\{v_0, h_0\}, \{v_1, h_1\},...\}$.
EXAMPLE if `number_of_components` is 3 and `subsampling_code` is 0x12345, then $v_0 = 5$, $h_0 = 4$, $v_1 = 3$, $h_1 = 2$, $v_2 = 1$, $h_2 = 0$.
For each element of `stream.components` its `v_samp_factor` and `h_samp_factor members` are updated to be $v_i + 1$ and $h_i + 1$ correspondingly, where `i` is the index of the element.
If any resulting value of `v_samp_factor` and `h_samp_factor` members is greater than 15, the codestream is ill-formed.
After that, the derivative values shall be calculated (see Aliases).
For each `component`, an element of `stream.components`, `component.coeffs` member shall be resized to have exactly `component.width_in_blocks` columns and `component.height_in_blocks` rows.

#### M.4.4.1 Width subsection

JPEG1 image width, in pixels.
`marker == 0x08` (`tag` is 1, `wire_type` is "Varint").
Payload: `width`.

#### M.4.4.2 Height subsection

JPEG1 image height, in pixels.
`marker == 0x10` (`tag` is 2, `wire_type` is "Varint").
Payload: `height`.

#### M.4.4.3 Version and component count subsection

Encoded recompression format compressor version and colour component count
`marker == 0x18` (`tag` is 3, `wire_type` is "Varint").
Payload: `version_and_component_count_code`.

#### M.4.4.4 Subsampling subsection

Encoded JPEG1 components sampling factors
`marker == 0x20` (`tag` is 4, `wire_type` is "Varint").
Payload: `subsampling_code`.

### M.4.5 Metadata section

The Metadata section contains auxiliary JPEG1 sections.
`marker == 0x1A` (`tag` is 3, `wire_type` is "Length-delimited").
Parsing of this section is done in two stages:

- first `encoded_data` byte array is extracted
- next, `encoded_data` is transformed into an ordered list of `stream.app_data`, `stream.com_data` elements and `stream.tail_data` value

First stage, depending on the payload length, works as follows:

- if the payload length is 0, then `encoded_data` is empty
- if the payload length is 1, then `encoded_data` is payload (1 byte) itself
- otherwise:
  - `uncompressed_size = Varint()`
  - remaining payload is a Brotli stream (see RFC 7932); `encoded_data` is obtained as a result of decompression; if `len(encoded_data) != uncompressed_size`, the codestream is ill-formed

On the second stage `encoded_data` is parsed as a concatenation of 1-, 2-, and multi-byte sequences. The first byte every of sequence is `type`. Transformation procedures for sequences of different classes are described in subclauses.
Sequences of different `type` are classified as follows:

- if `type` is in the range [0x00, 0x40), then sequence length is 1; Common APP0 sequence shall be generated (from type) and appended to `stream.app_data`
- if `type` is in the range [0x80, 0x83), then sequence length is 2; second byte of sequence is `code`; Common APPn sequence shall be generated (from `type` and `code`) and appended to `stream.app_data`
- if `type == 0xD9`, then `stream.tail_data` is assigned the remaining part of `encoded_data`
- if `type` is in the range [0xE0, 0xF0), then that is a multi-byte sequence, representing the Raw APPn sequence; its content is appended to `stream.app_data`
- if `type == 0xFE`, then that is a multi-byte sequence, representing the Raw COM sequence; its content, except the first byte, is appended to `stream.com_data`
- otherwise the codestream is ill-formed

### M.4.5.1 Common APP0 sequence

Common variants of "JFIF" (APP0) marker are generated using a `template` array provided in Table M.4 and 6-bit `type` value.
`(type & 1) + 1` is translated to the minor part of "version" (byte 9 in `template`).
Bits 1 and 2 of `type` become bits 0 and 1 of "density units" (byte 10 byte in `template`).
The remaining 3 bits (third to fifth) of type encode "xdensity" and "ydensity" (bytes at {11, 12, 13, 14} positions in `template`). Both "xdensity" and "ydensity" has the same `density` stored in big-endian order (highest 8 bits are stored in earlier byte). `density` is translated from `type` using the Table M.5.
After substitutions are applied to `template` the resulting array is appended to `stream.app_data`.

**Table M.4 – APP0 template**

Defined in `app0.md` (see Annex R).
NOTE template length is 17, MD5 checksum is `AE788A59D4774F869C5FB48D2CED6B26`.

**Table M.5 – type to density translation**

| `type >> 3` | 0 | 1 | 2 | 3 | 4 | 5 | 6 | 7 |
|---|---|---|---|---|---|---|---|---|
| `density` | 1 | 72 | 96 | 100 | 150 | 180 | 240 | 300 |

### M.4.5.2 Common APPn sequence

Variants of 3 APPn markers are efficiently encoded with 2 bytes – `type` and `code`.
`type` values correspond to 3 templates:

- 0x80 stands for common ICC profile; content is generated using `template` from Table M.6 with byte at index 84 substituted with `code`
- 0x81 stands for "Ducky" marker; content is generated using `template` from Table M.7 with byte at index 15 (quality) substituted with `code`
- 0x82 stands for "Adobe" marker; content is generated using `template` from Table M.8 with byte at index 10 (flags) substituted with `code`

After substitutions are applied, the resulting content is appended to `stream.app_data`.

**Table M.6 – common ICC profile template**

Defined in `icc.md` (see Annex R).
NOTE template length is 3161, MD5 checksum is `C02BFC5B3730AC5B6E26C943ACC0F651`.

**Table M.7 – "Ducky" marker template**

Define in `ducky.md` (see Annex R).
NOTE template length is 18, MD5 checksum is `36262F60927E01B4CD40885F6052A10E`.

**Table M.8 – "Adobe" marker template**

Defined in `adobe.md` (see Annex R).
NOTE template length is 15, MD5 checksum is `AC7C549589B20B9FC21FEC1675F85134`.

### M.4.5.3 Raw APPn sequence

This sequence is a JPEG1 APPn marker without the first byte (0xFF).

That means that the first byte of the sequence is 0xE*n* and the following 2 bytes encode the length of the remaining part of sequence, including themselves. Length is a 16-bit unsigned integer stored in big-endian order (highest 8 bits are stored in earlier byte).
EXAMPLE {0xE9, 0x00, 0x04, 0xCA, 0xFE}.

#### M.4.5.4 Raw COM sequence

This sequence is a JPEG1 COM marker without the first byte (0xFF).
That means that the first byte of the sequence is 0xFE and the following 2 bytes encode the length of the remaining part of sequence, including themselves. Length is a 16-bit unsigned integer stored in big-endian order (highest 8 bits are stored in earlier byte).
EXAMPLE {0xFE, 0x00, 0x05, 0xDE, 0xC0, 0xDE}.

### M.4.6 JPEG1 Internals section

JPEG1 internals section contains auxiliary information required for byte-exact original JPEG1 stream reconstruction.
`marker == 0x22`. (`tag` is 4, `wire_type` is "Length-delimited").
Decoding of section is performed in stages.
On the first stage `stream.marker_order` elements are decoded in the loop:

- `m = 0xC0 + u(6)`
- append `m` to `stream.marker_order`
- if `m == 0xD9`, then loop is finished

After that, if at least one `stream.marker_order` element is 0xDD (DRI marker), then:

- `stream.restart_interval = u(16)`

On the second stage `stream.huffman_code` elements are decoded in the loop:

- `is_last = Bool()`
- `HuffmanCode` entity is decoded using `is_last` value (see subclause), and appended to `stream.huffman_code`
- if `is_last == true`, then loop is finished
- if `len(stream.huffman_code) == 512`, the codestream is ill-formed

After that correctness check is performed: if the number of 0xC4 elements in `stream.marker_order` is not equal to the number of `stream.huffman_code` elements with `is_last` member equal to 1, the codestream is ill-formed.
On the third stage `stream.scan_infos` elements are decoded:

- `num_scans` is assigned the number of 0xDA elements of `stream.marker_order`
- `num_scans` times:
    - `ScanInfo` entity is decoded (see subclause), and appended to `stream.scan_infos`

On the fourth stage `stream.quant` elements are partially decoded:

- `num_quant = u(2) + 1`
- `num_quant` times:
    - initialize `q` as `QuantTable`
    - `q.index = u(2)`
    - `q.is_last = (len(stream.quant) == (num_quant - 1)) ? 1 : u(1)`
    - `q.precision = u(4)`
    - if `q.precision > 1`, the codestream is ill-formed
    - append `q` to `stream.quant`

On the fifth stage `stream.components` elements `id` members are decoded:

- for every `component` of `stream.components`:
    - `component.id = 0`
- `colour_code = u(2)`
- if `colour_code == 0`, then:
    - if `len(stream.components) < 3`, the codestream is ill-formed
    - `stream.components[0] = 1`
    - `stream.components[1] = 2`

- `stream.components[2] = 3`
- if `colour_code == 1`, then:
    - `stream.components[0] = 1`
- if `colour_code == 2`, then:
    - if `len(components) < 3` field, the codestream is ill-formed
    - `stream.components[0] = 0x52` ('R')
    - `stream.components[1] = 0x47` ('G')
    - `stream.components[2] = 0x42` ('B')
- if `colour_code == 3`, then:
    - for every `component` of `stream.components`:
        - `component.id = u(8)`

On the sixth stage `stream.padding_bits` elements are decoded:

- `num_padding_bits = GL(8, 4)`
- `num_padding_bits` times:
    - append `u(1)` to `stream.padding_bits`

After that `state.has_zero_padding_bit` shall be calculated.
Before the seventh stage decoder shall read `pu()` and discard the value.
On the seventh stage `stream.nter_marker_data` elements are decoded:

- `num_intermarker` is assigned the number of 0xFF elements of `stream.marker_order`
- `num_intermarker` times:
    - `length = Varint()`
    - `data` is initialized as an empty array of bytes
    - `length` times:
        - `u(8)` is appended to data
    - `data` is appended to `stream.nter_marker_data`

#### M.4.6.1 HuffmanCode stream

The HuffmanCode stream contains data required for symbol histogram reconstruction.
`is_last` flag (calculated in the outer loop) is used in computations.
Before decoding `hc` is initialized as HuffmanCode entity.
At first stage:

- `slot = u(2)`
- `is_ac = Bool()`
- `hc.slot_id = slot + (is_ac ? 0x10 : 0)`
- `hc.is_last = is_last ? 1 : u(1)`
- `hc.counts` and `hc.values` are initialized with 0 elements
- `is_stock = Bool()`

If `is_stock` is set, then:

- `stock_index = u(1)`
- first elements of `hc.counts` and `hc.values` are copied from Table M.9 and Table M.10 respectively
- after the decoding of `HuffmanCode` stream is finished, `hc` is returned as a result

**Table M.9 – stock counts arrays**

Defined in `stock_counts.md` (see Annex R).

**Table M.10 – stock values arrays**

Defined in `stock_values.md` (see Annex R).
Otherwise (`is_stock == false`) decoding of `hc.counts` and `hc.values` is performed in the second and third stages.
On the second stage `hc.counts` elements are decoded:

- `total_count = 0`
- `max_count = is_ac ? 256 : 12`
- `space = 65536`
- `max_len = u(4) + 1`
- `hc.counts[max_len] += 1`

- `space -= 1 << (16 - max_len)`
- for every unsigned integer `l` in the range [`1`, `max_len + 1`):
    - `remaining_count = max_count - total_count`
    - `allowed_count = space >> (16 - l)`
    - `required_count = min(remaining_count, allowed_count)`
    - if `required_count > 0`:
        - `count = u(ilog2(required_count))`
        - if `count > required_count`, the codestream is ill-formed
        - `counts[l] += count`
        - `space -= count << (16 - max_len)`
        - `total_count += count`

After all `hc.counts` elements are calculated it is known that first `total_count + 1` elements of `hc.values` shall be filled; "terminal" symbol is not encoded, and processed explicitly.
On the third stage `hc.values` elements are decoded:

- `count = 0`
- `order` list is filled with predefined values, from Table M.11
- `total_count` times:
- `order_length = len(order)`
- `required = (order_length == 1) ? 0 : ilog2(order_length - 1)`
- `required = ceil(required / 2)`
- `code = N(2, required)`
- if `code >= order_length`, the codestream is ill-formed
- `hc.values[count] = order[code]`
- `count += 1`
- The element of `order` with index `code` is removed from the list (i.e. all elements after it will get their index decremented by 1, and `order` will become shorter by 1)
- `hc.values[count] = 256`

**Table M.11 – predefined symbol order**

Defined in `symbol_order.md` (see Annex R).

### M.4.6.2 ScanInfo stream

`ScanInfo` entities contains data about progressive encoding scans.
Before decoding is started, `si` is initialized as `ScanInfo`.
On the first stage scalar members of `si` are decoded:

- `si.Ss = u(6)`
- `si.Se = u(6)`
- `si.Ah = u(4)`
- `si.Al = u(4)`

On the second stage elements of `si.component_scans` are decoded:

- `num_components = u(2) + 1`
- `num_components` times:
    - `csi` is initialized as `ComponentScanInfo`
    - `csi.comp_idx = u(2)`
    - `csi.dc_tbl_idx = u(2)`
    - `csi.ac_tbl_idx = u(2)`
    - `csi` is appended to `si.component_scans`

On the third stage elements of `si.reset_points` is decoded:

- `block_index = -1`
- loop:
    - `has_hext = Bool()`
    - if `has_hext == false`, then loop is finished
    - `block_index += SL(28) + 1`
    - `block_index` is put to `si.reset_points`

On the fourth stage elements of `si.extra_zero_runs` are decoded:

- `zero_runs` is initialized as empty array of unsigned integers
- `block_index = 0`
- loop:
  - `has_hext = Bool()`
  - if `has_hext == false`, then the loop is finished
  - `block_index += SL(28)`
  - `block_index` is appended to `zero_runs`

After that `zero_runs` is converted to `si.extra_zero_runs` elements: distinct values become `block_idx` member values, repetition counts become `num_extra_zero_runs` member values.

EXAMPLE {1, 1, 2, 2, 2, 3} becomes to 3 elements $z_1$, $z_2$ and $z_3$; $z_1$.block_idx == 1, $z_1$.num_extra_zero_runs == 2, $z_2$.block_idx == 1, $z_2$.num_extra_zero_runs == 3, $z_3$.block_idx == 3, $z_3$.num_extra_zero_runs == 1

### M.4.7 Quant Data section

The Quant data section contains quantization tables values and quantization table mapping.
`marker == 0x2A` (`tag` is 5, `wire_type` is "Length-delimited").
On the first stage integrity check is performed:

- `num_quant_tables = u(2) + 1`
- if `num_quant_tables != len(stream.quant)`, the codestream is ill-formed

On the second stage `stream.quant` elements are refined:

- for every integer `i` in the range [0, `num_quant_tables`):
  - `is_luma = (i == 0) ? true : false`
  - let `q` reference `stream.quant[i]`
  - `is_stock = u(1)`
  - if `is_stock == true`:
    - `stock_index = u(3)`
    - corresponding array from Table M.12 is copied to `q.values`
    - loop iteration is finished
  - `scale = u(6)`
  - let `template` reference corresponding array from Table M.13
  - `delta = 0`
  - for every integer `k` in the range [0, 64):
    - `has_correction = Bool()`
    - if `has_correction == true`:
      - `correction_sign = Bool()`
      - `abs_correction = SL(16) + 1`
      - `delta += abs_correction × (correction_sign ? -1 : 1)`
    - `approximation = (template[zigzag[k]] × scale + 32) >> 6`
    - `value = min(1, max(approximation, 255)) + delta`
    - if `value < 1`, the codestream is ill-formed
    - if `q.precision == 0` and `value >= 256`, the codestream is ill-formed
    - `q.values[zigzag[k]] = value`

On the third stage `stream.component` elements are refined:

- for every integer `i` in the range [0, `len(stream.component)`):
  - `quant_idx = u(2)`
  - if `quant_idx >= num_quant_tables`, the codestream is ill-formed
  - `stream.component[i].quant_idx = quant_idx`

After that, the decoder shall read `pu()` and discard the value.

**Table M.12 – stock quant tables**

Defined in `stock_quant.md` (see Annex R).

**Table M.13 – template quant tables**

Defined in `quant.md` (see Annex R).

### M.4.8 Histogram Data section

Histogram data section contains data describing the entropy sources used for AC / DC coefficient decoding.
`marker == 0x32` (`tag` is 6, `wire_type` is "Length-delimited").
On the first stage `state.schemes` elements are decoded:

- `len(stream.components)` times:
  - `scheme = u(3)`
  - if `scheme >= 7`, the codestream is ill-formed
  - `scheme` is appended to `state.schemes`
  - `num_contexts += context_count[scheme]` (see Table M.14)
- `num_contexts *= num_avg_ctx`

**Table M.14 – context_count**

```
8, 15, 31, 61, 120, 231, 412
```

On the second stage `state.context_map` elements are decoded:

- `num_histograms = U8() + 1`
- `num_contexts` times:
  - append 0 to `state.context_map`
- if `num_histograms > 1`:
  - decode `state.context_map` as described in the subclause

On the third stage `state.entropy_codes` elements are decoded:

- `num_histograms` times:
  - decode `d` – "sANS distribution stream" with `alphabet_size == 18`
  - append `d` to `state.entropy_codes`

#### M.4.8.1 ContextMap stream

ContextMap stream encodes `num_contexts` unsigned integer values, and each value is in the range [0, `num_histograms`).
On the first, stage auxiliary variables are decoded / initialized:

- `use_rle_for_zeros = Bool()`
- `max_run_length_prefix = use_rle_for_zeros ? (u(4) + 1) : 0`
- `alphabet_size = num_histograms + max_run_length_prefix`
- decode `huffman_code` as Huffman Histogram stream with specified `alphabet_size`

On the second, stage an array of intermediate values `map` is decoded:

- `map` is initialized as an empty array of unsigned integers
- `zero_count = 0`
- `num_contexts` times:
  - if `zero_count > 0`, then:
    - 0 is appended to `map`
    - `zero_count -= 1`
    - loop iteration is finished
  - `symbol` is decoded using `huffman_code`
  - if `symbol == 0`, then:
    - 0 is appended to `map`
    - loop iteration is finished
  - if `symbol > max_run_length_prefix`, then:
    - `symbol - max_run_length_prefix` is appended to `map`
    - loop iteration is finished
  - `extra = u(symbol)`
  - 0 is appended to `map`
  - `zero_count = (1 << symbol) + extra - 1`
- if `zero_count != 0`, the codestream is ill-formed

On the third stage, map is transformed to resulting values:

- apply_inverse_move_to_front = Bool()
- values list is initialized with consecutive integers from 0 to 255
- if apply_inverse_move_to_front == true then:
  - for each unsigned integer i in the range [0, len(map)):
    - m = map[i]
    - v = values[m]
    - element with index m is removed from values
    - v is prepended to values
    - map[i] = v
- map is returned as result

### M.4.9 DC Data section

DC data section contains values of lowest frequency DCT coefficients, and "is empty block" flags.
marker == 0x3A (tag is 7, wire_type is "Length-delimited").
Items (DC coefficients of DCT blocks) are decoded in DCT Block Order. For each entry, let block_index be the index of element in the described order, let c be the colour component index being processed, let bx and by be the indices of column and row of DCT block being processed.
Before decoding, sANS Symbol Decoder ans, bu(n) Decoder and 8-bit Binary Arithmetic Decoder bits are initialized; dc_states is initialized as an array of len(stream.components) ComponentStateDC entities.
Whenever in this subclause is_zero, is_empty, sign, first_extra, prev_is_nonempty, prev_abs, prev_sign are mentioned, it means the members of dc_states[c]; block_coeffs means the reference to state.components[c].coeffs.
For each DCT block decoding shall be done as follows:

- is_empty_ctx = prev_is_nonempty[bx - 1, by] + prev_is_nonempty[bx, by - 1]
- is_empty_bit = A(bits, is_empty[is_empty_ctx]) ? true : false
- state.block_state[block_index] = is_empty_bit
- abs_val = 0
- sign_bit = 0
- if is_empty_bit == 0, then:
  - is_zero_bit = A(bits, is_zero)
  - if is_zero_bit == 0, then:
    - sign_ctx = prev_sign[bx - 1, by] + 3 × prev_sign[bx, by - 1]
    - sign_bit = A(bits, sign[sign_ctx])
    - avg = 1 + prev_abs[bx - 2, by] + prev_abs[bx - 1, by] + prev_abs[bx, by - 1] + prev_abs[bx + 1, by - 1]
    - avg_ctx = min(num_avg_ctx - 1, floor(log2(avg))
    - code = E(ans, entropy_codes[num_avg_ctx × component_index + avg_ctx])
    - if code < 8, then abs_val = code + 1
    - if code >= 8:
      - num_extra = code - 8
      - first_extra_bit = A(bits, first_extra[num_extra])
      - exta_bits = bu(num_extra)
      - abs_val = 7 + (1 << (num_extra + 1)) + (first_extra_bit << num_extra) + extra_bits
- prev_abs[bx, by] = abs_val
- prev_sign[bx, by] = (abs_val == 0) ? 0 : (sign_bit + 1)
- delta = ((1 - 2 × sign_bit) × abs_val
- predicted_dc = 0
- if by == 0 and bx > 0, then predicted_dc = block_coeffs[bx - 1, by][0]
- if by > 0 and bx == 0, then predicted_dc = block_coeffs[bx, by - 1][0]
- if by > 0 and bx == 0:

- `n = block_coeffs[bx, by - 1][0]`
    - `w = block_coeffs[bx - 1, by][0]`
    - `nw = block_coeffs[bx - 1, by - 1][0]`
    - `predicted_dc = n + w - nw`
    - if `nw > max(n, w)`, then `predicted_dc = min(n, w)`
    - if `nw < min(n, w)`, then `predicted_dc = max(n, w)`
- `block_coeffs[bx, by][0] = predicted_dc + delta`

After all DC coefficients are decoded, `ans` is finalized (checked for integrity).

### M.4.10 AC Data section

AC data section contains contains values of higher frequency DCT coefficients.
`marker == 0x42` (`tag` is 8, `wire_type` is "Length-delimited").

Before decoding, sANS Symbol Decoder `ans`, `bu(n)` Decoder and 8-bit Binary Arithmetic Decoder `bits` are initialized.
On the first stage `ac_states` is initialized:

- `num_c = len(stream.components)`
- `ac_states` is initialized as an array of `num_c` `ComponentStateAC` entities
- `context_offset = num_components`
- for every unsigned integer `c` in the range [0, `num_c`):
    - let `acs` reference `ac_states[c]`
    - let `q` reference `stream.quant[stream.component[c].quant_idx]`
    - `acs.context_offset = num_avg_ctx × context_offset`
    - `context_offset += context_count[state.schemes[c]]` (see Table M.14)
    - `multiplier = floor(sqrt(2) × 8192))`
    - for every integer `y` in the range [0, 8):
        - for every integer `x` in the range [0, 8):
            - `acs.mult_row[x + 8 × y] = (q[x + 8 × y] × multiplier) Idiv q[y × 8]`
            - `acs.mult_col[x + 8 × y] = (q[x + 8 × y] × multiplier) Idiv q[x]`
    - `code` is initialized as empty array
    - while `len(code) < 64`:
        - `is_nonzero_span = bu(1)`
        - if `is_nonzero_span == 0`, then:
            - 16 0-valued elements are appended to `code`
            - loop iteration is finished
        - 16 times:
            - `v = 0`
            - if `len(code) > 0`:
                - while `v < 64`:
                    - `delta = bu(3)`
                    - `v += delta`
                    - if `delta < 7`, then loop is finished
                - if `v > 64`, the codestream is ill-formed
            - `v` is appended to `code`
    - if at least one element of `code` is not 0:
        - let `end` be the index of the last element of `code` not equal to 0
        - if `code[end] == 1`, the codestream is ill-formed
        - for every integer `i` in the range [`1`, `end + 1`):
            - if `code[i] == 0`, the codestream is ill-formed
            - `code[i] -= 1`
    - `items` is initialized as list with 64 sequential integers in the range [0, 64)

- `acs.order` is initialized as empty array
- for each `code` element `offset`:
    - if `offset >= len(items)`, the codestream is ill-formed
    - `zigzag[items[offset]]` is appended to `acs.order`
    - element of `items` with index `offset` is removed from the list (i.e. all elements after it will get their index decremented by 1, and `items` will become shorter by 1)

On the second stage, AC coefficients are decoded.

Items (AC coefficients of DCT blocks) are decoded in DCT Block Order. For each entry, let `block_index` be the index of element in the described order, let `c` be the colour component index being processed, let `bx` and `by` be the indices of column and row of DCT block being processed; `block_coeffs` means the reference to `state.components[c].coeffs`, `scheme` means `state.schemes[c]`; `freq_context` and `num_nonzero_context` mean the references to the corresponding arrays from Table M.15 and Table M.16; `nonzero_buckets` means array from Table M.17.

In this subclause, `context_offset`, `is_zero`, `sign`, `first_extra`, `num_nonzeros`, `prev_is_nonempty`, `prev_num_nonzeros`, `prev_abs`, `prev_sign`, `order`, `mult_row` and `mult_col` denote the corresponding members of `ac_states[c]`.

For each DCT block decoding shall be done as follows:

- `last_nonzero = 0`
- `num_nonzero = 0`
- if `state.block_state[block_index] == false`:
    - `nonzero_ctx = prev_num_nonzeros[bx - 1, by] + prev_num_nonzeros[bx, by - 1]`
    - if `bx > 0` and `bx > 0`, then `nonzero_ctx = nonzero_ctx Idiv 2`
    - if `bx == 0` or `by == 0`, then `nonzero_ctx = (nonzero_ctx + 1) Idiv 4`
    - `num_nonzeros_probs` is assigned a reference to `num_nonzeros[nonzero_ctx]`
    - for `b` in range [0, 6):
        - `bit = A(bits, num_nonzeros_probs[(1 << b) - 1 + last_nonzero])`
        - `last_nonzero = 2 × last_nonzero + bit`
- for every integer `m` in the range [1, 64):
    - `k = 64 - m`
    - `pos = order[k]`
    - if `k > last_nonzero`, then:
        - `prev_abs[bx, by][k] = 0`
        - `prev_sign[bx, by][k] = 0`
        - `block_coeffs[bx, by][pos] = 0`
        - iteration is finished
    - if `k == last_nonzero`, then `is_zero_coeff = false`
    - if `k < last_nonzero`:
        - `nonzero_bucket = nonzero_buckets[num_nonzero]`
        - `is_zero_coeff = A(bits, is_zero[nonzero_bucket][k]) ? true : false`
    - if `is_zero_coeff == true`, then:
        - `prev_abs[bx, by][k] = 0`
        - `prev_sign[bx, by][k] = 0`
        - `block_coeffs[bx, by][pos] = 0`
        - iteration is finished
    - `dct_row = pos Idiv 8`
    - `dct_col = pos Umod 8`
    - `avg_ctx = 0`
    - `sign_ctx = 8`
    - if `dct_row == 0` and `by > 0`:
        - `delta = 0`
        - for every integer `i` in the range [1, 8):

- `delta += mult_col[dct_col + i × 8] × (block_coeffs[bx, by][dct_col + i × 8] + (2 × (i & 1) - 1) × coeffs[bx, by - 1][dct_col + i × 8])`
        - `prediction = coeffs[bx, by - 1][dct_col] - truncate(delta / 8192)`
        - `avg_ctx = max(num_avg_ctx - 1, ilog2(2 × abs(prediction) + 1) - 1)`
        - `sign_ctx = num_avg_ctx - 1 + copysign(avg_ctx, prediction)`
    - if `dct_row > 0` and `dct_col == 0` and `bx > 0`:
        - `delta = 0`
        - for every integer `i` in the range [1, 8):
            - `delta += mult_row[i + dct_row × 8] × (block_coeffs[bx, by][i + dct_row × 8] + (2 × (i & 1) - 1) × coeffs[bx - 1, by][i + dct_row × 8])`
        - `prediction = coeffs[bx - 1, by][pos] - truncate(delta / 8192)`
        - `avg_ctx = max(num_avg_ctx - 1, ilog2(2 × abs(prediction) + 1) - 1)`
        - `sign_ctx = num_avg_ctx - 1 + copysign(avg_ctx, prediction)`
    - if `dct_row > 0` and `dct_col > 0`:
        - `weight = 4 + coeffs[bx, by - 2][k] + 2 × coeffs[bx - 1, by][k] + 2 × coeffs[bx, by - 1][k] + coeffs[bx - 2, by][k] + coeffs[bx - 1, by - 1][k] + coeffs[bx + 1, by - 1][k]`
        - `avg_ctx = min(num_avg_ctx - 1, ilog2(weight) - 3)`
        - `sign_ctx = prev_sign[bx - 1, by] + 3 × prev_sign[bx, by - 1]`
    - `sign_bit = A(bits, sign[sign_ctx][k])`
    - `entropy_ctx = freq_context[num_nonzero] + freq_context[k]`
    - `code = E(ans, entropy_codes[context_offset + num_avg_ctx × entropy_ctx + avg_ctx])`
    - if `code < 8`, then `abs_val = code + 1`
    - if `code >= 8`:
        - `num_extra = code - 8`
        - `first_extra_bit = A(ans, first_extra[num_extra][k])`
        - `exta_bits = bu(num_extra)`
        - `abs_val = 7 + (1 << (num_extra + 1)) + (first_extra_bit << num_extra) + extra_bits`
    - `num_nonzero += 1`
    - `prev_abs[bx, by][k] = abs_val`
    - `prev_sign[bx, by][k] = sign_bit + 1`
    - `block_coeffs[bx, by][pos] = ((1 - 2 × sign_bit) × abs_val`

After all AC coefficients are decoded `ans` is finalized (checked for integrity).

**Table M.15 – freq_context**

Defined in `freq_context.md` (see Annex R).

**Table M.16 – num_nonzero_context**

Defined in `num_nonzero_context.md` (see Annex R).

**Table M.17 – nonzero_buckets**

Defined in `nonzero_buckets.md` (see Annex R).

### M.4.11 Fallback section

The Fallback section contains original JPEG1 file contents. It may be used, e.g. when it is otherwise impossible to convert the original JPEG1 stream contents to recompressed stream.

`marker == 0x4A` (`tag` is 9, wire_type is "Length-delimited").

Payload: original_jpg (array of bytes).

Payload shall be transferred directly to `stream.original_jpg`.

## M.5 Reconstruction of JPEG1

`stream` contains all the necessary information for byte-precise reconstruction of JPEG1 streams.
The stream is designed to allow most JPEG1 streams to be reconstructed in Regular mode. For JPEG1 streams that are impossible to be synthesized in Regular mode, the Fallback mode allows storing the raw stream without any changes.

### M.5.1 Fallback mode

If `stream.version == 1`, then JPEG1 stream is reconstructed in fallback mode.
The output is copied from `stream.original_jpg` without any transformations. Output stream may be non-JPEG1 compliant.

### M.5.2 Regular mode

If `stream.version == 0`, then JPEG1 stream is reconstructed using other members of `stream`.
Output stream is produced as a concatenation of {0xFF, 0xD8} bytes (implicit SOI segment) and a sequence of segments, generated on the basis of `stream.marker_order` elements. Element value determines the `type` of segment. Detailed instructions for various `type` values are specified in subclauses. If encountered `type` does not match any instruction, then the codestream is ill-formed.
`stream.app_data`, `stream.com_data`, `stream.inter_marker_data`, `stream.huffman_code`, `stream.quant` and `stream.scan_infos` members are accessed through "iterators"; this means that during JPEG1 reconstruction every element (usually referenced as "**next**") shall be requested only once.

#### M.5.2.1 SOF segment

"Start Of Frame" segment.
`type` is one of {0xC0, 0xC1, 0xC2, 0xC9, 0xCA}.
Segment starts with 10 bytes: {0xFF, `type`, `len_hi`, `len_lo`, 8, `height_hi`, `height_lo`, `width_hi`, `width_lo`, `number_of_components`}, where `*_hi` and `*_lo` are the highest and lowest 8 bits of corresponding values, `len` is the final length of this segment minus 2, `height` and `width` are `stream.height` and `stream.width` correspondingly, and `number_of_components = len(stream.components)`.
For each element of `components c`, 3 more bytes are appended to segment: {`c.id`, `(c.h_samp_factor << 4) | c.v_samp_factor`, `c.quant_idx`}.

#### M.5.2.2 DHT segment

"Define Huffman Table(s)" segment.
`type` is 0xC4.
In this segment a series of `HuffmanCode` entities are serialized. Entities are fetched from the *next* `stream.huffman_code` until the element with set `is_last` member is met.
Segment starts with 4 bytes: {0xFF, 0xC4, `len_hi`, `len_lo`}, where `len_hi` and `len_lo` are the highest and lowest 8 bits of the final length of this segment minus 2.
For each `HuffmanCode` entity `hc`, segment content is defined in ISO/IEC 10918-1:1993(E) / B.2.4.2, with the following mapping:

- Tc together with Th, encoded as a single byte, is `hc.slot_id`
- $L_i$ are corresponding `hc.counts` elements, except for the last non-zero element, which shall be decremented by 1, before storing
- (flattened) $V_{i,j}$ are corresponding `hc.values` elements

#### M.5.2.3 RSTn segment

"Restart" segment.
`type` is in the range [0xD0, 0xD8).
Segment contains 2 bytes: {0xFF, `type`}.

### M.5.2.4 EOI segment

"End Of Image" segment; shall be the last segment, otherwise the codestream is ill-formed.
`type` is 0xD9.
Segment starts with {0xFF, 0xD9} bytes. The rest of the segment is copied from `stream.tail_data`.

### M.5.2.5 SOS segment

"Start Of Scan" segment.
`type` is 0xDA.
Let `scan_info` be the next `stream.scan_infos` element, `num_scans = len(scan_info.component_scans)`, and `len = 6 + 2 × num_scans`.
Segment starts with {0xFF, 0xDA, `len_hi`, `len_lo`, `num_scans`} bytes, where `len_hi` and `len_lo` are the highest and lowest 8 bits of `len`.
For each `scan_info.component_scans` element `csi`, the segment content is defined in ISO/IEC 10918-1:1993(E) / B.2.3, with the following mapping:

- $Cs_j$ is `stream.components[csi.comp_idx].id`
- $Td_j$ is `csi.dc_tbl_idx`
- $Ta_j$ is `csi.ac_tbl_idx`

Next 3 bytes of segment are also defined in ISO/IEC 10918-1:1993(E) / B.2.3, with the following mapping:

- Ss is `scan_info.Ss`
- Se is `scan_info.Se`
- Ah is `scan_info.Ah`
- Al is `scan_info.Al`

DCT coefficient data is encoded according to ISO/IEC 10918-1:1993(E) with the following changes to enable unambiguous bit-precise JPEG1 stream reconstruction:

- if `stream.has_zero_padding_bit == false` then, making entropy-coded segments (ISO/IEC 10918-1:1993(E) / B.1.1.5) an integer number of bytes is performed as follows: *next* bits from `stream.padding_bits` are used, if necessary, to pad to the end of the compressed data to complete the final byte of a segment
- when encoding AC coefficients for sequential DCT (ISO/IEC 10918-1:1993(E) / F.1.2.2.1) `ezr.num_extra_zero_runs` extra "ZRL" symbols shall be emitted before processing the end-of-block, where `ezr` is `scan_info.extra_zero_runs` element, whose `block_idx` member matches the currently serialized block index (in the current scan); final "EOB" symbol is emitted only if the number of 0 coefficients remaining to be encoded is non-negative
- when encoding AC coefficients for progressive DCT (ISO/IEC 10918-1:1993(E) / G.1.2.2) `ezr.num_extra_zero_runs` extra "ZRL" symbols shall be emitted before updating "EOBRUN", where `ezr` is `scn_info.extra_zero_runs` element, whose `block_idx` member matches the currently serialized block index (in the current scan); "EOBRUN" is updated only if the number of 0 coefficients remaining to be encoded is non-negative
- before encoding the block, if the block the currently serialized block index (in the current scan) is present in `scan_info.reset_points` set, then "Encode_EOBRUN" is invoked (ISO/IEC 10918-1:1993(E) / G.1.2.2)

### M.5.2.6 DQT segment

"Define Quantization Table(s)" segment.
`type` is 0xDB.
In this segment a series of `QuantTable` entities are serialized. Series of items are fetched from the *next* `stream.quant` until the element with set `is_last` member is met.
Segment starts with {0xFF, 0xDB, `len_hi`, `len_lo`} bytes, where `len_hi` and `len_lo` are the highest and lowest 8 bits of the final length of this segment minus 2.

For each `QuantTable` element `q`, the serialized content is defined in ISO/IEC 10918-1:1993(E) / B.2.4.1, with the following mapping:

- Pq is `q.precision`
- Tq is `q.index`
- $Q_k$ are corresponding `q.values` elements

#### M.5.2.7 DRI segment

"Define Restart Interval" segment.
`type` is 0xDD.
Segment contains {0xFF, 0xDD, 0x00, 0x04, `hi`, `lo`} bytes, where `hi` and `lo` are the highest and lowest 8 bits of `stream.restart_interval`.

#### M.5.2.8 APPn segment

"Application-specific" segment.
`type` is in the range [0xE0, 0xF0).
Segment starts with {0xFF} byte. The rest of the segment is copied from the *next* `stream.app_data` element.
The second byte of generated segment shall be equal to `type`, otherwise the codestream is ill-formed.

#### M.5.2.9 COM segment

"Comment" segment.
`type` is 0xFE.
Segment starts with {0xFF, 0xFE} bytes. The rest of the segment is copied from the *next* `stream.com_data` element.

#### M.5.2.10 Unrecognized data segment

Unrecognized JPEG1 pieces are stored as raw unrecognized data segments.
`type` is 0xFF.
Segment contents are copied from *next* `stream.inter_marker_data` element.

## M.6 Direct rasterization

After the recompressed stream is decoded (M.4), the decoder may produce an image output directly, rather than serializing the intermediate JPEG1 stream first (M.5).

### M.6.1 Dequantization

Let `num_c = len(stream.component)`. For each `c` in the range [0, `num_c`) `stream.component[c].coeffs` is transformed into `grid` of `dct_block` of coefficients as follows:

- let `component = stream.component[c]`
- let `xsize = component.width_in_blocks`
- let `ysize = component.height_in_blocks`
- `dequantized` is initialized as `grid` of `dct_block` of coefficients with `xsize` columns and `ysize` rows
- let `quant_table = stream.quant[component.quant_idx].values`
- each `dct_block src` of `component.coeffs` is transformed to a corresponding `dct_block dst` of `dequantized`:
  - for every integer `u` in the range [0, 8):
    - for every integer `v` in the range [0, 8):
      - `index = 8 × u + v`
      - `dst[index] = src[index] × quant_table[index] × I(8, u) × I(8, v)` (see I.2.6.1)

#### M.6.1.1 DCT coefficients conditioning

DCT coefficients can be conditioned in a way that does not contradict the quantized values. The conditioning is controlled by 2 parameters: `max_gap`, `min_step`, signaled via an optional "side-channel" stored as an APP9 section with the following structure:

- {0xFF, 0xE9} – APP9 marker
- {`hi`, `lo`} – chunk size (similar to ICC profile section encoding)
- "BRN:DCC" – section identifier
- {0x01, 0x01} – part 1 of 1 (similar to ICC profile)
- `payload` – codestream of encoded parameters

If this side-channel information is present, the decoder shall read 2 parameters (from `payload`) for each colour component as follows, and then perform the conditioning as specified below, separately for every `stream.component` element `component`.

- `max_gap = u(8) / 64`
- `min_step = u(8) / 64`

NOTE If the side-channel information is not present, the decoder can choose whether to perform the conditioning, for example with default parameter values `max_gap = 0.5`, `min_step = 0.5`.

Let `xsize = component.width_in_blocks`, `ysize = component.height_in_blocks`.
Let `dc(u, v) = component.coeffs[u, v][0]`, `ac_h(u, v) = component.coeffs[u, v][1]`, `ac_v(u, v) = component.coeffs[u, v][8]`.
Let `q_ac_dc_ h = quant_table[1] / quant_table[0]` and `q_ac_dc_ v = quant_table[8] / quant_table[0]`.
Let `h_ranges` be the collection of all the triplets {`y`, `start`, `end`} fulfilling all the following requirements:

- 0 <= `start` <= `end` < `xsize`
- either `start == 0`, or `dc(start - 1, y) != dc(start, y)`
- either `end == xsize - 1`, or `dc(end + 1, y) != dc(end, y)`
- if `start == 0` then `end != xsize - 1`
- `dc(x, y) == dc(end, y)` for all integer `x` in the range [`start`, `end + 1`)
- `ac_h(x, y) == 0` for all integer `x` in the range [`start`, `end + 1`)

Let `v_ranges` be the collection of all triplets {`x`, `start`, `end`} fulfilling all the following requirements:

- 0 <= `start` <= `end` < `ysize`
- either `start == 0`, or `dc(x, start - 1) != dc(x, start)`
- either `end == ysize - 1`, or `dc(x, end + 1) != dc(x, end)`
- if `start == 0` then `end != ysize - 1`
- `dc(x, y) == dc(x, end)` for all integer `y` in the range [`start`, `end + 1`)
- `ac_v(x, y) == 0` for all integer `x` in the range [`start`, `end + 1`)

After the ranges have been identified, the approximations and slopes are determined:

- let `h_approx` be a `grid` with `xsize` columns and `ysize` rows
- let `h_slopes` be a `grid` with `xsize` columns and `ysize` rows
- let `has_h` be a `grid` of `boolean` with `xsize` columns and `ysize` rows, initialized with `false` values
- for each triplet {`y`, `start`, `end`} in `h_ranges`:
  - let `current = dc(y, start)`
  - if `start == 0`, then `from = current`
  - if `start > 0`, then `from = 0.5 x (current + dc(y, start - 1) + q_ac_dc_h x ac_h(x, start - 1))`
  - if `end == xsize - 1`, then `to = current`
  - if `end < xsize - 1`, then `to = 0.5 x (current + dc(y, end + 1) + q_ac_dc_h x ac_h(x, end + 1))`
  - `is_sane` is assigned `true` if all the following requirements are met (otherwise, `false`):
    - `abs(to - current) <= max_gap`
    - `abs(current - from) <= max_gap`

- `abs(to - from) > min_step`
    - if `is_sane == true`:
        - let `span = end - start + 1`
        - let `delta = to - from`
        - for every integer `x` in the range `[start, end]`:
            - let `i = x - start`
            - `h_approx[x, y] = from + (0.5 + i) x (delta / span)`
            - `h_slopes[x, y] = -0.5 x (delta / span)`
            - `has_h[x, y] = true`

What is done above with `h_ranges` shall also be done to `v_ranges`, resulting in a grid `v_approx`, `v_slopes` and `has_v`, with the following variable swaps: x* <-> y*, h* <-> v*, *_h <-> *_v.
Let `dc_multiplier = quant_table[0] × I(8, u) × I(8, v)` (see I.2.6.1).
Once the vertical and horizontal ranges of plateaus are identified, the three most important DCT coefficients are replaced with approximations, which are linear interpolations between the averages of the end points and the neighboring values:

- for every integer `y` in the range [0, `ysize`):
    - for every integer `x` in the range [0, `xsize`):
        - let `approx_dc = dc(x, y)`
        - if `has_h[x, y] && has_v[x, y]`:
            - `approx_dc = 0.5 x (h_approx[x, y] + v_approx[x, y])`
        - else if `has_h[x, y]`:
            - `approx_dc = h_approx[x, y]`
        - else if `has_v[v, u]`:
            - `approx_dc = v_approx[x, y]`
        - `dequantized[x, y][0] = approx_dc x dc_multiplier`
        - if `has_h[x, y]`:
            - `dequantized[x, y][1] = h_slopes[x, y] x dc_multiplier`
        - if `has_v[x, y]`:
            - `dequantized[x, y][8] = v_slopes[x, y] x dc_multiplier`

### M.6.2 IDCT

To convert DCT coefficients to pixels, the decoder shall apply `IDCT_2D` (I.2.2) to each `dct_block` of `dequantized`.

### M.6.3 Gaborish loop filter

The decoder shall parse "side-channel" and apply corresponding loop filter as follows.
"Side-channel" information contains the parameters of Gaborish filter and criteria of block-wise applicability. It is stored as APP9 section with the following structure:

- {0xFF, 0xE9} – APP9 marker
- {`hi`, `lo`} – chunk size (similar to ICC profile section encoding)
- "BRN:GAB" – section identifier
- {0x01, 0x01} – part 1 of 1 (similar to ICC profile)
- `payload` – codestream of encoded parameters

For each colour component, 4 parameters are decoded (from `payload`) the following way:

- `gab_weight1 = u(8) / 1024`
- `gab_weight2 = u(8) / 1024`
- `threshold = u(8) / 16`
- `limit = u(8)`

If "side-channel" is present in JPEG1 stream, then after pixel `grid pixels` is obtained for every colour component (M.6.2), the following is applied:

- derive `smooth_pixels` from `pixels` by using Gaborish convolution (J.4) with parameters `gab_weight1` and `gab_weight2` (as if they were stored in `loop_filter`)

- for each 8x8 block of colour plane:
  - let `count` be the number of pairs {`u`, `v`} such that:
    - `1 <= u < 7`
    - `1 <= v < 7`
    - let (`x`, `y`) be the coordinates of the pixel within the block with the offset (`u`, `v`)
    - `abs(pixels[x, y] - smooth_pixels[x, y]) > threshold`
  - if `count < limit` then block of pixels is copied from `smooth_pixels` to `pixels`

#### M.6.4 Upsampling

For each subsampled colour component `component` (i.e. when `component.h_samp_factor != state.max_h_samp_factor` or `component.v_samp_factor != state.max_v_samp_factor`), the decoder shall perform upsampling. If both horizontal and vertical resampling is required, then horizontal resampling is performed first.
Horizontal upsampling is performed row-wise, vertical upsampling is performed column-wise. In both cases linear upsampling is used, with the respect of pixel centers as described in "JPEG File Interchange Format, Version 1.02" / "Spatial Relationship of Components", with the boundary element mirroring (i.e. element at index -1 is the same as at index 1).
EXAMPLE array {0, 4, 12, 8} after 2x upsampling becomes {1, 1, 3, 6, 10, 11, 9, 9}; calculation: 1 == ¾ × 0 + ¼ × 4; 3 == ¾ × 4 + ¼ × 0; 6 == ¾ × 4 + ¼ × 12; 10 == ¾ × 12 + ¼ × 4; 11 == ¾ × 12 + ¼ × 8; 9 == ¾ × 8 + ¼ × 12

#### M.6.5 Colour space transformation

If the rasterized image is in the YCbCr colour space (`state.component[i].id = i + 1` for `i` in range [0, 3)) and requested colour space is RGB, then the pixel values are transformed, using the following equations:
- R = 1.402 × Cr + (Y + 128)
- G = -0.71414 × Cr - 0.34414 × Cb + (Y + 128)
- B = 1.772 × Cb + (Y + 128)

where R, G, B, Y, Cr, Cb are the corresponding pixel values from previous stage.

##### M.6.5.1 HDR colour space transformation

"User-friendly HDR" JPEG1 images are produced by encoding HDR images using a specific colour profile, that maps both dark and bright sample values into a single range. Regular JPEG1 decoders will produce an image looking less saturated (due to more dense colour space). Decoder shall use "side-channel" to restore the original colour space for proper HDR image rasterization.
"Side-channel" information is a pair of colour spaces that specifies the desired colour transform. It is stored as APP9 section with the following structure:
- {0xFF, 0xE9} – APP9 marker
- {`hi`, `lo`} – chunk size (similar to ICC profile section encoding)
- "BRN:HDR" – section identifier
- {0x01, 0x01} – part 1 of 1 (similar to ICC profile)
- `descriptor` – string with the contents like "orig_colourspace>storage_colourspace", where *_colourspace is a description of the colour encoding (e.g. "RGB_D65_SRG_Rel_Lin", or "RGB_D65_202_Rel_PeQ"), specified below.

During rasterization, if "side-channel" is provided, then embedded ICC profile (APP2 / "ICC_PROFILE\0") shall be ignored. After pixel values are obtained, the specified colour transform is applied (i.e. values are considered to be in storage_colourspace and translated back to orig_colourspace).

The colour encoding description consists of 4 codes concatenated with "_":
- colourspace, one of the 3 values {"RGB", "Gra", "XYZ"}, corresponding to A.4.3.2 {"kRGB", "kGrey", "kXYZ"} constants
- whitepoint, one of the 3 values {"D65", "EER", "DCI"}, corresponding to A.4.3.3 {"kD65", "kE", "kDCI"} constants

- primaries, one of the 3 values {"SRG", "202", "DCI"}, corresponding to A.4.3.4 {"kSRGB", "k2100", "kP3"} constants
- transfer function, one of the 7 values {"SRG", "Lin", "709", "PeQ", "HLG", "DCI"}, corresponding to A.4.3.4 {"kSRGB", "kLinear", "k709", "kPQ", "kHLG", "kDCI"} constants

### M.6.6 Cropping

As a final step, the excess rows from the bottom and excess columns from the right are removed to make the resulting image exactly the same size as the original one.

# Annex N
(informative)
**Intentionally left blank**

## Annex O
(normative)
**Lossless mode**

### O.1 Overview

Lossless mode is enabled if `frame_header.encoding == kLossless`. In this case, the decoder shall interpret differently the contents of the `AcGlobal` and `PassGroup` sections (see table C.1), as specified below.

The lossless decoder shall support 4 modes, given by LosslessMode indicated in the `frame_header` as defined in C.2.4:

- greyscale 8-bit, for a greyscale image with 8 bits per sample
- coloured 8-bit, for a coloured image with 8 bits per sample
- greyscale 16-bit, for a greyscale image with 16 bits per sample
- coloured 16-bit, for a coloured image with 16 bits per sample

The `AcGlobal` section shall contain an encoding of the per-channel palette, as specified in O.2.

Each `PassGroup` section shall contain the pixel data for the corresponding group of pixels, as specified in O.3.

### O.2 Decoding the per-channel palette

The per-channel palette indicates, independently per colour channel, a palette_size and a series of bits. The bits indicate whether each sample value is present in this channel (value 1) or not (value 0). If the LosslessMode is greyscale, only a single channel is present, otherwise respectively the red, green and blue channels.

In this subclause, “no palette present for a channel” means that:

- applying the per-channel palette for that channel shall produce no effects
- `palette_size` is `256` for `8`-bit images, and `65536` for `16`-bit images.

If the total number of pixels in the image (computed as xsize × ysize, using the dimensions of the full image, not just a PassGroup) is less than `257`, no values shall be read from the bitstream, and no palette shall be present for any channel.

Otherwise, the decoder shall read a `u(8)` indicating whether the palette is encoded or not. If this value is `0`, no further values shall be read from the bitstream and no palette shall present for any channel.

Otherwise, the per-channel palette shall be decoded from the stream in the following way.

In `8`-bit greyscale mode, the decoder shall read a `u(1)` for each value in `[0, 256)`. When applying the palette, `i` shall be mapped to the value `k` such that exactly `i` bits with value `1` were read for the values in `[0, k)`; the `palette_size` shall be the total number of `1` bits that were read with this procedure.

In `8`-bit colour mode, the decoder shall read three `u(1)`, indicating whether channels `0`, `1` and `2` respectively shall have a per-channel palette. It shall then read a `pu0()`.
The decoder shall then read three `u(8)`, which, after incrementing by `1`, shall indicate `n_read_1` for channels `0`, `1` and `2` respectively. For each channel, a corresponding value `n_read_0` shall be set to `256 - n_read_1`.
For each channel that has a per-channel palette, the decoder shall read a `u(1)` for each value in `[0, 256)`. After reading `n_read_1` bits with value `1`, the decoder shall stop reading bits from the bitstream and treat all the skipped bits as `0` bits, then read a `pu0()`. Alternatively, after reading `n_read_0` bits with value `0`, the decoder shall stop reading bits from the bitstream and treat all the skipped bits as `1` bits, then read a `pu0()`.

In `16`-bit mode, the decoder shall read a `u(1)` for each channel, indicating whether the palette shall be present for that channel. It shall then read a `pu0()`.
The decoder shall then read a `u(16)` for each channel, which, after incrementing by `1`, shall indicate `n_read_1` for each channel respectively. For each channel, a corresponding value `n_read_0` shall be set to `65536 - n_read_1`.

For each channel for which a palette is present, the decoder shall read a `u(1)` value `method` and a `u(15)` value $p_s$, which shall be used as follows:

- If `method` is `0`, the decoder shall read a `u(1)` for each value in `[0, 65536)`. After reading `n_read_1` bits with value `1`, the decoder shall stop reading bits from the bitstream and treat all the skipped bits as `0` bits, then read a `pu0()`. Alternatively, after reading `n_read_0` bits with value `0`, the decoder shall stop reading bits from the bitstream and treat all the skipped bits as `1` bits, then read a `pu0()`. The decoded palette shall be applied in the same way as the `8`-bit greyscale palette is applied.
- If `method` is `1`, the decoder shall use the procedure listed below to decode the palette, where `palette[i]` represents the actual value of the channel for which the decoded value is `i`.

```
// The special case decoder for the 16-bit per channel palette
DecodePalette16() {
  uint16_t smt0[64] = {
      0x2415, 0x1d7d, 0x1f71, 0x46fe, 0x24f1, 0x3f15, 0x4a65, 0x6236,
      0x242c, 0x34ce, 0x4872, 0x5cf6, 0x4857, 0x64fe, 0x6745, 0x7986,
      0x24ad, 0x343c, 0x499a, 0x5fb5, 0x49a9, 0x61e8, 0x6e1f, 0x78ae,
      0x4ba3, 0x6332, 0x6c8b, 0x7ccd, 0x6819, 0x8247, 0x83f2, 0x8cce,
      0x247e, 0x3277, 0x391f, 0x5ea3, 0x4694, 0x5168, 0x67e3, 0x784b,
      0x474b, 0x5072, 0x666b, 0x6cb3, 0x6514, 0x7ba6, 0x83e4, 0x8cef,
      0x48bf, 0x6363, 0x6677, 0x7b76, 0x67f9, 0x7e0d, 0x826f, 0x8a52,
      0x659f, 0x7d6f, 0x7f8e, 0x8f66, 0x7ed6, 0x9169, 0x9269, 0x90e4,
  };
  uint64_t pos = 0;
  uint32_t smt[64];
  int x1 = 0;
  int x2 = 0xffffffff;
  int xr = 0;
  int context6 = 0;
  uint64_t sumv = 0;
  for (i = 0; i < 4; ++i) {
    xr = (xr << 8) + (pos >= p_s ? 0xFF : u(8));
    pos++;
  }
  for (i = 0; i < 64; ++i) {
    smt[i] = smt0[i] << 11;
  }
  for (x = 0; x < 0x10000;) {
    int v;
    uint32_t pr = smt[context6] >> 11;
    uint32_t xmid = x1 + ((x2 - x1) >> 16) × pr +
      ((((x2 - x1) & 0xffff) × pr) >> 16);
    if (xr <= xmid) {
      x2 = xmid;
      v = 1;
    } else {
```

```
      x1 = xmid + 1;
      v = 0;
    }
    while (((x1 ^ x2) & 0xff000000) == 0) {
      xr = (xr << 8) + (pos >=  p_s ? 0xFF : u(8));
      pos++;
      x1 <<= 8;
      x2 = (x2 << 8) + 255;
    }
    int p0 = smt[context6];
    p0 += ((v << (16 + 11)) - p0) × 5 >> 7;
    smt[context6] = p0;
    context6 = (context6 × 2 + v) & 0x3f;
    palette[sumv] = x++;
    sumv += v;
    if (sumv == n_read_1) break;
    if (sumv - x == 0x10000 - n_read_1) {
      while (x < 0x10000) palette[sumv++] = x++;
    }
  }
}
```

## O.3 Decoding samples

Decoding of samples of a PassGroup shall be performed given a bitstream and the xsize and ysize dimensions of this PassGroup (not necessarily the entire image), the colour and bit depth mode as described in O.1, and the per-channel palette decoded earlier as described in O.2.

If the LosslessMode is coloured, the decoder shall first read a u(8) representing `colour_transform`. If `colour_transform` is not in the range `[0, 30)`, the codestream is ill-formed.

Let `area` be defined as `xsize × ysize` and `max_err_shift`, `max_err_round` and `num_contexts` (the amount of contexts per colour channel) be defined as follows, depending on the LosslessMode:

greyscale 8-bit and coloured 8-bit:

```
max_err_shift = (area > 25600 ? 0 : area > 12800 ? 1 : area > 4000 ? 2 : area > 400
? 3 : 4);
max_err_round = 0;
num_contexts = ((17 - 1) >> max_err_shift) + 1;
```

greyscale 16-bit and coloured 16-bit:

```
max_err_shift = (area > 25600 ? 0 : area > 12800 ? 1 : area > 2800 ? 2 : area > 512
? 3 : 4);
max_err_round = (1 << max_err_shift) - 1;
num_contexts = ((18 - 1 + max_err_round) >> max_err_shift) + 1;
```

If the LosslessMode is 8-bit, the decoder shall read a u(8) `P` representing the `prediction_mode` to be used for each channel. If the LosslessMode is greyscale, the `prediction_mode` is `P`, if the LosslessMode is coloured, the prediction_mode for the first channel is `P & 3`, the `prediction_mode` for the second channel is `(p >> 2) & 3` and the `prediction_mode` for the third channel is `P >> 4`. If any `prediction_mode` is not one of `PM_Regular (0)`, `PM_West (1)`, or `PM_North (2)`, the codestream is ill-formed.

The decoder shall then read `num_contexts × num_channels` clustered distributions, according to D.3.5, where `num_channels` is 1 if the LosslessMode is greyscale, 3 if it is coloured.

Then, for each channel of the image, the decoder shall apply the following procedure to compute the corresponding samples.

Let `channel_index` be 0 if this is the first or only channel, 1 if this is the second channel or 2 if this is the third channel.

Let `max_value` for the current channel be defined as follows:

- if a per-channel palette is enabled for this channel, and the current channel is not affected by the colour transform given by `colour_transform`, or the mode is a greyscale mode, then `max_value` shall be `palette_size - 1`. The current channel is not affected by the colour transform if its corresponding R', G' or B' value are equal to respectively R, G or B after applying the transform given in O.4.
- otherwise, when the value can be modified for this channel due to the colour transform, or no per-channel palette is enabled for this channel, `max_value` for this channel shall be 255 for 8-bit modes, 65535 for 16-bit modes

The decoder shall then use the clustered distributions and the parameters defined above to obtain sample values for the current channel, using the weighted predictor as specified in E.3.2 on each channel.

After decoding all channels, if the LosslessMode is coloured, a colour transform shall be applied depending on `colour_transform`, as specified in O.4.

Finally, if per-channel palettes were used (decoded as described in O.2.), the decoded values of each pixel of each channel shall be expanded from consecutive (compacted) range to original full range. Each sample value `v` shall be replaced with the index of the `v`-th bit that has value `1` in the per-channel palette (where the first bit has index 0). If a sample value is larger than or equal to the amount of 1-bits (the `palette_size`), then the codestream is ill-formed.

## O.4 Colour transforms

The coloured lossless decoder shall apply the following colour transforms after computing sample values, but before applying the per-channel palette. In the following description: `R`, `G`, `B` shall represent the values of each channel (`0`, `1`, `2` respectively) before the transform, `R'`, `G'`, `B'` the values after (in the same order) and C the center point: `128` for `8`-bit, `32768` for `16`-bit. `mask` shall be `255` for `8`-bit, `65535` for `16`-bit. Values shall be implicitly masked at the end of the computation by computing a bitwise-and with the `mask` value, and masking operations are explicitly written here when values are reused for follow up computations. If a `R'`, `G'`, or `B'` value is not set by a colour transform, it shall be assumed to be equal to the corresponding `R`, `G`, or `B` value.

The `colour_transform` value from O.3 shall decide which of the following 30 possible colour transforms shall be applied by the decoder.

| **`formula`** | **`transform`** |
|---|---|
| none | `0` |
| `G' = R + G + C;` | `1` |
| `B' = R + B + C;` | `2` |
| `G' = R + G + C; B' = R + B + C;` | 3 |
| `G' = G + B + C;` | 4 |
| `B' = G - B + C;` | 5 |
| `G' = (G + R + C) & mask; B' = B + ((R + G') >> 1) + C;` | 6 |
| `B' = (B + R + C) & mask; G' = G + ((R + B') >> 1) + C;` | 7 |
| `B' = B + ((R + G) >> 1) + C;` | 8 |
| `G' = G + ((R + B) >> 1) + C;` | 9 |
| same as 0 | 10 |
| `R' = G + R + C;` | 11 |
| `B' = G + B + C;` | 12 |
| `R' = R + G + C; B' = B + G + C;` | 13 |
| `R' = R + B + C;` | 14 |
| `B' = R - B + C;` | 15 |
| `R' = (R + G + C) & mask; B' = B + ((G + R') >> 1) + C;` | 16 |
| `B' = (B + G + C) & mask; R' = R + ((G + B') >> 1) + C;` | 17 |
| `B' = B + ((R + G) >> 1) + C;` | 18 |
| `R' = R + ((B + G) >> 1) + C;` | 19 |
| same as 0 | 20 |
| same as 14 | 21 |
| same as 4 | 22 |
| `R' = R + B + C; G' = G + B + C;` | 23 |
| same as 11 | 24 |
| `G' = R - G + C;` | 25 |
| `R' = (R + B + C) & mask; G' = G + ((B + R') >> 1) + C;` | 26 |
| `G' = (G + B + C) & mask; R' = R + ((B + G') >> 1) + C;` | 27 |
| `G' = G + ((R + B) >> 1) + C;` | 28 |
| `R' = R + ((G + B) >> 1) + C;` | 29 |

After applying the colour transform, the decoder shall expand the channel values according to the per-channel palette (O.2), if any is present.

# Annex P
(informative)
**Encoder overview**

An encoder is an embodiment of the encoding process. This document does not specify an encoding process, and any encoding process is acceptable as long as the codestream is as specified in Annex A. This Annex gives an overview of the steps that may be involved.

The encoder can choose the `FrameEncoding` based on input type and acceptable loss: JPEG transcoding if JPEG input and lossless; `kLossless` for photographic inputs and lossless; `kModular` for maximum responsiveness or non-photographic inputs; and `kPasses` for lossy.

For `kModular`, the encoder can apply a palette transform if the image contains few colours, a colour transform to decorrelate the channels, and the Haar-wavelet-like squeeze transform to obtain a multiresolution representation. The latter recursively replaces each channel with two channels: one downsampled in horizontal or vertical direction, and the other containing residual values with respect to a predictor. The downsampling averages two pixels, rounding up if the left/top one is greater, otherwise down (this prevents bias).

For `kPasses`, the encoder first transforms pixels into the XYB colour space. It can then perform the inverse operation (sharpening) to compensate for the blurring caused by the decoder loop filter's convolution.

To partition the image into varblocks, the encoder can compute a DCT for naturally-aligned 32 × 32, 16 × 16, and 8 × 8 regions, and then choose the varblock sizes such that entropy of their coefficients is minimized.

After computing varblock DCTs, encoders can compute a DC image by taking LLF and computing the IDCT of size **N** / 8 of them to create a low resolution image (conceptually a ⅛ × ⅛ downscale of the original image).

The encoder can search for the X/BFromY that result in the largest number of quantized zero residuals, and an adaptive quantization map that minimizes a distance metric between the input and the expected decoder output.

The X, Y, B predictor residuals are quantized by rounding the value after multiplying by the quantization matrix times the adaptive quantization entry for the 8 × 8 block.

# Annex Q
(informative)

# Annex R
(normative)
# Electronic inserts

The electronic inserts and tables that reference this Annex are appended to this document and appear after the current page.

**Electronic Insert I.1 – DCT-II / DCT-III code generator**

```
########################################################################
# DCT-II / DCT-III generator
#
# Based on:
#  "A low multiplicative complexity fast recursive DCT-2 algorithm"
#  by Maxim Vashkevich and Alexander Petrovsky / arXiv / 20 Jul 2012
########################################################################

import math
import sys
N = 8

########################################################################
# Base transforms / generators
########################################################################

CNTR = 0
def makeTmp():
  global CNTR
  result = "t{:02d}".format(CNTR)
  CNTR = CNTR + 1
  return result

def makeVar(i):
  return "i{:02d}".format(i)

def add(x, y):
  tmp = makeTmp()
  print(tmp + " = " + x + " + " + y + ";")
  return tmp

def sub(x, y):
  tmp = makeTmp()
  print(tmp + " = " + x + " - " + y + ";")
  return tmp

def mul(x, c):
  tmp = makeTmp()
  print(tmp + " = " + x + " * " + c + ";")
  return tmp

# 2.0 * math.cos((a + 0.0) / (b + 0.0) * math.pi)
```

```
def C2(a, b):
  return "c_c2_" + str(a) + "_" + str(b)

# 1.0 / C2(a, b)
def iC2(a, b):
  return "c_ic2_" + str(a) + "_" + str(b)

########################################################################
# Utilities
########################################################################

# Generate identity matrix. Usually this matrix is passed to
# DCT algorithm to generate "basis" vectors of the transform.
def makeVars():
  return [makeVar(i) for i in range(N)]

# Split list of variables info halves.
def split(x):
  m = len(x)
  m2 = m // 2
  return (x[0 : m2], x[m2 : m])

# Make a list of variables in a reverse order.
def reverse(varz):
  m = len(varz)
  result = [0] * m
  for i in range(m):
    result[i] = varz[m - 1 - i]
  return result

# Apply permutation
def permute(x, p):
 return [x[p[i]] for i in range(len(p))]

def transposePermutation(p):
  n = len(p)
  result = [0] * n
  for i in range(n):
    result[p[i]] = i
  return result

# See paper. Split even-odd elements.
def P(n):
  if n == 1:
    return [0]
```

```
  n2 = n // 2
  return [2 * i for i in range(n2)] + [2 * i + 1 for i in range(n2)]

# See paper. Interleave first and second half.
def Pt(n):
  return transposePermutation(P(n))

########################################################################
# Scheme
########################################################################

def B2(x):
  n = len(x)
  n2 = n // 2
  if n == 1:
    raise "ooops"
  (top, bottom) = split(x)
  bottom = reverse(bottom)
  t = [add(top[i], bottom[i]) for i in range(n2)]
  b = [sub(top[i], bottom[i]) for i in range(n2)]
  return t + b

def iB2(x):
  n = len(x)
  n2 = n // 2
  if n == 1:
    raise "ooops"
  (top, bottom) = split(x)
  t = [add(top[i], bottom[i]) for i in range(n2)]
  b = [sub(top[i], bottom[i]) for i in range(n2)]
  return t + reverse(b)

def B4(x, rn):
  n = len(x)
  n2 = n // 2
  if n == 1:
    raise "ooops"
  (top, bottom) = split(x)
  rbottom = reverse(bottom)
  t = [sub(top[i], rbottom[i]) for i in range(n2)]
  b = [mul(bottom[i], C2(rn, 2 * N)) for i in range(n2)]
  top = [add(t[i], b[i]) for i in range(n2)]
  bottom = [sub(t[i], b[i]) for i in range(n2)]
  return top + bottom
```

```
def iB4(x, rn):
  n = len(x)
  n2 = n // 2
  if n == 1:
    raise "ooops"
  (top, bottom) = split(x)
  t = [add(top[i], bottom[i]) for i in range(n2)]
  b = [sub(top[i], bottom[i]) for i in range(n2)]
  bottom = [mul(b[i], iC2(rn, 2 * N)) for i in range(n2)]
  rbottom = reverse(bottom)
  top = [add(t[i], rbottom[i]) for i in range(n2)]
  return top + bottom

def P4(n):
  if n == 1:
    return [0]
  if n == 2:
    return [0, 1]
  n2 = n // 2
  result = [0] * n
  tc = 0
  bc = 0
  i = 0
  result[i] = tc; tc = tc + 1; i = i + 1
  turn = True
  while i < n - 1:
    if turn:
      result[i] = n2 + bc; bc = bc + 1; i = i + 1
      result[i] = n2 + bc; bc = bc + 1; i = i + 1
    else:
      result[i] = tc; tc = tc + 1; i = i + 1
      result[i] = tc; tc = tc + 1; i = i + 1
    turn = not turn
  result[i] = tc; tc = tc + 1; i = i + 1
  return result

def iP4(n):
  return transposePermutation(P4(n))

def d2n(x):
  n = len(x)
  if n == 1:
    return x
  y = B2(x)
  (top, bottom) = split(y)
```

```
  return permute(d2n(top) + d4n(bottom, N // 2), Pt(n))

def id2n(x):
  n = len(x)
  if n == 1:
    return x
  (top, bottom) = split(permute(x, P(n)))
  return iB2(id2n(top) + id4n(bottom, N // 2))

def d4n(x, rn):
  n = len(x)
  if n == 1:
    return x
  y = B4(x, rn)
  (top, bottom) = split(y)
  rn2 = rn // 2
  return permute(d4n(top, rn2) + d4n(bottom, N - rn2), P4(n))

def id4n(x, rn):
  n = len(x)
  if n == 1:
    return x
  (top, bottom) = split(permute(x, iP4(n)))
  rn2 = rn // 2
  y = id4n(top, rn2) + id4n(bottom, N -rn2)
  return iB4(y, rn)

########################################################################
# Main.
########################################################################

def help():
  print("Usage: %s [N [T]]" % sys.argv[0])
  print("  N should be the power of 2, default is 8")
  print("  T is one of {2, 3}, default is 2")
  sys.exit()

def parseInt(s):
  try:
    return int(s)
  except ValueError:
    help()

if __name__ == "__main__":
  if len(sys.argv) < 1 or len(sys.argv) > 3: help()
```

```
  if len(sys.argv) >= 2:
    N = parseInt(sys.argv[1])
    if (N & (N - 1)) != 0: help()
  type = 0
  if len(sys.argv) >= 3:
    typeOption = sys.argv[2]
    if len(typeOption) != 1: help()
    type = "23".index(typeOption)
    if type == -1: help()
  if type == 0:
    vars = d2n(makeVars())
  else:  # type == 1
    vars = id2n(makeVars())
  print("Output vector: " + str(vars))
```

**Table M.1 – is_zero_base table**

```
228, 216, 216, 195, 192, 189, 182, 184, 179, 176, 171, 168, 166, 159,
156, 151, 151, 150, 150, 146, 144, 138, 138, 137, 135, 131, 127, 126,
124, 123, 124, 123, 122, 121, 118, 117, 114, 115, 116, 116, 115, 115,
114, 111, 111, 111, 112, 111, 110, 110, 110, 111, 111, 114, 110, 111,
112, 113, 116, 120, 126, 131, 147, 160
```

**Table M.2 – num_nonzeros_base table**

```
251, 252, 117, 249, 161, 136,  83, 238, 184, 126, 137, 129, 140, 119,
 70, 213, 160, 175, 174, 130, 166, 134, 122, 125, 131, 144, 136, 133,
139, 123,  79, 216, 128, 128, 128, 128, 128, 128, 128, 128, 128, 128,
128, 128, 128, 128, 128, 128, 128, 128, 128, 128, 128, 128, 128, 128,
128, 128, 128, 128, 128, 128, 128
```

```
254, 252, 174, 232, 189, 155, 122, 177, 204, 173, 146, 149, 141, 133,
103, 109, 167, 187, 168, 142, 154, 147, 125, 139, 144, 138, 138, 153,
141, 133,  90, 121, 128, 128, 128, 128, 128, 128, 128, 128, 128, 128,
128, 128, 128, 128, 128, 128, 128, 128, 128, 128, 128, 128, 128, 128,
128, 128, 128, 128, 128, 128, 128
```

```
251, 240, 197, 176, 184, 177, 114,  89, 194, 165, 153, 161, 158, 136,
 92,  95, 123, 171, 160, 140, 148, 136, 129, 139, 145, 136, 143, 134,
138, 124,  92, 154, 128, 128, 128, 128, 128, 128, 128, 128, 128, 128,
128, 128, 128, 128, 128, 128, 128, 128, 128, 128, 128, 128, 128, 128,
128, 128, 128, 128, 128, 128, 128
```

```
247, 220, 201, 110, 194, 176, 147,  59, 175, 171, 156, 157, 152, 146,
115, 114,  88, 151, 164, 141, 153, 135, 141, 131, 146, 139, 140, 145,
138, 137, 112, 184, 128, 128, 128, 128, 128, 128, 128, 128, 128, 128,
128, 128, 128, 128, 128, 128, 128, 128, 128, 128, 128, 128, 128, 128,
128, 128, 128, 128, 128, 128, 128
```

```
238, 179, 203,  63, 194, 173, 149,  71, 139, 169, 154, 159, 150, 146,
117, 143,  78, 122, 152, 137, 149, 138, 138, 133, 134, 142, 142, 142,
148, 128, 118, 199, 128, 128, 128, 128, 128, 128, 128, 128, 128, 128,
128, 128, 128, 128, 128, 128, 128, 128, 128, 128, 128, 128, 128, 128,
128, 128, 128, 128, 128, 128, 128
```

```
227, 127, 200,  44, 192, 170, 148, 100, 102, 161, 156, 153, 148, 149,
124, 160,  88, 101, 134, 132, 149, 145, 134, 134, 136, 141, 138, 142,
144, 137, 116, 208, 128, 128, 128, 128, 128, 128, 128, 128, 128, 128,
128, 128, 128, 128, 128, 128, 128, 128, 128, 128, 128, 128, 128, 128,
128, 128, 128, 128, 128, 128, 128
```

```
214,  86, 195,  44, 187, 163, 148, 126,  81, 147, 156, 152, 150, 144,
121, 172,  96,  95, 117, 122, 145, 152, 136, 133, 135, 135, 131, 142,
141, 135, 114, 217, 128, 128, 128, 128, 128, 128, 128, 128, 128, 128,
128, 128, 128, 128, 128, 128, 128, 128, 128, 128, 128, 128, 128, 128,
128, 128, 128, 128, 128, 128, 128
```

```
198,  56, 191,  54, 171, 162, 147, 144,  74, 128, 152, 149, 150, 142,
119, 177, 101, 100, 106, 111, 135, 154, 136, 137, 136, 132, 133, 142,
144, 130, 117, 222, 128, 128, 128, 128, 128, 128, 128, 128, 128, 128,
128, 128, 128, 128, 128, 128, 128, 128, 128, 128, 128, 128, 128, 128,
128, 128, 128, 128, 128, 128, 128
```

```
176,  40, 189,  73, 147, 159, 148, 152,  79, 106, 147, 149, 151, 139,
123, 188, 108, 110, 106,  97, 125, 151, 137, 138, 135, 135, 134, 136,
140, 131, 116, 221, 128, 128, 128, 128, 128, 128, 128, 128, 128, 128,
128, 128, 128, 128, 128, 128, 128, 128, 128, 128, 128, 128, 128, 128,
128, 128, 128, 128, 128, 128, 128
```

```
148,  33, 185,  88, 117, 158, 145, 163,  95,  91, 137, 146, 150, 140,
120, 197, 115, 116, 114,  92, 114, 144, 130, 133, 132, 133, 129, 140,
138, 130, 111, 224, 128, 128, 128, 128, 128, 128, 128, 128, 128, 128,
128, 128, 128, 128, 128, 128, 128, 128, 128, 128, 128, 128, 128, 128,
128, 128, 128, 128, 128, 128, 128
```

```
117,  31, 180, 104,  93, 150, 143, 166,  99,  85, 124, 139, 148, 142,
118, 201, 105, 120, 120,  90, 107, 135, 127, 130, 131, 131, 132, 140,
142, 133, 114, 229, 128, 128, 128, 128, 128, 128, 128, 128, 128, 128,
128, 128, 128, 128, 128, 128, 128, 128, 128, 128, 128, 128, 128, 128,
128, 128, 128, 128, 128, 128, 128
```

```
 87,  35, 170, 110,  78, 141, 144, 176, 106,  90, 112, 132, 143, 138,
119, 204, 111, 121, 125,  90, 105, 131, 124, 122, 129, 128, 129, 137,
138, 133, 114, 227, 128, 128, 128, 128, 128, 128, 128, 128, 128, 128,
128, 128, 128, 128, 128, 128, 128, 128, 128, 128, 128, 128, 128, 128,
128, 128, 128, 128, 128, 128, 128
```

```
 63,  42, 159, 123,  73, 127, 142, 191, 105,  91, 105, 123, 139, 137,
120, 209, 117, 110, 122,  98, 110, 125, 115, 123, 122, 126, 128, 134,
141, 129, 113, 229, 128, 128, 128, 128, 128, 128, 128, 128, 128, 128,
128, 128, 128, 128, 128, 128, 128, 128, 128, 128, 128, 128, 128, 128,
128, 128, 128, 128, 128, 128, 128
```

```
 45,  53, 146, 135,  71, 114, 138, 193, 100,  98,  98, 113, 133, 135,
118, 222, 113, 111, 139, 103, 107, 126, 111, 119, 121, 122, 127, 135,
141, 128, 114, 242, 128, 128, 128, 128, 128, 128, 128, 128, 128, 128,
128, 128, 128, 128, 128, 128, 128, 128, 128, 128, 128, 128, 128, 128,
128, 128, 128, 128, 128, 128, 128
```

```
 33,  60, 132, 138,  75, 100, 134, 203, 112,  99,  98, 105, 126, 131,
115, 229, 107,  93, 121, 106, 108, 122, 106, 109, 114, 116, 127, 133,
143, 128, 110, 242, 128, 128, 128, 128, 128, 128, 128, 128, 128, 128,
128, 128, 128, 128, 128, 128, 128, 128, 128, 128, 128, 128, 128, 128,
128, 128, 128, 128, 128, 128, 128
```

```
 24,  70, 118, 134,  76,  87, 130, 201, 110,  96,  99,  97, 119, 130,
111, 229,  97, 104, 125, 102, 112, 125, 101, 109, 113, 114, 125, 129,
142, 127, 112, 241, 128, 128, 128, 128, 128, 128, 128, 128, 128, 128,
128, 128, 128, 128, 128, 128, 128, 128, 128, 128, 128, 128, 128, 128,
128, 128, 128, 128, 128, 128, 128
```

```
 17,  65, 100, 121,  80,  75, 124, 174, 117, 100,  94,  93, 114, 128,
110, 216, 103,  94, 113, 122, 118, 126, 113, 108, 105, 108, 122, 128,
141, 125, 113, 238, 128, 128, 128, 128, 128, 128, 128, 128, 128, 128,
128, 128, 128, 128, 128, 128, 128, 128, 128, 128, 128, 128, 128, 128,
128, 128, 128, 128, 128, 128, 128
```

```
 12,  70,  82, 132,  78,  65, 118, 155, 136, 103,  97,  89, 106, 124,
111, 215, 115, 123, 129,  99, 104, 127, 110, 108, 101, 109, 118, 126,
136, 123, 110, 233, 128, 128, 128, 128, 128, 128, 128, 128, 128, 128,
128, 128, 128, 128, 128, 128, 128, 128, 128, 128, 128, 128, 128, 128,
128, 128, 128, 128, 128, 128, 128
```

```
  8,  66,  61, 117,  91,  59, 108, 195, 101, 112,  99,  99,  99, 116,
106, 230, 127,  99, 144, 101, 118, 137, 117, 111, 106, 104, 116, 121,
134, 122, 110, 223, 128, 128, 128, 128, 128, 128, 128, 128, 128, 128,
128, 128, 128, 128, 128, 128, 128, 128, 128, 128, 128, 128, 128, 128,
128, 128, 128, 128, 128, 128, 128
```

```
  6,  78,  42, 146, 101,  54,  94, 201, 116, 102, 110,  94,  92, 108,
103, 214, 108, 111, 127, 102, 121, 132, 120, 121,  95,  98, 110, 121,
129, 117, 107, 235, 128, 128, 128, 128, 128, 128, 128, 128, 128, 128,
128, 128, 128, 128, 128, 128, 128, 128, 128, 128, 128, 128, 128, 128,
128, 128, 128, 128, 128, 128, 128
```

```
  5,  93,  29, 145, 102,  52,  77, 216, 108, 115, 108, 102,  89,  97,
 94, 229,  89, 103, 139, 120, 103, 151, 102, 100,  97,  96,  99, 111,
125, 116, 104, 242, 128, 128, 128, 128, 128, 128, 128, 128, 128, 128,
128, 128, 128, 128, 128, 128, 128, 128, 128, 128, 128, 128, 128, 128,
128, 128, 128, 128, 128, 128, 128
```

```
  4, 105,  21, 145, 100,  54,  64, 217, 100, 122, 128,  87,  88,  91,
 87, 230, 112,  80, 148,  95, 146, 123,  96, 140,  90,  91,  98, 106,
122, 111, 100, 249, 128, 128, 128, 128, 128, 128, 128, 128, 128, 128,
128, 128, 128, 128, 128, 128, 128, 128, 128, 128, 128, 128, 128, 128,
128, 128, 128, 128, 128, 128, 128
```

```
  4, 130,  14, 142, 104,  56,  51, 208, 116, 135, 100,  89,  82,  84,
 75, 239,  85,  85, 122, 125,  94, 144, 151, 136,  92,  97, 104, 109,
113, 110,  91, 246, 128, 128, 128, 128, 128, 128, 128, 128, 128, 128,
128, 128, 128, 128, 128, 128, 128, 128, 128, 128, 128, 128, 128, 128,
128, 128, 128, 128, 128, 128, 128
```

```
  3, 126,   9, 172, 105,  57,  39, 219,  95, 120, 118,  96,  93,  75,
 66, 241, 102, 134,  96, 156, 146, 162, 130, 112,  82,  89,  97, 101,
116, 103,  82, 254, 128, 128, 128, 128, 128, 128, 128, 128, 128, 128,
128, 128, 128, 128, 128, 128, 128, 128, 128, 128, 128, 128, 128, 128,
128, 128, 128, 128, 128, 128, 128
```

```
  3, 149,   7, 182, 122,  54,  29, 224, 103, 100, 113,  96,  90,  74,
 55, 250, 127,  94, 118,  93, 135, 160, 113, 130,  95, 117, 106,  96,
111,  97,  77, 242, 128, 128, 128, 128, 128, 128, 128, 128, 128, 128,
128, 128, 128, 128, 128, 128, 128, 128, 128, 128, 128, 128, 128, 128,
128, 128, 128, 128, 128, 128, 128
```

```
  3, 150,   4, 170, 138,  59,  20, 229,  91, 150, 107,  98,  92,  68,
 48, 245, 113,  64, 114, 111, 134, 127, 102, 104,  85, 118, 103, 107,
102,  91,  72, 245, 128, 128, 128, 128, 128, 128, 128, 128, 128, 128,
128, 128, 128, 128, 128, 128, 128, 128, 128, 128, 128, 128, 128, 128,
128, 128, 128, 128, 128, 128, 128
```

```
  3, 171,   3, 165, 137,  62,  14, 211,  96, 127, 132, 121,  95,  62,
 37, 248, 102,  57, 144,  85, 127, 191, 102,  97, 127, 104,  91, 102,
107,  81,  64, 254, 128, 128, 128, 128, 128, 128, 128, 128, 128, 128,
128, 128, 128, 128, 128, 128, 128, 128, 128, 128, 128, 128, 128, 128,
128, 128, 128, 128, 128, 128, 128
```

```
  2, 166,   2, 196, 122,  65,  10, 243, 102,  93, 117,  92,  96,  63,
 29, 251, 169, 159, 149,  96,  91, 139, 157,  40, 100,  89, 120,  92,
109,  79,  58, 247, 128, 128, 128, 128, 128, 128, 128, 128, 128, 128,
128, 128, 128, 128, 128, 128, 128, 128, 128, 128, 128, 128, 128, 128,
128, 128, 128, 128, 128, 128, 128
```

```
  2, 176,   2, 189, 118,  48,   7, 219,  68,  43, 109,  96, 129,  75,
 19, 254,   2,   3, 185,   6, 102, 127, 127, 127,   1, 131,  83,  99,
107,  80,  45, 254, 128, 128, 128, 128, 128, 128, 128, 128, 128, 128,
128, 128, 128, 128, 128, 128, 128, 128, 128, 128, 128, 128, 128, 128,
128, 128, 128, 128, 128, 128, 128
```

```
  1, 205,   2, 208,  64,  89,   4, 223,  29, 169,  29, 123, 118,  76,
 11, 240, 202, 243,  65,   6,  12, 243,  96,  55, 102, 102, 114, 102,
107,  74,  31, 247, 128, 128, 128, 128, 128, 128, 128, 128, 128, 128,
128, 128, 128, 128, 128, 128, 128, 128, 128, 128, 128, 128, 128, 128,
128, 128, 128, 128, 128, 128, 128
```

```
  1, 216,   1, 214, 127,  94,   2, 234, 145,   3, 127, 106, 155,  80,
  4, 247,   4,  65,  86, 127, 127, 127, 127, 102, 127, 143, 143, 108,
113,  80,  16, 216, 128, 128, 128, 128, 128, 128, 128, 128, 128, 128,
128, 128, 128, 128, 128, 128, 128, 128, 128, 128, 128, 128, 128, 128,
128, 128, 128, 128, 128, 128, 128
```

```
  2, 199,   1, 222,  93,  94,   1, 232,   2,  65,  74, 139, 201,  48,
  2, 254, 169, 127,  52, 243, 251, 249, 102,  86, 202, 153,  65,  65,
146,  69,   8, 238, 128, 128, 128, 128, 128, 128, 128, 128, 128, 128,
128, 128, 128, 128, 128, 128, 128, 128, 128, 128, 128, 128, 128, 128,
128, 128, 128, 128, 128, 128, 128
```

**Table M.3 – Protocol Buffer descriptor of top-level structure of losslessly compressed JPEG stream**

```
message Header {
  optional uint64 width = 1;
  optional uint64 height = 2;
  required uint64 version_and_component_count_code = 3;
  optional uint64 subsampling_code = 4;
}

message Jpeg {
  required bytes signature = 1;
  required Header header = 2;
  optional bytes meta_data = 3;
  optional bytes jpeg1_internals = 4;
  optional bytes quant_data = 5;
  optional bytes histogram_data = 6;
  optional bytes dc_data = 7;
  optional bytes ac_data = 8;
  optional bytes original_jpg = 9;
}
```

**Table M.4 – APP0 template**

```
0xE0, 0x00, 0x10, 0x4A, 0x46, 0x49, 0x46, 0x00, 0x01, 0x01, 0x00, 0x00,
0x01, 0x00, 0x01, 0x00, 0x00
```

**Table M.6 – common ICC profile template**

```
0xE2, 0x0C, 0x58, 0x49, 0x43, 0x43, 0x5F, 0x50, 0x52, 0x4F, 0x46, 0x49,
0x4C, 0x45, 0x00, 0x01, 0x01, 0x00, 0x00, 0x0C, 0x48, 0x4C, 0x69, 0x6E,
0x6F, 0x02, 0x10, 0x00, 0x00, 0x6D, 0x6E, 0x74, 0x72, 0x52, 0x47, 0x42,
0x20, 0x58, 0x59, 0x5A, 0x20, 0x07, 0xCE, 0x00, 0x02, 0x00, 0x09, 0x00,
0x06, 0x00, 0x31, 0x00, 0x00, 0x61, 0x63, 0x73, 0x70, 0x4D, 0x53, 0x46,
0x54, 0x00, 0x00, 0x00, 0x00, 0x49, 0x45, 0x43, 0x20, 0x73, 0x52, 0x47,
0x42, 0x00, 0x00, 0x00, 0x00, 0x00, 0x00, 0x00, 0x00, 0x00, 0x00, 0x00,
0x01, 0x00, 0x00, 0xF6, 0xD6, 0x00, 0x01, 0x00, 0x00, 0x00, 0x00, 0xD3,
```

```
0x2D, 0x48, 0x50, 0x20, 0x20, 0x00, 0x00, 0x00, 0x00, 0x00, 0x00, 0x00,
0x00, 0x00, 0x00, 0x00, 0x00, 0x00, 0x00, 0x00, 0x00, 0x00, 0x00, 0x00,
0x00, 0x00, 0x00, 0x00, 0x00, 0x00, 0x00, 0x00, 0x00, 0x00, 0x00, 0x00,
0x00, 0x00, 0x00, 0x00, 0x00, 0x00, 0x00, 0x00, 0x00, 0x00, 0x00, 0x00,
0x00, 0x00, 0x00, 0x00, 0x11, 0x63, 0x70, 0x72, 0x74, 0x00, 0x00, 0x01,
0x50, 0x00, 0x00, 0x00, 0x33, 0x64, 0x65, 0x73, 0x63, 0x00, 0x00, 0x01,
0x84, 0x00, 0x00, 0x00, 0x6C, 0x77, 0x74, 0x70, 0x74, 0x00, 0x00, 0x01,
0xF0, 0x00, 0x00, 0x00, 0x14, 0x62, 0x6B, 0x70, 0x74, 0x00, 0x00, 0x02,
0x04, 0x00, 0x00, 0x00, 0x14, 0x72, 0x58, 0x59, 0x5A, 0x00, 0x00, 0x02,
0x18, 0x00, 0x00, 0x00, 0x14, 0x67, 0x58, 0x59, 0x5A, 0x00, 0x00, 0x02,
0x2C, 0x00, 0x00, 0x00, 0x14, 0x62, 0x58, 0x59, 0x5A, 0x00, 0x00, 0x02,
0x40, 0x00, 0x00, 0x00, 0x14, 0x64, 0x6D, 0x6E, 0x64, 0x00, 0x00, 0x02,
0x54, 0x00, 0x00, 0x00, 0x70, 0x64, 0x6D, 0x64, 0x64, 0x00, 0x00, 0x02,
0xC4, 0x00, 0x00, 0x00, 0x88, 0x76, 0x75, 0x65, 0x64, 0x00, 0x00, 0x03,
0x4C, 0x00, 0x00, 0x00, 0x86, 0x76, 0x69, 0x65, 0x77, 0x00, 0x00, 0x03,
0xD4, 0x00, 0x00, 0x00, 0x24, 0x6C, 0x75, 0x6D, 0x69, 0x00, 0x00, 0x03,
0xF8, 0x00, 0x00, 0x00, 0x14, 0x6D, 0x65, 0x61, 0x73, 0x00, 0x00, 0x04,
0x0C, 0x00, 0x00, 0x00, 0x24, 0x74, 0x65, 0x63, 0x68, 0x00, 0x00, 0x04,
0x30, 0x00, 0x00, 0x00, 0x0C, 0x72, 0x54, 0x52, 0x43, 0x00, 0x00, 0x04,
0x3C, 0x00, 0x00, 0x08, 0x0C, 0x67, 0x54, 0x52, 0x43, 0x00, 0x00, 0x04,
0x3C, 0x00, 0x00, 0x08, 0x0C, 0x62, 0x54, 0x52, 0x43, 0x00, 0x00, 0x04,
0x3C, 0x00, 0x00, 0x08, 0x0C, 0x74, 0x65, 0x78, 0x74, 0x00, 0x00, 0x00,
0x00, 0x43, 0x6F, 0x70, 0x79, 0x72, 0x69, 0x67, 0x68, 0x74, 0x20, 0x28,
0x63, 0x29, 0x20, 0x31, 0x39, 0x39, 0x38, 0x20, 0x48, 0x65, 0x77, 0x6C,
0x65, 0x74, 0x74, 0x2D, 0x50, 0x61, 0x63, 0x6B, 0x61, 0x72, 0x64, 0x20,
0x43, 0x6F, 0x6D, 0x70, 0x61, 0x6E, 0x79, 0x00, 0x00, 0x64, 0x65, 0x73,
0x63, 0x00, 0x00, 0x00, 0x00, 0x00, 0x00, 0x00, 0x12, 0x73, 0x52, 0x47,
0x42, 0x20, 0x49, 0x45, 0x43, 0x36, 0x31, 0x39, 0x36, 0x36, 0x2D, 0x32,
0x2E, 0x31, 0x00, 0x00, 0x00, 0x00, 0x00, 0x00, 0x00, 0x00, 0x00, 0x00,
0x00, 0x12, 0x73, 0x52, 0x47, 0x42, 0x20, 0x49, 0x45, 0x43, 0x36, 0x31,
0x39, 0x36, 0x36, 0x2D, 0x32, 0x2E, 0x31, 0x00, 0x00, 0x00, 0x00, 0x00,
0x00, 0x00, 0x00, 0x00, 0x00, 0x00, 0x00, 0x00, 0x00, 0x00, 0x00, 0x00,
0x00, 0x00, 0x00, 0x00, 0x00, 0x00, 0x00, 0x00, 0x00, 0x00, 0x00, 0x00,
0x00, 0x00, 0x00, 0x00, 0x00, 0x00, 0x00, 0x00, 0x00, 0x00, 0x00, 0x00,
0x00, 0x00, 0x00, 0x00, 0x00, 0x00, 0x00, 0x00, 0x00, 0x58, 0x59, 0x5A,
0x20, 0x00, 0x00, 0x00, 0x00, 0x00, 0x00, 0xF3, 0x51, 0x00, 0x01, 0x00,
0x00, 0x00, 0x01, 0x16, 0xCC, 0x58, 0x59, 0x5A, 0x20, 0x00, 0x00, 0x00,
0x00, 0x00, 0x00, 0x00, 0x00, 0x00, 0x00, 0x00, 0x00, 0x00, 0x00, 0x00,
0x00, 0x58, 0x59, 0x5A, 0x20, 0x00, 0x00, 0x00, 0x00, 0x00, 0x00, 0x6F,
0xA2, 0x00, 0x00, 0x38, 0xF5, 0x00, 0x00, 0x03, 0x90, 0x58, 0x59, 0x5A,
0x20, 0x00, 0x00, 0x00, 0x00, 0x00, 0x00, 0x62, 0x99, 0x00, 0x00, 0xB7,
0x85, 0x00, 0x00, 0x18, 0xDA, 0x58, 0x59, 0x5A, 0x20, 0x00, 0x00, 0x00,
0x00, 0x00, 0x00, 0x24, 0xA0, 0x00, 0x00, 0x0F, 0x84, 0x00, 0x00, 0xB6,
0xCF, 0x64, 0x65, 0x73, 0x63, 0x00, 0x00, 0x00, 0x00, 0x00, 0x00, 0x00,
0x16, 0x49, 0x45, 0x43, 0x20, 0x68, 0x74, 0x74, 0x70, 0x3A, 0x2F, 0x2F,
```

```
0x77, 0x77, 0x77, 0x2E, 0x69, 0x65, 0x63, 0x2E, 0x63, 0x68, 0x00, 0x00,
0x00, 0x00, 0x00, 0x00, 0x00, 0x00, 0x00, 0x00, 0x00, 0x16, 0x49, 0x45,
0x43, 0x20, 0x68, 0x74, 0x74, 0x70, 0x3A, 0x2F, 0x2F, 0x77, 0x77, 0x77,
0x2E, 0x69, 0x65, 0x63, 0x2E, 0x63, 0x68, 0x00, 0x00, 0x00, 0x00, 0x00,
0x00, 0x00, 0x00, 0x00, 0x00, 0x00, 0x00, 0x00, 0x00, 0x00, 0x00, 0x00,
0x00, 0x00, 0x00, 0x00, 0x00, 0x00, 0x00, 0x00, 0x00, 0x00, 0x00, 0x00,
0x00, 0x00, 0x00, 0x00, 0x00, 0x00, 0x00, 0x00, 0x00, 0x00, 0x00, 0x00,
0x00, 0x00, 0x00, 0x00, 0x00, 0x64, 0x65, 0x73, 0x63, 0x00, 0x00, 0x00,
0x00, 0x00, 0x00, 0x00, 0x2E, 0x49, 0x45, 0x43, 0x20, 0x36, 0x31, 0x39,
0x36, 0x36, 0x2D, 0x32, 0x2E, 0x31, 0x20, 0x44, 0x65, 0x66, 0x61, 0x75,
0x6C, 0x74, 0x20, 0x52, 0x47, 0x42, 0x20, 0x63, 0x6F, 0x6C, 0x6F, 0x75,
0x72, 0x20, 0x73, 0x70, 0x61, 0x63, 0x65, 0x20, 0x2D, 0x20, 0x73, 0x52,
0x47, 0x42, 0x00, 0x00, 0x00, 0x00, 0x00, 0x00, 0x00, 0x00, 0x00, 0x00,
0x00, 0x2E, 0x49, 0x45, 0x43, 0x20, 0x36, 0x31, 0x39, 0x36, 0x36, 0x2D,
0x32, 0x2E, 0x31, 0x20, 0x44, 0x65, 0x66, 0x61, 0x75, 0x6C, 0x74, 0x20,
0x52, 0x47, 0x42, 0x20, 0x63, 0x6F, 0x6C, 0x6F, 0x75, 0x72, 0x20, 0x73,
0x70, 0x61, 0x63, 0x65, 0x20, 0x2D, 0x20, 0x73, 0x52, 0x47, 0x42, 0x00,
0x00, 0x00, 0x00, 0x00, 0x00, 0x00, 0x00, 0x00, 0x00, 0x00, 0x00, 0x00,
0x00, 0x00, 0x00, 0x00, 0x00, 0x00, 0x00, 0x00, 0x00, 0x64, 0x65, 0x73,
0x63, 0x00, 0x00, 0x00, 0x00, 0x00, 0x00, 0x00, 0x2C, 0x52, 0x65, 0x66,
0x65, 0x72, 0x65, 0x6E, 0x63, 0x65, 0x20, 0x56, 0x69, 0x65, 0x77, 0x69,
0x6E, 0x67, 0x20, 0x43, 0x6F, 0x6E, 0x64, 0x69, 0x74, 0x69, 0x6F, 0x6E,
0x20, 0x69, 0x6E, 0x20, 0x49, 0x45, 0x43, 0x36, 0x31, 0x39, 0x36, 0x36,
0x2D, 0x32, 0x2E, 0x31, 0x00, 0x00, 0x00, 0x00, 0x00, 0x00, 0x00, 0x00,
0x00, 0x00, 0x00, 0x2C, 0x52, 0x65, 0x66, 0x65, 0x72, 0x65, 0x6E, 0x63,
0x65, 0x20, 0x56, 0x69, 0x65, 0x77, 0x69, 0x6E, 0x67, 0x20, 0x43, 0x6F,
0x6E, 0x64, 0x69, 0x74, 0x69, 0x6F, 0x6E, 0x20, 0x69, 0x6E, 0x20, 0x49,
0x45, 0x43, 0x36, 0x31, 0x39, 0x36, 0x36, 0x2D, 0x32, 0x2E, 0x31, 0x00,
0x00, 0x00, 0x00, 0x00, 0x00, 0x00, 0x00, 0x00, 0x00, 0x00, 0x00, 0x00,
0x00, 0x00, 0x00, 0x00, 0x00, 0x00, 0x00, 0x00, 0x00, 0x00, 0x00, 0x00,
0x00, 0x76, 0x69, 0x65, 0x77, 0x00, 0x00, 0x00, 0x00, 0x00, 0x13, 0xA4,
0xFE, 0x00, 0x14, 0x5F, 0x2E, 0x00, 0x10, 0xCF, 0x14, 0x00, 0x03, 0xED,
0xCC, 0x00, 0x04, 0x13, 0x0B, 0x00, 0x03, 0x5C, 0x9E, 0x00, 0x00, 0x00,
0x01, 0x58, 0x59, 0x5A, 0x20, 0x00, 0x00, 0x00, 0x00, 0x00, 0x4C, 0x09,
0x56, 0x00, 0x50, 0x00, 0x00, 0x00, 0x57, 0x1F, 0xE7, 0x6D, 0x65, 0x61,
0x73, 0x00, 0x00, 0x00, 0x00, 0x00, 0x00, 0x00, 0x01, 0x00, 0x00, 0x00,
0x00, 0x00, 0x00, 0x00, 0x00, 0x00, 0x00, 0x00, 0x00, 0x00, 0x00, 0x00,
0x00, 0x00, 0x00, 0x02, 0x8F, 0x00, 0x00, 0x00, 0x02, 0x73, 0x69, 0x67,
0x20, 0x00, 0x00, 0x00, 0x00, 0x43, 0x52, 0x54, 0x20, 0x63, 0x75, 0x72,
0x76, 0x00, 0x00, 0x00, 0x00, 0x00, 0x00, 0x04, 0x00, 0x00, 0x00, 0x00,
0x05, 0x00, 0x0A, 0x00, 0x0F, 0x00, 0x14, 0x00, 0x19, 0x00, 0x1E, 0x00,
0x23, 0x00, 0x28, 0x00, 0x2D, 0x00, 0x32, 0x00, 0x37, 0x00, 0x3B, 0x00,
0x40, 0x00, 0x45, 0x00, 0x4A, 0x00, 0x4F, 0x00, 0x54, 0x00, 0x59, 0x00,
0x5E, 0x00, 0x63, 0x00, 0x68, 0x00, 0x6D, 0x00, 0x72, 0x00, 0x77, 0x00,
0x7C, 0x00, 0x81, 0x00, 0x86, 0x00, 0x8B, 0x00, 0x90, 0x00, 0x95, 0x00,
```

```
0x9A, 0x00, 0x9F, 0x00, 0xA4, 0x00, 0xA9, 0x00, 0xAE, 0x00, 0xB2, 0x00,
0xB7, 0x00, 0xBC, 0x00, 0xC1, 0x00, 0xC6, 0x00, 0xCB, 0x00, 0xD0, 0x00,
0xD5, 0x00, 0xDB, 0x00, 0xE0, 0x00, 0xE5, 0x00, 0xEB, 0x00, 0xF0, 0x00,
0xF6, 0x00, 0xFB, 0x01, 0x01, 0x01, 0x07, 0x01, 0x0D, 0x01, 0x13, 0x01,
0x19, 0x01, 0x1F, 0x01, 0x25, 0x01, 0x2B, 0x01, 0x32, 0x01, 0x38, 0x01,
0x3E, 0x01, 0x45, 0x01, 0x4C, 0x01, 0x52, 0x01, 0x59, 0x01, 0x60, 0x01,
0x67, 0x01, 0x6E, 0x01, 0x75, 0x01, 0x7C, 0x01, 0x83, 0x01, 0x8B, 0x01,
0x92, 0x01, 0x9A, 0x01, 0xA1, 0x01, 0xA9, 0x01, 0xB1, 0x01, 0xB9, 0x01,
0xC1, 0x01, 0xC9, 0x01, 0xD1, 0x01, 0xD9, 0x01, 0xE1, 0x01, 0xE9, 0x01,
0xF2, 0x01, 0xFA, 0x02, 0x03, 0x02, 0x0C, 0x02, 0x14, 0x02, 0x1D, 0x02,
0x26, 0x02, 0x2F, 0x02, 0x38, 0x02, 0x41, 0x02, 0x4B, 0x02, 0x54, 0x02,
0x5D, 0x02, 0x67, 0x02, 0x71, 0x02, 0x7A, 0x02, 0x84, 0x02, 0x8E, 0x02,
0x98, 0x02, 0xA2, 0x02, 0xAC, 0x02, 0xB6, 0x02, 0xC1, 0x02, 0xCB, 0x02,
0xD5, 0x02, 0xE0, 0x02, 0xEB, 0x02, 0xF5, 0x03, 0x00, 0x03, 0x0B, 0x03,
0x16, 0x03, 0x21, 0x03, 0x2D, 0x03, 0x38, 0x03, 0x43, 0x03, 0x4F, 0x03,
0x5A, 0x03, 0x66, 0x03, 0x72, 0x03, 0x7E, 0x03, 0x8A, 0x03, 0x96, 0x03,
0xA2, 0x03, 0xAE, 0x03, 0xBA, 0x03, 0xC7, 0x03, 0xD3, 0x03, 0xE0, 0x03,
0xEC, 0x03, 0xF9, 0x04, 0x06, 0x04, 0x13, 0x04, 0x20, 0x04, 0x2D, 0x04,
0x3B, 0x04, 0x48, 0x04, 0x55, 0x04, 0x63, 0x04, 0x71, 0x04, 0x7E, 0x04,
0x8C, 0x04, 0x9A, 0x04, 0xA8, 0x04, 0xB6, 0x04, 0xC4, 0x04, 0xD3, 0x04,
0xE1, 0x04, 0xF0, 0x04, 0xFE, 0x05, 0x0D, 0x05, 0x1C, 0x05, 0x2B, 0x05,
0x3A, 0x05, 0x49, 0x05, 0x58, 0x05, 0x67, 0x05, 0x77, 0x05, 0x86, 0x05,
0x96, 0x05, 0xA6, 0x05, 0xB5, 0x05, 0xC5, 0x05, 0xD5, 0x05, 0xE5, 0x05,
0xF6, 0x06, 0x06, 0x06, 0x16, 0x06, 0x27, 0x06, 0x37, 0x06, 0x48, 0x06,
0x59, 0x06, 0x6A, 0x06, 0x7B, 0x06, 0x8C, 0x06, 0x9D, 0x06, 0xAF, 0x06,
0xC0, 0x06, 0xD1, 0x06, 0xE3, 0x06, 0xF5, 0x07, 0x07, 0x07, 0x19, 0x07,
0x2B, 0x07, 0x3D, 0x07, 0x4F, 0x07, 0x61, 0x07, 0x74, 0x07, 0x86, 0x07,
0x99, 0x07, 0xAC, 0x07, 0xBF, 0x07, 0xD2, 0x07, 0xE5, 0x07, 0xF8, 0x08,
0x0B, 0x08, 0x1F, 0x08, 0x32, 0x08, 0x46, 0x08, 0x5A, 0x08, 0x6E, 0x08,
0x82, 0x08, 0x96, 0x08, 0xAA, 0x08, 0xBE, 0x08, 0xD2, 0x08, 0xE7, 0x08,
0xFB, 0x09, 0x10, 0x09, 0x25, 0x09, 0x3A, 0x09, 0x4F, 0x09, 0x64, 0x09,
0x79, 0x09, 0x8F, 0x09, 0xA4, 0x09, 0xBA, 0x09, 0xCF, 0x09, 0xE5, 0x09,
0xFB, 0x0A, 0x11, 0x0A, 0x27, 0x0A, 0x3D, 0x0A, 0x54, 0x0A, 0x6A, 0x0A,
0x81, 0x0A, 0x98, 0x0A, 0xAE, 0x0A, 0xC5, 0x0A, 0xDC, 0x0A, 0xF3, 0x0B,
0x0B, 0x0B, 0x22, 0x0B, 0x39, 0x0B, 0x51, 0x0B, 0x69, 0x0B, 0x80, 0x0B,
0x98, 0x0B, 0xB0, 0x0B, 0xC8, 0x0B, 0xE1, 0x0B, 0xF9, 0x0C, 0x12, 0x0C,
0x2A, 0x0C, 0x43, 0x0C, 0x5C, 0x0C, 0x75, 0x0C, 0x8E, 0x0C, 0xA7, 0x0C,
0xC0, 0x0C, 0xD9, 0x0C, 0xF3, 0x0D, 0x0D, 0x0D, 0x26, 0x0D, 0x40, 0x0D,
0x5A, 0x0D, 0x74, 0x0D, 0x8E, 0x0D, 0xA9, 0x0D, 0xC3, 0x0D, 0xDE, 0x0D,
0xF8, 0x0E, 0x13, 0x0E, 0x2E, 0x0E, 0x49, 0x0E, 0x64, 0x0E, 0x7F, 0x0E,
0x9B, 0x0E, 0xB6, 0x0E, 0xD2, 0x0E, 0xEE, 0x0F, 0x09, 0x0F, 0x25, 0x0F,
0x41, 0x0F, 0x5E, 0x0F, 0x7A, 0x0F, 0x96, 0x0F, 0xB3, 0x0F, 0xCF, 0x0F,
0xEC, 0x10, 0x09, 0x10, 0x26, 0x10, 0x43, 0x10, 0x61, 0x10, 0x7E, 0x10,
0x9B, 0x10, 0xB9, 0x10, 0xD7, 0x10, 0xF5, 0x11, 0x13, 0x11, 0x31, 0x11,
0x4F, 0x11, 0x6D, 0x11, 0x8C, 0x11, 0xAA, 0x11, 0xC9, 0x11, 0xE8, 0x12,
```

```
0x07, 0x12, 0x26, 0x12, 0x45, 0x12, 0x64, 0x12, 0x84, 0x12, 0xA3, 0x12,
0xC3, 0x12, 0xE3, 0x13, 0x03, 0x13, 0x23, 0x13, 0x43, 0x13, 0x63, 0x13,
0x83, 0x13, 0xA4, 0x13, 0xC5, 0x13, 0xE5, 0x14, 0x06, 0x14, 0x27, 0x14,
0x49, 0x14, 0x6A, 0x14, 0x8B, 0x14, 0xAD, 0x14, 0xCE, 0x14, 0xF0, 0x15,
0x12, 0x15, 0x34, 0x15, 0x56, 0x15, 0x78, 0x15, 0x9B, 0x15, 0xBD, 0x15,
0xE0, 0x16, 0x03, 0x16, 0x26, 0x16, 0x49, 0x16, 0x6C, 0x16, 0x8F, 0x16,
0xB2, 0x16, 0xD6, 0x16, 0xFA, 0x17, 0x1D, 0x17, 0x41, 0x17, 0x65, 0x17,
0x89, 0x17, 0xAE, 0x17, 0xD2, 0x17, 0xF7, 0x18, 0x1B, 0x18, 0x40, 0x18,
0x65, 0x18, 0x8A, 0x18, 0xAF, 0x18, 0xD5, 0x18, 0xFA, 0x19, 0x20, 0x19,
0x45, 0x19, 0x6B, 0x19, 0x91, 0x19, 0xB7, 0x19, 0xDD, 0x1A, 0x04, 0x1A,
0x2A, 0x1A, 0x51, 0x1A, 0x77, 0x1A, 0x9E, 0x1A, 0xC5, 0x1A, 0xEC, 0x1B,
0x14, 0x1B, 0x3B, 0x1B, 0x63, 0x1B, 0x8A, 0x1B, 0xB2, 0x1B, 0xDA, 0x1C,
0x02, 0x1C, 0x2A, 0x1C, 0x52, 0x1C, 0x7B, 0x1C, 0xA3, 0x1C, 0xCC, 0x1C,
0xF5, 0x1D, 0x1E, 0x1D, 0x47, 0x1D, 0x70, 0x1D, 0x99, 0x1D, 0xC3, 0x1D,
0xEC, 0x1E, 0x16, 0x1E, 0x40, 0x1E, 0x6A, 0x1E, 0x94, 0x1E, 0xBE, 0x1E,
0xE9, 0x1F, 0x13, 0x1F, 0x3E, 0x1F, 0x69, 0x1F, 0x94, 0x1F, 0xBF, 0x1F,
0xEA, 0x20, 0x15, 0x20, 0x41, 0x20, 0x6C, 0x20, 0x98, 0x20, 0xC4, 0x20,
0xF0, 0x21, 0x1C, 0x21, 0x48, 0x21, 0x75, 0x21, 0xA1, 0x21, 0xCE, 0x21,
0xFB, 0x22, 0x27, 0x22, 0x55, 0x22, 0x82, 0x22, 0xAF, 0x22, 0xDD, 0x23,
0x0A, 0x23, 0x38, 0x23, 0x66, 0x23, 0x94, 0x23, 0xC2, 0x23, 0xF0, 0x24,
0x1F, 0x24, 0x4D, 0x24, 0x7C, 0x24, 0xAB, 0x24, 0xDA, 0x25, 0x09, 0x25,
0x38, 0x25, 0x68, 0x25, 0x97, 0x25, 0xC7, 0x25, 0xF7, 0x26, 0x27, 0x26,
0x57, 0x26, 0x87, 0x26, 0xB7, 0x26, 0xE8, 0x27, 0x18, 0x27, 0x49, 0x27,
0x7A, 0x27, 0xAB, 0x27, 0xDC, 0x28, 0x0D, 0x28, 0x3F, 0x28, 0x71, 0x28,
0xA2, 0x28, 0xD4, 0x29, 0x06, 0x29, 0x38, 0x29, 0x6B, 0x29, 0x9D, 0x29,
0xD0, 0x2A, 0x02, 0x2A, 0x35, 0x2A, 0x68, 0x2A, 0x9B, 0x2A, 0xCF, 0x2B,
0x02, 0x2B, 0x36, 0x2B, 0x69, 0x2B, 0x9D, 0x2B, 0xD1, 0x2C, 0x05, 0x2C,
0x39, 0x2C, 0x6E, 0x2C, 0xA2, 0x2C, 0xD7, 0x2D, 0x0C, 0x2D, 0x41, 0x2D,
0x76, 0x2D, 0xAB, 0x2D, 0xE1, 0x2E, 0x16, 0x2E, 0x4C, 0x2E, 0x82, 0x2E,
0xB7, 0x2E, 0xEE, 0x2F, 0x24, 0x2F, 0x5A, 0x2F, 0x91, 0x2F, 0xC7, 0x2F,
0xFE, 0x30, 0x35, 0x30, 0x6C, 0x30, 0xA4, 0x30, 0xDB, 0x31, 0x12, 0x31,
0x4A, 0x31, 0x82, 0x31, 0xBA, 0x31, 0xF2, 0x32, 0x2A, 0x32, 0x63, 0x32,
0x9B, 0x32, 0xD4, 0x33, 0x0D, 0x33, 0x46, 0x33, 0x7F, 0x33, 0xB8, 0x33,
0xF1, 0x34, 0x2B, 0x34, 0x65, 0x34, 0x9E, 0x34, 0xD8, 0x35, 0x13, 0x35,
0x4D, 0x35, 0x87, 0x35, 0xC2, 0x35, 0xFD, 0x36, 0x37, 0x36, 0x72, 0x36,
0xAE, 0x36, 0xE9, 0x37, 0x24, 0x37, 0x60, 0x37, 0x9C, 0x37, 0xD7, 0x38,
0x14, 0x38, 0x50, 0x38, 0x8C, 0x38, 0xC8, 0x39, 0x05, 0x39, 0x42, 0x39,
0x7F, 0x39, 0xBC, 0x39, 0xF9, 0x3A, 0x36, 0x3A, 0x74, 0x3A, 0xB2, 0x3A,
0xEF, 0x3B, 0x2D, 0x3B, 0x6B, 0x3B, 0xAA, 0x3B, 0xE8, 0x3C, 0x27, 0x3C,
0x65, 0x3C, 0xA4, 0x3C, 0xE3, 0x3D, 0x22, 0x3D, 0x61, 0x3D, 0xA1, 0x3D,
0xE0, 0x3E, 0x20, 0x3E, 0x60, 0x3E, 0xA0, 0x3E, 0xE0, 0x3F, 0x21, 0x3F,
0x61, 0x3F, 0xA2, 0x3F, 0xE2, 0x40, 0x23, 0x40, 0x64, 0x40, 0xA6, 0x40,
0xE7, 0x41, 0x29, 0x41, 0x6A, 0x41, 0xAC, 0x41, 0xEE, 0x42, 0x30, 0x42,
0x72, 0x42, 0xB5, 0x42, 0xF7, 0x43, 0x3A, 0x43, 0x7D, 0x43, 0xC0, 0x44,
0x03, 0x44, 0x47, 0x44, 0x8A, 0x44, 0xCE, 0x45, 0x12, 0x45, 0x55, 0x45,
```

```
0x9A, 0x45, 0xDE, 0x46, 0x22, 0x46, 0x67, 0x46, 0xAB, 0x46, 0xF0, 0x47,
0x35, 0x47, 0x7B, 0x47, 0xC0, 0x48, 0x05, 0x48, 0x4B, 0x48, 0x91, 0x48,
0xD7, 0x49, 0x1D, 0x49, 0x63, 0x49, 0xA9, 0x49, 0xF0, 0x4A, 0x37, 0x4A,
0x7D, 0x4A, 0xC4, 0x4B, 0x0C, 0x4B, 0x53, 0x4B, 0x9A, 0x4B, 0xE2, 0x4C,
0x2A, 0x4C, 0x72, 0x4C, 0xBA, 0x4D, 0x02, 0x4D, 0x4A, 0x4D, 0x93, 0x4D,
0xDC, 0x4E, 0x25, 0x4E, 0x6E, 0x4E, 0xB7, 0x4F, 0x00, 0x4F, 0x49, 0x4F,
0x93, 0x4F, 0xDD, 0x50, 0x27, 0x50, 0x71, 0x50, 0xBB, 0x51, 0x06, 0x51,
0x50, 0x51, 0x9B, 0x51, 0xE6, 0x52, 0x31, 0x52, 0x7C, 0x52, 0xC7, 0x53,
0x13, 0x53, 0x5F, 0x53, 0xAA, 0x53, 0xF6, 0x54, 0x42, 0x54, 0x8F, 0x54,
0xDB, 0x55, 0x28, 0x55, 0x75, 0x55, 0xC2, 0x56, 0x0F, 0x56, 0x5C, 0x56,
0xA9, 0x56, 0xF7, 0x57, 0x44, 0x57, 0x92, 0x57, 0xE0, 0x58, 0x2F, 0x58,
0x7D, 0x58, 0xCB, 0x59, 0x1A, 0x59, 0x69, 0x59, 0xB8, 0x5A, 0x07, 0x5A,
0x56, 0x5A, 0xA6, 0x5A, 0xF5, 0x5B, 0x45, 0x5B, 0x95, 0x5B, 0xE5, 0x5C,
0x35, 0x5C, 0x86, 0x5C, 0xD6, 0x5D, 0x27, 0x5D, 0x78, 0x5D, 0xC9, 0x5E,
0x1A, 0x5E, 0x6C, 0x5E, 0xBD, 0x5F, 0x0F, 0x5F, 0x61, 0x5F, 0xB3, 0x60,
0x05, 0x60, 0x57, 0x60, 0xAA, 0x60, 0xFC, 0x61, 0x4F, 0x61, 0xA2, 0x61,
0xF5, 0x62, 0x49, 0x62, 0x9C, 0x62, 0xF0, 0x63, 0x43, 0x63, 0x97, 0x63,
0xEB, 0x64, 0x40, 0x64, 0x94, 0x64, 0xE9, 0x65, 0x3D, 0x65, 0x92, 0x65,
0xE7, 0x66, 0x3D, 0x66, 0x92, 0x66, 0xE8, 0x67, 0x3D, 0x67, 0x93, 0x67,
0xE9, 0x68, 0x3F, 0x68, 0x96, 0x68, 0xEC, 0x69, 0x43, 0x69, 0x9A, 0x69,
0xF1, 0x6A, 0x48, 0x6A, 0x9F, 0x6A, 0xF7, 0x6B, 0x4F, 0x6B, 0xA7, 0x6B,
0xFF, 0x6C, 0x57, 0x6C, 0xAF, 0x6D, 0x08, 0x6D, 0x60, 0x6D, 0xB9, 0x6E,
0x12, 0x6E, 0x6B, 0x6E, 0xC4, 0x6F, 0x1E, 0x6F, 0x78, 0x6F, 0xD1, 0x70,
0x2B, 0x70, 0x86, 0x70, 0xE0, 0x71, 0x3A, 0x71, 0x95, 0x71, 0xF0, 0x72,
0x4B, 0x72, 0xA6, 0x73, 0x01, 0x73, 0x5D, 0x73, 0xB8, 0x74, 0x14, 0x74,
0x70, 0x74, 0xCC, 0x75, 0x28, 0x75, 0x85, 0x75, 0xE1, 0x76, 0x3E, 0x76,
0x9B, 0x76, 0xF8, 0x77, 0x56, 0x77, 0xB3, 0x78, 0x11, 0x78, 0x6E, 0x78,
0xCC, 0x79, 0x2A, 0x79, 0x89, 0x79, 0xE7, 0x7A, 0x46, 0x7A, 0xA5, 0x7B,
0x04, 0x7B, 0x63, 0x7B, 0xC2, 0x7C, 0x21, 0x7C, 0x81, 0x7C, 0xE1, 0x7D,
0x41, 0x7D, 0xA1, 0x7E, 0x01, 0x7E, 0x62, 0x7E, 0xC2, 0x7F, 0x23, 0x7F,
0x84, 0x7F, 0xE5, 0x80, 0x47, 0x80, 0xA8, 0x81, 0x0A, 0x81, 0x6B, 0x81,
0xCD, 0x82, 0x30, 0x82, 0x92, 0x82, 0xF4, 0x83, 0x57, 0x83, 0xBA, 0x84,
0x1D, 0x84, 0x80, 0x84, 0xE3, 0x85, 0x47, 0x85, 0xAB, 0x86, 0x0E, 0x86,
0x72, 0x86, 0xD7, 0x87, 0x3B, 0x87, 0x9F, 0x88, 0x04, 0x88, 0x69, 0x88,
0xCE, 0x89, 0x33, 0x89, 0x99, 0x89, 0xFE, 0x8A, 0x64, 0x8A, 0xCA, 0x8B,
0x30, 0x8B, 0x96, 0x8B, 0xFC, 0x8C, 0x63, 0x8C, 0xCA, 0x8D, 0x31, 0x8D,
0x98, 0x8D, 0xFF, 0x8E, 0x66, 0x8E, 0xCE, 0x8F, 0x36, 0x8F, 0x9E, 0x90,
0x06, 0x90, 0x6E, 0x90, 0xD6, 0x91, 0x3F, 0x91, 0xA8, 0x92, 0x11, 0x92,
0x7A, 0x92, 0xE3, 0x93, 0x4D, 0x93, 0xB6, 0x94, 0x20, 0x94, 0x8A, 0x94,
0xF4, 0x95, 0x5F, 0x95, 0xC9, 0x96, 0x34, 0x96, 0x9F, 0x97, 0x0A, 0x97,
0x75, 0x97, 0xE0, 0x98, 0x4C, 0x98, 0xB8, 0x99, 0x24, 0x99, 0x90, 0x99,
0xFC, 0x9A, 0x68, 0x9A, 0xD5, 0x9B, 0x42, 0x9B, 0xAF, 0x9C, 0x1C, 0x9C,
0x89, 0x9C, 0xF7, 0x9D, 0x64, 0x9D, 0xD2, 0x9E, 0x40, 0x9E, 0xAE, 0x9F,
0x1D, 0x9F, 0x8B, 0x9F, 0xFA, 0xA0, 0x69, 0xA0, 0xD8, 0xA1, 0x47, 0xA1,
0xB6, 0xA2, 0x26, 0xA2, 0x96, 0xA3, 0x06, 0xA3, 0x76, 0xA3, 0xE6, 0xA4,
```

```
0x56, 0xA4, 0xC7, 0xA5, 0x38, 0xA5, 0xA9, 0xA6, 0x1A, 0xA6, 0x8B, 0xA6,
0xFD, 0xA7, 0x6E, 0xA7, 0xE0, 0xA8, 0x52, 0xA8, 0xC4, 0xA9, 0x37, 0xA9,
0xA9, 0xAA, 0x1C, 0xAA, 0x8F, 0xAB, 0x02, 0xAB, 0x75, 0xAB, 0xE9, 0xAC,
0x5C, 0xAC, 0xD0, 0xAD, 0x44, 0xAD, 0xB8, 0xAE, 0x2D, 0xAE, 0xA1, 0xAF,
0x16, 0xAF, 0x8B, 0xB0, 0x00, 0xB0, 0x75, 0xB0, 0xEA, 0xB1, 0x60, 0xB1,
0xD6, 0xB2, 0x4B, 0xB2, 0xC2, 0xB3, 0x38, 0xB3, 0xAE, 0xB4, 0x25, 0xB4,
0x9C, 0xB5, 0x13, 0xB5, 0x8A, 0xB6, 0x01, 0xB6, 0x79, 0xB6, 0xF0, 0xB7,
0x68, 0xB7, 0xE0, 0xB8, 0x59, 0xB8, 0xD1, 0xB9, 0x4A, 0xB9, 0xC2, 0xBA,
0x3B, 0xBA, 0xB5, 0xBB, 0x2E, 0xBB, 0xA7, 0xBC, 0x21, 0xBC, 0x9B, 0xBD,
0x15, 0xBD, 0x8F, 0xBE, 0x0A, 0xBE, 0x84, 0xBE, 0xFF, 0xBF, 0x7A, 0xBF,
0xF5, 0xC0, 0x70, 0xC0, 0xEC, 0xC1, 0x67, 0xC1, 0xE3, 0xC2, 0x5F, 0xC2,
0xDB, 0xC3, 0x58, 0xC3, 0xD4, 0xC4, 0x51, 0xC4, 0xCE, 0xC5, 0x4B, 0xC5,
0xC8, 0xC6, 0x46, 0xC6, 0xC3, 0xC7, 0x41, 0xC7, 0xBF, 0xC8, 0x3D, 0xC8,
0xBC, 0xC9, 0x3A, 0xC9, 0xB9, 0xCA, 0x38, 0xCA, 0xB7, 0xCB, 0x36, 0xCB,
0xB6, 0xCC, 0x35, 0xCC, 0xB5, 0xCD, 0x35, 0xCD, 0xB5, 0xCE, 0x36, 0xCE,
0xB6, 0xCF, 0x37, 0xCF, 0xB8, 0xD0, 0x39, 0xD0, 0xBA, 0xD1, 0x3C, 0xD1,
0xBE, 0xD2, 0x3F, 0xD2, 0xC1, 0xD3, 0x44, 0xD3, 0xC6, 0xD4, 0x49, 0xD4,
0xCB, 0xD5, 0x4E, 0xD5, 0xD1, 0xD6, 0x55, 0xD6, 0xD8, 0xD7, 0x5C, 0xD7,
0xE0, 0xD8, 0x64, 0xD8, 0xE8, 0xD9, 0x6C, 0xD9, 0xF1, 0xDA, 0x76, 0xDA,
0xFB, 0xDB, 0x80, 0xDC, 0x05, 0xDC, 0x8A, 0xDD, 0x10, 0xDD, 0x96, 0xDE,
0x1C, 0xDE, 0xA2, 0xDF, 0x29, 0xDF, 0xAF, 0xE0, 0x36, 0xE0, 0xBD, 0xE1,
0x44, 0xE1, 0xCC, 0xE2, 0x53, 0xE2, 0xDB, 0xE3, 0x63, 0xE3, 0xEB, 0xE4,
0x73, 0xE4, 0xFC, 0xE5, 0x84, 0xE6, 0x0D, 0xE6, 0x96, 0xE7, 0x1F, 0xE7,
0xA9, 0xE8, 0x32, 0xE8, 0xBC, 0xE9, 0x46, 0xE9, 0xD0, 0xEA, 0x5B, 0xEA,
0xE5, 0xEB, 0x70, 0xEB, 0xFB, 0xEC, 0x86, 0xED, 0x11, 0xED, 0x9C, 0xEE,
0x28, 0xEE, 0xB4, 0xEF, 0x40, 0xEF, 0xCC, 0xF0, 0x58, 0xF0, 0xE5, 0xF1,
0x72, 0xF1, 0xFF, 0xF2, 0x8C, 0xF3, 0x19, 0xF3, 0xA7, 0xF4, 0x34, 0xF4,
0xC2, 0xF5, 0x50, 0xF5, 0xDE, 0xF6, 0x6D, 0xF6, 0xFB, 0xF7, 0x8A, 0xF8,
0x19, 0xF8, 0xA8, 0xF9, 0x38, 0xF9, 0xC7, 0xFA, 0x57, 0xFA, 0xE7, 0xFB,
0x77, 0xFC, 0x07, 0xFC, 0x98, 0xFD, 0x29, 0xFD, 0xBA, 0xFE, 0x4B, 0xFE,
0xDC, 0xFF, 0x6D, 0xFF, 0xFF
```

**Table M.7 – "Ducky" marker template**

```
0xEC, 0x00, 0x11, 0x44, 0x75, 0x63, 0x6B, 0x79, 0x00, 0x01, 0x00, 0x04,
0x00, 0x00, 0x00, 0x64, 0x00, 0x00
```

**Table M.8 – "Adobe" marker template**

```
0xEE, 0x00, 0x0E, 0x41, 0x64, 0x6F, 0x62, 0x65, 0x00, 0x64, 0x00, 0x00,
0x00, 0x00, 0x01
```

**Table M.9 – stock counts arrays**

is_ac == 0,stock_index == 0:

```
0, 0, 3, 1, 1, 1, 1, 1, 1, 1, 1, 2, 0, 0, 0, 0, 0
```

is_ac == 0,stock_index == 1:

```
0, 0, 1, 5, 1, 1, 1, 1, 1, 2, 0, 0, 0, 0, 0, 0, 0
```

is_ac == 1,stock_index == 0:

```
0, 0, 2, 1, 3, 3, 2, 4, 3, 5, 5, 4, 4, 0, 0, 1, 126
```

is_ac == 1,stock_index == 1:

```
0, 0, 2, 1, 2, 4, 4, 3, 4, 7, 5, 4, 4, 0, 1, 2, 120
```

**Table M.10 – stock values arrays**

is_ac == 0,stock_index == 0:

```
0, 1, 2, 3, 4, 5, 6, 7, 8, 9, 10, 11, 256
```

is_ac == 0,stock_index == 1:

```
0, 1, 2, 3, 4, 5, 6, 7, 8, 9, 10, 11, 256
```

is_ac == 1,stock_index == 0:

```
  1,   2,   3,   0,   4,  17,   5,  18,  33,  49,  65,   6,  19,  81,
 97,   7,  34, 113,  20,  50, 129, 145, 161,   8,  35,  66, 177, 193,
 21,  82, 209, 240,  36,  51,  98, 114, 130,   9,  10,  22,  23,  24,
 25,  26,  37,  38,  39,  40,  41,  42,  52,  53,  54,  55,  56,  57,
 58,  67,  68,  69,  70,  71,  72,  73,  74,  83,  84,  85,  86,  87,
 88,  89,  90,  99, 100, 101, 102, 103, 104, 105, 106, 115, 116, 117,
118, 119, 120, 121, 122, 131, 132, 133, 134, 135, 136, 137, 138, 146,
147, 148, 149, 150, 151, 152, 153, 154, 162, 163, 164, 165, 166, 167,
168, 169, 170, 178, 179, 180, 181, 182, 183, 184, 185, 186, 194, 195,
196, 197, 198, 199, 200, 201, 202, 210, 211, 212, 213, 214, 215, 216,
217, 218, 225, 226, 227, 228, 229, 230, 231, 232, 233, 234, 241, 242,
243, 244, 245, 246, 247, 248, 249, 250, 256
```

is_ac == 1,stock_index == 1:

```
  0,   1,   2,   3,  17,   4,   5,  33,  49,   6,  18,  65,  81,   7,
 97, 113,  19,  34,  50, 129,   8,  20,  66, 145, 161, 177, 193,   9,
 35,  51,  82, 240,  21,  98, 114, 209,  10,  22,  36,  52, 225,  37,
241,  23,  24,  25,  26,  38,  39,  40,  41,  42,  53,  54,  55,  56,
 57,  58,  67,  68,  69,  70,  71,  72,  73,  74,  83,  84,  85,  86,
 87,  88,  89,  90,  99, 100, 101, 102, 103, 104, 105, 106, 115, 116,
117, 118, 119, 120, 121, 122, 130, 131, 132, 133, 134, 135, 136, 137,
138, 146, 147, 148, 149, 150, 151, 152, 153, 154, 162, 163, 164, 165,
166, 167, 168, 169, 170, 178, 179, 180, 181, 182, 183, 184, 185, 186,
194, 195, 196, 197, 198, 199, 200, 201, 202, 210, 211, 212, 213, 214,
215, 216, 217, 218, 226, 227, 228, 229, 230, 231, 232, 233, 234, 242,
243, 244, 245, 246, 247, 248, 249, 250, 256
```

**Table M.11 – predefined symbol order**

`is_ac == 0`:

```
0, 1, 2, 3, 4, 5, 6, 7, 8, 9, 10, 11, 12, 13, 14, 15
```

`is_ac == 1`:

```
  1,   0,   2,   3,  17,   4,   5,  33,  18,  49,  65,   6,  81,  19,
 97,   7,  34, 113,  50, 129,  20, 145, 161,   8,  35,  66, 177, 193,
 21,  82, 209, 240,  36,  51,  98, 114,   9, 130,  10,  22,  52, 225,
 23,  37, 241,  24,  25,  26,  38,  39,  40,  41,  42,  53,  54,  55,
 56,  57,  58,  67,  68,  69,  70,  71,  72,  73,  74,  83,  84,  85,
 86,  87,  88,  89,  90,  99, 100, 101, 102, 103, 104, 105, 106, 115,
116, 117, 118, 119, 120, 121, 122, 131, 132, 133, 134, 135, 136, 137,
138, 146, 147, 148, 149, 150, 151, 152, 153, 154, 162, 163, 164, 165,
166, 167, 168, 169, 170, 178, 179, 180, 181, 182, 183, 184, 185, 186,
194, 195, 196, 197, 198, 199, 200, 201, 202, 210, 211, 212, 213, 214,
215, 216, 217, 218, 226, 227, 228, 229, 230, 231, 232, 233, 234, 242,
243, 244, 245, 246, 247, 248, 249, 250,  16,  32,  48,  64,  80,  96,
112, 128, 144, 160, 176, 192, 208,  11,  12,  13,  14,  15,  27,  28,
 29,  30,  31,  43,  44,  45,  46,  47,  59,  60,  61,  62,  63,  75,
 76,  77,  78,  79,  91,  92,  93,  94,  95, 107, 108, 109, 110, 111,
123, 124, 125, 126, 127, 139, 140, 141, 142, 143, 155, 156, 157, 158,
159, 171, 172, 173, 174, 175, 187, 188, 189, 190, 191, 203, 204, 205,
206, 207, 219, 220, 221, 222, 223, 224, 235, 236, 237, 238, 239, 251,
252, 253, 254, 255
```

**Table M.12 – stock quant tables**

is_luma == true, stock_index == 0:

```
 3,  2,  2,  3,  5,  8, 10, 12,  2,  2,  3,  4,  5, 12, 12, 11,  3,  3,
 3,  5,  8, 11, 14, 11,  3,  3,  4,  6, 10, 17, 16, 12,  4,  4,  7, 11,
14, 22, 21, 15,  5,  7, 11, 13, 16, 21, 23, 18, 10, 13, 16, 17, 21, 24,
24, 20, 14, 18, 19, 20, 22, 20, 21, 20
```

is_luma == true, stock_index == 1:

```
 8,  6,  5,  8, 12, 20, 26, 31,  6,  6,  7, 10, 13, 29, 30, 28,  7,  7,
 8, 12, 20, 29, 35, 28,  7,  9, 11, 15, 26, 44, 40, 31,  9, 11, 19, 28,
34, 55, 52, 39, 12, 18, 28, 32, 41, 52, 57, 46, 25, 32, 39, 44, 52, 61,
60, 51, 36, 46, 48, 49, 56, 50, 52, 50
```

is_luma == true, stock_index == 2:

```
 6,  4,  4,  6, 10, 16, 20, 24,  5,  5,  6,  8, 10, 23, 24, 22,  6,  5,
 6, 10, 16, 23, 28, 22,  6,  7,  9, 12, 20, 35, 32, 25,  7,  9, 15, 22,
27, 44, 41, 31, 10, 14, 22, 26, 32, 42, 45, 37, 20, 26, 31, 35, 41, 48,
48, 40, 29, 37, 38, 39, 45, 40, 41, 40
```

is_luma == true, stock_index == 3:

```
 5,  3,  3,  5,  7, 12, 15, 18,  4,  4,  4,  6,  8, 17, 18, 17,  4,  4,
 5,  7, 12, 17, 21, 17,  4,  5,  7,  9, 15, 26, 24, 19,  5,  7, 11, 17,
20, 33, 31, 23,  7, 11, 17, 19, 24, 31, 34, 28, 15, 19, 23, 26, 31, 36,
36, 30, 22, 28, 29, 29, 34, 30, 31, 30
```

is_luma == true, stock_index == 4:

```
 1,  1,  1,  1,  1,  1,  1,  1,  1,  1,  1,  1,  1,  1,  1,  1,  1,  1,
 1,  1,  1,  1,  1,  1,  1,  1,  1,  1,  1,  1,  1,  1,  1,  1,  1,  1,
 1,  1,  1,  1,  1,  1,  1,  1,  1,  1,  1,  1,  1,  1,  1,  1,  1,  1,
 1,  1,  1,  1,  1,  1,  1,  1,  1,  1
```

is_luma == true, stock_index == 5:

```
 2,  1,  1,  2,  2,  4,  5,  6,  1,  1,  1,  2,  3,  6,  6,  6,  1,  1,
 2,  2,  4,  6,  7,  6,  1,  2,  2,  3,  5,  9,  8,  6,  2,  2,  4,  6,
 7, 11, 10,  8,  2,  4,  6,  6,  8, 10, 11,  9,  5,  6,  8,  9, 10, 12,
12, 10,  7,  9, 10, 10, 11, 10, 10, 10
```

is_luma == true, stock_index == 6:

```
 1,  1,  1,  1,  1,  1,  1,  1,  1,  1,  1,  1,  1,  1,  1,  1,  1,  1,
 1,  1,  1,  1,  1,  2,  1,  1,  1,  1,  1,  1,  2,  2,  1,  1,  1,  1,
 1,  2,  2,  3,  1,  1,  1,  1,  2,  2,  3,  3,  1,  1,  1,  2,  2,  3,
 3,  3,  1,  1,  2,  2,  3,  3,  3,  3
```

`is_luma == true`, `stock_index == 7`:

```
10,  7,  6, 10, 14, 24, 31, 37,  7,  7,  8, 11, 16, 35, 36, 33,  8,  8,
10, 14, 24, 34, 41, 34,  8, 10, 13, 17, 31, 52, 48, 37, 11, 13, 22, 34,
41, 65, 62, 46, 14, 21, 33, 38, 49, 62, 68, 55, 29, 38, 47, 52, 62, 73,
72, 61, 43, 55, 57, 59, 67, 60, 62, 59
```

`is_luma == false`, `stock_index == 0`:

```
 9,  9,  9, 12, 11, 12, 24, 13, 13, 24, 50, 33, 28, 33, 50, 50, 50, 50,
50, 50, 50, 50, 50, 50, 50, 50, 50, 50, 50, 50, 50, 50, 50, 50, 50, 50,
50, 50, 50, 50, 50, 50, 50, 50, 50, 50, 50, 50, 50, 50, 50, 50, 50, 50,
50, 50, 50, 50, 50, 50, 50, 50, 50, 50
```

`is_luma == false`, `stock_index == 1`:

```
 3,  4,  5,  9, 20, 20, 20, 20,  4,  4,  5, 13, 20, 20, 20, 20,  5,  5,
11, 20, 20, 20, 20, 20,  9, 13, 20, 20, 20, 20, 20, 20, 20, 20, 20, 20,
20, 20, 20, 20, 20, 20, 20, 20, 20, 20, 20, 20, 20, 20, 20, 20, 20, 20,
20, 20, 20, 20, 20, 20, 20, 20, 20, 20
```

`is_luma == false`, `stock_index == 2`:

```
 9,  9, 12, 24, 50, 50, 50, 50,  9, 11, 13, 33, 50, 50, 50, 50, 12, 13,
28, 50, 50, 50, 50, 50, 24, 33, 50, 50, 50, 50, 50, 50, 50, 50, 50, 50,
50, 50, 50, 50, 50, 50, 50, 50, 50, 50, 50, 50, 50, 50, 50, 50, 50, 50,
50, 50, 50, 50, 50, 50, 50, 50, 50, 50
```

`is_luma == false`, `stock_index == 3`:

```
 5,  5,  7, 14, 30, 30, 30, 30,  5,  6,  8, 20, 30, 30, 30, 30,  7,  8,
17, 30, 30, 30, 30, 30, 14, 20, 30, 30, 30, 30, 30, 30, 30, 30, 30, 30,
30, 30, 30, 30, 30, 30, 30, 30, 30, 30, 30, 30, 30, 30, 30, 30, 30, 30,
30, 30, 30, 30, 30, 30, 30, 30, 30, 30
```

`is_luma == false`, `stock_index == 4`:

```
 7,  7, 10, 19, 40, 40, 40, 40,  7,  8, 10, 26, 40, 40, 40, 40, 10, 10,
22, 40, 40, 40, 40, 40, 19, 26, 40, 40, 40, 40, 40, 40, 40, 40, 40, 40,
40, 40, 40, 40, 40, 40, 40, 40, 40, 40, 40, 40, 40, 40, 40, 40, 40, 40,
40, 40, 40, 40, 40, 40, 40, 40, 40, 40
```

`is_luma == false`, `stock_index == 5`:

```
 1,  1,  1,  1,  1,  1,  1,  1,  1,  1,  1,  1,  1,  1,  1,  1,  1,  1,
 1,  1,  1,  1,  1,  1,  1,  1,  1,  1,  1,  1,  1,  1,  1,  1,  1,  1,
 1,  1,  1,  1,  1,  1,  1,  1,  1,  1,  1,  1,  1,  1,  1,  1,  1,  1,
 1,  1,  1,  1,  1,  1,  1,  1,  1,  1
```

`is_luma == false`, `stock_index == 6`:

```
 2,  2,  2,  5, 10, 10, 10, 10,  2,  2,  3,  7, 10, 10, 10, 10,  2,  3,
 6, 10, 10, 10, 10, 10,  5,  7, 10, 10, 10, 10, 10, 10, 10, 10, 10, 10,
10, 10, 10, 10, 10, 10, 10, 10, 10, 10, 10, 10, 10, 10, 10, 10, 10, 10,
10, 10, 10, 10, 10, 10, 10, 10, 10, 10
```

`is_luma == false`, `stock_index == 7`:

```
10, 11, 14, 28, 59, 59, 59, 59, 11, 13, 16, 40, 59, 59, 59, 59, 14, 16,
34, 59, 59, 59, 59, 59, 28, 40, 59, 59, 59, 59, 59, 59, 59, 59, 59, 59,
59, 59, 59, 59, 59, 59, 59, 59, 59, 59, 59, 59, 59, 59, 59, 59, 59, 59,
59, 59, 59, 59, 59, 59, 59, 59, 59, 59
```

**Table M.13 – template quant tables**

`is_luma == true`:

```
 16,  11,  10,  16, 24, 40, 51,  61,  12,  12,  14, 19, 26, 58, 60,
 55,  14,  13,  16, 24, 40, 57,  69,  56,  14,  17, 22, 29, 51, 87,
 80,  62,  18,  22, 37, 56, 68, 109, 103,  77,  24, 35, 55, 64, 81,
104, 113,  92,  49, 64, 78, 87, 103, 121, 120, 101, 72, 92, 95, 98,
112, 100, 103,  99
```

`is_luma == false`:

```
17, 18, 24, 47, 99, 99, 99, 99, 18, 21, 26, 66, 99, 99, 99, 99, 24, 26,
56, 99, 99, 99, 99, 99, 47, 66, 99, 99, 99, 99, 99, 99, 99, 99, 99, 99,
99, 99, 99, 99, 99, 99, 99, 99, 99, 99, 99, 99, 99, 99, 99, 99, 99, 99,
99, 99, 99, 99, 99, 99, 99, 99, 99, 99
```

**Table M.15 – freq_context**

`scheme == 0`:

```
0, 0, 0, 0, 0, 0, 0, 0, 0, 0, 0, 0, 0, 0, 0, 0, 0, 0, 0, 0, 0, 0, 0, 0,
0, 0, 0, 0, 0, 0, 0, 0, 0, 0, 0, 0, 0, 0, 0, 0, 0, 0, 0, 0, 0, 0, 0, 0,
0, 0, 0, 0, 0, 0, 0, 0, 0, 0, 0, 0, 0, 0, 0, 0
```

`scheme == 1`:

```
0, 0, 0, 0, 0, 0, 0, 0, 0, 0, 0, 0, 0, 0, 0, 0, 0, 0, 0, 0, 0, 0, 0, 0,
1, 1, 1, 1, 1, 1, 1, 1, 1, 1, 1, 1, 1, 1, 1, 1, 1, 1, 1, 1, 1, 1, 1, 1,
1, 1, 1, 1, 1, 1, 1, 1, 1, 1, 1, 1, 1, 0, 0, 0
```

`scheme == 2`:

```
0, 1, 1, 1, 1, 1, 1, 1, 1, 1, 2, 2, 2, 2, 2, 2, 2, 2, 2, 2, 2, 2, 2, 2,
2, 2, 3, 3, 3, 3, 3, 3, 3, 3, 3, 3, 3, 3, 3, 3, 3, 3, 3, 3, 3, 3, 3, 3,
3, 3, 3, 3, 3, 3, 3, 3, 3, 3, 3, 3, 3, 1, 1, 1
```

`scheme == 3`:

```
0, 1, 1, 2, 2, 2, 3, 3, 3, 3, 4, 4, 4, 4, 4, 4, 5, 5, 5, 5, 5, 5, 5, 5,
6, 6, 6, 6, 6, 6, 6, 6, 6, 6, 6, 6, 6, 6, 6, 6, 7, 7, 7, 7, 7, 7, 7, 7,
7, 7, 7, 7, 7, 7, 7, 7, 7, 7, 7, 7, 7, 2, 2, 2
```

`scheme == 4`:

```
 0,  1,  2,  3,  4,  4,  5,  5,  6,  6,  7,  7,  8,  8,  8,  8,  9,  9,
 9,  9, 10, 10, 10, 10, 11, 11, 11, 11, 12, 12, 12, 12, 13, 13, 13, 13,
13, 13, 13, 13, 14, 14, 14, 14, 14, 14, 14, 14, 15, 15, 15, 15, 15, 15,
15, 15, 15, 15, 15, 15, 15, 15, 15, 15
```

`scheme == 5`:

```
 0,  1,  2,  3,  4,  5,  6,  7,  8,  9, 10, 11, 12, 13, 14, 15, 16, 16,
17, 17, 18, 18, 19, 19, 20, 20, 21, 21, 22, 22, 23, 23, 24, 24, 24, 24,
25, 25, 25, 25, 26, 26, 26, 26, 27, 27, 27, 27, 28, 28, 28, 28, 29, 29,
29, 29, 30, 30, 30, 30, 31, 31, 31, 31
```

`scheme == 6`:

```
```

```
 0,  1,  2,  3,  4,  5,  6,  7,  8,  9, 10, 11, 12, 13, 14, 15, 16, 17,
18, 19, 20, 21, 22, 23, 24, 25, 26, 27, 28, 29, 30, 31, 32, 33, 34, 35,
36, 37, 38, 39, 40, 41, 42, 43, 44, 45, 46, 47, 48, 49, 50, 51, 52, 53,
54, 55, 56, 57, 58, 59, 60, 61, 62, 63
```

**Table M.16 – num_nonzero_context**

`scheme == 0`:

```
0, 1, 1, 2, 2, 2, 3, 3, 3, 3, 4, 4, 4, 4, 4, 4, 5, 5, 5, 5, 5, 5, 5, 5,
6, 6, 6, 6, 6, 6, 6, 6, 6, 6, 6, 6, 6, 6, 6, 6, 7, 7, 7, 7, 7, 7, 7, 7,
7, 7, 7, 7, 7, 7, 7, 7, 7, 7, 7, 7, 7, 7, 7, 7
```

`scheme == 1`:

```
 0,  2,  2,  4,  4,  4,  6,  6,  6,  6,  8,  8,  8,  8,  8,  8, 10, 10,
10, 10, 10, 10, 10, 10, 12, 12, 12, 12, 12, 12, 12, 12, 12, 12, 12, 12,
12, 12, 12, 12, 14, 14, 14, 14, 14, 14, 14, 14, 14, 14, 14, 14, 14, 14,
14, 14, 14, 14, 14, 14, 14, 14, 14, 14
```

`scheme == 2`:

```
 0,  4,  4,  8,  8,  8, 12, 12, 12, 12, 16, 16, 16, 16, 16, 16, 20, 20,
20, 20, 20, 20, 20, 20, 24, 24, 24, 24, 24, 24, 24, 24, 24, 24, 24, 24,
24, 24, 24, 24, 28, 28, 28, 28, 28, 28, 28, 28, 28, 28, 28, 28, 28, 28,
28, 28, 28, 28, 28, 28, 28, 28, 28, 28
```

`scheme == 3`:

```
 0,  8,  8, 16, 16, 16, 24, 24, 24, 24, 32, 32, 32, 32, 32, 32, 40, 40,
40, 40, 40, 40, 40, 40, 48, 48, 48, 48, 48, 48, 48, 48, 48, 48, 48, 48,
48, 48, 48, 48, 55, 55, 55, 55, 55, 55, 55, 55, 55, 55, 55, 55, 55, 55,
55, 55, 55, 55, 55, 55, 55, 55, 55, 55
```

`scheme == 4`:

```
  0,  16,  16,  32,  32,  32,  48,  48,  48,  48,  64,  64,  64,  64,
 64,  64,  80,  80,  80,  80,  80,  80,  80,  80,  95,  95,  95,  95,
 95,  95,  95,  95,  95,  95,  95,  95,  95,  95,  95,  95, 109, 109,
109, 109, 109, 109, 109, 109, 109, 109, 109, 109, 109, 109, 109, 109,
109, 109, 109, 109, 109, 109, 109, 109
```

`scheme == 5`:

```
  0,  32,  32,  64,  64,  64,  96,  96,  96,  96, 127, 127, 127, 127,
127, 127, 157, 157, 157, 157, 157, 157, 157, 157, 185, 185, 185, 185,
185, 185, 185, 185, 185, 185, 185, 185, 185, 185, 185, 185, 211, 211,
211, 211, 211, 211, 211, 211, 211, 211, 211, 211, 211, 211, 211, 211,
211, 211, 211, 211, 211, 211, 211, 211
```

scheme == 6:

```
  0,  64,  64, 127, 127, 127, 188, 188, 188, 188, 246, 246, 246, 246,
246, 246, 300, 300, 300, 300, 300, 300, 300, 300, 348, 348, 348, 348,
348, 348, 348, 348, 348, 348, 348, 348, 348, 348, 348, 348, 388, 388,
388, 388, 388, 388, 388, 388, 388, 388, 388, 388, 388, 388, 388, 388,
388, 388, 388, 388, 388, 388, 388, 388
```

**Table M.17 – nonzero_buckets**

```
  0,  1,  2,  3,  4,  4,  5,  5,  5,  6,  6,  6,  6,  7,  7,  7,  7,
  7,  7,  7,  7,  8,  8,  8,  8,  8,  8,  8,  8,  8,  8,  8,  9,  9,
  9,  9,  9,  9,  9,  9,  9,  9,  9,  9,  9, 10, 10, 10, 10, 10, 10,
 10, 10, 10, 10, 10, 10, 10, 10, 10, 10, 10, 10, 10
```